\documentclass[12pt]{article}
\usepackage[utf8]{inputenc}
\usepackage{extsizes}
\usepackage{amsmath}
\usepackage[left=0.65in,right=0.65in,top = 0.7in, bottom = 0.7in]{geometry}
\usepackage{graphicx}
\usepackage[toc,page]{appendix}
\usepackage[inline]{enumitem}
\usepackage{multicol}
\usepackage[T1]{fontenc}
\usepackage{multirow}
\usepackage{amssymb}
\usepackage{setspace}
\usepackage{titling}
\usepackage{bigints}
\usepackage{bibentry}
\usepackage{import}
\usepackage{subcaption}
\usepackage{upgreek}
\usepackage{longtable}
\usepackage{bm}
\usepackage{bbm}
\usepackage{algorithm}
\usepackage{siunitx}
\usepackage{algpseudocode}
\usepackage{mathtools, nccmath}
\usepackage{setspace}
\usepackage[mathlines]{lineno}
\usepackage[most]{tcolorbox}
\usepackage{tabularray}
\usepackage{esint}
\usepackage[unicode=true]{hyperref}
\usepackage[nameinlink]{cleveref}
\usepackage{nameref}
\usepackage{booktabs}
\usepackage{collcell}
\usepackage{listings}
\usepackage{pgf,tikz,pgfplots}
\usepackage{asymptote}
\usepackage{fancyhdr}
\usepackage{lscape}
\usepackage{authblk}
\usepackage{pdfpages}
\usepackage{soul}
\pgfplotsset{compat=1.15}
\usepackage{amsthm}
\usepackage{mathrsfs}
\numberwithin{equation}{section}
\usepackage{float}
\usepackage{longfigure}
\usepackage{adjustbox}
\usepackage[english]{babel}
\usetikzlibrary{arrows}
\usetikzlibrary{shapes}
\newcommand*{\currentname}{\@currentlabelname}
\usepackage{titlesec}
\titleclass{\subsubsubsection}{straight}[\subsection]
\newcounter{subsubsubsection}[subsubsection]
\renewcommand\thesubsubsubsection{\thesubsubsection.\arabic{subsubsubsection}}
\titleformat{\subsubsubsection}
  {\normalfont\normalsize\bfseries}{\thesubsubsubsection}{1em}{}
\titlespacing*{\subsubsubsection}
{0pt}{3.25ex plus 1ex minus .2ex}{1.5ex plus .2ex}
\def\toclevel@subsubsubsection{4}
\def\l@subsubsubsection{\@dottedtocline{4}{7em}{4em}}
\makeatother

\usepackage{mathtools}
\usepackage{upquote}
\usepackage{amsthm}
\usepackage{chngcntr}
\usepackage{apptools}
\usepackage{etoolbox}
\AtAppendix{\counterwithin{theorem}{section}}

\newcommand\mydots{\ifmmode\ldots\else\makebox[1em][c]{.\hfil.\hfil.}\thinspace\fi}

\usepackage{scalerel,stackengine}
\stackMath
\newcommand\reallywidehat[1]{%
\savestack{\tmpbox}{\stretchto{%
  \scaleto{%
    \scalerel*[\widthof{\ensuremath{#1}}]{\kern-.6pt\bigwedge\kern-.6pt}%
    {\rule[-\textheight/2]{1ex}{\textheight}}
  }{\textheight}%
}{0.5ex}}%
\stackon[1pt]{#1}{\tmpbox}%
}
\definecolor{qqwuqq}{rgb}{0.,0.39215686274509803,0.}
\definecolor{ududff}{rgb}{0.30196078431372547,0.30196078431372547,1.}
\newcommand{\splitfigurealt}[2]{%
  \ifnum#1=6
    \adjincludegraphics[trim={0cm {0.8333333333333334\height} 0cm {0.0\height}},clip,width=0.95\textwidth]{#2}\\
    \adjincludegraphics[trim={0cm {0.6666666666666666\height} 0cm {0.16666666666666666\height}},clip,width=0.95\textwidth]{#2}\\
    \adjincludegraphics[trim={0cm {0.5\height} 0cm {0.3333333333333333\height}},clip,width=0.95\textwidth]{#2}\\
    \adjincludegraphics[trim={0cm {0.33\height} 0cm {0.5\height}},clip,width=0.95\textwidth]{#2}\\
    \adjincludegraphics[trim={0cm {0.16\height} 0cm {0.68\height}},clip,width=0.95\textwidth]{#2}\\
    \adjincludegraphics[trim={0cm {0.0\height} 0cm {0.85\height}},clip,width=0.95\textwidth]{#2}\\
  \fi
}
\newcommand{\splitfigure}[2]{%
  \ifnum#1=3
    \adjincludegraphics[trim={0cm {0.6666666666666666\height} 0cm {0.0\height}},clip,width=0.95\textwidth]{#2}\\
    \adjincludegraphics[trim={0cm {0.3333333333333333\height} 0cm {0.3333333333333333\height}},clip,width=0.95\textwidth]{#2}\\
    \adjincludegraphics[trim={0cm {0.0\height} 0cm {0.6666666666666666\height}},clip,width=0.95\textwidth]{#2}\\
  \fi
  \ifnum#1=4
    \adjincludegraphics[trim={0cm {0.75\height} 0cm {0.0\height}},clip,width=0.95\textwidth]{#2}\\
    \adjincludegraphics[trim={0cm {0.5\height} 0cm {0.25\height}},clip,width=0.95\textwidth]{#2}\\
    \adjincludegraphics[trim={0cm {0.25\height} 0cm {0.5\height}},clip,width=0.95\textwidth]{#2}\\
    \adjincludegraphics[trim={0cm {0.0\height} 0cm {0.75\height}},clip,width=0.95\textwidth]{#2}\\
  \fi
  \ifnum#1=5
    \adjincludegraphics[trim={0cm {0.8\height} 0cm {0.0\height}},clip,width=0.95\textwidth]{#2}\\
    \adjincludegraphics[trim={0cm {0.6\height} 0cm {0.2\height}},clip,width=0.95\textwidth]{#2}\\
    \adjincludegraphics[trim={0cm {0.4\height} 0cm {0.4\height}},clip,width=0.95\textwidth]{#2}\\
    \adjincludegraphics[trim={0cm {0.2\height} 0cm {0.6\height}},clip,width=0.95\textwidth]{#2}\\
    \adjincludegraphics[trim={0cm {0.0\height} 0cm {0.8\height}},clip,width=0.95\textwidth]{#2}\\
  \fi
  \ifnum#1=6
    \adjincludegraphics[trim={0cm {0.8333333333333334\height} 0cm {0.0\height}},clip,width=0.95\textwidth]{#2}\\
    \adjincludegraphics[trim={0cm {0.6666666666666666\height} 0cm {0.16666666666666666\height}},clip,width=0.95\textwidth]{#2}\\
    \adjincludegraphics[trim={0cm {0.5\height} 0cm {0.3333333333333333\height}},clip,width=0.95\textwidth]{#2}\\
    \adjincludegraphics[trim={0cm {0.3333333333333333\height} 0cm {0.5\height}},clip,width=0.95\textwidth]{#2}\\
    \adjincludegraphics[trim={0cm {0.16666666666666666\height} 0cm {0.6666666666666666\height}},clip,width=0.95\textwidth]{#2}\\
    \adjincludegraphics[trim={0cm {0.0\height} 0cm {0.8333333333333334\height}},clip,width=0.95\textwidth]{#2}\\
  \fi
}
\newcommand{\splitatcommas}[1]{%
  \begingroup
  \begingroup\lccode`~=`, \lowercase{\endgroup
    \edef~{\mathchar\the\mathcode`, \penalty0 \noexpand\hspace{0pt plus 1em}}%
  }\mathcode`,="8000 #1%
  \endgroup
}
\newcommand{\ignore}[1]{}
\newcommand{\nobibentry}[1]{{\let\nocite\ignore\bibentry{#1}}}

\newtcbtheorem[auto counter]{theo}%
  {Theorem}{fonttitle=\bfseries\upshape,
     arc=0mm, colback=black!5!white,colframe=black!75!black}{theorem}
\hypersetup{
    colorlinks=true,
    linkcolor=blue, 
    citecolor=blue,        
    filecolor=magenta,      
    urlcolor=cyan           
    }

\date{}
\makeatother
\title{Sensitivity analysis-guided model reduction of a mathematical model of pembrolizumab therapy for de novo metastatic MSI-H/dMMR colorectal cancer}
\author[1, *]{Georgio Hawi}
\author[1, $\dagger$]{Peter S. Kim}
\author[2, $\dagger$]{Peter P. Lee}
\affil[1]{School of Mathematics and Statistics, University of Sydney, Sydney, Australia}
\affil[2]{Department of Immuno-Oncology, Beckman Research Institute, City of Hope, Duarte, California, USA}
\affil[*]{Corresponding author: \href{mailto:georgio.hawi@sydney.edu.au}{georgio.hawi@sydney.edu.au}}
\affil[$\dagger$]{These authors contributed comparably to this work}
\usepackage{float}
\floatstyle{plaintop}
\restylefloat{table}

\theoremstyle{definition}

\newcommand{\set}[1]{\left \{ #1 \right \}}

\allowdisplaybreaks
\usepackage{longtable}
\usepackage[numbers,square,sort&compress]{natbib}
\usepackage{bibunits}
\usepackage{hypernat}
\makeatletter
\pretocmd{\@startbibunit}{%
  \begingroup
    \count@\@bibunitauxcnt
    \advance\count@\@ne
    \xdef\@extra@binfo{.bu\the\count@}%
    \xdef\@extra@b@citeb{.bu\the\count@}%
  \endgroup
}{}{}
\apptocmd{\endbibunit}{%
  \gdef\@extra@binfo{}%
  \gdef\@extra@b@citeb{}%
}{}{}
\makeatother
\defaultbibliographystyle{vancouver}%
\begin{document}
\maketitle
\begin{bibunit}[vancouver]
\begin{abstract}
Colorectal cancer (CRC) is the third most commonly diagnosed cancer worldwide and the leading cause of cancer-related deaths in adults under 55, involving a complex interplay of biological processes such as dendritic cell (DC) maturation and migration, T cell activation and proliferation, cytokine production, and T cell and natural killer (NK) cell-mediated cancer cell killing. Microsatellite instability-high (MSI-H) CRC and deficient mismatch repair (dMMR) CRC constitute 15\% of all CRC and 4\% of metastatic CRC, and exhibit remarkable responsiveness to immunotherapy, especially with PD-1 inhibitors such as pembrolizumab. Mathematical models of the underlying immunobiology and the interactions underpinning immune checkpoint blockade offer mechanistic insights into tumour--immune dynamics and provide avenues for treatment optimisation and the identification of novel therapeutic targets. We used our data-driven model of de novo metastatic MSI-H/dMMR CRC (dnmMCRC) and performed sensitivity analysis-guided model reduction using the Fourier amplitude sensitivity testing (FAST) and extended FAST (EFAST) methods. In this work, we constructed two simplified models of dnmMCRC: one that faithfully reproduces all of the original model's trajectories, and a second, minimal model that accurately replicates the original dynamics while being highly extensible for future inclusion of additional components to explore various aspects of the anti-tumour immune response. Together, these resulting models offer a tractable foundation for future theoretical and computational studies of immune checkpoint blockade, avoiding unnecessary complexity while preserving mechanistic interpretability.
\end{abstract}
\section{Introduction}
Colorectal cancer (CRC) is the third most common cancer worldwide, with more than 1.85 million cases and 850,000 deaths annually \citep{Biller2021}. Despite CRC being diagnosed mostly in adults aged 65 and older, there has been an increase in the incidence rate of CRC amongst younger populations \citep{SifakiPistolla2022, Siegel2023, Siegel2024} since the mid-1990s, with CRC being the leading cause of cancer-related deaths in adults under 55 \citep{Siegel2024}. In the United States, the 5-year survival rates for stage IIIA, stage IIIB, and stage IIIC colon cancer are 90\%, 72\%, and 53\%, respectively, whilst stage IV CRC has a 5-year survival of only 12\% \citep{Rawla2019}. Of new CRC diagnoses, 20\% of patients present with de novo metastatic disease, with an estimated 75\%--90\% of these patients presenting with unresectable metastatic lesions \citep{Cook2005}. However, patients with the hypermutant microsatellite instability-high (MSI-H) or deficient mismatch repair (dMMR) phenotype, hereafter referred to as the MSI-H/dMMR phenotype, who have reached metastasis are less responsive to conventional chemotherapy and have a poorer prognosis compared to patients with microsatellite stable (MSS) CRC \citep{Shulman2018}. \\~\\
That being said, patients with MSI-H/dMMR CRC have a much better prognosis with immunotherapy, in particular with immune checkpoint inhibitors (ICIs). Immune checkpoints, such as programmed cell death protein 1 (PD-1), normally downregulate immune responses after antigen activation \citep{Topalian2012}. PD-1, a cell membrane receptor that is expressed on a variety of cell types, including activated T cells, activated B cells and monocytes, has been extensively researched in the context of cancer such as MSI-H/dMMR CRC \citep{Sarshekeh2018, Yaghoubi2019}. When PD-1 interacts with its ligands, programmed death-ligand 1 (PD-L1) and programmed death-ligand 2 (PD-L2), effector T cell activity is inhibited, resulting in the downregulation of pro-inflammatory cytokine secretion and the upregulation of immunosuppressive regulatory T cells (Tregs) \citep{Lin2024, han2020pd}. Cancers can exploit this by expressing PD-L1 themselves, evading immunosurveillance, and impairing the proliferation and activity of cytotoxic T lymphocytes (CTLs) \citep{Oliveira2019}. Blockade of PD-1/PD-L1 complex formation reinvigorates effector T cell activity, resulting in enhanced anti-tumour immunity and responses, leading to improved clinical outcomes in cancer patients \citep{Lee2015, Zhang2020nature}. In particular, we refer to the FDA-approved regimen for the first-line treatment of metastatic MSI-H/dMMR CRC (mMCRC) in adults---200 mg of pembrolizumab, an anti-PD-1 antibody, administered by intravenous infusion over 30 minutes every 3 weeks until disease progression or unacceptable toxicity \citep{fda_pembrolizumab}---as the standard regimen for the remainder of this work. \\~\\
Mathematical models provide a powerful framework for analysing the immunobiology underpinning diseases, which can then be used to investigate the dynamics of relevant biological components---improving theoretical understanding---especially when approached from a mechanistic perspective. Additionally, such modelling provides avenues for comparing and optimising treatment regimens, whilst avoiding the significant time and financial costs associated with human clinical trials. However, to effectively model complex biological systems, one must master the balance between model complexity and model accuracy. Traditional models incorporate multiple biological processes and usually result in models containing many input variables and model parameters \citep{Zhang2015}. In addition to the challenge of properly estimating and calibrating these parameters, one must also consider the resources and computational costs needed to repeatedly run the model \citep{Morris1991}. As such, it is ideal to reduce models such that their key dynamics are maintained, but parameters and inputs with minimal influence on the outputs are identified and appropriately removed. There are numerous ways to do this reduction, including multiple timescale methods \citep{Bertram2017}, singular value decomposition \citep{Padoan2021}, and sensitivity analysis (SA).\\~\\
SA aims to determine the robustness of a system to fluctuations in its input parameters and components, and can be used to eliminate the least significant components in a model \citep{Snowden2017}. One can classify SA techniques into two categories: local and global. Local SA concerns slightly perturbing a model's parameters in a predetermined region in parameter space and exploring the model outputs. As such, the results attained by local SA are only valid in a small neighbourhood of the perturbed region \citep{Marzban2013}. The main drawback of local SA is its inapplicability to nonlinear models, as it fails to consider interaction effects between variables, resulting in an underestimation of variables' true significance \citep{Qian2020}. Nonetheless, due to their transparency, ease of understanding and computational speed, local SA methods such as the partial rank correlation coefficient (PRCC) \citep{Marino2008}, and derivative-based local methods have been widely used in analysing the sensitivity of parameters in mathematical models of CRC and ICI therapy \citep{Lai2017, Lai2019, Liao2024, Siewe2021, Kirshtein2020, Budithi2021}. On the other hand, global SA methods allow for exploration of the entire parameter space and can examine the effects of large changes of model inputs on the model output---even in nonlinear models \citep{https://doi.org/10.5281/zenodo.6110623}. Whilst being computationally intensive, these methods account for higher-order interactions between variables, unlike local SA methods. There exist many types of global SA methods, such as the elementary effects methods \citep{Morris1991, Campolongo2007}, high-dimensional model representation \citep{Li2001}, and metamodel-based analyses \citep{Gratiet2015}. We focus on variance-based SA, which features model-independent methods that have been used to analyse models in immunology, biophysics, and engineering \citep{Jayachandran2014, Dela2022, VazquezCruz2014, Qian2020}.\\~\\
In this work, we consider our two-compartmental model of pembrolizumab therapy in de novo metastatic MSI-H/dMMR CRC (dnmMCRC) \citep{Hawi2026meta}, which, to the authors' knowledge, is the only immunobiological model of ICI therapy in metastatic CRC, and extends \citep{Hawi2025local} to the metastatic setting. We perform global variance-based SA on this model to evaluate the sensitivity of individual parameters to the model and use this to guide model reduction and facilitate the development of two simplified models: one that faithfully replicates all model trajectories, and a minimal model that accurately replicates trajectories of its state variables.
\section{Variance-Based Sensitivity Analysis Background\label{SAsection}}
It is prudent for us to detail the mathematical background for variance-based SA, beginning with the Sobol' method. The rationale behind the Sobol' method, and variance-based SA in general, is to quantify the impact of each input parameter of a model on the total variance of its output \citep{Cannav2012}. For simplicity, we consider a model $Y=f(\mathbf{X})$, where $Y$ is a scalar and $\mathbf{X} = (X_1,X_2,\dots,X_n)$ are the $n$ model input parameters. Via rescaling and appropriate transformations, we can consider each $X_i$ as a random variable to be independently and uniformly distributed on $[0,1]$. Furthermore, we assume that $f\in L^2([0,1]^n)$.
\subsection{Sobol' Method}
The key idea behind variance-based SA is that $f(\mathbf{X})$ can be uniquely decomposed via a Sobol-Hoeffding decomposition \citep{sobol1993sensitivity,Hoeffding1948} given by 
\begin{align}
    f(\mathbf{X}) &= f_0 + \sum_{i=1}^{n} f_i(X_i) + \sum_{i=1}^{n}\sum_{j=1}^{i-1} f_{ij}(X_i,X_j) + \dots + f_{1,2,\dots,n}(X_1,X_2,\dots,X_n), \label{ANOVadecomp}
\intertext{where}
f_0 &= \mathbb{E}[f(\mathbf{X})] = \int_{[0,1]^{n}} f(\mathbf{X}) \ \mathrm{d}\mathbf{X}, \\ 
f_i(X_i) &= \mathbb{E}[f(\mathbf{X})|X_i]-f_0 = \int_{[0,1]^{n-1}} f(\mathbf{X}) \ \prod_{\substack{k=1 \\ k\neq i}}^{n} \mathrm{d}X_k - f_0 ,\\
f_{ij}(X_i,X_j) &= \mathbb{E}[f(\mathbf{X})|X_i,X_j]-f_0-f_i(X_i)-f_j(X_j) = \int_{[0,1]^{n-2}} f(\mathbf{X}) \ \prod_{\substack{k=1 \\ k\neq i,j}}^{n} \mathrm{d}X_k - f_0 -f_i-f_j,
\end{align}
and so forth. From this, it is easy to see that $f_i$ represents the effect of varying just $X_i$, denoted the main effect of $X_i$, $f_{ij}$ represents the effect of varying $X_i$ and $X_j$ simultaneously, etc. We note that this decomposition is only unique when all terms in \eqref{ANOVadecomp} are orthogonal, i.e.
\begin{equation}
    \mathbb{E}[f_I(X_I)|X_J] = 0,
\end{equation}
for all $I\subseteq \set{1,2,\dots,n}$ and all $J\subseteq \set{1,2,\dots,n}$ where $J \subsetneq I$. If $I = \set{i_1,\dots,i_s}$ for $i_1,\dots,i_s\in \set{1,2,\dots,n}$, then we define $X_I$ as the set $\set{X_{i_1},\dots,X_{i_s}}$ and $f_I$ as $f_{i_1,\dots,i_s}$ and analogously for $X_J$. Squaring \eqref{ANOVadecomp} and integrating, with multiple applications of the orthogonality condition, renders
\begin{equation}
    \int_{[0,1]^n} f^2(\mathbf{X}) \ \mathrm{d}\mathbf{X} - f_0^2 = \sum_{s=1}^{n}\sum_{i_1<\dots<i_s} \int_{[0,1]^s} f^2_{i_1\dots i_s}(X_{i_1},\dots,X_{i_s}) \ \prod_{k=1}^{|I|} \mathrm{d}X_{i_k}. \label{squareanov}
\end{equation}
By definition, the left-hand side of \eqref{squareanov} is $\operatorname{Var}(Y)=\operatorname{Var}(f(\mathbf{X}))$ and the right-hand side is a sum of partial variances over all subsets of $\set{X_1, X_2,\dots, X_n}$. Thus, this leads to the analysis of variance (ANOVA) decomposition
\begin{align}
    \operatorname{Var}(f(\mathbf{X})) &= \sum_{i=1}^n V_i + \sum_{i=1}^{n}\sum_{j=1}^{i-1} V_{ij} + \dots + V_{12\dots n}, \label{vardecomp}
    \intertext{where}
    V_I &= \int_{[0,1]^{|I|}} f_I^2(X_I) \ \prod_{k=1}^{|I|} \mathrm{d}X_{i_k},
\end{align}
for all $I\subseteq \set{1,2,\dots,n}$. We consider the first-order sensitivity index, $S_i$, as the main contribution of a
single parameter to the output variance, neglecting interactions with other input parameters. This is given by \citep{sobol1993sensitivity}
\begin{equation}
    S_i = \frac{V_i}{\operatorname{Var}(Y)} \label{firstorderformulasobol}.
\end{equation}
In addition to $S_i$ being a way to measure the importance of $X_i$ via determining the sensitivity of the model output to $X_i$, it also represents the expected percentage reduction in $\operatorname{Var}(Y)$ when $X_i$ has zero uncertainty \citep{Cannav2012, Borgonovo2007}. Second-order sensitivity indices incorporate the main and second-order contributions---the extent of $\operatorname{Var}(Y)$ due to two input parameters that cannot be explained by the individual effects of the parameters alone and are given by
\begin{equation}
    S_{ij}=\frac{V_{ij}}{\operatorname{Var}(Y)}.
\end{equation}
This notion continues for higher-order indices and is notated similarly. By \eqref{vardecomp}, one has that
\begin{equation}
    \sum_{i=1}^{n} S_i +\sum_{i=1}^{n}\sum_{j=1}^{i-1} S_{ij} + \dots + S_{12\dots n}= 1.
\end{equation}
One particularly useful index is the total-order sensitivity index for an input parameter, $S_{Ti}$, which is the proportion of the total output variance that is caused by an input variable and its interactions with any of the other input variables \citep{Homma1996}. One can express this as
\begin{equation}
    S_{Ti} = S_i + \sum_{\substack{j=1\\j\neq i}}^n S_{ij} + \dots + S_{1\dots i \dots n}.
\end{equation}
We note that the sum of all the total-order indices is greater than or equal to $1$, with equality if and only if the model is purely additive. Since each sensitivity index must clearly be non-negative, it is easy to see that the closer an index is to $1$, the more important the corresponding variables are to the model output and vice versa. \\~\\
Whilst calculating each $S_i$ only involves $n$ integrals, to calculate all sensitivity indices (including the total-order indices), one needs $\sum_{k=0}^{n} \binom{n}{k} = 2^n$ integral evaluations. For low-dimensional models, this does not pose an issue, but for larger ones, this is computationally expensive.

\subsection{Fourier Amplitude Sensitivity Testing}
To remedy this, one can calculate the first-order Sobol' indices via optimisations such as Fourier amplitude sensitivity testing (FAST) \citep{Cukier1973, Saltelli1998}. As before, we consider the model $Y=f(\mathbf{X})$, where, after the necessary transformations, we can consider $\mathbf{X}$ as a random variable independently and uniformly distributed on the $n$-dimensional unit cube, $[0,1]^n$. Since the complex exponentials of multidimensional Fourier series form an orthonormal basis, there exist unique $C_{k_1,k_2,\dots,k_n}$ such that $f(\mathbf{X})$ can be represented as 
\begin{align}
    f(\mathbf{X})&=\sum_{k_1=-\infty}^{\infty} \sum_{k_2=-\infty}^{\infty} \cdots \sum_{k_n=-\infty}^{\infty} C_{k_1 k_2 \cdots k_n} \exp \left[2 \pi i\left(k_1 X_1+k_2 X_2+\cdots+k_n X_n\right)\right],
    \intertext{where the Fourier coefficients $C_{k_1 k_2\dots k_n}$ are given by}
    C_{k_1 k_2\dots k_n}&=\int_{[0,1]^n} f(\mathbf{X}) \exp \left[-2 \pi i\left(k_1 X_1+k_2 X_2+\cdots+k_n X_n\right)\right] \ d\mathbf{X}.
\end{align}
This immediately gives the decomposition in \eqref{ANOVadecomp}, where
\begin{align*}
    f_0 &= C_{00\dots 0}, \\
    f_i(X_i)&=\sum_{\substack{k_i=-\infty \\ k_i\neq 0}}^{\infty} C_{0\dots k_i \dots 0 }\exp\left[2\pi i(k_i X_i)\right], \\
    f_{ij}(X_i,X_j)&=\sum_{\substack{k_i=-\infty \\ k_i\neq 0}}^{\infty} \sum_{\substack{k_j=-\infty \\ k_j\neq 0}}^{\infty} C_{0\dots k_i \dots k_j \dots 0}\exp\left[2\pi i(k_i X_i+k_j X_j)\right], 
\end{align*}
and so forth. We thus see that
\begin{align}
    V_i &=\int_0^1 f_{i}^2(X_i) \ dX_i =\sum_{\substack{k_i=-\infty \\ k_i\neq 0}}^{\infty} \left|C_{0\dots k_i \dots 0 }\right|^2 =2\sum_{k_i=1}^\infty \left( A_{k_i}^2+B_{k_i}^2\right), \label{parsevalform1}
    \intertext{where $A_{k_i}$ and $B_{k_i}$ are the Fourier cosine and sine coefficients, respectively, given by}
    A_{k_i}&=\int_{[0,1]^n} f(\mathbf{X})\cos(2\pi k_i X_i) \ d\mathbf{X}, \quad B_{k_i}=\int_{[0,1]^n} f(\mathbf{X})\sin(2\pi k_i X_i) \ d\mathbf{X},
\end{align}
where we have used Parseval's identity. We immediately see a major problem: the computation of the Fourier coefficients requires an infinite number of multidimensional integrals to be calculated, which is very computationally expensive. To remedy this, we consider the set of transformations
\begin{equation}
    X_i(s) = \frac{1}{2}+\frac{1}{\pi}\arcsin(\sin(\omega_i s)),
\end{equation}
from \citep{Saltelli1999}, where $s$ is a new independent variable. The key idea is to choose the set of frequencies $\{\omega_i\}$ such that the $n$-dimensional Fourier integrals are converted to $1$-dimensional integrals with respect to $s$ \citep{Cannav2012}. If $\{\omega_i\}$ are chosen to be incommensurate, so that each $\omega_i$ cannot be expressed as a linear combination of the other frequencies with integer coefficients, then as $s$ varies from $-\infty$ to $\infty$, $\mathbf{X}(s)=(X_1(s), X_2(s),\dots, X_n(s))$ passes through every point in $[0,1]^n$. As such, the curve is space-filling, and by Weyl's ergodic theorem \citep{Weyl1938} we can replace the $n$-dimensional integrals with the appropriate $1$-dimensional counterparts. More precisely, for any integrable $g\in L^1([0,1]^n)$, Weyl's ergodic theorem yields
\begin{equation}
\int_{[0,1]^n} g(\mathbf{X}) \ d\mathbf{X}=\lim_{L\to\infty}\frac{1}{2L}\int_{-L}^{L} g(\mathbf{X}(s)) \ ds.
\end{equation}
For example, we have that
\begin{align}
    \mathbb{E}[f(\mathbf{X})]&=\lim_{L\to\infty}\frac{1}{2L}\int_{-L}^{L} f(\mathbf{X}(s)) \ ds ,\label{fv2}\\
    A_{k_i}&=\lim_{L\to\infty}\frac{1}{2L}\int_{-L}^{L} f(\mathbf{X}(s))\cos(2\pi k_i X_i(s)) \ ds , \label{Ak1v2}\\
    B_{k_i}&=\lim_{L\to\infty}\frac{1}{2L}\int_{-L}^{L} f(\mathbf{X}(s))\sin(2\pi k_i X_i(s)) \ ds , \label{Bk1v2}
\end{align}
and similarly for other integrals. Recalling that
\begin{equation}
    \operatorname{Var}(f(\mathbf{X}))=\mathbb{E}[f(\mathbf{X})^2]-\mathbb{E}[f(\mathbf{X})]^2, \label{varformula}
\end{equation}
in conjunction with \eqref{firstorderformulasobol}, \eqref{parsevalform1}, \eqref{fv2}, \eqref{Ak1v2}, and \eqref{Bk1v2} gives a means to calculate the first-order sensitivity indices.\\~\\
However, it is obvious that at most one of the $\omega_i$ can be rational, whilst the others must be irrational. Due to the finite precision of computers, irrational numbers cannot be stored exactly on computers, and so $\{\omega_i\}$ can only be incommensurate up to some level of precision. Thus, the resultant search curve is not truly space-filling, and a rational approximation of $\{\omega_i\}$ is needed. Without loss of generality, one can assume that each frequency is integral, and choose $\{\omega_i\}$ to be incommensurate to the order $M$, so that each $\omega_i$ cannot be expressed as a linear combination of the other frequencies with integer coefficients when the sums of the absolute values of those coefficients are less than or equal to $M+1$. $M$ is known as the interference number \citep{Cukier1973}, and is usually set to $4$ or higher \citep{Saltelli1999}, with $\{\omega_i\}$ becoming exactly incommensurate as $M\to\infty$. Since $\{\omega_i\}$ is not truly incommensurate for finite $M$, there exists some positive rational number $T$, such that $f(\mathbf{X}(s))=f(\mathbf{X}(s+T))$. In the case of integral $\omega_i$, it was shown in \citep{Cukier1973} that each $X_i(s)$ is $2\pi$-periodic, and so $f(\mathbf{X}(s))$ is also $2\pi$-periodic. Therefore, the long-time averages in \eqref{fv2}--\eqref{Bk1v2} can be approximated by averages over one period, i.e.
\begin{equation}
\lim_{L\to\infty}\frac{1}{2L}\int_{-L}^{L} g(\mathbf{X}(s)) \ ds=\frac{1}{2\pi}\int_{-\pi}^{\pi} g(\mathbf{X}(s)) \ ds,
\end{equation}
for appropriate integrands $g(\mathbf{X}(s))$. Thus, if we consider $f(\mathbf{X}(s))$ for $s\in [-\pi,\pi]$, we have that \eqref{Ak1v2}, \eqref{Bk1v2}, and \eqref{varformula} become
\begin{align}
    A_{k_i}&\approx\frac{1}{2\pi}\int_{-\pi}^{\pi} f(\mathbf{X}(s))\cos(2\pi k_i X_i(s)) \ ds, \label{Ak1v3}\\
    B_{k_i}&\approx\frac{1}{2\pi}\int_{-\pi}^{\pi} f(\mathbf{X}(s))\sin(2\pi k_i X_i(s)) \ ds, \label{Bk1v3}\\
    \operatorname{Var}(f(\mathbf{X}))&\approx\frac{1}{2\pi}\int_{-\pi}^{\pi} f^2(\mathbf{X}(s)) \ ds -\left(\frac{1}{2\pi}\int_{-\pi}^{\pi} f(\mathbf{X}(s)) \ ds \right)^2.
    \intertext{If we express $f(\mathbf{X}(s))$ as a 1-dimensional Fourier series in $s$, via the expansion}
    f(\mathbf{X}(s))&=\sum_{k=-\infty}^\infty \left(A_k\cos(ks)+B_k\sin(ks)\right),
    \intertext{then by Parseval's identity, we have that}
    \operatorname{Var}(f(\mathbf{X}))&\approx 2\sum_{k=1}^\infty \left(A_k^2 + B_k^2\right), \label{parsevalidentity}
    \intertext{where}
    A_k&=\frac{1}{2\pi}\int_{-\pi}^{\pi} f(\mathbf{X}(s))\cos(ks) \ ds, \quad B_k=\frac{1}{2\pi}\int_{-\pi}^{\pi} f(\mathbf{X}(s))\sin(ks) \ ds.
    \intertext{Thus, analogous to \eqref{firstorderformulasobol}, one has that the first-order FAST sensitivity indices are given by}
    S_i^\mathrm{FAST}&=\frac{\sum\limits_{k_i=1}^\infty \left(A_{k_i}^2+B_{k_i}^2\right)}{\sum\limits_{k=1}^\infty \left(A_{k}^2+B_{k}^2\right)}.
\end{align}
These integrals can be numerically determined using $N$ sampling points, uniformly chosen on $[-\pi,\pi]$, in conjunction with predetermined sets of frequencies (which have been calculated up to $n=50$) \citep{McRae1982}. \\~\\
There are a few important things to note. One important thing to consider is that since a finite number of sampling points are used, $N$ must be chosen so that aliasing is avoided, and that it is not possible for more than one frequency to contribute to each Fourier coefficient \citep{Cukier1973}. By the Nyquist--Shannon sampling theorem, this is equivalent to choosing $N$ such that
\begin{equation}
    N\geq 2M\omega_\text{max}+1,
\end{equation}
where $\omega_\text{max}$ is the maximum value of $\{\omega_i\}$. Furthermore, $S_i^\mathrm{FAST}$ is independent of the choice of $\{\omega_i\}$ \citep{Schaibly1973}, and it was shown in \citep{Saltelli1998} that $S_i^\mathrm{FAST}\to S_i$ as $N\to\infty$. There exist optimisations that can be made to FAST, such as that due to \citep{Koda1979}, in which if $\{\omega_i\}$ are all chosen to be odd, then $f(\mathbf{X}(s))$ is symmetric about $s=\pm \pi/2$, so that one can restrict the interval of integration from $[-\pi,\pi]$ to $[-\pi/2,\pi/2]$, and halve the number of model evaluations.
\subsection{Extended Fourier Amplitude Sensitivity Test (EFAST)}
The FAST methodology can be extended to estimate total-order sensitivity indices using the extended Fourier amplitude sensitivity test (EFAST) algorithm \citep{Saltelli1999}. In EFAST, one assigns parameter $X_i$ a high base frequency $\omega_i$ and assigns all other parameters to lower frequencies, so that the variance due to all factors except $X_i$ can be estimated from Fourier modes not associated with the harmonics $k=m\omega_i$, $m=1,\dots, M$. Using \eqref{parsevalidentity}, we define the complementary variance
\begin{equation}
V_{\sim i}\approx 2\sum_{\substack{k=1\\ k\notin\{m\omega_i\}_{m=1}^M}}^\infty \left(A_k^2 + B_k^2\right),
\end{equation}
and then estimate the total-order index by
\begin{equation}
    S_{Ti}^\mathrm{EFAST}\approx 1-\frac{V_{\sim i}}{\operatorname{Var}(f(\mathbf{X}))}.
\end{equation}
One point of comparison between FAST-based and Sobol' methods is that, in general, the Sobol' method requires $N$ times more model evaluations than FAST for a given $N$ \citep{Saltelli1998}. As such, FAST-based methods are computationally less expensive, with faster convergence to the true sensitivity indices.
\numberwithin{equation}{subsection}
\renewcommand{\theequation}{\thesubsection.\arabic{equation}}
\section{Mathematical Models}
\subsection{Model Assumptions}
For simplicity, we ignore spatial effects in the model, including diffusion, advection, and chemotaxis for all species. We instead assume that model species are well-mixed, which is a common approach in differential equation-based models of tumour--immune interactions \citep{Marzban2024}. We assume the system has two spatial compartments: one at the tumour site (TS), located in the colon or rectum, and one at the tumour-draining lymph node (TDLN), which receives lymphatic drainage from the primary tumour and constitutes the principal site of tumour antigen presentation and T cell priming. This is a simplification since metastatic CRC typically involves multiple tumour-draining lymph nodes \citep{Wang2024}; however, for simplicity, we focus on the sentinel node and refer to it as the TDLN for the purposes of the model \citep{Li2022}. In a similar fashion to nearly all models of de novo metastatic cancer, we primarily focus on the growth of the primary tumour, using it as a proxy to infer the progression of lymph node and distant metastases. We assume that cytokines in the TS are produced only by effector or activated cells \citep{Hoekstra2021} and that damage-associated molecular patterns (DAMPs) in the TS are only produced by necrotic cancer cells \citep{Murao2021}. We assume that all mature DCs modelled in the TDLN are cancer-antigen-bearing and that all T cells modelled in the TS are primed with cancer antigens. Furthermore, we assume that all activated T cells considered in the TDLN are activated with cancer antigens and that T cell proliferation/division follows a deterministic program \citep{Marchingo2014, Kaech2001}. We ignore CD4+ and CD8+ memory T cells and assume that naive CD4+ T cells differentiate immediately upon activation. We also assume that all Tregs in the TS are natural Tregs (nTregs), ignoring induced Tregs (iTregs), since most tumour-infiltrating Tregs are nTregs \citep{Adeegbe2013}. We assume, for simplicity, that activated macrophages polarise into the M1/M2 dichotomy, where M1 macrophages are pro-inflammatory, and M2 macrophages are immunosuppressive \citep{Viola2019, Mills2012, Han2021}. We also assume that the duration of pembrolizumab infusion is negligible compared to the timescale of the model \citep{fda_pembrolizumab}. Therefore, we treat its infusion as an intravenous bolus so that drug absorption occurs immediately after infusion. Finally, we assume a constant solution history, where the history for each species is set to its respective initial condition.\\~\\
We assume that all species, $X_i$, degrade/die at a rate proportional to their concentration, with decay constant $d_{X_i}$. We assume that the rate of activation/polarisation of a species $X_i$ by a species $X_j$ follows the Michaelis-Menten kinetic law $\lambda_{X_i X_j}X_i\frac{X_j}{K_{X_i X_j}+X_j}$, for rate constant $\lambda_{X_i X_j}$, and half-saturation constant $K_{X_i X_j}$. Similarly, we model the rate of inhibition of a species $X_i$ by a species $X_j$ using a term of the form $\lambda_{X_i X_j}\frac{X_i}{1+X_j/K_{X_i X_j}}$ for rate constant $\lambda_{X_i X_j}$, and half-saturation constant $K_{X_i X_j}$. Production of $X_i$ by $X_j$ is modelled using mass-action kinetics unless otherwise specified, so that the rate at which $X_i$ is formed is given by $\lambda_{X_iX_j}X_j$ for some positive constant $\lambda_{X_iX_j}$. Finally, we assume that the rate of lysis of $X_i$ by $X_j$ follows mass-action kinetics in the case where $X_j$ is a cell and follows Michaelis-Menten kinetics in the case where $X_j$ is a cytokine.
\subsection{Model Summary}
We also outline some of the main processes accounted for in the models.
\begin{enumerate}
    \item Effector CD8+ T cells and NK cells induce apoptosis of cancer cells, with this being inhibited by TGF-$\upbeta$ and the PD-1/PD-L1 complex. However, TNF and IFN-$\upgamma$ induce necroptosis of cancer cells, causing them to become necrotic before they are removed.
    \item Necrotic cancer cells release DAMPs such as high mobility group box 1 (HMGB1) and calreticulin, which stimulate immature DCs to mature.
    \item Some mature DCs migrate to the T cell zone of the TDLN and activate naive CD8+ and CD4+ T cells (including Tregs), with CD8+ T cell and CD4+ T helper 1 (Th1) cell activation being inhibited by Tregs and the PD-1/PD-L1 complex.
    \item Activated T cells undergo clonal expansion and proliferate rapidly in the TDLN, with CD8+ T cell and Th1 cell proliferation being inhibited by Tregs and the PD-1/PD-L1 complex.
    \item T cells that have completed proliferation migrate to the TS and perform effector functions including the production of pro-inflammatory cytokines such as interleukin-2 (IL-2), interferon-$\upgamma$ (IFN-$\upgamma$), and tumour necrosis factor (TNF), and immunosuppressive cytokines such as transforming growth factor-$\upbeta$ (TGF-$\upbeta$) and interleukin-10 (IL-10). Extended exposure to the cancer antigen can lead CD8+ T cells to become exhausted; however, this exhaustion can be reversed by pembrolizumab.
    \item In addition, mature DCs, NK cells and macrophages secrete cytokines that can activate NK cells and polarise and repolarise macrophages into pro-inflammatory and immunosuppressive phenotypes.
    \item Pembrolizumab infusion promotes the binding of unbound PD-1 receptors to pembrolizumab, forming the PD-1/pembrolizumab complex instead of the PD-1/PD-L1 complex. This reduces the inhibition of pro-inflammatory CD8+ and Th1 cell activation and proliferation while also reducing the inhibition of cancer cell lysis.
\end{enumerate}
\subsection{Full Model}
\subsubsection{Model Variables}
The variables and their units in the full model are shown in \autoref{modelvars}. 
\begin{table}[H]
\centering
\resizebox{\columnwidth}{!}{%
\begin{tabular}{|lp{7.7cm}|lp{7.7cm}|}
\hline
\textbf{Var} & \textbf{Description} & \textbf{Var} & \textbf{Description} \\ 
\hline
$V_\mathrm{TS}$ & Primary tumour volume & & \\
\hline
$C$ & Viable cancer cell density & $N_c$ & Necrotic cell density \\
$D_0$ & Immature DC density & $D$ & Mature DC density in the TS \\
$D^\mathrm{LN}$ & Mature DC density in the TDLN & $T_0^8$ & Naive CD8+ T cell density in the TDLN \\
$T_A^8$ & Effector CD8+ T cell density in the TDLN & $T_8$ & Effector CD8+ T cell density in the TS \\
$T_{\mathrm{ex}}$ & Exhausted CD8+ T cell density in the TS & $T_0^4$ & Naive CD4+ T cell density in the TDLN \\
$T_A^1$ & Effector Th1 cell density in the TDLN & $T_1$ & Effector Th1 cell density in the TS \\
$T_0^r$ & Naive Treg density in the TDLN & $T_A^r$ & Effector Treg density in the TDLN \\ 
$T_r$ & Effector Treg density in the TS & $M_0$ & Naive macrophage density \\
$M_1$ & M1 macrophage density & $M_2$ & M2 macrophage density \\
$K_0$ & Resting NK cell density & $K$ & Activated NK cell density\\
\hline
$H$ & HMGB1 concentration & $S$ & Calreticulin concentration \\
$I_2$ & IL-2 concentration & $I_\upgamma$ & IFN-$\upgamma$ concentration \\
$I_\upalpha$ & TNF concentration & $I_\upbeta$ & TGF-$\upbeta$ concentration \\
$I_{10}$ & IL-10 concentration & & \\
\hline
$P_D^{T_8}$ & Unbound PD-1 receptor concentration on effector CD8+ T cells in the TS & $P_D^{T_1}$ & Unbound PD-1 receptor concentration on effector Th1 cells in the TS \\
$P_D^{K}$ & Unbound PD-1 receptor concentration on activated NK cells & $Q_A^{T_8}$ & PD-1/pembrolizumab complex concentration on effector CD8+ T cells in the TS \\
$Q_A^{T_1}$ & PD-1/pembrolizumab complex concentration on effector Th1 cells in the TS & $Q_A^{K}$ & PD-1/pembrolizumab complex concentration on activated NK cells \\
$P_L$ & Unbound PD-L1 concentration in the TS & $Q^{T_8}$ & PD-1/PD-L1 complex concentration on effector CD8+ T cells in the TS \\
$Q^{T_1}$ & PD-1/PD-L1 complex concentration on effector Th1 cells in the TS & $Q^{K}$ & PD-1/PD-L1 complex concentration on activated NK cells \\
$A_{1}$ & Concentration of pembrolizumab in the TS & & \\
\hline 
$P_D^{8\mathrm{LN}}$ & Unbound PD-1 receptor concentration on effector CD8+ T cells in the TDLN & $P_D^{1\mathrm{LN}}$ & Unbound PD-1 receptor concentration on effector Th1 cells in the TDLN \\
$Q_A^{8\mathrm{LN}}$ & PD-1/pembrolizumab complex concentration on effector CD8+ T cells in the TDLN & $Q_A^{1\mathrm{LN}}$ & PD-1/pembrolizumab complex concentration on effector Th1 cells in the TDLN \\
$P_L^\mathrm{LN}$ & Unbound PD-L1 concentration in the TDLN & $Q^{8\mathrm{LN}}$ & PD-1/PD-L1 complex concentration on effector CD8+ T cells in the TDLN \\
$Q^{1\mathrm{LN}}$ & PD-1/PD-L1 complex concentration on effector Th1 cells in the TDLN & $A_{1}^\mathrm{LN}$ & Concentration of pembrolizumab in the TDLN \\
\hline
\end{tabular}%
}
\caption{\label{modelvars}Variables used in the full model. Quantities in the top box are in units of $\mathrm{cm^3}$, quantities in the second box are in units of $\mathrm{cell/{cm}^3}$, quantities in the third box are in units of $\mathrm{g/{cm}^3}$, and all other quantities are in units of $\mathrm{molec/{cm}^3}$. All quantities pertain to the tumour site unless otherwise specified. TS denotes the tumour site, whilst TDLN denotes the tumour-draining lymph node.}
\end{table}
\subsubsection{Model Equations}
The model equations are identical to those from \citep{Hawi2026meta}; however, for completeness, we provide the mathematical model below. A full derivation of the model is available in Appendix A, and the parameter values used---listed in Table A.1---are estimated as described therein.
\paragraph{Equations for Cancer Cells, DAMPs, and DCs}
\begin{align}
  \begin{split}
  \frac{dC}{dt} &= \underbrace{\lambda_{C}C\left(1-\frac{C}{C_0}\right)}_{\text{growth}} - \underbrace{\lambda_{CT_8}T_8 \frac{1}{1+I_{\upbeta}/K_{CI_{\upbeta}}}\frac{1}{1+Q^{T_8}/K_{CQ^{T_8}}}C}_{\substack{\text{elimination by $T_8$} \\ \text{inhibited by $I_{\upbeta}$ and $Q^{T_8}$}}} - \underbrace{\lambda_{CK}K \frac{1}{1+I_{\upbeta}/K_{CI_{\upbeta}}}\frac{1}{1+Q^K/K_{CQ^K}}C}_{\substack{\text{elimination by $K$} \\ \text{inhibited by $I_\upbeta$ and $Q^K$}}} \\
  &- \underbrace{\lambda_{CI_{\upalpha}}\frac{I_{\upalpha}}{K_{CI_{\upalpha}}+I_{\upalpha}}C}_{\text{elimination by $I_{\upalpha}$}}- \underbrace{\lambda_{CI_{\upgamma}}\frac{I_{\upgamma}}{K_{CI_{\upgamma}}+I_{\upgamma}}C}_{\text{elimination by $I_{\upgamma}$}},
  \end{split} \label{cancereqn} \\
  \begin{split}
    \frac{dN_c}{dt}&= \underbrace{\lambda_{CI_{\upalpha}}\frac{I_{\upalpha}}{K_{CI_{\upalpha}}+I_{\upalpha}}C}_{\text{elimination by $I_{\upalpha}$}} + \underbrace{\lambda_{CI_{\upgamma}}\frac{I_{\upgamma}}{K_{CI_{\upgamma}}+I_{\upgamma}}C}_{\text{elimination by $I_{\upgamma}$}} -\underbrace{d_{N_c}N_c}_{\text{removal}}.
    \end{split}\label{necroticcelleqn} \\
\begin{split}
\frac{dV_\mathrm{TS}}{dt} &= \frac{1}{f_{C}+f_{N_c}}\left[\lambda_{C}f_{C}V_\mathrm{TS}\left(1-\frac{f_C V_\mathrm{TS}}{C_0}\right) - \lambda_{CT_8}T_8 \frac{1}{1+I_{\upbeta}/K_{CI_{\upbeta}}}\frac{1}{1+Q^{T_8}/K_{CQ^{T_8}}}f_C V_\mathrm{TS} \right. \\
&\left.- \lambda_{CK}K \frac{1}{1+I_\upbeta/K_{CI_\upbeta}}\frac{1}{1+Q^K/K_{CQ^K}}f_C V_\mathrm{TS} -d_{N_c}f_{N_c}V_\mathrm{TS}\right],
\end{split} \label{Veqn}
\end{align}
where $C(t) = f_C V_\mathrm{TS}(t)$ and $N_c(t) = f_{N_c}V_\mathrm{TS}(t)$.
\begin{align}
 \frac{dH}{dt} &= \underbrace{\lambda_{HN_c} N_c}_{\text{production by $N_c$}} - \underbrace{d_{H} H}_{\text{degradation}}, \label{Heqn} \\
\frac{dS}{dt} &= \underbrace{\lambda_{SN_c} N_c}_{\text{production by $N_c$}}- \underbrace{d_{S} S}_{\text{degradation}}, \label{Seqn} \\
\begin{split}
    \frac{dD_0}{dt} &= \underbrace{\mathcal{A}_{D_0}}_{\text{source}} - \underbrace{\lambda_{DH}D_0\frac{H}{K_{DH}+H}}_{\text{$D_0 \to D$ by $H$}} - \underbrace{\lambda_{DS}D_0\frac{S}{K_{DS}+S}}_{\text{$D_0 \to D$ by $S$}}-\underbrace{\lambda_{D_0K} D_0K\frac{1}{1+I_\upbeta/K_{D_0I_\upbeta}}}_{\substack{\text{elimination by $K$} \\ \text{inhibited by $I_\upbeta$}}} - \underbrace{d_{D_0}D_0}_{\text{death}},
    \end{split} \label{D0eqn} \\
  \frac{dD}{dt} &= \underbrace{\lambda_{DH}D_0\frac{H}{K_{DH}+H}}_{\text{$D_0 \to D$ by $H$}} +   \underbrace{\lambda_{DS}D_0\frac{S}{K_{DS}+S}}_{\text{$D_0 \to D$ by $S$}} - \underbrace{\lambda_{DD^\mathrm{LN}}D}_{\substack{\text{$D$ migration} \\ \text{to TDLN}}} - \underbrace{d_{D}D}_{\text{death}}, \label{Deqn} \\
\frac{dD^\mathrm{LN}}{dt} &= \frac{V_\mathrm{TS}}{V_\mathrm{LN}}\underbrace{\lambda_{DD^\mathrm{LN}}\exp\left(-d_D \tau_m\right)D(t-\tau_m)}_{\text{$D$ migration to TDLN}} - \underbrace{d_{D}D^\mathrm{LN}}_{\text{death}}. \label{DLNeqn}
\end{align}
\paragraph{Equations for T Cells}
\begin{align}
  \frac{dT_0^8}{dt}&=\underbrace{\mathcal{A}_{T_0^8}}_{\text{source}} -\underbrace{R^8(t)}_{\substack{\text{CD8+ T cell} \\ \text{activation}}} - \underbrace{d_{T_0^8}T_0^8}_{\text{death}}, \label{naivecd8eqn}
\intertext{where $R^8(t)$ is defined as}
    R^8(t) &:= \underbrace{\frac{\lambda_{T_0^8 T_A^8}\exp\left(-d_{T_0^8}\tau_8^\mathrm{act}\right)D^\mathrm{LN}(t-\tau_8^\mathrm{act})T_0^8(t-\tau_8^\mathrm{act})}{\left(1+\int_{t-\tau_8^\mathrm{act}}^{t} T_A^{r}(s) \ ds/K_{T_0^8T_A^r}\right)\left(1+\int_{t-\tau_8^\mathrm{act}}^{t} Q^{8\mathrm{LN}}(s) \ ds/K_{T_0^8Q^{8\mathrm{LN}}}\right)}}_{\text{CD8+ T cell activation inhibited by $T_A^r$ and $Q^{8\mathrm{LN}}$}}. \label{R8eqn}\\
    \frac{dT_A^8}{dt} &= \underbrace{\frac{2^{n^8_\mathrm{max}}\exp\left(-d_{T_0^8}\tau_{T_A^8}\right)R^8(t- \tau_{T_A^8})}{\left(1+\int_{t- \tau_{T_A^8}}^{t} T_A^{r}(s) \ ds/K_{T_A^8 T_A^{r}}\right)\left(1+\int_{t- \tau_{T_A^8}}^{t} Q^{8\mathrm{LN}}(s) \ ds/K_{T_A^8 Q^{8\mathrm{LN}}}\right)}}_{\text{CD8+ T cell proliferation inhibited by $T_A^r$ and $Q^{8\mathrm{LN}}$}} - \underbrace{\lambda_{T_A^8T_8}T_A^8}_{\substack{\text{$T_A^8$ migration} \\ \text{to the TS}}}-\underbrace{d_{T_8} T_A^8}_\text{death}, \label{TA8n8maxeqn}
\intertext{where $\tau_{T_A^8}$ is defined as}
    \tau_{T_A^8} &:=\Delta_8^0 + (n^8_\mathrm{max}-1)\Delta_8.\\
\begin{split}
    \frac{d T_8}{dt} &= \frac{V_\mathrm{LN}}{V_\mathrm{TS}}\underbrace{\lambda_{T_A^8T_8}\exp\left(-d_{T_8} \tau_a\right)T_A^8(t - \tau_a)}_{\text{$T_A^8$ migration to the TS}} + \underbrace{\lambda_{T_8 I_{2}}\frac{T_8 I_{2}}{K_{T_8 I_{2}}+ I_{2}}\frac{1}{1+T_r/K_{T_8T_r}}}_{\text{growth by $I_2$ inhibited by $T_r$}} \\
    &- \underbrace{\lambda_{T_8C}\frac{T_8\int_{t-\tau_{l}}^t C(s) \ ds}{K_{T_8C}+\int_{t-\tau_{l}}^t C(s) \ ds}}_{\text{$T_8 \to T_\mathrm{ex}$ from $C$ exposure}} + \underbrace{\lambda_{T_\mathrm{ex}A_1}\frac{T_\mathrm{ex}A_1}{K_{T_\mathrm{ex}A_1} + A_1}}_{\text{$T_\mathrm{ex} \to T_8$ by $A_1$}} - \underbrace{\frac{d_{T_8} T_8}{1+I_{10}/K_{T_8I_{10}}}}_{\substack{\text{death} \\ \text{inhibited by $I_{10}$}}},
    \end{split}\label{t8eqn}\\
    \frac{dT_\mathrm{ex}}{dt} &= \underbrace{\lambda_{T_8C}\frac{T_8\int_{t-\tau_{l}}^t C(s) \ ds}{K_{T_8C}+\int_{t-\tau_{l}}^t C(s) \ ds}}_{\text{$T_8 \to T_\mathrm{ex}$ from $C$ exposure}} - \underbrace{\lambda_{T_\mathrm{ex}A_1}\frac{T_\mathrm{ex}A_1}{K_{T_\mathrm{ex}A_1} + A_1}}_{\text{$T_\mathrm{ex} \to T_8$ by $A_1$}} - \underbrace{\frac{d_{T_\mathrm{ex}} T_\mathrm{ex}}{1+I_{10}/K_{T_\mathrm{ex}I_{10}}}}_{\substack{\text{death} \\ \text{inhibited by $I_{10}$}}}. \label{Texeqn}\\
\frac{dT_0^4}{dt} &=\underbrace{\mathcal{A}_{T_0^4}}_{\text{source}} - \underbrace{R^1(t)}_{\text{Th1 cell activation}} - \underbrace{d_{T_0^4}T_0^4}_{\text{death}}, \label{naivecd4eqn}
\intertext{where $R^1(t)$ is defined as}
  R^1(t) &:= \underbrace{\frac{\lambda_{T_0^4 T_A^1}\exp\left(-d_{T_0^4}\tau_4^\mathrm{act}\right) D^\mathrm{LN}(t-\tau_4^\mathrm{act})T_0^4(t-\tau_4^\mathrm{act})}{\left(1+\int_{t-\tau_4^\mathrm{act}}^{t} T_A^r(s) \ ds/K_{T_0^4 T_A^r}\right)\left(1+\int_{t-\tau_4^\mathrm{act}}^{t} Q^{1\mathrm{LN}}(s) \ ds/K_{T_0^4 Q^{1\mathrm{LN}}}\right)}}_{\text{Th1 cell activation inhibited by $T_A^r$ and $Q^{1\mathrm{LN}}$}}. \label{R1eqn}\\
    \frac{dT_A^1}{dt} &= \underbrace{\frac{2^{n^1_\mathrm{max}}\exp\left(-d_{T_0^4}\tau_{T_A^1}\right) R^1(t- \tau_{T_A^1})}{\left(1+\int_{t- \tau_{T_A^1}}^{t} Q^{1\mathrm{LN}}(s) \ ds/K_{T_A^1 Q^{1\mathrm{LN}}}\right)\left(1+\int_{t- \tau_{T_A^1}}^{t} T_A^r(s) \ ds/K_{T_A^1 T_A^r}\right)}}_{\text{Th1 cell proliferation inhibited by $T_A^r$ and $Q^{1\mathrm{LN}}$}} - \underbrace{\lambda_{T_A^1T_1}T_A^1}_{\substack{\text{$T_A^1$ migration} \\ \text{to the TS}}}-\underbrace{d_{T_1} T_A^1}_\text{death}, \label{TA1n1maxeqn}
\intertext{where $\tau_{T_A^1}$ is defined as}
\tau_{T_A^1} &:=\Delta_1^0 + (n^1_\mathrm{max}-1)\Delta_1.\\
\begin{split}
\frac{dT_1}{dt} &= \frac{V_\mathrm{LN}}{V_\mathrm{TS}}\underbrace{\lambda_{T_A^1T_1}\exp\left(-d_{T_1} \tau_a\right)T_A^1(t-\tau_a)}_{\text{$T_A^1$ migration to the TS}} +\underbrace{\lambda_{T_1 I_{2}}\frac{T_1 I_{2}}{K_{T_1 I_{2}}+ I_{2}}\frac{1}{1+T_r/K_{T_1T_r}}}_{\text{growth by $I_2$ inhibited by $T_r$}} - \underbrace{\lambda_{T_1T_r}T_1\frac{Q^{T_1}}{K_{T_1Q^{T_1}} + Q^{T_1}}}_{\text{$T_1 \to T_r$ by $Q^{T_1}$}} \\
&- \underbrace{d_{T_1}T_1}_{\text{death}} \label{th1eqn},
\end{split}\\
  \frac{dT_0^r}{dt} &=\underbrace{\mathcal{A}_{T_0^r}}_{\text{source}} - \underbrace{R^r(t)}_{\text{Treg activation}} - \underbrace{d_{T_0^r}T_0^r}_{\text{death}}, \label{naivetregeqn}
\intertext{where $R^r(t)$ is defined as}
R^r(t) &:= \underbrace{\lambda_{T_0^r T_A^r}\exp\left(-d_{T_0^r}\tau_r^\mathrm{act}\right) D^\mathrm{LN}(t-\tau_r^\mathrm{act})T_0^r(t-\tau_r^\mathrm{act})}_{\text{Treg activation}}. \label{Rreqn}\\
\frac{dT_A^r}{dt} &= \underbrace{2^{n^r_\mathrm{max}}\exp\left(-d_{T_0^r}\tau_{T_A^r}\right) R^r(t- \tau_{T_A^r})}_{\text{Treg proliferation}} - \underbrace{\lambda_{T_A^rT_r}T_A^r}_{\substack{\text{$T_A^r$ migration} \\ \text{to the TS}}}-\underbrace{d_{T_r} T_A^r}_\text{death}, \label{TArnrmaxeqn}
\intertext{where $\tau_{T_A^r}$ is defined as}
\tau_{T_A^r} &:=\Delta_r^0 + (n^r_\mathrm{max}-1)\Delta_r. \\
\frac{dT_r}{dt} &= \frac{V_\mathrm{LN}}{V_\mathrm{TS}}\underbrace{\lambda_{T_A^rT_r}\exp\left(-d_{T_r} \tau_a\right)T_A^r(t-\tau_a)}_{\text{$T_A^r$ migration to the TS}} + \underbrace{\lambda_{T_1T_r}T_1\frac{Q^{T_1}}{K_{T_1Q^{T_1}} + Q^{T_1}}}_{\text{$T_1 \to T_r$ by $Q^{T_1}$}} - \underbrace{d_{T_r}T_r}_{\text{death}}. \label{tregeqn}
\end{align}
\paragraph{Equations for Macrophages and NK Cells}
\begin{align}
 \begin{split}
 \frac{dM_0}{dt} &=\underbrace{\mathcal{A}_{M_0}}_{\text{source}} - \underbrace{\lambda_{M_1I_\upalpha}M_0\frac{I_\upalpha}{K_{M_1 I_\upalpha}+I_\upalpha}}_{\text{$M_0 \to M_1$ by $I_{\upalpha}$}} - \underbrace{\lambda_{M_1I_\upgamma}M_0\frac{I_\upgamma}{K_{M_1 I_\upgamma}+I_\upgamma}}_{\text{$M_0 \to M_1$ by $I_\upgamma$}} - \underbrace{\lambda_{M_2I_{10}}M_0\frac{I_{10}}{K_{M_2 I_{10}}+I_{10}}}_{\text{$M_0 \to M_2$ by $I_{10}$}} \\
 &- \underbrace{\lambda_{M_2I_\upbeta}M_0\frac{I_\upbeta}{K_{M_2 I_\upbeta}+I_\upbeta}}_{\text{$M_0 \to M_2$ by $I_\upbeta$}} - \underbrace{d_{M_0}M_0}_{\text{degradation}},
 \end{split} \label{M0eqn} \\
 \begin{split}
 \frac{dM_1}{dt} &= \underbrace{\lambda_{M_1I_\upalpha}M_0\frac{I_\upalpha}{K_{M_1 I_\upalpha}+I_\upalpha}}_{\text{$M_0 \to M_1$ by $I_{\upalpha}$}} + \underbrace{\lambda_{M_1I_\upgamma}M_0\frac{I_\upgamma}{K_{M_1 I_\upgamma}+I_\upgamma}}_{\text{$M_0 \to M_1$ by $I_\upgamma$}} + \underbrace{\lambda_{MI_{\upgamma}}M_2\frac{I_{\upgamma}}{K_{MI_{\upgamma}}+I_{\upgamma}}}_{\text{$M_2 \to M_1$ by $I_{\upgamma}$}} + \underbrace{\lambda_{MI_{\upalpha}}M_2 \frac{I_{\upalpha}}{K_{MI_{\upalpha}}+I_{\upalpha}}}_{\text{$M_2 \to M_1$ by $I_{\upalpha}$}} \\ 
 &- \underbrace{\lambda_{MI_{\upbeta}}M_1 \frac{I_{\upbeta}}{K_{MI_{\upbeta}}+I_{\upbeta}}}_{\text{$M_1 \to M_2$ by $I_{\upbeta}$}} - \underbrace{d_{M_1}M_1}_{\text{degradation}},     
 \end{split} \label{M1eqn} \\
\begin{split}
 \frac{dM_2}{dt} &= \underbrace{\lambda_{M_2I_{10}}M_0\frac{I_{10}}{K_{M_2 I_{10}}+I_{10}}}_{\text{$M_0 \to M_2$ by $I_{10}$}} + \underbrace{\lambda_{M_2I_\upbeta}M_0\frac{I_\upbeta}{K_{M_2 I_\upbeta}+I_\upbeta}}_{\text{$M_0 \to M_2$ by $I_\upbeta$}} - \underbrace{\lambda_{MI_{\upgamma}}M_2\frac{I_{\upgamma}}{K_{MI_{\upgamma}}+I_{\upgamma}}}_{\text{$M_2 \to M_1$ by $I_{\upgamma}$}} - \underbrace{\lambda_{MI_{\upalpha}}M_2 \frac{I_{\upalpha}}{K_{MI_{\upalpha}}+I_{\upalpha}}}_{\text{$M_2 \to M_1$ by $I_{\upalpha}$}} \\
 &+ \underbrace{\lambda_{MI_{\upbeta}}M_1 \frac{I_{\upbeta}}{K_{MI_{\upbeta}}+I_{\upbeta}}}_{\text{$M_1 \to M_2$ by $I_{\upbeta}$}} - \underbrace{d_{M_2}M_2}_{\text{degradation}},     
 \end{split} \label{M2eqn} \\
\frac{dK_0}{dt} &= \underbrace{\mathcal{A}_{K_0}}_{\text{source}} - \left(\underbrace{\lambda_{KI_2}K_0\frac{I_2}{K_{KI_2}+I_2}}_{\text{$K_0 \to K$ by $I_2$}} +\underbrace{\lambda_{KD_{0}}K_0\frac{D_{0}}{K_{KD_0}+D_0}}_{\text{$K_0 \to K$ by $D_0$}}+\underbrace{\lambda_{KD}K_0\frac{D}{K_{KD}+D}}_{\text{$K_0 \to K$ by $D$}}\right)\underbrace{\frac{1}{1+I_\upbeta/K_{KI_\upbeta}}}_{\substack{\text{activation}\\ \text{inhibited by $I_\upbeta$}}} - \underbrace{d_{K_0}K_0}_{\text{degradation}}, \label{K0eqn}\\
\frac{dK}{dt} &= \left(\underbrace{\lambda_{KI_2}K_0\frac{I_2}{K_{KI_2}+I_2}}_{\text{$K_0 \to K$ by $I_2$}} + \underbrace{\lambda_{KD_{0}}K_0\frac{D_{0}}{K_{KD_0}+D_0}}_{\text{$K_0 \to K$ by $D_0$}}+\underbrace{\lambda_{KD}K_0\frac{D}{K_{KD}+D}}_{\text{$K_0 \to K$ by $D$}}\right)\underbrace{\frac{1}{1+I_\upbeta/K_{KI_\upbeta}}}_{\substack{\text{activation}\\ \text{inhibited by $I_\upbeta$}}} - \underbrace{d_{K}K}_{\text{degradation}}. \label{Keqn}
\end{align}
\paragraph{Equations for Cytokines}
\begin{align}
\frac{dI_2}{dt} &= \underbrace{\lambda_{I_2 T_8}T_8}_{\text{production by $T_8$}} + \underbrace{\lambda_{I_2 T_1}T_1}_{\text{production by $T_1$}} - \underbrace{d_{I_2}I_2}_{\text{degradation}}. \label{il2eqn}
\intertext{After applying a quasi-steady-state approximation (QSSA), this becomes}
I_2 &= \frac{1}{d_{I_2}}\left(\lambda_{I_2 T_8}T_8 + \lambda_{I_2 T_1}T_1\right). \label{il2qssa}\\
 \frac{dI_{\upgamma}}{dt} &=\left(\underbrace{\lambda_{I_{\upgamma} T_8}T_8}_{\text{production by $T_8$}} + \underbrace{\lambda_{I_{\upgamma} T_1}T_1}_{\text{production by $T_1$}}\right)\underbrace{\frac{1}{1+T_r/K_{I_{\upgamma}T_r}}}_{\text{inhibition by $T_r$}} + \underbrace{\lambda_{I_{\upgamma} K}K}_{\text{production by $K$}} - \underbrace{d_{I_\upgamma}I_\upgamma}_{\text{degradation}}. \label{ifngammaeqn}
\intertext{After applying a QSSA, this becomes}
I_\upgamma &= \frac{1}{d_{I_\upgamma}}\left[\left(\lambda_{I_{\upgamma} T_8}T_8 + \lambda_{I_{\upgamma} T_1}T_1\right)\frac{1}{1+T_r/K_{I_{\upgamma}T_r}} + \lambda_{I_{\upgamma} K}K\right]. \label{ifngammaqssa}\\
 \frac{dI_\upalpha}{dt} &= \underbrace{\lambda_{I_{\upalpha}T_8}T_8}_{\text{production by $T_8$}} + \underbrace{\lambda_{I_{\upalpha}T_1}T_1}_{\text{production by $T_1$}} + \underbrace{\lambda_{I_{\upalpha}M_1}M_1}_{\text{production by $M_1$}} + \underbrace{\lambda_{I_{\upalpha}K}K}_{\text{production by $K$}}- \underbrace{d_{I_{\upalpha}}I_{\upalpha}}_{\text{degradation}}. \label{tnfeqn}
 \intertext{After applying a QSSA, this becomes}
 I_\upalpha &= \frac{1}{d_{I_{\upalpha}}}\left(\lambda_{I_{\upalpha}T_8}T_8 + \lambda_{I_{\upalpha}T_1}T_1 + \lambda_{I_{\upalpha}M_1}M_1 + \lambda_{I_{\upalpha}K}K\right). \label{tnfqssa}\\
 \frac{dI_\upbeta}{dt}&= \underbrace{\lambda_{I_{\upbeta}C}C}_{\text{production by $C$}} + \underbrace{\lambda_{I_{\upbeta}T_r}T_r}_{\text{production by $T_r$}} + \underbrace{\lambda_{I_{\upbeta}M_2}M_2}_{\text{production by $M_2$}} - \underbrace{d_{I_{\upbeta}}I_{\upbeta}}_{\text{degradation}}. \label{tgfbetaeqn}
\intertext{After applying a QSSA, this becomes}
I_\upbeta &= \frac{1}{d_{I_{\upbeta}}}\left(\lambda_{I_{\upbeta}C}C + \lambda_{I_{\upbeta}T_r}T_r + \lambda_{I_{\upbeta}M_2}M_2\right). \label{ibetaqssa}\\
\frac{dI_{10}}{dt} &= \underbrace{\lambda_{I_{10}C}C}_{\text{production by $C$}} + \underbrace{\lambda_{I_{10}M_2}M_2}_{\text{production by $M_2$}} + \underbrace{\lambda_{I_{10}T_{r}}T_r\left(1+\lambda_{I_{10}I_{2}}\frac{I_{2}}{K_{I_{10}I_{2}}+I_{2}}\right)}_{\text{production by $T_r$ enhanced by $I_2$}} - \underbrace{d_{I_{10}}I_{10}}_{\text{degradation}}. \label{il10eqn}
\end{align}
\paragraph{Equations for Immune Checkpoint-Associated Components in the TS}
\begin{align}
\frac{dP_D^{T_8}}{dt} &= \underbrace{\lambda_{P_D^{T_8}}T_8}_{\text{synthesis}} + \underbrace{\lambda_{Q_A}Q_A^{T_8}}_{\substack{\text{dissociation} \\ \text{of $Q_A^{T_8}$}}} - \underbrace{\lambda_{P_DA_1}P_D^{T_8}A_1}_{\text{binding to $A_1$}} - \underbrace{d_{P_D}P_D^{T_8}}_{\text{degradation}}, \label{PD8eqn} \\
    \frac{dP_D^{T_1}}{dt} &= \underbrace{\lambda_{P_D^{T_1}}T_1}_{\text{synthesis}} + \underbrace{\lambda_{Q_A}Q_A^{T_1}}_{\substack{\text{dissociation} \\ \text{of $Q_A^{T_1}$}}} - \underbrace{\lambda_{P_DA_1}P_D^{T_1}A_1}_{\text{binding to $A_1$}} - \underbrace{d_{P_D}P_D^{T_1}}_{\text{degradation}}, \label{PD1eqn} \\
    \frac{dP_D^{K}}{dt} &= \underbrace{\lambda_{P_D^{K}}K}_{\text{synthesis}} + \underbrace{\lambda_{Q_A}Q_A^{K}}_{\substack{\text{dissociation} \\ \text{of $Q_A^{K}$}}} - \underbrace{\lambda_{P_DA_1}P_D^{K}A_1}_{\text{binding to $A_1$}} - \underbrace{d_{P_D}P_D^{K}}_{\text{degradation}}, \label{PDKeqn} \\
    \frac{dQ_A^{T_8}}{dt} &= \underbrace{\lambda_{P_DA_1}P_D^{T_8}A_1}_{\text{formation of $Q_A^{T_8}$}} - \underbrace{\lambda_{Q_A}Q_A^{T_8}}_{\text{dissociation of $Q_A^{T_8}$}} - \underbrace{d_{Q_A}Q_A^{T_8}}_{\text{internalisation}}, \label{QA8eqn} \\
    \frac{dQ_A^{T_1}}{dt} &= \underbrace{\lambda_{P_DA_1}P_D^{T_1}A_1}_{\text{formation of $Q_A^{T_1}$}} - \underbrace{\lambda_{Q_A}Q_A^{T_1}}_{\text{dissociation of $Q_A^{T_1}$}} - \underbrace{d_{Q_A}Q_A^{T_1}}_{\text{internalisation}}, \label{QA1eqn} \\
    \frac{dQ_A^{K}}{dt} &= \underbrace{\lambda_{P_DA_1}P_D^{K}A_1}_{\text{formation of $Q_A^{K}$}} - \underbrace{\lambda_{Q_A}Q_A^{K}}_{\text{dissociation of $Q_A^{K}$}} - \underbrace{d_{Q_A}Q_A^{K}}_{\text{internalisation}}, \label{QAKeqn} \\
    \frac{dA_{1}}{dt} &=\underbrace{\sum_{j=1}^{n} \xi_{j}f_\mathrm{pembro}\delta\left(t-t_j \right)}_{\text{infusion}} + \underbrace{\lambda_{Q_A}\left(Q_A^{T_8}+ Q_A^{T_1} + Q_A^{K}\right)}_{\text{dissociation of $Q_A^{T_8}$, $Q_A^{T_1}$, and $Q_A^{K}$}} - \underbrace{\lambda_{P_DA_1}\left(P_D^{T_8} + P_D^{T_1} + P_D^{K}\right)A_1}_{\text{formation of $Q_A^{T_8}$, $Q_A^{T_1}$, and $Q_A^{K}$}} - \underbrace{d_{A_1}A_{1}}_{\text{elimination}}, \label{A1eqn} \\
    \frac{dP_{L}}{dt} &= \underbrace{\sum_{X \in \mathcal{X}}\lambda_{P_L X}X}_{\text{synthesis}} - \underbrace{d_{P_L}P_L}_{\text{degradation}},\label{PLeqn}
\intertext{where $\mathcal{X}:=\set{C, D, T_8, T_1, T_r, M_2}$.}
Q^{T_8} &= \frac{\lambda_{P_{D}P_{L}}}{\lambda_{Q}} P_{D}^{T_8}P_{L}, \label{Q8eqn} \\
Q^{T_1} &= \frac{\lambda_{P_{D}P_{L}}}{\lambda_{Q}} P_{D}^{T_1}P_{L}, \label{QT1eqn} \\ 
Q^{K} &= \frac{\lambda_{P_{D}P_{L}}}{\lambda_{Q}} P_{D}^{K}P_{L}. \label{QKeqn}
\end{align}
\paragraph{Equations for Immune Checkpoint-Associated Components in the TDLN}
\begin{align}
\frac{dP_D^{8\mathrm{LN}}}{dt} &= \underbrace{\lambda_{P_D^{8\mathrm{LN}}}T_A^8}_{\text{synthesis}} + \underbrace{\lambda_{Q_A}Q_A^{8\mathrm{LN}}}_{\substack{\text{dissociation} \\ \text{of $Q_A^{8\mathrm{LN}}$}}} - \underbrace{\lambda_{P_DA_1}P_D^{8\mathrm{LN}}A_1^\mathrm{LN}}_{\text{binding to $A_1^\mathrm{LN}$}} - \underbrace{d_{P_D}P_D^{8\mathrm{LN}}}_{\text{degradation}}, \label{PD8LNeqn} \\
    \frac{dP_D^{1\mathrm{LN}}}{dt} &= \underbrace{\lambda_{P_D^{1\mathrm{LN}}}T_A^1}_{\text{synthesis}} + \underbrace{\lambda_{Q_A}Q_A^{1\mathrm{LN}}}_{\substack{\text{dissociation} \\ \text{of $Q_A^{1\mathrm{LN}}$}}} - \underbrace{\lambda_{P_DA_1}P_D^{1\mathrm{LN}}A_1^\mathrm{LN}}_{\text{binding to $A_1^\mathrm{LN}$}} - \underbrace{d_{P_D}P_D^{1\mathrm{LN}}}_{\text{degradation}}, \label{PD1LNeqn} \\
    \frac{dQ_A^{8\mathrm{LN}}}{dt} &= \underbrace{\lambda_{P_DA_1}P_D^{8\mathrm{LN}}A_1^\mathrm{LN}}_{\text{formation of $Q_A^{8\mathrm{LN}}$}} - \underbrace{\lambda_{Q_A}Q_A^{8\mathrm{LN}}}_{\text{dissociation of $Q_A^{8\mathrm{LN}}$}} - \underbrace{d_{Q_A}Q_A^{8\mathrm{LN}}}_{\text{internalisation}}, \label{QA8LNeqn} \\
    \frac{dQ_A^{1\mathrm{LN}}}{dt} &= \underbrace{\lambda_{P_DA_1}P_D^{1\mathrm{LN}}A_1^\mathrm{LN}}_{\text{formation of $Q_A^{1\mathrm{LN}}$}} - \underbrace{\lambda_{Q_A}Q_A^{1\mathrm{LN}}}_{\text{dissociation of $Q_A^{1\mathrm{LN}}$}} - \underbrace{d_{Q_A}Q_A^{1\mathrm{LN}}}_{\text{internalisation}}, \label{QA1LNeqn} \\
    \frac{dA_{1}^\mathrm{LN}}{dt} &=\underbrace{\sum_{j=1}^{n} \xi_{j}f_\mathrm{pembro}\delta\left(t-t_j \right)}_{\text{infusion}} + \underbrace{\lambda_{Q_A}\left(Q_A^{8\mathrm{LN}} + Q_A^{1\mathrm{LN}}\right)}_{\text{dissociation of $Q_A^{8\mathrm{LN}}$ and $Q_A^{1\mathrm{LN}}$}} - \underbrace{\lambda_{P_DA_1}\left(P_D^{8\mathrm{LN}} + P_D^{1\mathrm{LN}}\right)A_1^\mathrm{LN}}_{\text{formation of $Q_A^{8\mathrm{LN}}$ and $Q_A^{1\mathrm{LN}}$}} - \underbrace{d_{A_1}A_{1}^\mathrm{LN}}_{\text{elimination}}, \label{A1LNeqn} \\
    \frac{dP_{L}^\mathrm{LN}}{dt} &= \underbrace{\sum_{Y \in \mathcal{Y}} \lambda_{P_L^\mathrm{LN}Y}Y}_{\text{synthesis}} - \underbrace{d_{P_L}P_L^\mathrm{LN}}_{\text{degradation}}, \label{PLLNeqn}
\intertext{where $\mathcal{Y}:=\set{D^\mathrm{LN}, T_A^8, T_A^1, T_A^r}$.}
Q^{8\mathrm{LN}} &= \frac{\lambda_{P_{D}P_{L}}}{\lambda_{Q}} P_{D}^{8\mathrm{LN}}P_{L}^\mathrm{LN}, \label{Q8LNeqn} \\
Q^{1\mathrm{LN}} &= \frac{\lambda_{P_{D}P_{L}}}{\lambda_{Q}} P_{D}^{1\mathrm{LN}}P_{L}^\mathrm{LN}. \label{Q1LNeqn}
\end{align}
\subsubsection{Initial Conditions \label{initcond section}}
For all species in the model, we assume a constant solution history, where the history for each species is set to its respective initial condition. In particular, all initial conditions are estimated as in \citep{Hawi2026meta}, but are repeated here for completeness, using data from a combination of pharmacokinetic, bioanalytical, and radiographic studies, as well as digital cytometry and tumour deconvolution of bulk RNA-sequencing data.

\begin{table}[H]
\centering
\resizebox{\columnwidth}{!}{%
\begin{tabular}{|c|c|c|c|c|c|}
\hline
\textbf{Variable} & \textbf{Initial Condition} & \textbf{Unit} & \textbf{Variable} & \textbf{Initial Condition} & \textbf{Unit} \\
\hline
$C$ & $3.90 \times 10^{7}$ & $\mathrm{cell/cm^3}$ & $N_c$ & $2.05 \times 10^{6}$ & $\mathrm{cell/cm^3}$ \\
$D_0$ & $1.04 \times 10^{6}$ & $\mathrm{cell/cm^3}$ & $D$ & $1.04 \times 10^{6}$ & $\mathrm{cell/cm^3}$ \\
$D^\mathrm{LN}$ & $1.78 \times 10^{7}$ & $\mathrm{cell/cm^3}$ & $T_0^8$ & $1.20 \times 10^{7}$ & $\mathrm{cell/cm^3}$ \\
$T_A^8$ & $8.60 \times 10^{5}$ & $\mathrm{cell/cm^3}$ & $T_8$ & $1.61 \times 10^{5}$ & $\mathrm{cell/cm^3}$ \\
$T_{\mathrm{ex}}$ & $1.93 \times 10^{5}$ & $\mathrm{cell/cm^3}$ & $T_0^4$ & $4.31 \times 10^{6}$ & $\mathrm{cell/cm^3}$ \\
$T_A^1$ & $7.76 \times 10^{6}$ & $\mathrm{cell/cm^3}$ & $T_1$ & $1.01 \times 10^{5}$ & $\mathrm{cell/cm^3}$ \\
$T_0^r$ & $1.72 \times 10^{5}$ & $\mathrm{cell/cm^3}$ & $T_A^r$ & $7.81 \times 10^{5}$ & $\mathrm{cell/cm^3}$ \\
$T_r$ & $2.02 \times 10^{5}$ & $\mathrm{cell/cm^3}$ & $M_0$ & $2.50 \times 10^{5}$ & $\mathrm{cell/cm^3}$ \\
$M_1$ & $2.09 \times 10^{5}$ & $\mathrm{cell/cm^3}$ & $M_2$ & $1.29 \times 10^{6}$ & $\mathrm{cell/cm^3}$ \\
$K_0$ & $4.47 \times 10^{6}$ & $\mathrm{cell/cm^3}$ & $K$ & $4.47 \times 10^{5}$ & $\mathrm{cell/cm^3}$ \\
\hline
$H$ & $1.33 \times 10^{-8}$ & $\mathrm{g/cm^3}$ & $S$ & $3.25 \times 10^{-8}$ & $\mathrm{g/cm^3}$ \\
\hline
$I_2$ & $1.87 \times 10^{-12}$ & $\mathrm{g/cm^3}$ & $I_\upgamma$ & $1.10 \times 10^{-10}$ & $\mathrm{g/cm^3}$ \\
$I_\upalpha$ & $1.25 \times 10^{-10}$ & $\mathrm{g/cm^3}$ & $I_\upbeta$ & $9.20 \times 10^{-7}$ & $\mathrm{g/cm^3}$ \\
$I_{10}$ & $1.15 \times 10^{-10}$ & $\mathrm{g/cm^3}$ &  &  &  \\
\hline
$P_D^{T_8}$ & $4.44 \times 10^{8}$ & $\mathrm{molec/cm^3}$ & $P_D^{T_1}$ & $2.07 \times 10^{8}$ & $\mathrm{molec/cm^3}$ \\
$P_D^{K}$ & $2.46 \times 10^{8}$ & $\mathrm{molec/cm^3}$ & $Q_A^{T_8}$ & $0$ & $\mathrm{molec/cm^3}$ \\
$Q_A^{T_1}$ & $0$ & $\mathrm{molec/cm^3}$ & $Q_A^{K}$ & $0$ & $\mathrm{molec/cm^3}$ \\
$P_L$ & $7.40 \times 10^{12}$ & $\mathrm{molec/cm^3}$ & $Q^{T_8}$ & $6.99 \times 10^{5}$ & $\mathrm{molec/cm^3}$ \\
$Q^{T_1}$ & $3.26 \times 10^{5}$ & $\mathrm{molec/cm^3}$ & $Q^{K}$ & $3.88 \times 10^{5}$ & $\mathrm{molec/cm^3}$ \\
$A_{1}$ & $0$ & $\mathrm{molec/cm^3}$ &  &  &  \\
\hline
$P_D^{8\mathrm{LN}}$ & $2.37 \times 10^{9}$ & $\mathrm{molec/cm^3}$ & $P_D^{1\mathrm{LN}}$ & $1.59 \times 10^{10}$ & $\mathrm{molec/cm^3}$ \\
$Q_A^{8\mathrm{LN}}$ & $0$ & $\mathrm{molec/cm^3}$ & $Q_A^{1\mathrm{LN}}$ & $0$ & $\mathrm{molec/cm^3}$ \\
$P_L^\mathrm{LN}$ & $3.34 \times 10^{11}$ & $\mathrm{molec/cm^3}$ & $Q^{8\mathrm{LN}}$ & $1.69 \times 10^{5}$ & $\mathrm{molec/cm^3}$ \\
$Q^{1\mathrm{LN}}$ & $1.13 \times 10^{6}$ & $\mathrm{molec/cm^3}$ & $A_{1}^\mathrm{LN}$ & $0$ & $\mathrm{molec/cm^3}$ \\
\hline
\end{tabular}%
}
\caption{Initial conditions for all model components in the full model, with model variables as in \autoref{modelvars}.}
\end{table}
\subsubsection{Sensitivity Analysis}
\begin{figure}[H]
    \centering
    \includegraphics[width=\textwidth]{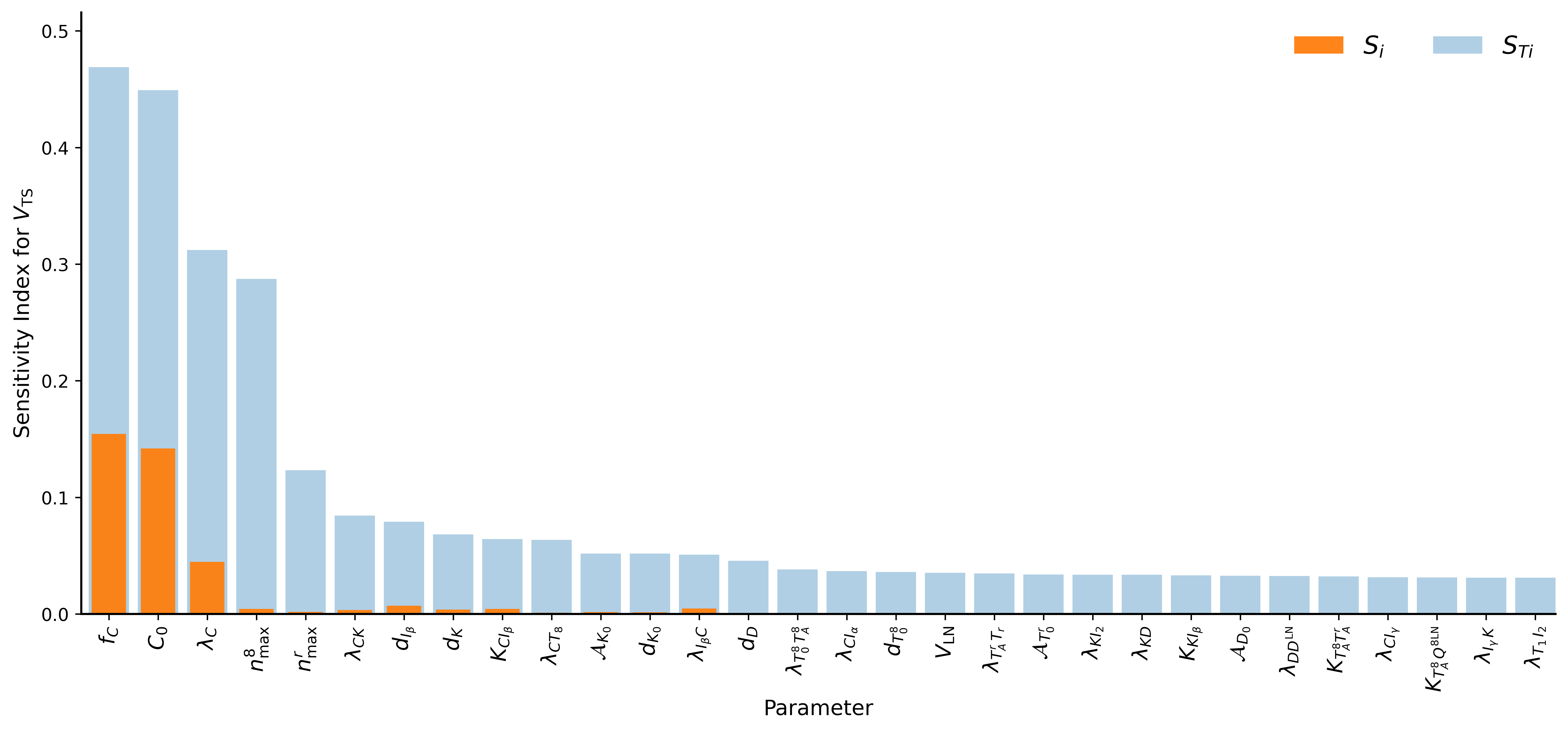}
    \caption{\label{fullmodelsensindexplot}First-order indices ($S_i$ in orange) and total-order indices ($S_{Ti}$ in blue) for the full model for the top 30 parameters associated with the RMSRE of $V_\mathrm{TS}$ at 180.9 days under the standard regimen, sorted in descending order of $S_{Ti}$.}
\end{figure}

We now perform a sensitivity analysis on the full model in order to assess the significance of its parameters and identify avenues for model reduction. We simulated the standard regimen using the initial conditions in \Cref{initcond section} for 180.9 days---the duration of time needed for the cancer concentrations to reach their steady-state value without treatment as explained in \citep{Hawi2026meta}. However, whilst we have gone into detail about variance-based sensitivity analysis methods for a function with single-valued output in \Cref{SAsection}, the full model is multidimensional and time-dependent in its output. As such, we perform a sensitivity analysis on the constant vector defined by the root mean squared relative error (RMSRE) between the trajectories of the full model using parameters from Table A.1, and those generated from a given set of parameter inputs, at 180.9 days. To define the RMSRE, we first denote the trajectory of a state variable, $X$, using parameters from Table A.1 as $X(t)$, and the trajectory using a given set of parameters, $\bm{p}$, as $X_{\bm{p}}(t)$. The RMSRE at time $t$ is then defined by
\begin{equation}
    \operatorname{RMSRE}^X_{\bm{p}}(t) := \sqrt{\frac{1}{t}\int_{0}^{t} \left(\frac{X_{\bm{p}}(s) - X(s)}{X(s)}\right)^2 \ ds} \label{RMSRESAeqn}.
\end{equation}
In particular, the RMSRE represents the overall accuracy of the reduction for a given variable by aggregating relative errors across the entire period of integration and penalising large deviations more heavily. We used the FAST method to compute the first-order sensitivity indices and the EFAST method for the total-order indices, with the model simulations carried out using the dde23 integrator. We then aggregated the output via \eqref{RMSRESAeqn} for each state variable, and a separate sensitivity analysis was performed for each state variable to obtain the corresponding indices. \\~\\
One question that needs to be answered is regarding the number of parameters sampled. Denoting $N$ as the number of samples per parameter, and noting that the full model contains $157$ independent parameters, this corresponds to $157N$ combinations of parameters to be evaluated as part of the sensitivity analysis. Whilst in an ideal world one could have an arbitrarily large sample size, limits on the memory available to computers, as well as the time available for the computation to execute, restrict $N$. As such, we chose $N = 10,000$, corresponding to $1,570,000$ parameter combinations in total, observing that the sensitivity indices showed convergence as the sample size approached this value, as indicated by minimal deviations in the indices. Similarly, choosing the value of $M$ is a compromise between accuracy and computational expense. We chose $M=4$ as this is the value that is usually used and is less susceptible to numerical error in the characteristic amplitudes, which tend to be larger for higher values of $M$ \citep{Xu2008}. Additionally, we considered parameter ranges corresponding to a $\pm 50\%$ deviation from the nominal calibrated values of each parameter, noting that $n^8_\mathrm{max}$, $n^1_\mathrm{max}$, and $n^r_\mathrm{max}$ are integers and were therefore rounded to the nearest integer. While biological parameters can globally vary over several orders of magnitude, this linear $\pm 50\%$ range was chosen to assess the robustness of the dynamics in the neighbourhood of the calibrated parameter set---a standard methodology for quantifying parameter influence and verifying model stability under significant perturbations without entering biologically implausible regimes where the model assumptions may no longer hold.\\~\\
The SALib Python library \citep{Iwanaga2022} was used to perform FAST and EFAST analysis on the full model with the aforementioned configuration, leading to the indices in Table F.1. The sensitivity indices for the full model for the top 30 parameters associated with the RMSRE of $V_\mathrm{TS}$ at 180.9 days under the standard regimen, sorted in descending order of the total-order indices, are shown in \autoref{fullmodelsensindexplot}.
We observe from the results in Table F.1 and \autoref{fullmodelsensindexplot} that the most significant parameters underpinning tumour growth over time, and by extension the viable and necrotic cancer cell density over time, are $f_C$, $\lambda_C$, $\lambda_{CT_8}$, $\lambda_{CK}$, $C_0$, $K_{CI_\upbeta}$, $n^8_\mathrm{max}$, and $n^r_\mathrm{max}$. We must also note that cancer cell necrosis via TNF is significant; however, only appears in the equations for $C$ and $N_c$, and does not appear in the equation for $V_\mathrm{TS}$. We note that the parameters $\lambda_{CQ^{T_8}}$ and $\lambda_{CQ^{K}}$ are also important; however, their exact values are less critical, as the concentration of the PD-1/PD-L1 complex on effector CD8+ T cells and NK cells will be orders of magnitude lower under treatment than without treatment. Consequently, a $\pm 50\%$ deviation in these parameters does not lead to major deviations in model trajectories, which is consistent with their low first-order and total-order sensitivity indices for $V_\mathrm{TS}$.
\subsection{Reduced Model\label{reducedmodelsec}}
We now aim to use the sensitivity analysis results of the full model found in Table F.1 to remove as many parameters as possible while still replicating the original trajectories and dynamics for each state variable. Each parameter in the model corresponds to a biochemical process; however, some processes are more influential on the model than others. In particular, by considering parameters with low maximal first-order and total-order sensitivity indices, we can assess whether the corresponding process has a sufficiently low impact. If so, we can simplify the model by eliminating the process---in most cases by setting the associated parameter to zero \citep{Snowden2017}. Indeed, we performed this for all insensitive parameters, with the exception of insensitive half-saturation and inhibition constants, which were treated separately. It should be noted that setting the associated term to population-average constants would lead to a decrease in mechanistic interpretability, which is an important property of any desired reduction.\\~\\
To rigorise this notion, we used the sensitivity indices in Table F.1 to define an explicit screening criterion for process elimination. We let $S_i^{(X)}$ and $S_{Ti}^{(X)}$ be the first-order and total-order indices of a parameter $p$ for the RMSRE of a state variable $X$ at $t=180.9$ days, as defined in \eqref{RMSRESAeqn}. We classified a parameter, and thus the corresponding process, as potentially insensitive if 
\begin{equation}
\max_{X\in\mathcal{X}_\mathrm{full}} S_i^{(X)} \leq 2.77\times 10^{-3}\text{,}\quad S_{i}^{(V_\mathrm{TS})} \leq 5.9\times 10^{-5},\quad\text{and}\quad
S_{Ti}^{(V_\mathrm{TS})} \leq 3.16\times 10^{-2}, \label{fulltoreducedcriterion}
\end{equation}
where $\mathcal{X}_\mathrm{full}$ is the set of all model state variables in \autoref{modelvars}. In particular, the first part of the condition enforces that the main effect of $p$ is uniformly negligible across all outputs. The conditions in \eqref{fulltoreducedcriterion} were used as a screening criterion to define a set of eligible candidate parameters/processes for removal. We selected the numerical thresholds in \eqref{fulltoreducedcriterion} empirically by performing a grid search over candidate values, and then, among the eligible candidates, removed as many processes as possible while keeping deviations in the trajectories of all model variables sufficiently small under the standard regimen.\\~\\
Candidate insensitive processes were then removed, noting that insensitive half-saturation and inhibition constants were not eliminated purely on the basis of their own indices; instead, they were removed only when the entire associated process term was removed. In addition, before we proceeded, we made two key observations.
\begin{enumerate}
    \item [1)] Regardless of the set of parameters eliminated, the degradation terms, notated as $d_{X}$ for state variable $X$, must remain. This is obvious from a biological standpoint and is well-supported by Table F.1. This makes sense from a mathematical standpoint as well---since we expect the state variables to be bounded.
    \item [2)] These degradation terms cannot be the only terms present for any differential equation. If this were the case for a particular state variable, then the system would decouple, with that state variable exponentially decaying over time, and approaching $0$ as $t\to \infty$. However, we expect the system to be persistent, with none of the biological state variables $\to 0$, as $t\to \infty$, implying that each differential equation must have both a positive and negative contribution---that is, at least $2$ terms.
\end{enumerate}
The SA revealed that approximately one quarter of parameters/processes in the full model could be appropriately removed without significantly affecting the model trajectories. As a result, the following parameters were set to $0$: $\lambda_{CI_{\upgamma}}$, $\lambda_{DS}$, $\lambda_{T_8 I_2}$, $\lambda_{T_1 I_2}$, $\lambda_{MI_{\upgamma}}$, $\lambda_{MI_{\upalpha}}$, $\lambda_{MI_{\upbeta}}$, $\lambda_{KD_0}$, $\lambda_{I_{\upgamma} T_8}$, $\lambda_{I_{\upgamma}T_1}$, $\lambda_{I_{10}T_r}$, $\lambda_{I_{10}I_{2}}$, $\lambda_{P_LD}$, $\lambda_{P_LT_8}$, $\lambda_{P_LT_1}$, $\lambda_{P_LT_r}$, $\lambda_{P_L^\mathrm{LN}T_A^8}$, $\lambda_{P_L^\mathrm{LN}T_A^r}$, $\tau_8^\mathrm{act}$, $\tau_a$, $\tau_4^\mathrm{act}$, and $\tau_r^\mathrm{act}$. In particular, this also results in the elimination of $K_{CI_{\upgamma}}$, $K_{DS}$, $K_{T_8I_2}$, $K_{T_1I_2}$, $K_{MI_{\upgamma}}$, $K_{MI_{\upalpha}}$, $K_{MI_{\upbeta}}$, $K_{KD_0}$, $K_{I_{10}I_{2}}$, and $K_{T_8T_r}$. Additionally, we eliminate $\tau_l$ as well as the inhibition constant, $K_{I_\upgamma T_r}$, by removing the entire $\frac{1}{1+T_r/K_{I_\upgamma T_r}}$ term from \eqref{ifngammaqssa}.\\~\\
The elimination of $\tau_l$ and the T cell activation times requires special care, as the associated integrals will degenerate, and the corresponding half-saturation and inhibition constants must be redefined. Removing these times is equivalent to considering the corresponding processes as being instantaneous; as such, we replace all affected integrals with point estimate values evaluated at the upper bound of integration. As such, the following replacements are made:
\begin{align*}
    \frac{1}{\tau_l}\int_{t-\tau_l}^t C(s) \ ds &\mapsto C(t) \text{ in \eqref{t8eqn} and \eqref{Texeqn},} \\
    \frac{1}{\tau_8^\mathrm{act}}\int_{t-\tau_8^\mathrm{act}}^{t} T_A^{r}(s) \ ds & \mapsto T_A^r(t) \text{ in \eqref{R8eqn},} \\
    \frac{1}{\tau_8^\mathrm{act}}\int_{t-\tau_8^\mathrm{act}}^{t} Q^{8\mathrm{LN}}(s) \ ds & \mapsto Q^{8\mathrm{LN}}(t) \text{ in \eqref{R8eqn},} \\
    \frac{1}{\tau_4^\mathrm{act}}\int_{t-\tau_4^\mathrm{act}}^{t} T_A^{r}(s) \ ds & \mapsto T_A^r(t) \text{ in \eqref{R1eqn},} \\
    \frac{1}{\tau_4^\mathrm{act}}\int_{t-\tau_4^\mathrm{act}}^{t} Q^{1\mathrm{LN}}(s) \ ds & \mapsto Q^{1\mathrm{LN}}(t) \text{ in \eqref{R1eqn}.}
\intertext{We also perform a similar simplification for the T cell proliferation integrals, noting that since $\tau_{T_A^8}$ and $\tau_{T_A^1}$ have not been eliminated, we preserve mechanistic consistency and, by causality, replace all affected integrals with point estimate values evaluated at the lower bound of integration. As such,}
 \frac{1}{\tau_{T_A^8}}\int_{t- \tau_{T_A^8}}^{t} T_A^{r}(s) \ ds & \mapsto T_A^{r}(t-\tau_{T_A^8}) \text{ in \eqref{TA8n8maxeqn}, } \\
   \frac{1}{\tau_{T_A^8}}\int_{t- \tau_{T_A^8}}^{t} Q^{8\mathrm{LN}}(s) \ ds & \mapsto Q^{8\mathrm{LN}}(t-\tau_{T_A^8}) \text{ in \eqref{TA8n8maxeqn}, } \\
   \frac{1}{\tau_{T_A^1}}\int_{t- \tau_{T_A^1}}^{t} T_A^{r}(s) \ ds & \mapsto T_A^{r}(t-\tau_{T_A^1}) \text{ in \eqref{TA1n1maxeqn}, } \\
   \frac{1}{\tau_{T_A^1}}\int_{t- \tau_{T_A^1}}^{t} Q^{1\mathrm{LN}}(s) \ ds & \mapsto Q^{1\mathrm{LN}}(t-\tau_{T_A^1}) \text{ in \eqref{TA1n1maxeqn}.}
\end{align*}
It is worth noting that, by performing the model reduction in this manner, the biological interpretation of the individual parameters is preserved, as the approach does not rely on any transformations.
\subsubsection{Model Equations}
The model variables for the reduced models are the same as those in \autoref{modelvars}. 
\begin{align}
  \begin{split}
  \frac{dC}{dt} &= \underbrace{\lambda_{C}C\left(1-\frac{C}{C_0}\right)}_{\text{growth}} - \underbrace{\lambda_{CT_8}T_8 \frac{1}{1+I_{\upbeta}/K_{CI_{\upbeta}}}\frac{1}{1+Q^{T_8}/K_{CQ^{T_8}}}C}_{\substack{\text{elimination by $T_8$} \\ \text{inhibited by $I_{\upbeta}$ and $Q^{T_8}$}}} - \underbrace{\lambda_{CK}K \frac{1}{1+I_{\upbeta}/K_{CI_{\upbeta}}}\frac{1}{1+Q^K/K_{CQ^K}}C}_{\substack{\text{elimination by $K$} \\ \text{inhibited by $I_\upbeta$ and $Q^K$}}} \\
  &- \underbrace{\lambda_{CI_{\upalpha}}\frac{I_{\upalpha}}{K_{CI_{\upalpha}}+I_{\upalpha}}C}_{\text{elimination by $I_{\upalpha}$}},
  \end{split} \label{reducedcancereqn} \\
  \begin{split}
    \frac{dN_c}{dt}&= \underbrace{\lambda_{CI_{\upalpha}}\frac{I_{\upalpha}}{K_{CI_{\upalpha}}+I_{\upalpha}}C}_{\text{elimination by $I_{\upalpha}$}} -\underbrace{d_{N_c}N_c}_{\text{removal}},
    \end{split}\label{reducednecroticcelleqn} \\
  \begin{split}
  \frac{dV_\mathrm{TS}}{dt} &= \frac{1}{f_{C}+f_{N_c}}\left[\lambda_{C}f_{C}V_\mathrm{TS}\left(1-\frac{f_C V_\mathrm{TS}}{C_0}\right) - \lambda_{CT_8}T_8 \frac{1}{1+I_{\upbeta}/K_{CI_{\upbeta}}}\frac{1}{1+Q^{T_8}/K_{CQ^{T_8}}}f_C V_\mathrm{TS} \right. \\
  &\left.- \lambda_{CK}K \frac{1}{1+I_\upbeta/K_{CI_\upbeta}}\frac{1}{1+Q^K/K_{CQ^K}}f_C V_\mathrm{TS} -d_{N_c}f_{N_c}V_\mathrm{TS}\right],
  \end{split} \label{reducedVeqn}
  \intertext{where $C(t)=f_C V_\mathrm{TS}(t)$ and $N_c(t)=f_{N_c}V_\mathrm{TS}(t)$.}
  \frac{dH}{dt} &= \underbrace{\lambda_{HN_c} N_c}_{\text{production by $N_c$}} - \underbrace{d_{H} H}_{\text{degradation}}, \label{reducedHeqn}\\
  \frac{dS}{dt} &= \underbrace{\lambda_{SN_c} N_c}_{\text{production by $N_c$}}- \underbrace{d_{S} S}_{\text{degradation}}, \label{reducedSeqn} \\
  \begin{split}
  \frac{dD_0}{dt} &= \underbrace{\mathcal{A}_{D_0}}_{\text{source}} - \underbrace{\lambda_{DH}D_0\frac{H}{K_{DH}+H}}_{\text{$D_0 \to D$ by $H$}} -\underbrace{\lambda_{D_0K} D_0K\frac{1}{1+I_\upbeta/K_{D_0I_\upbeta}}}_{\substack{\text{elimination by $K$} \\ \text{inhibited by $I_\upbeta$}}} - \underbrace{d_{D_0}D_0}_{\text{death}},
  \end{split} \label{reducedD0eqn} \\
  \frac{dD}{dt} &= \underbrace{\lambda_{DH}D_0\frac{H}{K_{DH}+H}}_{\text{$D_0 \to D$ by $H$}} - \underbrace{\lambda_{DD^\mathrm{LN}}D}_{\substack{\text{$D$ migration} \\ \text{to TDLN}}} - \underbrace{d_{D}D}_{\text{death}}, \label{reducedDeqn} \\
  \frac{dD^\mathrm{LN}}{dt} &= \frac{V_\mathrm{TS}}{V_\mathrm{LN}}\underbrace{\lambda_{DD^\mathrm{LN}}\exp\left(-d_D \tau_m\right)D(t-\tau_m)}_{\text{$D$ migration to TDLN}} - \underbrace{d_{D}D^\mathrm{LN}}_{\text{death}}, \label{reducedDLNeqn} \\
  \frac{dT_0^8}{dt} &= \underbrace{\mathcal{A}_{T_0^8}}_{\text{source}} -\underbrace{R^8(t)}_{\substack{\text{CD8+ T cell} \\ \text{activation}}} - \underbrace{d_{T_0^8}T_0^8}_{\text{death}}, \label{reducednaivecd8eqn}
  \intertext{where $R^8(t)$ is defined as}
  R^8(t) &:= \underbrace{\frac{\lambda_{T_0^8 T_A^8}D^\mathrm{LN}(t)T_0^8(t)}{\left(1+T_A^r(t)/K_{T_0^8T_A^r}\right)\left(1+Q^{8\mathrm{LN}}(t)/K_{T_0^8Q^{8\mathrm{LN}}}\right)}}_{\text{CD8+ T cell activation inhibited by $T_A^r$ and $Q^{8\mathrm{LN}}$}}. \label{reducedR8eqn} \\
  \begin{split}
  \frac{dT_A^8}{dt} &= \underbrace{\frac{2^{n^8_\mathrm{max}}\exp\left(-d_{T_0^8}\tau_{T_A^8}\right)R^8(t- \tau_{T_A^8})}{\left(1+T_A^{r}(t- \tau_{T_A^8})/K_{T_A^8 T_A^{r}}\right)\left(1+ Q^{8\mathrm{LN}}(t- \tau_{T_A^8})/K_{T_A^8 Q^{8\mathrm{LN}}}\right)}}_{\text{CD8+ T cell proliferation inhibited by $T_A^r$ and $Q^{8\mathrm{LN}}$}} - \underbrace{\lambda_{T_A^8T_8}T_A^8}_{\substack{\text{$T_A^8$ migration} \\ \text{to the TS}}}-\underbrace{d_{T_8} T_A^8}_\text{death},
  \end{split} \label{reducedTA8n8maxeqn} 
  \intertext{where $\tau_{T_A^8}$ is defined as}
  \tau_{T_A^8} &:=\Delta_8^0 + (n^8_\mathrm{max}-1)\Delta_8.\\
  \frac{d T_8}{dt} &= \frac{V_\mathrm{LN}}{V_\mathrm{TS}}\underbrace{\lambda_{T_A^8T_8}T_A^8}_{\substack{\text{$T_A^8$ migration} \\ \text{to the TS}}} - \underbrace{\lambda_{T_8C}\frac{T_8 C}{K_{T_8C}+ C}}_{\text{$T_8 \to T_\mathrm{ex}$ from $C$ exposure}} + \underbrace{\lambda_{T_\mathrm{ex}A_1}\frac{T_\mathrm{ex}A_1}{K_{T_\mathrm{ex}A_1} + A_1}}_{\text{$T_\mathrm{ex} \to T_8$ by $A_1$}} - \underbrace{\frac{d_{T_8} T_8}{1+I_{10}/K_{T_8I_{10}}}}_{\substack{\text{death} \\ \text{inhibited by $I_{10}$}}}, \label{reducedt8eqn}\\
  \frac{dT_\mathrm{ex}}{dt} &= \underbrace{\lambda_{T_8C}\frac{T_8 C}{K_{T_8C}+ C}}_{\text{$T_8 \to T_\mathrm{ex}$ from $C$ exposure}} - \underbrace{\lambda_{T_\mathrm{ex}A_1}\frac{T_\mathrm{ex}A_1}{K_{T_\mathrm{ex}A_1} + A_1}}_{\text{$T_\mathrm{ex} \to T_8$ by $A_1$}} - \underbrace{\frac{d_{T_\mathrm{ex}} T_\mathrm{ex}}{1+I_{10}/K_{T_\mathrm{ex}I_{10}}}}_{\substack{\text{death} \\ \text{inhibited by $I_{10}$}}}, \label{reducedTexeqn} \\
  \frac{dT_0^4}{dt} &=\underbrace{\mathcal{A}_{T_0^4}}_{\text{source}} - \underbrace{R^1(t)}_{\text{Th1 cell activation}} - \underbrace{d_{T_0^4}T_0^4}_{\text{death}}, \label{reducednaivecd4eqn}
  \intertext{where $R^1(t)$ is defined as}
  R^1(t) &:= \underbrace{\frac{\lambda_{T_0^4 T_A^1}D^\mathrm{LN}(t)T_0^4(t)}{\left(1+T_A^r(t)/K_{T_0^4 T_A^r}\right)\left(1+Q^{1\mathrm{LN}}(t)/K_{T_0^4 Q^{1\mathrm{LN}}}\right)}}_{\text{Th1 cell activation inhibited by $T_A^r$ and $Q^{1\mathrm{LN}}$}}. \\
  \frac{dT_A^1}{dt} &= \underbrace{\frac{2^{n^1_\mathrm{max}}\exp\left(-d_{T_0^4}\tau_{T_A^1}\right)R^1(t- \tau_{T_A^1})}{\left(1+Q^{1\mathrm{LN}}(t- \tau_{T_A^1})/K_{T_A^1 Q^{1\mathrm{LN}}}\right)\left(1+ T_A^r(t- \tau_{T_A^1})/K_{T_A^1 T_A^r}\right)}}_{\text{Th1 cell proliferation inhibited by $T_A^r$ and $Q^{1\mathrm{LN}}$}} - \underbrace{\lambda_{T_A^1T_1}T_A^1}_{\substack{\text{$T_A^1$ migration} \\ \text{to the TS}}}-\underbrace{d_{T_1} T_A^1}_\text{death} \label{reducedTA1n1maxeqn},
  \intertext{where $\tau_{T_A^1}$ is defined as}
  \tau_{T_A^1} &:=\Delta_1^0 + (n^1_\mathrm{max}-1)\Delta_1.\\
  \frac{dT_1}{dt} &= \frac{V_\mathrm{LN}}{V_\mathrm{TS}}\underbrace{\lambda_{T_A^1T_1}T_A^1}_{\substack{\text{$T_A^1$ migration} \\ \text{to the TS}}} - \underbrace{\lambda_{T_1T_r}T_1\frac{Q^{T_1}}{K_{T_1Q^{T_1}} + Q^{T_1}}}_{\text{$T_1 \to T_r$ by $Q^{T_1}$}} - \underbrace{d_{T_1}T_1}_{\text{death}} \label{reducedth1eqn}, \\
  \frac{dT_0^r}{dt} &=\underbrace{\mathcal{A}_{T_0^r}}_{\text{source}} - \underbrace{R^r(t)}_{\text{Treg activation}} - \underbrace{d_{T_0^r}T_0^r}_{\text{death}}, \label{reducednaivetregeqn}
  \intertext{where $R^r(t)$ is defined as}
  R^r(t) &:= \underbrace{\lambda_{T_0^r T_A^r}D^\mathrm{LN}(t)T_0^r(t)}_{\text{Treg activation}}.\\
  \frac{dT_A^r}{dt} &= \underbrace{2^{n^r_\mathrm{max}}\exp\left(-d_{T_0^r}\tau_{T_A^r}\right)R^r(t- \tau_{T_A^r})}_{\text{Treg proliferation}} - \underbrace{\lambda_{T_A^rT_r}T_A^r}_{\substack{\text{$T_A^r$ migration} \\ \text{to the TS}}}-\underbrace{d_{T_r} T_A^r}_\text{death}, \label{reducedTArnrmaxeqn}
  \intertext{where $\tau_{T_A^r}$ is defined as}
  \tau_{T_A^r} &:=\Delta_r^0 + (n^r_\mathrm{max}-1)\Delta_r. \\
  \frac{dT_r}{dt} &= \frac{V_\mathrm{LN}}{V_\mathrm{TS}}\underbrace{\lambda_{T_A^rT_r}T_A^r}_{\substack{\text{$T_A^r$ migration} \\ \text{to the TS}}} + \underbrace{\lambda_{T_1T_r}T_1\frac{Q^{T_1}}{K_{T_1Q^{T_1}} + Q^{T_1}}}_{\text{$T_1 \to T_r$ by $Q^{T_1}$}} - \underbrace{d_{T_r}T_r}_{\text{death}}, \label{reducedtregeqn} \\
  \begin{split}
  \frac{dM_0}{dt} &=\underbrace{\mathcal{A}_{M_0}}_{\text{source}} - \underbrace{\lambda_{M_1I_\upalpha}M_0\frac{I_\upalpha}{K_{M_1 I_\upalpha}+I_\upalpha}}_{\text{$M_0 \to M_1$ by $I_{\upalpha}$}} - \underbrace{\lambda_{M_1I_\upgamma}M_0\frac{I_\upgamma}{K_{M_1 I_\upgamma}+I_\upgamma}}_{\text{$M_0 \to M_1$ by $I_\upgamma$}} - \underbrace{\lambda_{M_2I_{10}}M_0\frac{I_{10}}{K_{M_2 I_{10}}+I_{10}}}_{\text{$M_0 \to M_2$ by $I_{10}$}} \\
  &- \underbrace{\lambda_{M_2I_\upbeta}M_0\frac{I_\upbeta}{K_{M_2 I_\upbeta}+I_\upbeta}}_{\text{$M_0 \to M_2$ by $I_\upbeta$}} - \underbrace{d_{M_0}M_0}_{\text{degradation}},
  \end{split} \label{reducedM0eqn} \\
  \frac{dM_1}{dt} &= \underbrace{\lambda_{M_1I_\upalpha}M_0\frac{I_\upalpha}{K_{M_1 I_\upalpha}+I_\upalpha}}_{\text{$M_0 \to M_1$ by $I_{\upalpha}$}} + \underbrace{\lambda_{M_1I_\upgamma}M_0\frac{I_\upgamma}{K_{M_1 I_\upgamma}+I_\upgamma}}_{\text{$M_0 \to M_1$ by $I_\upgamma$}} - \underbrace{d_{M_1}M_1}_{\text{degradation}}, \label{reducedM1eqn} \\
  \frac{dM_2}{dt} &= \underbrace{\lambda_{M_2I_{10}}M_0\frac{I_{10}}{K_{M_2 I_{10}}+I_{10}}}_{\text{$M_0 \to M_2$ by $I_{10}$}} + \underbrace{\lambda_{M_2I_\upbeta}M_0\frac{I_\upbeta}{K_{M_2 I_\upbeta}+I_\upbeta}}_{\text{$M_0 \to M_2$ by $I_\upbeta$}} - \underbrace{d_{M_2}M_2}_{\text{degradation}}, \label{reducedM2eqn} \\
  \frac{dK_0}{dt} &= \underbrace{\mathcal{A}_{K_0}}_{\text{source}} - \left(\underbrace{\lambda_{KI_2}K_0\frac{I_2}{K_{KI_2}+I_2}}_{\text{$K_0 \to K$ by $I_2$}} +\underbrace{\lambda_{KD}K_0\frac{D}{K_{KD}+D}}_{\text{$K_0 \to K$ by $D$}}\right)\underbrace{\frac{1}{1+I_\upbeta/K_{KI_\upbeta}}}_{\substack{\text{activation}\\ \text{inhibited by $I_\upbeta$}}} - \underbrace{d_{K_0}K_0}_{\text{degradation}}, \label{reducedK0eqn}\\
  \frac{dK}{dt} &= \left(\underbrace{\lambda_{KI_2}K_0\frac{I_2}{K_{KI_2}+I_2}}_{\text{$K_0 \to K$ by $I_2$}} +\underbrace{\lambda_{KD}K_0\frac{D}{K_{KD}+D}}_{\text{$K_0 \to K$ by $D$}}\right)\underbrace{\frac{1}{1+I_\upbeta/K_{KI_\upbeta}}}_{\substack{\text{activation}\\ \text{inhibited by $I_\upbeta$}}} - \underbrace{d_{K}K}_{\text{degradation}}, \label{reducedKeqn} \\
  \frac{dI_2}{dt} &= \underbrace{\lambda_{I_2 T_8}T_8}_{\text{production by $T_8$}} + \underbrace{\lambda_{I_2 T_1}T_1}_{\text{production by $T_1$}} - \underbrace{d_{I_2}I_2}_{\text{degradation}}. \label{reducedil2eqn}
  \intertext{After applying a QSSA, this becomes}
  I_2 &= \frac{1}{d_{I_2}}\left(\lambda_{I_2 T_8}T_8 + \lambda_{I_2 T_1}T_1\right). \label{reducedil2qssa} \\
  \frac{dI_{\upgamma}}{dt} &=\underbrace{\lambda_{I_{\upgamma} K}K}_{\text{production by $K$}} - \underbrace{d_{I_\upgamma}I_\upgamma}_{\text{degradation}}. \label{reducedifngammaeqn}
  \intertext{After applying a QSSA, this becomes}
  I_\upgamma &= \frac{\lambda_{I_{\upgamma} K}}{d_{I_\upgamma}}K. \label{reducedifngammaqssa} \\
  \frac{dI_\upalpha}{dt}&= \underbrace{\lambda_{I_{\upalpha}T_8}T_8}_{\text{production by $T_8$}} + \underbrace{\lambda_{I_{\upalpha}T_1}T_1}_{\text{production by $T_1$}} + \underbrace{\lambda_{I_{\upalpha}M_1}M_1}_{\text{production by $M_1$}} + \underbrace{\lambda_{I_{\upalpha}K}K}_{\text{production by $K$}}- \underbrace{d_{I_{\upalpha}}I_{\upalpha}}_{\text{degradation}}. \label{reducedtnfeqn}
  \intertext{After applying a QSSA, this becomes}
  I_\upalpha &= \frac{1}{d_{I_{\upalpha}}}\left(\lambda_{I_{\upalpha}T_8}T_8 + \lambda_{I_{\upalpha}T_1}T_1 + \lambda_{I_{\upalpha}M_1}M_1 + \lambda_{I_{\upalpha}K}K\right). \label{reducedtnfqssa} \\
  \frac{dI_\upbeta}{dt} &= \underbrace{\lambda_{I_{\upbeta}C}C}_{\text{production by $C$}} + \underbrace{\lambda_{I_{\upbeta}T_r}T_r}_{\text{production by $T_r$}} + \underbrace{\lambda_{I_{\upbeta}M_2}M_2}_{\text{production by $M_2$}} - \underbrace{d_{I_{\upbeta}}I_{\upbeta}}_{\text{degradation}}. \label{reducedtgfbetaeqn}
  \intertext{After applying a QSSA, this becomes}
  I_\upbeta &= \frac{1}{d_{I_{\upbeta}}}\left(\lambda_{I_{\upbeta}C}C + \lambda_{I_{\upbeta}T_r}T_r + \lambda_{I_{\upbeta}M_2}M_2\right). \label{reducedibetaqssa} \\
  \frac{dI_{10}}{dt}&= \underbrace{\lambda_{I_{10}C}C}_{\text{production by $C$}} + \underbrace{\lambda_{I_{10}M_2}M_2}_{\text{production by $M_2$}} - \underbrace{d_{I_{10}}I_{10}}_{\text{degradation}}, \label{reducedil10eqn} \\
  \frac{dP_D^{T_8}}{dt} &= \underbrace{\lambda_{P_D^{T_8}}T_8}_{\text{synthesis}} + \underbrace{\lambda_{Q_A}Q_A^{T_8}}_{\substack{\text{dissociation} \\ \text{of $Q_A^{T_8}$}}} - \underbrace{\lambda_{P_DA_1}P_D^{T_8}A_1}_{\text{binding to $A_1$}} - \underbrace{d_{P_D}P_D^{T_8}}_{\text{degradation}}, \label{reducedPD8eqn} \\
  \frac{dP_D^{T_1}}{dt} &= \underbrace{\lambda_{P_D^{T_1}}T_1}_{\text{synthesis}} + \underbrace{\lambda_{Q_A}Q_A^{T_1}}_{\substack{\text{dissociation} \\ \text{of $Q_A^{T_1}$}}} - \underbrace{\lambda_{P_DA_1}P_D^{T_1}A_1}_{\text{binding to $A_1$}} - \underbrace{d_{P_D}P_D^{T_1}}_{\text{degradation}}, \label{reducedPD1eqn} \\
  \frac{dP_D^{K}}{dt} &= \underbrace{\lambda_{P_D^{K}}K}_{\text{synthesis}} + \underbrace{\lambda_{Q_A}Q_A^{K}}_{\substack{\text{dissociation} \\ \text{of $Q_A^{K}$}}} - \underbrace{\lambda_{P_DA_1}P_D^{K}A_1}_{\text{binding to $A_1$}} - \underbrace{d_{P_D}P_D^{K}}_{\text{degradation}}, \label{reducedPDKeqn} \\
  \frac{dQ_A^{T_8}}{dt} &= \underbrace{\lambda_{P_DA_1}P_D^{T_8}A_1}_{\text{formation of $Q_A^{T_8}$}} - \underbrace{\lambda_{Q_A}Q_A^{T_8}}_{\text{dissociation of $Q_A^{T_8}$}} - \underbrace{d_{Q_A}Q_A^{T_8}}_{\text{internalisation}}, \label{reducedQA8eqn} \\
  \frac{dQ_A^{T_1}}{dt} &= \underbrace{\lambda_{P_DA_1}P_D^{T_1}A_1}_{\text{formation of $Q_A^{T_1}$}} - \underbrace{\lambda_{Q_A}Q_A^{T_1}}_{\text{dissociation of $Q_A^{T_1}$}} - \underbrace{d_{Q_A}Q_A^{T_1}}_{\text{internalisation}}, \label{reducedQA1eqn} \\
  \frac{dQ_A^{K}}{dt} &= \underbrace{\lambda_{P_DA_1}P_D^{K}A_1}_{\text{formation of $Q_A^{K}$}} - \underbrace{\lambda_{Q_A}Q_A^{K}}_{\text{dissociation of $Q_A^{K}$}} - \underbrace{d_{Q_A}Q_A^{K}}_{\text{internalisation}}, \label{reducedQAKeqn} \\
  \frac{dA_{1}}{dt} &=\underbrace{\sum_{j=1}^{n} \xi_{j}f_\mathrm{pembro}\delta\left(t-t_j \right)}_{\text{infusion}} + \underbrace{\lambda_{Q_A}\left(Q_A^{T_8}+ Q_A^{T_1} + Q_A^{K}\right)}_{\text{dissociation of $Q_A^{T_8}$, $Q_A^{T_1}$, and $Q_A^{K}$}} - \underbrace{\lambda_{P_DA_1}\left(P_D^{T_8} + P_D^{T_1} + P_D^{K}\right)A_1}_{\text{formation of $Q_A^{T_8}$, $Q_A^{T_1}$, and $Q_A^{K}$}} - \underbrace{d_{A_1}A_{1}}_{\text{elimination}}, \label{reducedA1eqn} \\
  \frac{dP_{L}}{dt} &= \underbrace{\lambda_{P_L C}C + \lambda_{P_L M_2}M_2}_{\text{synthesis}} - \underbrace{d_{P_L}P_L}_{\text{degradation}}, \label{reducedPLeqn} \\
  Q^{T_8} &= \frac{\lambda_{P_{D}P_{L}}}{\lambda_{Q}} P_{D}^{T_8}P_{L}, \label{reducedQ8eqn} \\
  Q^{T_1} &= \frac{\lambda_{P_{D}P_{L}}}{\lambda_{Q}} P_{D}^{T_1}P_{L}, \label{reducedQT1eqn} \\ 
  Q^{K} &= \frac{\lambda_{P_{D}P_{L}}}{\lambda_{Q}} P_{D}^{K}P_{L}, \label{reducedQKeqn} \\
  \frac{dP_D^{8\mathrm{LN}}}{dt} &= \underbrace{\lambda_{P_D^{8\mathrm{LN}}}T_A^8}_{\text{synthesis}} + \underbrace{\lambda_{Q_A}Q_A^{8\mathrm{LN}}}_{\substack{\text{dissociation} \\ \text{of $Q_A^{8\mathrm{LN}}$}}} - \underbrace{\lambda_{P_DA_1}P_D^{8\mathrm{LN}}A_1^\mathrm{LN}}_{\text{binding to $A_1^\mathrm{LN}$}} - \underbrace{d_{P_D}P_D^{8\mathrm{LN}}}_{\text{degradation}}, \label{reducedPD8LNeqn} \\
  \frac{dP_D^{1\mathrm{LN}}}{dt} &= \underbrace{\lambda_{P_D^{1\mathrm{LN}}}T_A^1}_{\text{synthesis}} + \underbrace{\lambda_{Q_A}Q_A^{1\mathrm{LN}}}_{\substack{\text{dissociation} \\ \text{of $Q_A^{1\mathrm{LN}}$}}} - \underbrace{\lambda_{P_DA_1}P_D^{1\mathrm{LN}}A_1^\mathrm{LN}}_{\text{binding to $A_1^\mathrm{LN}$}} - \underbrace{d_{P_D}P_D^{1\mathrm{LN}}}_{\text{degradation}}, \label{reducedPD1LNeqn} \\
  \frac{dQ_A^{8\mathrm{LN}}}{dt} &= \underbrace{\lambda_{P_DA_1}P_D^{8\mathrm{LN}}A_1^\mathrm{LN}}_{\text{formation of $Q_A^{8\mathrm{LN}}$}} - \underbrace{\lambda_{Q_A}Q_A^{8\mathrm{LN}}}_{\text{dissociation of $Q_A^{8\mathrm{LN}}$}} - \underbrace{d_{Q_A}Q_A^{8\mathrm{LN}}}_{\text{internalisation}}, \label{reducedQA8LNeqn} \\
  \frac{dQ_A^{1\mathrm{LN}}}{dt} &= \underbrace{\lambda_{P_DA_1}P_D^{1\mathrm{LN}}A_1^\mathrm{LN}}_{\text{formation of $Q_A^{1\mathrm{LN}}$}} - \underbrace{\lambda_{Q_A}Q_A^{1\mathrm{LN}}}_{\text{dissociation of $Q_A^{1\mathrm{LN}}$}} - \underbrace{d_{Q_A}Q_A^{1\mathrm{LN}}}_{\text{internalisation}}, \label{reducedQA1LNeqn} \\
  \frac{dA_{1}^\mathrm{LN}}{dt} &=\underbrace{\sum_{j=1}^{n} \xi_{j}f_\mathrm{pembro}\delta\left(t-t_j \right)}_{\text{infusion}} + \underbrace{\lambda_{Q_A}\left(Q_A^{8\mathrm{LN}} + Q_A^{1\mathrm{LN}}\right)}_{\text{dissociation of $Q_A^{8\mathrm{LN}}$ and $Q_A^{1\mathrm{LN}}$}} - \underbrace{\lambda_{P_DA_1}\left(P_D^{8\mathrm{LN}} + P_D^{1\mathrm{LN}}\right)A_1^\mathrm{LN}}_{\text{formation of $Q_A^{8\mathrm{LN}}$ and $Q_A^{1\mathrm{LN}}$}} - \underbrace{d_{A_1}A_{1}^\mathrm{LN}}_{\text{elimination}}, \label{reducedA1LNeqn} \\
  \frac{dP_{L}^\mathrm{LN}}{dt} &= \underbrace{\lambda_{P_L^\mathrm{LN}D^\mathrm{LN}}D^\mathrm{LN} + \lambda_{P_L^\mathrm{LN} T_A^1}T_A^1}_{\text{synthesis}} - \underbrace{d_{P_L}P_L^\mathrm{LN}}_{\text{degradation}}, \label{reducedPLLNeqn} \\
  Q^{8\mathrm{LN}} &= \frac{\lambda_{P_{D}P_{L}}}{\lambda_{Q}} P_{D}^{8\mathrm{LN}}P_{L}^\mathrm{LN}, \label{reducedQ8LNeqn} \\
  Q^{1\mathrm{LN}} &= \frac{\lambda_{P_{D}P_{L}}}{\lambda_{Q}} P_{D}^{1\mathrm{LN}}P_{L}^\mathrm{LN}. \label{reducedQ1LNeqn}
\end{align}
The model parameter values are estimated in Appendix B and are listed in Table C.1.
\subsubsection{Initial Conditions\label{reducedmodelssic}}
All initial conditions are the same as those in \Cref{initcond section}, except for cytokines and the TS and TDLN immune checkpoint protein initial conditions, which are provided in \autoref{reducedcytokinetable}, \autoref{reducedTSICItable}, and \autoref{reducedTDLNICItable}, respectively. Justification for the choice of these values is given in Appendix B.3, Appendix B.7, and Appendix B.8, respectively.
\begin{table}[H]
    \centering
    \begin{tabular}{|c|c|}
\hline
\textbf{Cytokine} & \textbf{Initial Condition} \\
\hline 
$I_2$        & $1.87 \times 10^{-12}$\\
$I_\upgamma$ & $1.14 \times 10^{-10}$ \\
$I_\upalpha$ & $1.25 \times 10^{-10}$\\
$I_\upbeta$  & $9.20 \times 10^{-7}$ \\
$I_{10}$     & $1.15 \times 10^{-10}$\\
\hline
\end{tabular}
\caption{\label{reducedcytokinetable}Cytokine initial conditions for the reduced model. All values are in units of $\mathrm{g/cm^3}$.}
\end{table}

\begin{table}[H]
    \centering
    \begin{tabular}{|c|c|}
    \hline
    \textbf{Protein} & \textbf{Initial Condition} \\
    \hline 
    $P_D^{T_8}$ & $4.44 \times 10^8$ \\
    $P_D^{T_1}$ & $2.07 \times 10^8$ \\
    $P_D^{K}$   & $2.46 \times 10^8$ \\
    $P_L$       & $7.36 \times 10^{12}$ \\
    $Q^{T_8}$   & $6.96 \times 10^5$ \\
    $Q^{T_1}$   & $3.24 \times 10^5$ \\
    $Q^{K}$     & $3.86 \times 10^5$ \\
    \hline
    \end{tabular}
    \caption{\label{reducedTSICItable}TS immune checkpoint-associated component initial conditions for the reduced model. All values are in units of $\mathrm{molec/cm^3}$.}
\end{table}

\begin{table}[H]
    \centering
\begin{tabular}{|c|c|}
\hline
\textbf{Protein} & \textbf{Initial Condition} \\
\hline 
$P_{D}^{8\mathrm{LN}}$ & $2.37 \times 10^9$ \\
$P_{D}^{1\mathrm{LN}}$ & $1.59 \times 10^{10}$ \\
$P_{L}^{\mathrm{LN}}$  & $3.31 \times 10^{11}$ \\
$Q^{8\mathrm{LN}}$      & $1.67 \times 10^5$ \\
$Q^{1\mathrm{LN}}$      & $1.12 \times 10^6$ \\
\hline
\end{tabular}
    \caption{\label{reducedTDLNICItable}TDLN immune checkpoint-associated component initial conditions for the reduced model. All values are in units of $\mathrm{molec/cm^3}$.}
\end{table}
\subsubsection{Sensitivity Analysis}
\begin{figure}[H]
    \centering
    \includegraphics[width=\textwidth]{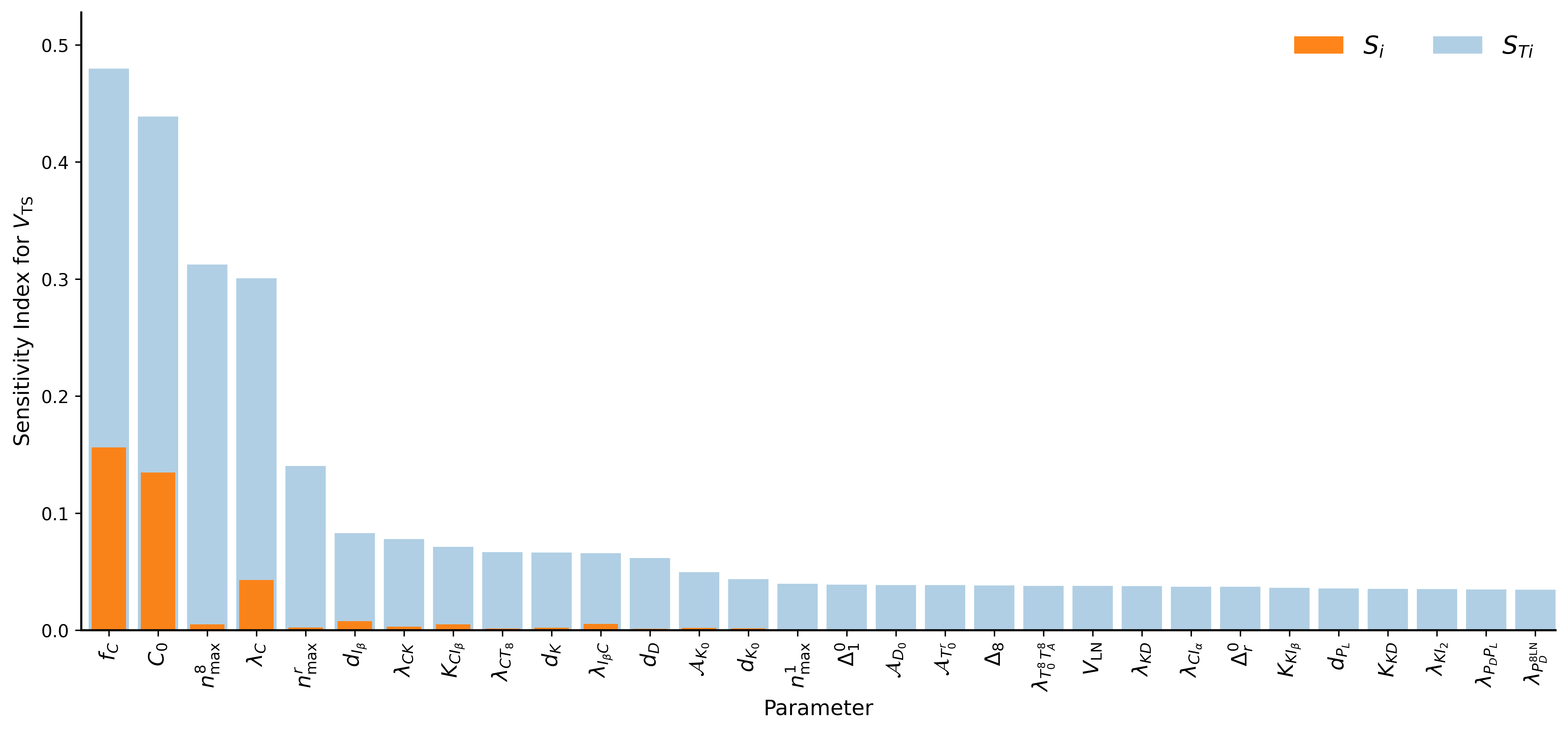}
    \caption{\label{reducedmodelsensindexplot}First-order indices ($S_i$ in orange) and total-order indices ($S_{Ti}$ in blue) for the reduced model for the top 30 parameters associated with the RMSRE of $V_\mathrm{TS}$ at 180.9 days under the standard regimen, sorted in descending order of $S_{Ti}$.}
\end{figure}

We now perform a sensitivity analysis on the reduced model, following the same procedure as for the full model. We simulated the standard regimen for 180.9 days using the initial conditions from \Cref{reducedmodelssic}, using the dde23 integrator and applying the FAST and EFAST methods to compute the first-order and total-order sensitivity indices, respectively. We sampled $10,000$ samples for each parameter, as this was sufficient to reach convergence in the indices, setting $M=4$, and considered parameter ranges corresponding to a $\pm 50\%$ deviation for each parameter, noting that $n^8_\mathrm{max}$, $n^1_\mathrm{max}$, and $n^r_\mathrm{max}$ are integers, and were therefore rounded to the nearest integer. \\~\\
The SALib Python library \citep{Iwanaga2022} was used to perform FAST and EFAST analysis on the reduced model with the aforementioned configuration, leading to the indices in Table G.1. The sensitivity indices for the reduced model for the top 30 parameters associated with the RMSRE of $V_\mathrm{TS}$ at 180.9 days under the standard regimen, sorted in descending order of the total-order indices, are shown in \autoref{reducedmodelsensindexplot}.
\subsection{Minimal Model}
We now aim to use the sensitivity analysis results of the reduced model, presented in Table G.1, to remove as many parameters and state variables as possible while still retaining the original trajectories and dynamics of the remaining state variables. \\~\\
To guide elimination of state variables, we distinguished between variables that exhibit significant coupling into the remaining tumour-control subsystem, from those with negligible influence over the 180.9-day window. To express this more precisely, we let $S_i^{(V_\mathrm{TS})}$ and $S_{Ti}^{(V_\mathrm{TS})}$ be the first-order and total-order indices of a parameter $p$ for the RMSRE of $V_\mathrm{TS}$ at $t=180.9$ days, as defined in \eqref{RMSRESAeqn}. For a particular set of state variables retained in the minimal model, denoted by $\mathcal{X}_\mathrm{min}$, for any candidate variable $X \in \mathcal{X}_\mathrm{min}$, we define the coupling parameter set, $\mathcal{P}_X$, via
\begin{equation}
    \mathcal{P}_X := \set{p | \exists Y\in \mathcal{X}_\mathrm{min} \text{ where }p \text{ appears in the equation for } \frac{dY}{dt} \text{ in a term that depends on }X }.
\end{equation}
If $\mathcal{P}_X=\emptyset$ or
\begin{equation}
\max_{p\in\mathcal{P}_X} S_i^{(V_\mathrm{TS})} \leq 1.11\times 10^{-4}
\quad\text{and}\quad
\max_{p\in\mathcal{P}_X} S_{Ti}^{(V_\mathrm{TS})} \leq 3.43\times 10^{-2},
\label{reducedtominimalscreen}
\end{equation}
then we removed $X$ by setting $X(t)\equiv 0$, with a grid search employed to derive these thresholds in a similar fashion to those used to derive \eqref{fulltoreducedcriterion}. To identify which variables could be eliminated, we sought the smallest subset of variables such that all parameters with large sensitivity indices are associated only with variables within this subset; equivalently, we aimed to minimise the number of state variables in $\mathcal{X}_\mathrm{min}$. Applying this filtering process led us to eliminate the following state variables by zeroing them: $S$, $T_0^4$, $T_A^1$, $T_1$, $M_0$, $M_1$, $M_2$, $I_\upgamma$, $P_D^{T_1}$, $Q_A^{T_1}$, $Q^{T_1}$, $P_D^{1\mathrm{LN}}$, $Q_A^{1\mathrm{LN}}$, and $Q^{1\mathrm{LN}}$. After substituting these zero values, any parameter that appeared only in either the eliminated differential equations or coupling terms nullified by $Y=0$ was removed from the parameter set; moreover, half-saturation and inhibition constants were removed only when the entire parent term vanished.\\~\\
We also eliminated $T_0^8$ and $K_0$ by treating them as constant precursor pools, with their baseline values absorbed into the corresponding kinetic constants. That is, we implicitly apply the transformations $\lambda_{T_0^8 T_A^8}\overline{T_0^8} \mapsto \lambda_{T_0^8 T_A^8}$, $\lambda_{KI_2}\overline{K_0} \mapsto \lambda_{KI_2}$, and $\lambda_{KD}\overline{K_0} \mapsto \lambda_{KD}$, where $\overline{T_0^8}$ and $\overline{K_0}$ are as in both the full and reduced models.
We can further simplify the model by performing a QSSA on $H$, by setting $\frac{dH}{dt}=0$, or equivalently
\begin{equation}
    H = \frac{\lambda_{HN_c}}{d_H}N_c. \label{minimalHqssa}
\end{equation}
We can eliminate $H$ from the model by substituting \eqref{minimalHqssa} into \eqref{reducedD0eqn} and \eqref{reducedDeqn} to get that
\begin{align*}
\begin{split}
    \frac{dD_0}{dt} &= \underbrace{\mathcal{A}_{D_0}}_{\text{source}} - \underbrace{\lambda_{DN_c}D_0\frac{N_c}{K_{DN_c}+N_c}}_{\text{$D_0 \to D$ by $N_c$}} -\underbrace{\lambda_{D_0K} D_0K\frac{1}{1+I_\upbeta/K_{D_0I_\upbeta}}}_{\substack{\text{elimination by $K$} \\ \text{inhibited by $I_\upbeta$}}} - \underbrace{d_{D_0}D_0}_{\text{death}},
    \end{split} \\
  \frac{dD}{dt} &= \underbrace{\lambda_{DN_c}D_0\frac{N_c}{K_{DN_c}+N_c}}_{\text{$D_0 \to D$ by $N_c$}} - \underbrace{\lambda_{DD^\mathrm{LN}}D}_{\substack{\text{$D$ migration} \\ \text{to TDLN}}} - \underbrace{d_{D}D}_{\text{death}},
 \end{align*}
where $\lambda_{DN_c} = \lambda_{DH}$, and $K_{DN_c} = d_H K_{DH}/\lambda_{HN_c}$.\\~\\
Similarly, we can perform a QSSA on $I_2$ and $I_{10}$, leading to 
\begin{align}
    I_2 &= \frac{\lambda_{I_2 T_8}}{d_{I_2}}T_8, \label{minimalil2qssa} \\
    I_{10} &= \frac{\lambda_{I_{10}C}}{d_{I_{10}}} C, \label{minimali10qssa}
\end{align}
respectively. We can thus similarly eliminate $I_2$ from the model by substituting \eqref{minimalil2qssa} into \eqref{reducedKeqn}, so that
\begin{equation*}
    \frac{dK}{dt} = \left(\underbrace{\lambda_{KT_8}\frac{T_8}{K_{KT_8}+T_8}}_{\text{$K_0 \to K$ by $T_8$}} +\underbrace{\lambda_{KD}\frac{D}{K_{KD}+D}}_{\text{$K_0 \to K$ by $D$}}\right)\underbrace{\frac{1}{1+I_\upbeta/K_{KI_\upbeta}}}_{\substack{\text{activation}\\ \text{inhibited by $I_\upbeta$}}} - \underbrace{d_{K}K}_{\text{degradation}},
  \end{equation*}
where $\lambda_{KT_8} = \lambda_{KI_2}$, and $K_{KT_8} = d_{I_2}K_{KI_2}/\lambda_{I_2 T_8}$. Repeating this process to eliminate $I_{10}$, by substituting \eqref{minimali10qssa} into \eqref{minimalt8eqn} and \eqref{minimalTexeqn}, leads to
\begin{align*}
    \frac{d T_8}{dt} &= \frac{V_\mathrm{LN}}{V_\mathrm{TS}}\underbrace{\lambda_{T_A^8T_8}T_A^8}_{\substack{\text{$T_A^8$ migration} \\ \text{to the TS}}} - \underbrace{\lambda_{T_8C}\frac{T_8 C}{K_{T_8T_\mathrm{ex}}+ C}}_{\text{$T_8 \to T_\mathrm{ex}$ from $C$ exposure}} + \underbrace{\lambda_{T_\mathrm{ex}A_1}\frac{T_\mathrm{ex}A_1}{K_{T_\mathrm{ex}A_1} + A_1}}_{\text{$T_\mathrm{ex} \to T_8$ by $A_1$}} - \underbrace{\frac{d_{T_8} T_8}{1+C/K_{T_8C}}}_{\substack{\text{death} \\ \text{inhibited by $C$}}}, \\
    \frac{dT_\mathrm{ex}}{dt} &= \underbrace{\lambda_{T_8C}\frac{T_8 C}{K_{T_8T_\mathrm{ex}}+ C}}_{\text{$T_8 \to T_\mathrm{ex}$ from $C$ exposure}} - \underbrace{\lambda_{T_\mathrm{ex}A_1}\frac{T_\mathrm{ex}A_1}{K_{T_\mathrm{ex}A_1} + A_1}}_{\text{$T_\mathrm{ex} \to T_8$ by $A_1$}} - \underbrace{\frac{d_{T_\mathrm{ex}} T_\mathrm{ex}}{1+C/K_{T_\mathrm{ex}C}}}_{\substack{\text{death} \\ \text{inhibited by $C$}}},
\end{align*}
where $K_{T_8C} = d_{I_{10}}K_{T_8I_{10}}/\lambda_{I_{10}C}$ and $K_{T_\mathrm{ex}C} = d_{I_{10}}K_{T_\mathrm{ex}I_{10}}/\lambda_{I_{10}C}$. We note that we have relabelled the half-saturation constant for CD8+ T cell exhaustion due to prolonged cancer antigen exposure as $K_{T_8T_\mathrm{ex}}$ to avoid confusion with the inhibition constant associated with CD8+ T cell death mediated by IL-10.\\~\\
Thus, with the application of QSSA to HMGB1, IL-2, and IL-10 concentrations, this analysis led to the removal of variables representing the concentrations of DAMPs, CD4+ T cells and associated variables, macrophages, resting NK and naive CD8+ T cells, IL-2, IFN-$\upgamma$, and IL-10, along with their corresponding parameters. The resulting minimal model achieves over a 50\% reduction in system dimensionality, along with more than 50\% and 33\% reductions in the number of parameters compared to the full and reduced models, respectively---representing a significant simplification of the original model.
\subsubsection{Model Variables}
The variables and their units in the minimal model are shown in \autoref{minimalmodelvars}.
\begin{table}[H]
\centering
\resizebox{\columnwidth}{!}{%
\begin{tabular}{|lp{7.7cm}|lp{7.7cm}|}
\hline
\textbf{Var} & \textbf{Description} & \textbf{Var} & \textbf{Description} \\ 
\hline
$V_\mathrm{TS}$ & Primary tumour volume & & \\
\hline
$C$ & Viable cancer cell density & $N_c$ & Necrotic cell density \\
$D_0$ & Immature DC density & $D$ & Mature DC density in the TS \\
$D^\mathrm{LN}$ & Mature DC density in the TDLN & $T_A^8$ & Effector CD8+ T cell density in the TDLN \\
$T_8$ & Effector CD8+ T cell density in the TS &
$T_{\mathrm{ex}}$ & Exhausted CD8+ T cell density in the TS \\
$T_0^r$ & Naive Treg density in the TDLN & $T_A^r$ & Effector Treg density in the TDLN \\ 
$T_r$ & Effector Treg density in the TS & $K$ & Activated NK cell density\\
\hline
$I_\upalpha$ & TNF concentration & $I_\upbeta$ & TGF-$\upbeta$ concentration \\
\hline
$P_D^{T_8}$ & Unbound PD-1 receptor concentration on effector CD8+ T cells in the TS & $P_D^{K}$ & Unbound PD-1 receptor concentration on activated NK cells \\
$Q_A^{T_8}$ & PD-1/pembrolizumab complex concentration on effector CD8+ T cells in the TS & $Q_A^{K}$ & PD-1/pembrolizumab complex concentration on activated NK cells \\
$P_L$ & Unbound PD-L1 concentration in the TS & $Q^{T_8}$ & PD-1/PD-L1 complex concentration on effector CD8+ T cells in the TS \\
$Q^{K}$ & PD-1/PD-L1 complex concentration on activated NK cells & $A_{1}$ & Concentration of pembrolizumab in the TS \\
\hline 
$P_D^{8\mathrm{LN}}$ & Unbound PD-1 receptor concentration on effector CD8+ T cells in the TDLN & $Q_A^{8\mathrm{LN}}$ & PD-1/pembrolizumab complex concentration on effector CD8+ T cells in the TDLN \\
$P_L^\mathrm{LN}$ & Unbound PD-L1 concentration in the TDLN & $Q^{8\mathrm{LN}}$ & PD-1/PD-L1 complex concentration on effector CD8+ T cells in the TDLN \\
$A_{1}^\mathrm{LN}$ & Concentration of pembrolizumab in the TDLN & & \\
\hline
\end{tabular}%
}
\caption{\label{minimalmodelvars}Variables used in the minimal model. Quantities in the top box are in units of $\mathrm{cm^3}$, quantities in the second box are in units of $\mathrm{cell/{cm}^3}$, quantities in the third box are in units of $\mathrm{g/{cm}^3}$, and all other quantities are in units of $\mathrm{molec/{cm}^3}$. All quantities pertain to the tumour site unless otherwise specified. TS denotes the tumour site, whilst TDLN denotes the tumour-draining lymph node.}
\end{table}
\subsubsection{Model Equations}
\begin{align}
  \begin{split}
  \frac{dC}{dt} &= \underbrace{\lambda_{C}C\left(1-\frac{C}{C_0}\right)}_{\text{growth}} - \underbrace{\lambda_{CT_8}T_8 \frac{1}{1+I_{\upbeta}/K_{CI_{\upbeta}}}\frac{1}{1+Q^{T_8}/K_{CQ^{T_8}}}C}_{\substack{\text{elimination by $T_8$} \\ \text{inhibited by $I_{\upbeta}$ and $Q^{T_8}$}}} - \underbrace{\lambda_{CK}K \frac{1}{1+I_{\upbeta}/K_{CI_{\upbeta}}}\frac{1}{1+Q^K/K_{CQ^K}}C}_{\substack{\text{elimination by $K$} \\ \text{inhibited by $I_\upbeta$ and $Q^K$}}} \\
  &- \underbrace{\lambda_{CI_{\upalpha}}\frac{I_{\upalpha}}{K_{CI_{\upalpha}}+I_{\upalpha}}C}_{\text{elimination by $I_{\upalpha}$}},
  \end{split} \label{minimalcancereqn} \\
  \begin{split}
    \frac{dN_c}{dt}&= \underbrace{\lambda_{CI_{\upalpha}}\frac{I_{\upalpha}}{K_{CI_{\upalpha}}+I_{\upalpha}}C}_{\text{elimination by $I_{\upalpha}$}} -\underbrace{d_{N_c}N_c}_{\text{removal}},
    \end{split}\label{minimalnecroticcelleqn} \\
  \begin{split}
  \frac{dV_\mathrm{TS}}{dt} &= \frac{1}{f_{C}+f_{N_c}}\left[\lambda_{C}f_{C}V_\mathrm{TS}\left(1-\frac{f_C V_\mathrm{TS}}{C_0}\right) - \lambda_{CT_8}T_8 \frac{1}{1+I_{\upbeta}/K_{CI_{\upbeta}}}\frac{1}{1+Q^{T_8}/K_{CQ^{T_8}}}f_C V_\mathrm{TS} \right. \\
  &\left.- \lambda_{CK}K \frac{1}{1+I_\upbeta/K_{CI_\upbeta}}\frac{1}{1+Q^K/K_{CQ^K}}f_C V_\mathrm{TS} -d_{N_c}f_{N_c}V_\mathrm{TS}\right],
  \end{split} \label{minimalVeqn}
  \intertext{where $C(t)=f_C V_\mathrm{TS}(t)$ and $N_c(t)=f_{N_c}V_\mathrm{TS}(t)$.}
  \begin{split}
    \frac{dD_0}{dt} &= \underbrace{\mathcal{A}_{D_0}}_{\text{source}} - \underbrace{\lambda_{DN_c}D_0\frac{N_c}{K_{DN_c}+N_c}}_{\text{$D_0 \to D$ by DAMPs}} -\underbrace{\lambda_{D_0K} D_0K\frac{1}{1+I_\upbeta/K_{D_0I_\upbeta}}}_{\substack{\text{elimination by $K$} \\ \text{inhibited by $I_\upbeta$}}} - \underbrace{d_{D_0}D_0}_{\text{death}},
    \end{split} \label{minimalD0eqn} \\
  \frac{dD}{dt} &= \underbrace{\lambda_{DN_c}D_0\frac{N_c}{K_{DN_c}+N_c}}_{\text{$D_0 \to D$ by DAMPs}} - \underbrace{\lambda_{DD^\mathrm{LN}}D}_{\substack{\text{$D$ migration} \\ \text{to TDLN}}} - \underbrace{d_{D}D}_{\text{death}}, \label{minimalDeqn} \\
  \frac{dD^\mathrm{LN}}{dt} &= \frac{V_\mathrm{TS}}{V_\mathrm{LN}}\underbrace{\lambda_{DD^\mathrm{LN}}\exp\left(-d_D \tau_m\right)D(t-\tau_m)}_{\text{$D$ migration to TDLN}} - \underbrace{d_{D}D^\mathrm{LN}}_{\text{death}}, \label{minimalDLNeqn} \\
  \begin{split}
  \frac{dT_A^8}{dt} &= \underbrace{\frac{2^{n^8_\mathrm{max}}\exp\left(-d_{T_0^8}\tau_{T_A^8}\right)R^8(t- \tau_{T_A^8})}{\left(1+T_A^{r}(t- \tau_{T_A^8})/K_{T_A^8 T_A^{r}}\right)\left(1+ Q^{8\mathrm{LN}}(t- \tau_{T_A^8})/K_{T_A^8 Q^{8\mathrm{LN}}}\right)}}_{\text{CD8+ T cell proliferation inhibited by $T_A^r$ and $Q^{8\mathrm{LN}}$}} - \underbrace{\lambda_{T_A^8T_8}T_A^8}_{\substack{\text{$T_A^8$ migration} \\ \text{to the TS}}}-\underbrace{d_{T_8} T_A^8}_\text{death},
  \end{split} \label{minimalTA8n8maxeqn} 
  \intertext{where $\tau_{T_A^8}$ is defined as}
  \tau_{T_A^8} &:=\Delta_8^0 + (n^8_\mathrm{max}-1)\Delta_8, 
  \intertext{and where $R^8(t)$ is defined as}
  R^8(t) &:= \underbrace{\frac{\lambda_{T_0^8 T_A^8}D^\mathrm{LN}(t)}{\left(1+T_A^r(t)/K_{T_0^8T_A^r}\right)\left(1+Q^{8\mathrm{LN}}(t)/K_{T_0^8Q^{8\mathrm{LN}}}\right)}}_{\text{CD8+ T cell activation inhibited by $T_A^r$ and $Q^{8\mathrm{LN}}$}}. \label{minimalR8eqn} \\
  \frac{d T_8}{dt} &= \frac{V_\mathrm{LN}}{V_\mathrm{TS}}\underbrace{\lambda_{T_A^8T_8}T_A^8}_{\substack{\text{$T_A^8$ migration} \\ \text{to the TS}}} - \underbrace{\lambda_{T_8C}\frac{T_8 C}{K_{T_8T_\mathrm{ex}}+ C}}_{\text{$T_8 \to T_\mathrm{ex}$ from $C$ exposure}} + \underbrace{\lambda_{T_\mathrm{ex}A_1}\frac{T_\mathrm{ex}A_1}{K_{T_\mathrm{ex}A_1} + A_1}}_{\text{$T_\mathrm{ex} \to T_8$ by $A_1$}} - \underbrace{\frac{d_{T_8} T_8}{1+C/K_{T_8C}}}_{\substack{\text{death} \\ \text{inhibited by IL-10}}}, \label{minimalt8eqn}\\
  \frac{dT_\mathrm{ex}}{dt} &= \underbrace{\lambda_{T_8C}\frac{T_8 C}{K_{T_8T_\mathrm{ex}}+ C}}_{\text{$T_8 \to T_\mathrm{ex}$ from $C$ exposure}} - \underbrace{\lambda_{T_\mathrm{ex}A_1}\frac{T_\mathrm{ex}A_1}{K_{T_\mathrm{ex}A_1} + A_1}}_{\text{$T_\mathrm{ex} \to T_8$ by $A_1$}} - \underbrace{\frac{d_{T_\mathrm{ex}} T_\mathrm{ex}}{1+C/K_{T_\mathrm{ex}C}}}_{\substack{\text{death} \\ \text{inhibited by IL-10}}}, \label{minimalTexeqn} \\
  \frac{dT_0^r}{dt} &=\underbrace{\mathcal{A}_{T_0^r}}_{\text{source}} - \underbrace{R^r(t)}_{\text{Treg activation}} - \underbrace{d_{T_0^r}T_0^r}_{\text{death}}, \label{minimalnaivetregeqn}
  \intertext{where $R^r(t)$ is defined as}
  R^r(t) &:= \underbrace{\lambda_{T_0^r T_A^r}D^\mathrm{LN}(t)T_0^r(t)}_{\text{Treg activation}}.\\
  \frac{dT_A^r}{dt} &= \underbrace{2^{n^r_\mathrm{max}}\exp\left(-d_{T_0^r}\tau_{T_A^r}\right)R^r(t- \tau_{T_A^r})}_{\text{Treg proliferation}} - \underbrace{\lambda_{T_A^rT_r}T_A^r}_{\substack{\text{$T_A^r$ migration} \\ \text{to the TS}}}-\underbrace{d_{T_r} T_A^r}_\text{death}, \label{minimalTArnrmaxeqn}
  \intertext{where $\tau_{T_A^r}$ is defined as}
  \tau_{T_A^r} &:=\Delta_r^0 + (n^r_\mathrm{max}-1)\Delta_r. \\
  \frac{dT_r}{dt} &= \frac{V_\mathrm{LN}}{V_\mathrm{TS}}\underbrace{\lambda_{T_A^rT_r}T_A^r}_{\substack{\text{$T_A^r$ migration} \\ \text{to the TS}}} - \underbrace{d_{T_r}T_r}_{\text{death}}, \label{minimaltregeqn} \\
  \frac{dK}{dt} &= \left(\underbrace{\lambda_{KT_8}\frac{T_8}{K_{KT_8}+T_8}}_{\text{$K_0 \to K$ by IL-2}} +\underbrace{\lambda_{KD}\frac{D}{K_{KD}+D}}_{\text{$K_0 \to K$ by $D$}}\right)\underbrace{\frac{1}{1+I_\upbeta/K_{KI_\upbeta}}}_{\substack{\text{activation}\\ \text{inhibited by $I_\upbeta$}}} - \underbrace{d_{K}K}_{\text{degradation}}, \label{minimalKeqn} \\
  \frac{dI_\upalpha}{dt}&= \underbrace{\lambda_{I_{\upalpha}T_8}T_8}_{\text{production by $T_8$}} + \underbrace{\lambda_{I_{\upalpha}K}K}_{\text{production by $K$}}- \underbrace{d_{I_{\upalpha}}I_{\upalpha}}_{\text{degradation}}. \label{minimaltnfeqn}
  \intertext{After applying a QSSA, this becomes}
  I_\upalpha &= \frac{1}{d_{I_{\upalpha}}}\left(\lambda_{I_{\upalpha}T_8}T_8 + \lambda_{I_{\upalpha}K}K\right). \label{minimaltnfqssa} \\
  \frac{dI_\upbeta}{dt} &= \underbrace{\lambda_{I_{\upbeta}C}C}_{\text{production by $C$}} + \underbrace{\lambda_{I_{\upbeta}T_r}T_r}_{\text{production by $T_r$}} - \underbrace{d_{I_{\upbeta}}I_{\upbeta}}_{\text{degradation}}. \label{minimaltgfbetaeqn}
  \intertext{After applying a QSSA, this becomes}
  I_\upbeta &= \frac{1}{d_{I_{\upbeta}}}\left(\lambda_{I_{\upbeta}C}C + \lambda_{I_{\upbeta}T_r}T_r\right). \label{minimalibetaqssa} \\
  \frac{dP_D^{T_8}}{dt} &= \underbrace{\lambda_{P_D^{T_8}}T_8}_{\text{synthesis}} + \underbrace{\lambda_{Q_A}Q_A^{T_8}}_{\substack{\text{dissociation} \\ \text{of $Q_A^{T_8}$}}} - \underbrace{\lambda_{P_DA_1}P_D^{T_8}A_1}_{\text{binding to $A_1$}} - \underbrace{d_{P_D}P_D^{T_8}}_{\text{degradation}}, \label{minimalPD8eqn} \\
  \frac{dP_D^{K}}{dt} &= \underbrace{\lambda_{P_D^{K}}K}_{\text{synthesis}} + \underbrace{\lambda_{Q_A}Q_A^{K}}_{\substack{\text{dissociation} \\ \text{of $Q_A^{K}$}}} - \underbrace{\lambda_{P_DA_1}P_D^{K}A_1}_{\text{binding to $A_1$}} - \underbrace{d_{P_D}P_D^{K}}_{\text{degradation}}, \label{minimalPDKeqn} \\
  \frac{dQ_A^{T_8}}{dt} &= \underbrace{\lambda_{P_DA_1}P_D^{T_8}A_1}_{\text{formation of $Q_A^{T_8}$}} - \underbrace{\lambda_{Q_A}Q_A^{T_8}}_{\text{dissociation of $Q_A^{T_8}$}} - \underbrace{d_{Q_A}Q_A^{T_8}}_{\text{internalisation}}, \label{minimalQA8eqn} \\
  \frac{dQ_A^{K}}{dt} &= \underbrace{\lambda_{P_DA_1}P_D^{K}A_1}_{\text{formation of $Q_A^{K}$}} - \underbrace{\lambda_{Q_A}Q_A^{K}}_{\text{dissociation of $Q_A^{K}$}} - \underbrace{d_{Q_A}Q_A^{K}}_{\text{internalisation}}, \label{minimalQAKeqn} \\
  \frac{dA_{1}}{dt} &=\underbrace{\sum_{j=1}^{n} \xi_{j}f_\mathrm{pembro}\delta\left(t-t_j \right)}_{\text{infusion}} + \underbrace{\lambda_{Q_A}\left(Q_A^{T_8}+ Q_A^{K}\right)}_{\text{dissociation of $Q_A^{T_8}$ and $Q_A^{K}$}} - \underbrace{\lambda_{P_DA_1}\left(P_D^{T_8} + P_D^{K}\right)A_1}_{\text{formation of $Q_A^{T_8}$ and $Q_A^{K}$}} - \underbrace{d_{A_1}A_{1}}_{\text{elimination}}, \label{minimalA1eqn} \\
  \frac{dP_{L}}{dt} &= \underbrace{\lambda_{P_L C}C}_{\text{synthesis}} - \underbrace{d_{P_L}P_L}_{\text{degradation}}, \label{minimalPLeqn} \\
  Q^{T_8} &= \frac{\lambda_{P_{D}P_{L}}}{\lambda_{Q}} P_{D}^{T_8}P_{L}, \label{minimalQ8eqn} \\
  Q^{K} &= \frac{\lambda_{P_{D}P_{L}}}{\lambda_{Q}} P_{D}^{K}P_{L}, \label{minimalQKeqn} \\
  \frac{dP_D^{8\mathrm{LN}}}{dt} &= \underbrace{\lambda_{P_D^{8\mathrm{LN}}}T_A^8}_{\text{synthesis}} + \underbrace{\lambda_{Q_A}Q_A^{8\mathrm{LN}}}_{\substack{\text{dissociation} \\ \text{of $Q_A^{8\mathrm{LN}}$}}} - \underbrace{\lambda_{P_DA_1}P_D^{8\mathrm{LN}}A_1^\mathrm{LN}}_{\text{binding to $A_1^\mathrm{LN}$}} - \underbrace{d_{P_D}P_D^{8\mathrm{LN}}}_{\text{degradation}}, \label{minimalPD8LNeqn} \\
  \frac{dQ_A^{8\mathrm{LN}}}{dt} &= \underbrace{\lambda_{P_DA_1}P_D^{8\mathrm{LN}}A_1^\mathrm{LN}}_{\text{formation of $Q_A^{8\mathrm{LN}}$}} - \underbrace{\lambda_{Q_A}Q_A^{8\mathrm{LN}}}_{\text{dissociation of $Q_A^{8\mathrm{LN}}$}} - \underbrace{d_{Q_A}Q_A^{8\mathrm{LN}}}_{\text{internalisation}}, \label{minimalQA8LNeqn} \\
  \frac{dA_{1}^\mathrm{LN}}{dt} &=\underbrace{\sum_{j=1}^{n} \xi_{j}f_\mathrm{pembro}\delta\left(t-t_j \right)}_{\text{infusion}} + \underbrace{\lambda_{Q_A}Q_A^{8\mathrm{LN}}}_{\text{dissociation of $Q_A^{8\mathrm{LN}}$}} - \underbrace{\lambda_{P_DA_1}P_D^{8\mathrm{LN}}A_1^\mathrm{LN}}_{\text{formation of $Q_A^{8\mathrm{LN}}$}} - \underbrace{d_{A_1}A_{1}^\mathrm{LN}}_{\text{elimination}}, \label{minimalA1LNeqn} \\
  \frac{dP_{L}^\mathrm{LN}}{dt} &= \underbrace{\lambda_{P_L^\mathrm{LN}D^\mathrm{LN}}D^\mathrm{LN}}_{\text{synthesis}} - \underbrace{d_{P_L}P_L^\mathrm{LN}}_{\text{degradation}}, \label{minimalPLLNeqn} \\
  Q^{8\mathrm{LN}} &= \frac{\lambda_{P_{D}P_{L}}}{\lambda_{Q}} P_{D}^{8\mathrm{LN}}P_{L}^\mathrm{LN}. \label{minimalQ8LNeqn}
\end{align}
The model parameter values are estimated in Appendix D and are listed in Table E.1.
\subsubsection{Initial Conditions\label{minimalmodelssic}}
All initial conditions are the same as those in \Cref{initcond section}, except for cytokines and the TS and TDLN immune checkpoint protein initial conditions, which are provided in \autoref{minimalcytokinetable}, \autoref{minimalTSICItable}, and \autoref{minimalTDLNICItable}, respectively. Justification for the choice of these values is given in Appendix D.3, Appendix D.6, and Appendix D.7, respectively. 
\begin{table}[H]
    \centering
    \begin{tabular}{|c|c|}
    \hline
    \textbf{Cytokine} & \textbf{Initial Condition} \\
    \hline 
    $I_\upalpha$ & $1.64 \times 10^{-10}$\\
    $I_\upbeta$  & $8.45 \times 10^{-7}$ \\
    \hline
    \end{tabular}
    \caption{\label{minimalcytokinetable}Cytokine initial conditions for the minimal model. All values are in units of $\mathrm{g/cm^3}$.}
\end{table}

\begin{table}[H]
    \centering
    \begin{tabular}{|c|c|}
    \hline
    \textbf{Protein} & \textbf{Initial Condition} \\
    \hline 
    $P_D^{T_8}$ & $4.44 \times 10^8$ \\
    $P_D^{K}$   & $2.46 \times 10^8$ \\
    $P_L$       & $7.02 \times 10^{12}$ \\
    $Q^{T_8}$   & $6.63 \times 10^5$ \\
    $Q^{K}$     & $3.68 \times 10^5$ \\
    \hline
    \end{tabular}
    \caption{\label{minimalTSICItable}TS immune checkpoint-associated component initial conditions for the minimal model. All values are in units of $\mathrm{molec/cm^3}$.}
\end{table}

\begin{table}[H]
    \centering
    \begin{tabular}{|c|c|}
    \hline
    \textbf{Protein} & \textbf{Initial Condition} \\
    \hline 
    $P_{D}^{8\mathrm{LN}}$  & $2.37 \times 10^9$ \\
    $P_{L}^{\mathrm{LN}}$  & $3.15 \times 10^{11}$ \\
    $Q^{8\mathrm{LN}}$      & $1.59 \times 10^5$ \\
    \hline
    \end{tabular}
    \caption{\label{minimalTDLNICItable}TDLN immune checkpoint-associated component initial conditions for the minimal model. All values are in units of $\mathrm{molec/cm^3}$.}
\end{table}
\subsubsection{Sensitivity Analysis}
\begin{figure}[H]
    \centering
    \includegraphics[width=\textwidth]{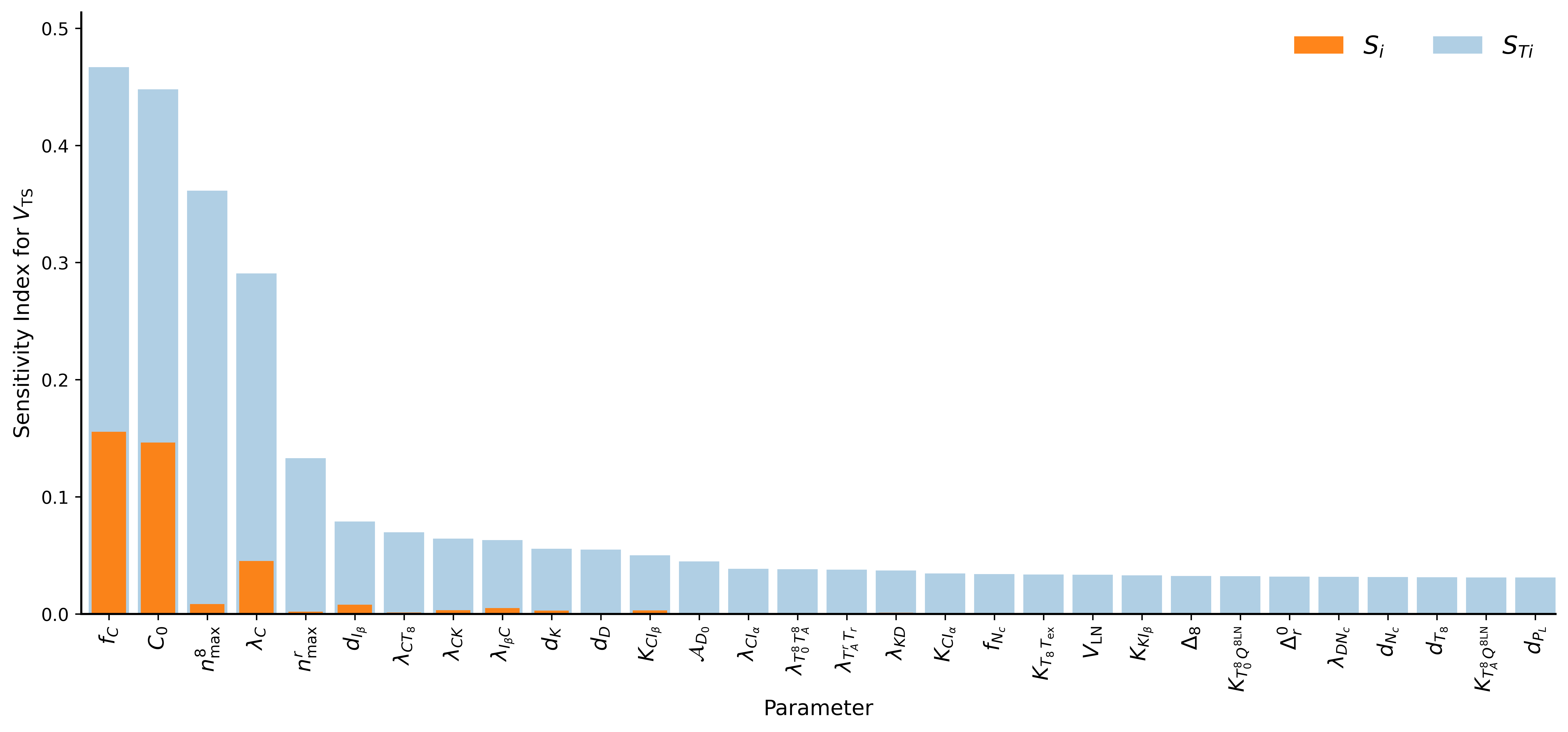}
    \caption{\label{minimalmodelsensindexplot}First-order indices ($S_i$ in orange) and total-order indices ($S_{Ti}$ in blue) for the minimal model for the top 30 parameters associated with the RMSRE of $V_\mathrm{TS}$ at 180.9 days under the standard regimen, sorted in descending order of $S_{Ti}$.}
\end{figure}

We now perform a sensitivity analysis on the minimal model, following the same procedure as for the full and reduced models. We simulated the standard regimen for 180.9 days using the initial conditions from \Cref{minimalmodelssic}, using the dde23 integrator and applying the FAST and EFAST methods to compute the first-order and total-order sensitivity indices, respectively. We sampled $10,000$ samples for each parameter, as this was sufficient to reach convergence in the indices, setting $M=4$, and considered parameter ranges corresponding to a $\pm 50\%$ deviation for each parameter, noting that $n^8_\mathrm{max}$ and $n^r_\mathrm{max}$ are integers, and were therefore rounded to the nearest integer. \\~\\
The SALib Python library \citep{Iwanaga2022} was used to perform FAST and EFAST analysis on the minimal model with the aforementioned configuration, leading to the indices in Table H.1. The sensitivity indices for the minimal model for the top 30 parameters associated with the RMSRE of $V_\mathrm{TS}$ at 180.9 days under the standard regimen, sorted in descending order of the total-order indices, are shown in \autoref{minimalmodelsensindexplot}.
We can verify that the minimal model cannot be simplified any further without losing substantial accuracy by considering the indices for its SA presented in Table H.1. Indeed, every parameter exhibits either a high first-order or total-order sensitivity index for at least one state variable, confirming that each retained process and parameter significantly influences the model's dynamics. In contrast to the full and reduced models, which possess greater dimensionality and display additional complexity, each component of the minimal model is functionally important, contributing meaningfully to the system's behaviour. This, in turn, not only enhances the model's interpretability but also increases the identifiability of individual parameters, positioning it well for future research.
\numberwithin{equation}{section}
\renewcommand{\theequation}{\thesection.\arabic{equation}}
\section{Results}
\begin{figure}[H]
    \centering
\includegraphics[width=0.6\textwidth]{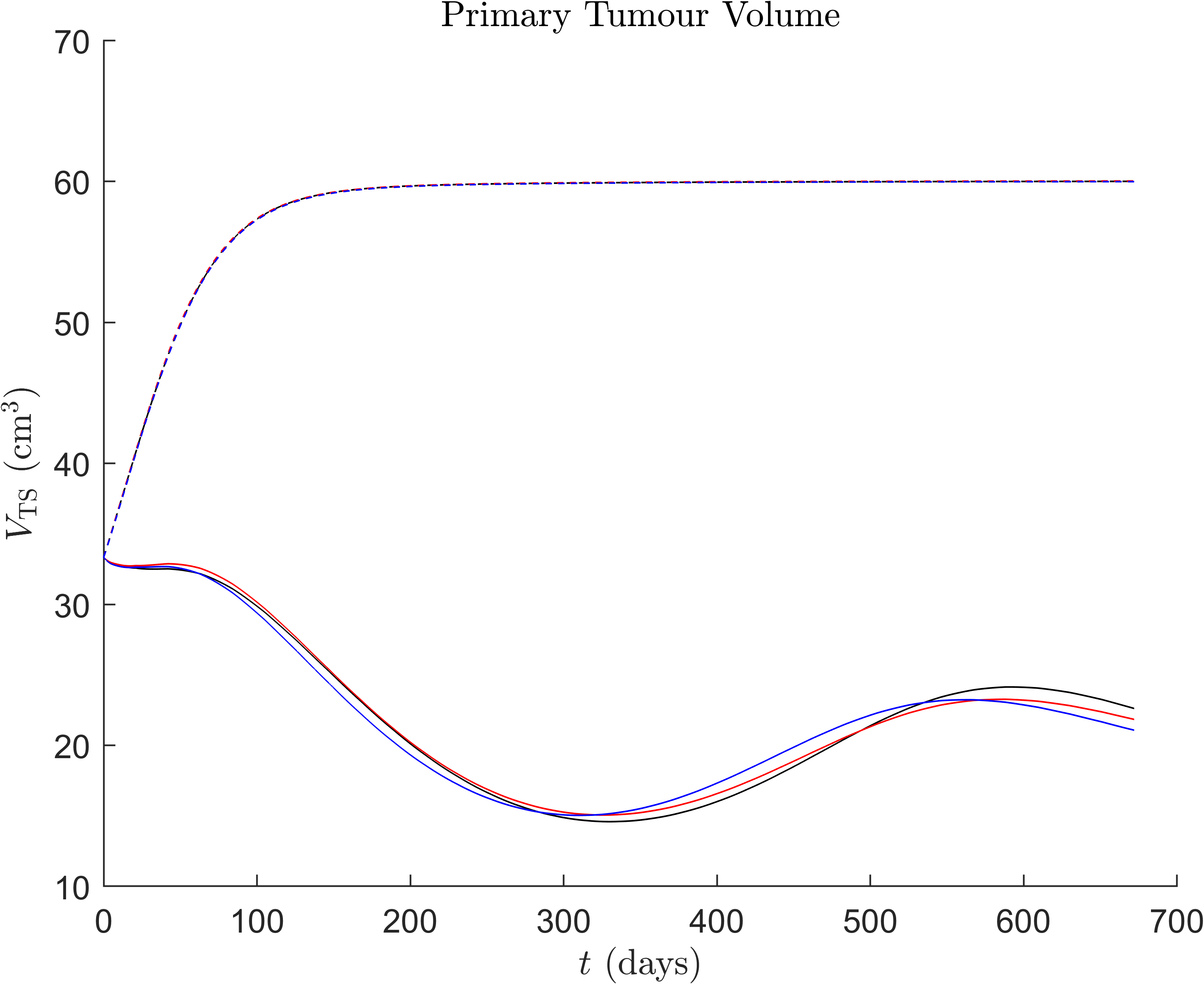}
    \caption{\label{pembrotvplot}Time traces of $V_\mathrm{TS}$ up to $672$ days from commencement in the minimal, reduced, and full models. Time traces from the full model are shown in black, from the reduced model in red, and from the minimal model in blue, with dashed lines indicating no treatment and solid lines indicating the standard regimen.}
\end{figure}

We compare the trajectories of the full, reduced, and minimal models by examining the time traces of model variables over 672 days, a time period long enough to reveal oscillatory dynamics \citep{Hawi2026meta}, in the cases of no treatment and the standard regimen, as shown in \autoref{minimalvsreducedvsfullcomparisonoftreatment}. Furthermore, time traces for the primary tumour volume, $V_\mathrm{TS}$, from the full, reduced, and minimal models in the cases of no treatment and the standard regimen are shown in \autoref{pembrotvplot}.
\begin{figure}[H]
\begin{subfigure}{\textwidth}
    \centering
    \includegraphics[width=\textwidth]{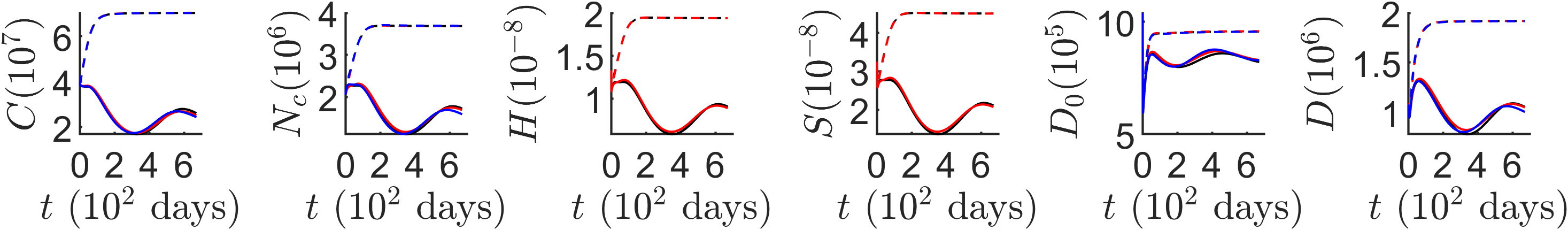}
\end{subfigure}
\end{figure}%

\begin{figure}[H]\ContinuedFloat
\begin{subfigure}{\textwidth}
    \centering
    \includegraphics[width=\textwidth]{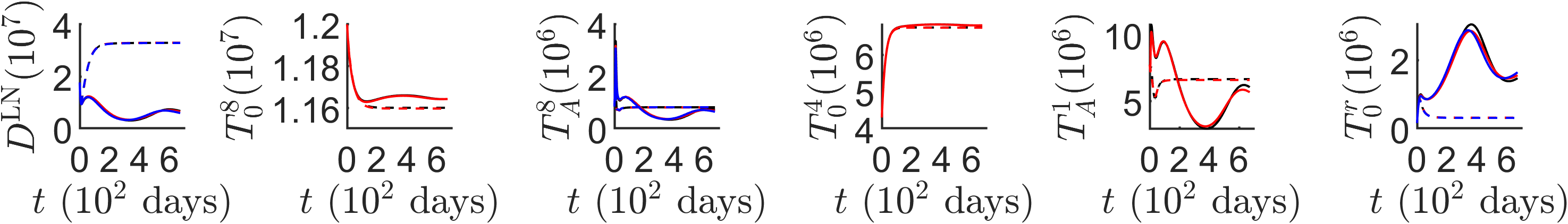}
\end{subfigure}
\end{figure}%

\begin{figure}[H]\ContinuedFloat
\begin{subfigure}{\textwidth}
    \centering
    \includegraphics[width=\textwidth]{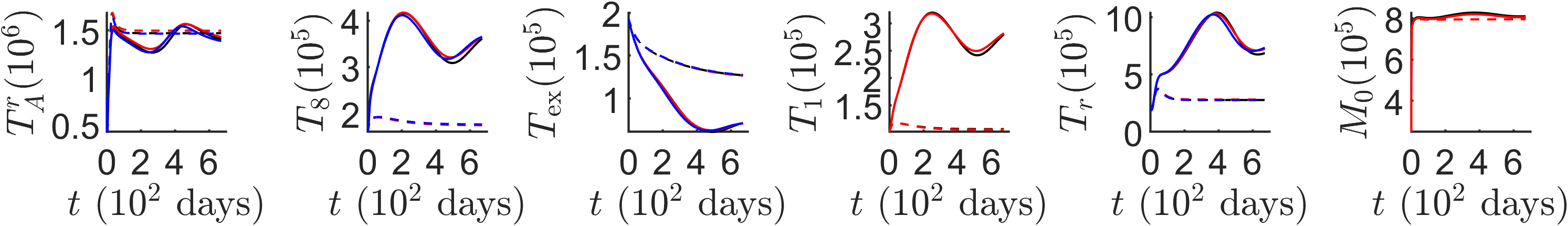}
\end{subfigure}
\end{figure}%

\begin{figure}[H]\ContinuedFloat
\begin{subfigure}{\textwidth}
    \centering
    \includegraphics[width=\textwidth]{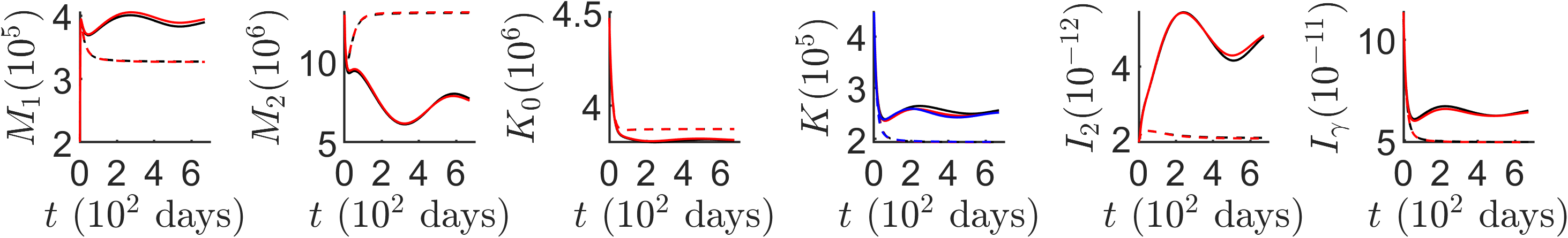}
\end{subfigure}
\end{figure}%

\begin{figure}[H]\ContinuedFloat
\begin{subfigure}{\textwidth}
    \centering
    \includegraphics[width=\textwidth]{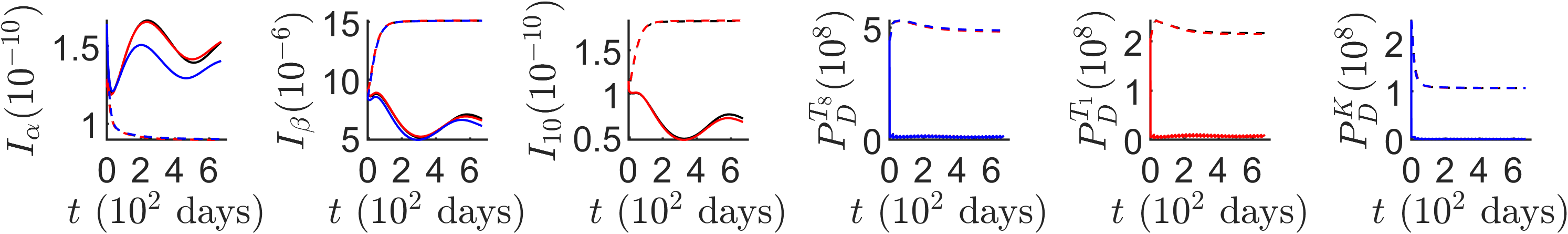}
\end{subfigure}
\end{figure}%

\begin{figure}[H]\ContinuedFloat
\begin{subfigure}{\textwidth}
    \centering
    \includegraphics[width=\textwidth]{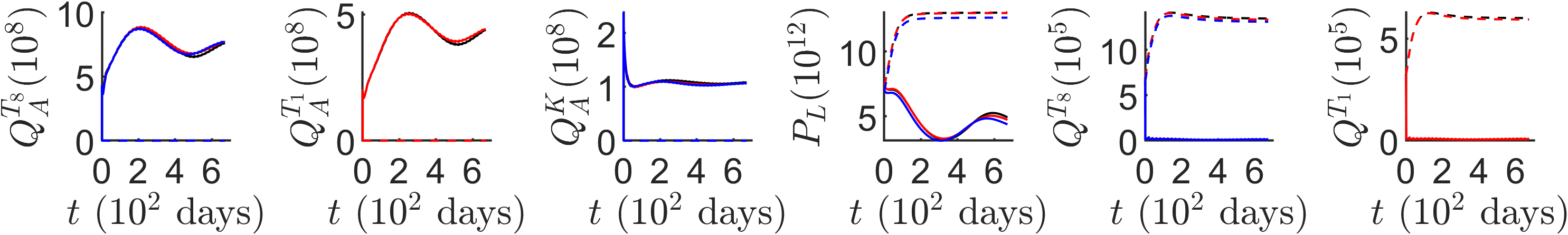}
\end{subfigure}
\end{figure}%

\begin{figure}[H]\ContinuedFloat
\begin{subfigure}{\textwidth}
    \centering
    \includegraphics[width=\textwidth]{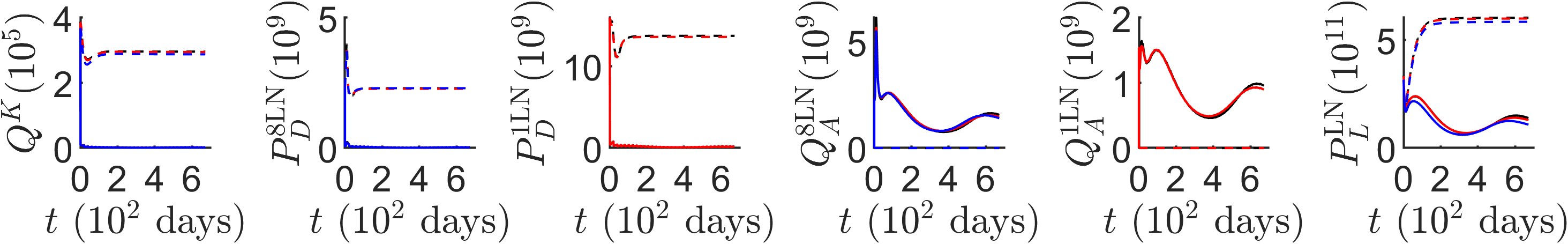}
\end{subfigure}
\end{figure}%

\begin{figure}[H]\ContinuedFloat
\begin{subfigure}{\textwidth}
    \centering
    \includegraphics[width=\textwidth]{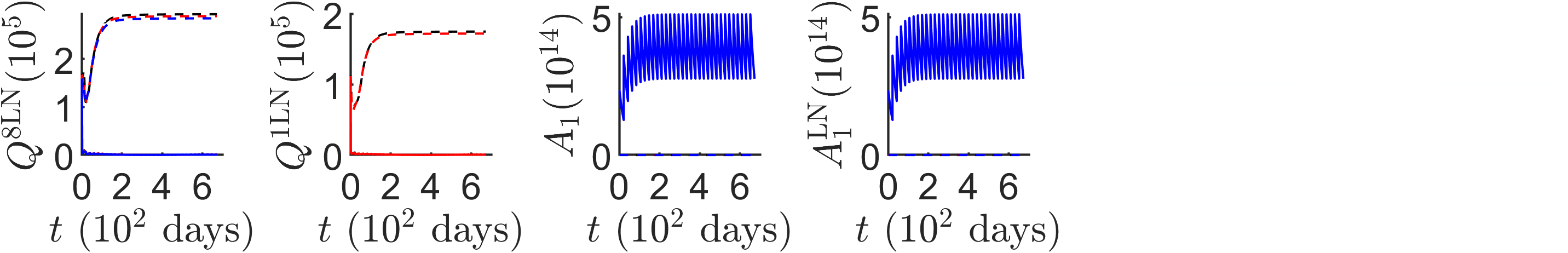}
\end{subfigure}
    \caption{\label{minimalvsreducedvsfullcomparisonoftreatment}Time traces of variables up to 672 days in the minimal, reduced, and full models, with the units of the variables as in \autoref{modelvars}. Time traces from the full model are shown in black, from the reduced model in red, and from the minimal model in blue, with dashed lines indicating no treatment and solid lines indicating the standard regimen.}
\end{figure}
\begin{figure}[H]
    \centering
\includegraphics[width=\textwidth]{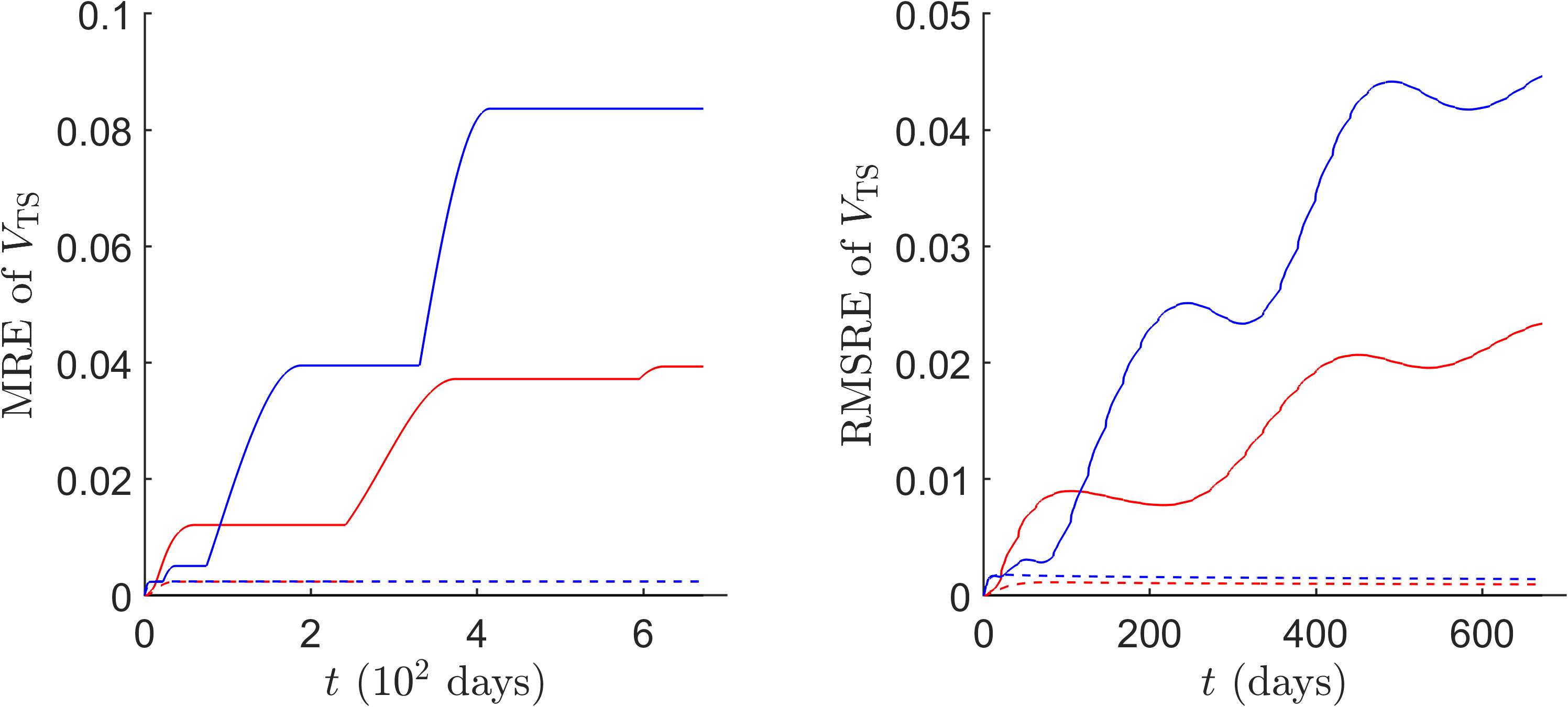}
    \caption{\label{metricplot}Time traces of MRE (left) and RMSRE (right) for $V_\mathrm{TS}$ between the reduced model and the full model (red) and the minimal model and full model (blue), from 0 days up to 672 days from commencement, with and without treatment. Dashed lines indicate no treatment, and solid lines indicate treatment with the standard regimen.}
\end{figure}

To quantitatively evaluate how well the reduced and minimal models replicate the trajectories of the full model, we consider two statistical measures: the maximum relative error (MRE) and the RMSRE. These measures are particularly well-suited for comparing models with dimensioned data and variables that span multiple orders of magnitude. \\~\\
To define these measures, we first denote $X_\mathrm{approx}$ as the variable in the reduced and minimal models corresponding to the variable $X_\mathrm{full}$ in the full model. The MRE at time $t$ is then defined as 
\begin{align}
    \operatorname{MRE}(t) &:= \max_{s \in [0,t]} \left|\frac{X_\mathrm{approx}(s) - X_\mathrm{full}(s)}{X_\mathrm{full}(s)}\right| \label{MREeqn},
    \intertext{whilst the RMSRE at time $t$ is defined as}
    \operatorname{RMSRE}(t) &:= \sqrt{\frac{1}{t}\int_{0}^{t} \left(\frac{X_\mathrm{approx}(s) - X_\mathrm{full}(s)}{X_\mathrm{full}(s)}\right)^2 \ ds} \label{RMSREeqn},
\end{align}
in identical fashion to \eqref{RMSRESAeqn}. In particular, the MRE represents the maximum deviation in a variable's trajectory between the full and simplified models. The MRE and RMSRE at 180.9 days and 672 days, both without treatment and under the standard regimen, are shown in Table I.1 and Table I.2 for variables in the reduced and minimal models, respectively. Time traces for the MRE and RMSRE of $V_\mathrm{TS}$, both without treatment and under the standard regimen, from 0 to 672 days for the reduced and minimal models, are shown in \autoref{metricplot}.
We observe that the reduced model, obtained by appropriately removing the aforementioned processes, faithfully replicates the trajectories of the full model under both treatment and no-treatment conditions. In both cases, the vast majority of MREs are below 5\%, with the exception of those corresponding to T cell concentrations in the TDLN and certain immune checkpoint and pembrolizumab-associated components. In particular, these exceptions primarily arise from very brief, transient dynamics immediately following the administration of the first pembrolizumab dose. However, these values represent only the maximum deviation over the entire timespan of integration and do not reflect the overall deviation, which is better captured by the RMSRE. The RMSRE more clearly demonstrates the validity of the reduction, with most values below 1\% regardless of whether treatment is applied. The largest RMSRE occurs for $T_A^r$ without treatment, reaching less than 10\% at 180.9 days. Furthermore, as shown in \autoref{minimalvsreducedvsfullcomparisonoftreatment}, the differences in trajectories across all state variables are minimal, confirming that the reduction performs as intended. This also demonstrates the robustness of the reduction, as the trajectories remain accurate over timespans significantly longer than those used for the sensitivity analysis, which was performed over 180.9 days.\\~\\
We can also verify the validity of the reduction for the minimal model by considering the MRE and RMSRE with and without treatment. As with the reduced model, we observe that, in both cases, the vast majority of MREs remain below 10\%, with the exception of variables related to the concentrations of the cytokines TNF and TGF-$\upbeta$, T cell concentrations in the TDLN, and immune checkpoint- and pembrolizumab-associated components. Notably, the RMSREs for most variables are very low under both treatment conditions, with the exceptions being the concentrations of TNF, TGF-$\upbeta$, and immune checkpoint-associated components. These deviations are expected due to the nature and application of QSSA, which results in different initial conditions for these variables compared to the full and reduced models. Additionally, minor differences in the steady states and initial conditions of certain immune checkpoint-associated components contribute to these deviations. However, since most of these components only influence the system through half-saturation or inhibition constants, their absolute concentrations are less critical, and their variation has minimal impact on the trajectories of other variables. Considering \autoref{minimalvsreducedvsfullcomparisonoftreatment}, we see that the minimal model successfully replicates the trajectories of its state variables well, in cases with and without treatment, even over timespans significantly longer than those used in the original sensitivity analysis, similar to the reduced model.
\section{Discussion\label{discussionsection}}
The results presented herein demonstrate that global variance-based sensitivity analysis (SA) serves a dual purpose: it quantifies parameter influence within a high-dimensional tumour--immune interaction space, and simultaneously guides a principled reduction to lower-dimensional systems. Crucially, the reduced and minimal models derived through this framework preserve the dominant dynamics under clinically relevant treatment protocols. Both models faithfully reproduce the trajectories of the full model under treatment and no-treatment conditions, while achieving substantial reductions in state-space dimensionality and parameter count.\\~\\
The full-model SA partitions the parameter space, identifying a small subset as the primary determinants of tumour growth and, consequently, of viable and necrotic cancer cell densities. These processes correspond to the logistic growth of cancer cells and their lysis by effector CD8+ T cells and NK cells, which is inhibited by TGF-$\upbeta$. Consequently, the reduction delineates the interactions essential for preserving emergent tumour--immune dynamics, distinguishing it from peripheral mechanisms that may be simplified without materially compromising the specific outputs of interest. This, in turn, not only enhances the model's interpretability but also increases the identifiability of individual parameters, positioning the framework well for future research. It is worth noting that while many parameters have low first-order sensitivity indices, they may still exhibit significant higher-order interactions, as indicated by their large total-order indices. This highlights the intrinsic nonlinearity of the full system, a characteristic that more commonly used SA methods fail to capture, thereby justifying the computational expense of the global variance-based approach performed.\\~\\
From a computational perspective, reducing the full system leads to a marked decrease in the cost of procedures such as variance-based SA, uncertainty quantification, and calibration. These tasks typically require many forward simulations; for instance, the full-model SA used $N=10,000$, corresponding to $1,570,000$ parameter combinations. Since the computational burden scales at least linearly with the number of uncertain parameters, reducing the parameter dimension therefore yields immediate savings in the number of model evaluations required for global analyses, while reducing system dimensionality decreases the time taken per simulation. Together, these effects enable substantially more extensive sensitivity and uncertainty analyses, parameter calibration, and therapeutic-regimen optimisation that would otherwise be computationally prohibitive.\\~\\
Beyond computational efficiency, the combination of reduced dimensionality and sensitivity-informed parsimony renders the minimal model amenable to rigorous mathematical analyses that are typically intractable in comprehensive immunotherapy models. These include equilibrium stability characterisation, bifurcation analysis, and structural and practical identifiability analyses. Such analyses can provide qualitative insights that are robust to parameter uncertainty, allowing for the explicit characterisation of parameter regimes associated with tumour control versus escape. In this sense, the minimal model functions as a bridge between data-calibrated numerical simulation and theory-driven mechanistic understanding.\\~\\
The reduction process also informs model-form choice. The validity of replacing the integrals in the full model with point estimates suggests that delay integro-differential equations may be unnecessarily complex for modelling the immunobiology of MSI-H/dMMR CRC, and that delay differential equations may be sufficient. Conversely, the analysis reveals that the equations governing the concentrations of unbound PD-1 receptors, PD-1/PD-L1 complexes, and PD-1/pembrolizumab complexes on effector CD8+ T cells, CD4+ T cells, and NK cells cannot be easily simplified or reduced without significantly affecting model trajectories. This implies that the current mass-action structure of these equations is already as close as possible to mechanistic fidelity while still preserving accuracy.\\~\\
Of additional interest is that many of the insights derived herein parallel observations in the locally advanced MSI-H/dMMR CRC (laMCRC) setting. Notably, a minimal model of neoadjuvant pembrolizumab therapy for laMCRC \citep{Hawi2026}, incorporating many of the reductions and eliminations presented here, demonstrates strong concordance with the corresponding full model \citep{Hawi2025local}. In particular, this framework achieved an approximate 50\% reduction in both system dimensionality and parameter count---representing a significant simplification while still accurately replicating model trajectories.\\~\\
Despite the utility of the reduction frameworks and models, several limitations exist that warrant note:
\begin{itemize}
\item The reduction is conditioned on the specific output metrics and parameter bounds defined in the global SA; regimes outside these biological priors may exhibit sensitivities not captured by the reduced and minimal models.
\item We ignored spatial effects in the models; however, their resolution can provide information about the distribution and clustering of different immune cell types in the tumour microenvironment and their clinical implications \citep{Barua2018, Maley2015}. 
\item The models do not explicitly account for additional anatomical compartments such as the spleen, nor do they directly model metastasis, which may impact the accuracy of systemic immune dynamics and tumour-specific responses.
\item The dnmMCRC models do not apply in many cases of recurrent mMCRC, as these cases often lack a primary tumour.
\end{itemize}
In this work, we have used global variance-based SA to guide model reduction of our model of dnmMCRC to construct reduced and minimal models that faithfully replicate the original model's trajectories. Moreover, the minimal model is fully self-contained and is highly extensible, offering a robust foundation for future experimentation and theoretical exploration. Importantly, the reduced model remains valuable as a simplified model for modelling and analysing the interactions and dynamics of many immune cell and cytokine types in MSI-H/dMMR CRC. Thus, one may use the reduced model as a guide for the sequential incorporation of additional variables, enabling targeted investigation of specific aspects of the tumour--immune response in MSI-H/dMMR CRC, as well as in other cancer types. This approach avoids introducing unnecessary complexity while enhancing the adaptability of the model through facilitating the straightforward integration of additional immune cell types, cytokines, or spatial compartments as needed.
\section{CRediT authorship contribution statement}
\textbf{Georgio Hawi}: conceptualisation, data curation, formal analysis, funding acquisition, investigation, methodology, project administration, resources, software, validation, visualisation, writing -- original draft, writing -- review \& editing. \\
\textbf{Peter S. Kim}: conceptualisation, formal analysis, funding acquisition, investigation, methodology, project administration, resources, supervision, validation, visualisation, writing -- original draft, writing -- review \& editing. \\
\textbf{Peter P. Lee}: conceptualisation, formal analysis, investigation, methodology, project administration, resources, supervision, validation, visualisation, writing -- original draft, writing -- review \& editing.
\section{Declaration of Competing Interests}
The authors declare that they have no known competing financial interests or personal relationships that could have appeared to influence the work reported in this paper.
\section{Data availability}
All data and procedures are available within the manuscript and its Supporting Information file.
\section{Acknowledgements}
This work was supported by an Australian Government Research Training Program Scholarship. PSK gratefully acknowledges support from the Australian Research Council Discovery Project (DP230100485).
\putbib[References.bib]
\end{bibunit}
\appendix
\counterwithin{figure}{section}
\counterwithin{table}{section}
\counterwithout{equation}{subsection}
\numberwithin{equation}{section}
\renewcommand{\theequation}{\thesection.\arabic{equation}}
\resetlinenumber
\nolinenumbers
\singlespacing
\title{Appendix A: Sensitivity analysis-guided model reduction of a mathematical model of pembrolizumab therapy for de novo metastatic MSI-H/dMMR colorectal cancer}
\maketitle
\setcounter{page}{1}
\begin{bibunit}[vancouver]
\section{Mathematical Model Derivation\label{fullmodelderivation}}
The derivation of the equations for the full model follows an almost identical approach to \cite{Hawi2025localsupp}, and are shown in \cite{Hawi2026metasupp}, but is repeated here for the sake of completeness.
\subsection{Equations for Cancer Cells, DAMPs, and DCs}
\subsubsection{Equations for Cancer Cells ($C$ and $N_c$)}
Viable cancer cells are killed by effector CD8+ T cells \cite{Raskov2020supp} and activated NK cells \cite{Zhang2020supp} through direct contact, whilst TNF and IFN-$\upgamma$ indirectly eliminate cancer cells via activating cell death pathways \cite{Josephs2018supp, Wang2008supp, Jorgovanovic2020supp}. In particular, TNF and IFN-$\upgamma$ induce the necroptosis, programmed necrotic cell death, of cancer cells \cite{Wang2008supp, Castro2018supp}. We note that TGF-$\upbeta$ and the PD-1/PD-L1 complex inhibit cancer cell lysis by CD8+ T cells \cite{Thomas2005supp, Juneja2017supp, Azuma2008supp}, and that TGF-$\upbeta$ and PD-1/PD-L1 have been shown to inhibit NK cell cytotoxicity \cite{Regis2020supp, Batlle2019supp, Hsu2018supp, Lin2024supp, Quatrini2020supp, Liu2017supp}. We assume that viable cancer cells grow logistically, as is done in many CRC models \cite{Kirshtein2020supp, dePillis2014supp, Budithi2021supp}, due to space and resource competition in the tumour microenvironment (TME). Combining these, we have
\begin{align}
  \begin{split}
  \frac{dC}{dt} &= \underbrace{\lambda_{C}C\left(1-\frac{C}{C_0}\right)}_{\text{growth}} - \underbrace{\lambda_{CT_8}T_8 \frac{1}{1+I_{\upbeta}/K_{CI_{\upbeta}}}\frac{1}{1+Q^{T_8}/K_{CQ^{T_8}}}C}_{\substack{\text{elimination by $T_8$} \\ \text{inhibited by $I_{\upbeta}$ and $Q^{T_8}$}}} - \underbrace{\lambda_{CK}K \frac{1}{1+I_{\upbeta}/K_{CI_{\upbeta}}}\frac{1}{1+Q^K/K_{CQ^K}}C}_{\substack{\text{elimination by $K$} \\ \text{inhibited by $I_\upbeta$ and $Q^K$}}} \\
  &- \underbrace{\lambda_{CI_{\upalpha}}\frac{I_{\upalpha}}{K_{CI_{\upalpha}}+I_{\upalpha}}C}_{\text{elimination by $I_{\upalpha}$}}- \underbrace{\lambda_{CI_{\upgamma}}\frac{I_{\upgamma}}{K_{CI_{\upgamma}}+I_{\upgamma}}C}_{\text{elimination by $I_{\upgamma}$}},
  \end{split} \label{cancereqnsupp} \\
  \begin{split}
    \frac{dN_c}{dt}&= \underbrace{\lambda_{CI_{\upalpha}}\frac{I_{\upalpha}}{K_{CI_{\upalpha}}+I_{\upalpha}}C}_{\text{Elimination by $I_{\upalpha}$}} + \underbrace{\lambda_{CI_{\upgamma}}\frac{I_{\upgamma}}{K_{CI_{\upgamma}}+I_{\upgamma}}C}_{\text{Elimination by $I_{\upgamma}$}} -\underbrace{d_{N_c}N_c}_{\text{Removal}}.
    \end{split}\label{necroticcelleqnsupp}
\end{align}
\subsubsection{Equation for Primary Tumour Volume ($V_\mathrm{TS}$)}
However, it is natural to consider the primary tumour volume, $V_\mathrm{TS}$, as it is experimentally easier to measure directly.
To convert between $C$, $N_c$, and $V_\mathrm{TS}$, we introduce scaling factors $f_C$ and $f_{N_c}$, so that $C(t) = f_C V_\mathrm{TS}(t)$ and $N_c(t) = f_{N_c}V_\mathrm{TS}(t)$. Hence $C(t)+N_c(t) = \left(f_{C}+f_{N_c}\right)V_\mathrm{TS}(t) \implies V_\mathrm{TS}(t) = \frac{C(t)+N_c(t)}{f_{C}+f_{N_c}}$. Thus, substituting these into the sum of \eqref{cancereqnsupp} and \eqref{necroticcelleqnsupp} leads to
\begin{equation}
\begin{split}
\frac{dV_\mathrm{TS}}{dt} &= \frac{1}{f_{C}+f_{N_c}}\left[\lambda_{C}f_{C}V_\mathrm{TS}\left(1-\frac{f_C V_\mathrm{TS}}{C_0}\right) - \lambda_{CT_8}T_8 \frac{1}{1+I_{\upbeta}/K_{CI_{\upbeta}}}\frac{1}{1+Q^{T_8}/K_{CQ^{T_8}}}f_C V_\mathrm{TS} \right. \\
&\left.- \lambda_{CK}K \frac{1}{1+I_\upbeta/K_{CI_\upbeta}}\frac{1}{1+Q^K/K_{CQ^K}}f_C V_\mathrm{TS} -d_{N_c}f_{N_c}V_\mathrm{TS}\right].
\end{split} \label{Veqnsupp}
\end{equation}
\subsubsection{Equation for HMGB1 ($H$)}
The molecule HMGB1 is released by necrotic cancer cells \cite{Fucikova2020supp} so that
\begin{equation}
 \frac{dH}{dt} = \underbrace{\lambda_{HN_c} N_c}_{\text{production by $N_c$}} - \underbrace{d_{H} H}_{\text{degradation}}. \label{Heqnsupp}
\end{equation}
\subsubsection{Equation for Calreticulin ($S$)}
Necrotic cancer cells release calreticulin \cite{Ahmed2020supp} so that
\begin{equation}
\frac{dS}{dt} = \underbrace{\lambda_{SN_c} N_c}_{\text{production by $N_c$}}- \underbrace{d_{S} S}_{\text{degradation}}. \label{Seqnsupp}
\end{equation}
\subsubsection{Equations for Immature and Mature DCs in the TS ($D_0$ and $D$)}
Immature DCs are stimulated to mature via DAMPs such as HMGB1 and calreticulin \cite{DelPrete2023supp}; however, we employ Michaelis-Menten kinetics to account for the limited rate of receptor recycling time \cite{Lai2017supp}. In addition, activated NK cells have been shown to efficiently kill immature DCs but not mature DCs; however, this is inhibited by TGF-$\upbeta$ \cite{Morandi2012supp, Vivier2008supp, Castriconi2003supp}. We also need to consider that some mature DCs migrate into the T cell zone of the TDLN and stimulate naive T cells, causing them to be activated \cite{Ruhland2020supp, Choi2017supp}. Assuming that immature DCs are supplied at a rate $\mathcal{A}_{D_0}$, we have that
\begin{align}
\begin{split}
    \frac{dD_0}{dt} &= \underbrace{\mathcal{A}_{D_0}}_{\text{source}} - \underbrace{\lambda_{DH}D_0\frac{H}{K_{DH}+H}}_{\text{$D_0 \to D$ by $H$}} - \underbrace{\lambda_{DS}D_0\frac{S}{K_{DS}+S}}_{\text{$D_0 \to D$ by $S$}}-\underbrace{\lambda_{D_0K} D_0K\frac{1}{1+I_\upbeta/K_{D_0I_\upbeta}}}_{\substack{\text{elimination by $K$} \\ \text{inhibited by $I_\upbeta$}}} - \underbrace{d_{D_0}D_0}_{\text{death}},
    \end{split} \label{D0eqnsupp} \\
  \frac{dD}{dt} &= \underbrace{\lambda_{DH}D_0\frac{H}{K_{DH}+H}}_{\text{$D_0 \to D$ by $H$}} +   \underbrace{\lambda_{DS}D_0\frac{S}{K_{DS}+S}}_{\text{$D_0 \to D$ by $S$}} - \underbrace{\lambda_{DD^\mathrm{LN}}D}_{\substack{\text{$D$ migration} \\ \text{to TDLN}}} - \underbrace{d_{D}D}_{\text{death}}. \label{Deqnsupp}
 \end{align}
\subsubsection{Equation for Mature DCs in the TDLN ($D^\mathrm{LN}$)}
\begin{figure}[ht]
    \centering
    \fbox{\includegraphics[width=0.75\textwidth]{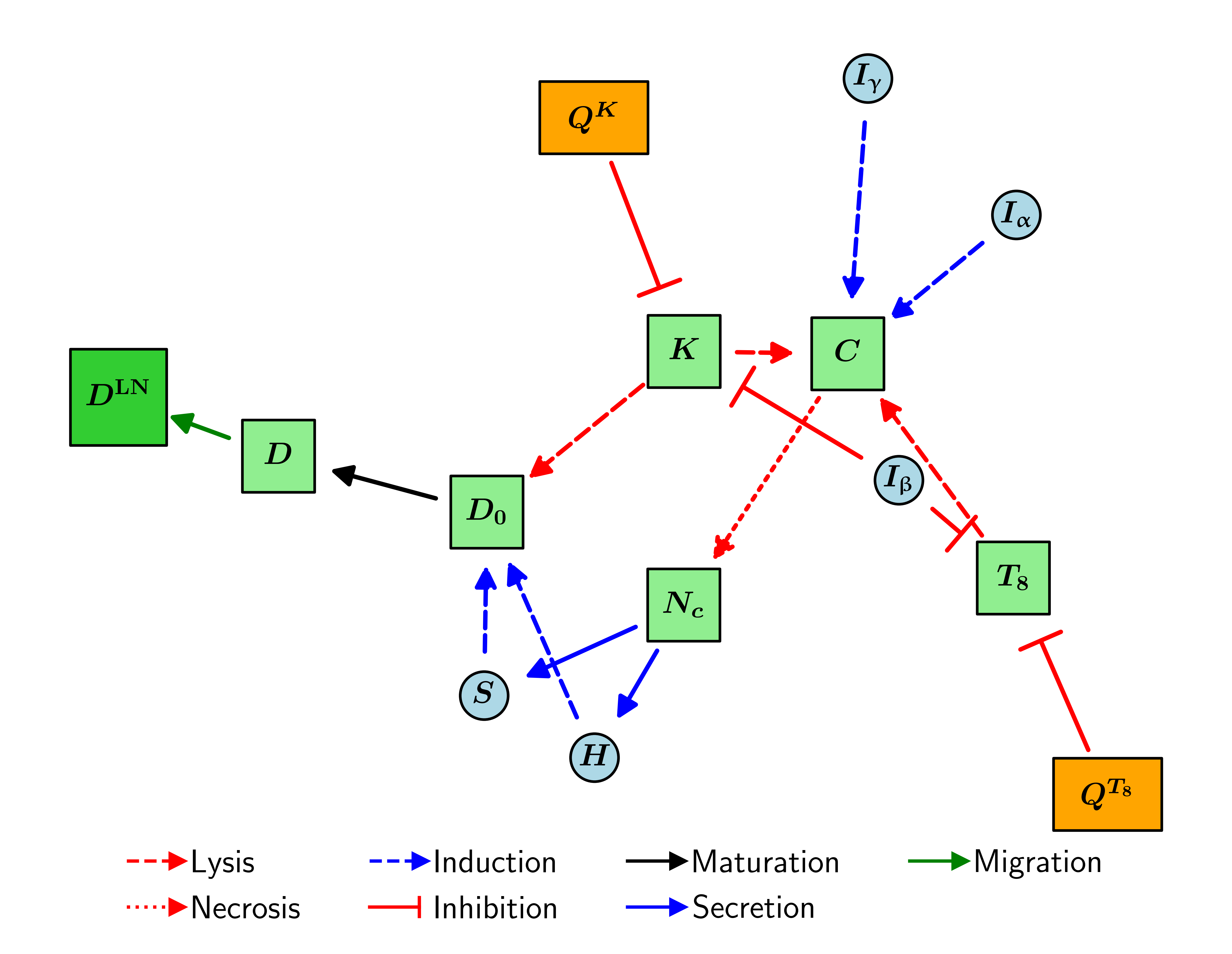}}
    \caption{\label{modeldiagramcancer}Schematic diagram of the interactions of cancer cells, DAMPs, and DCs in the full model.}
\end{figure}

We assume a fixed DC migration time of $\tau_m$ and also assume that only $e^{-d_{D}\tau_m}$ of the mature DCs that leave the TS survive migration. Taking into account the volume change between the TS and the TDLN, we have that
\begin{equation}
\frac{dD^\mathrm{LN}}{dt} = \frac{V_\mathrm{TS}}{V_\mathrm{LN}}\underbrace{\lambda_{DD^\mathrm{LN}}e^{-d_D \tau_m}D(t-\tau_m)}_{\text{$D$ migration to TDLN}} - \underbrace{d_{D}D^\mathrm{LN}}_{\text{death}}. \label{DLNeqnsupp}
 \end{equation}
A diagram encompassing the interactions of these components is shown in \autoref{modeldiagramcancer}.
\subsection{Equations for T Cells}
\subsubsection{Equation for Naive CD8+ T Cells in the TDLN ($T_0^8$)}
We assume that naive CD8+ T cells come into the TDLN at a constant rate and that they have not undergone cell division, nor will they until their activation. For simplicity, we do not consider cytokines in the TDLN, absorbing their influence into $\lambda_{T_0^8 T_A^8}$. We do, however, explicitly take into account the influence of effector Tregs and the PD-1/PD-L1 complex in the TDLN, which have been shown to inhibit T cell activation via mechanisms including limiting naive T cells from binding to mature DCs \cite{Tadokoro2006supp, Chen2022tregsupp, Li2020supp, Sakaguchi2008supp, BrunnerWeinzierl2018supp, Mizuno2019supp, Chen2023immunsupp, Peng2020supp, Arasanz2017supp}. Recalling that T cells that have become activated by mature DCs are no longer naive, and taking this all into account, leads to
\begin{equation}
  \frac{dT_0^8}{dt}=\underbrace{\mathcal{A}_{T_0^8}}_{\text{source}} -\underbrace{R^8(t)}_{\substack{\text{CD8+ T cell} \\ \text{activation}}} - \underbrace{d_{T_0^8}T_0^8}_{\text{death}}, \label{naivecd8eqnsupp}
\end{equation}
where $R^8(t)$ is defined as
\begin{equation}
    R^8(t) := \underbrace{\frac{\lambda_{T_0^8 T_A^8}e^{-d_{T_0^8}\tau_8^\mathrm{act}}D^\mathrm{LN}(t-\tau_8^\mathrm{act})T_0^8(t-\tau_8^\mathrm{act})}{\left(1+\int_{t-\tau_8^\mathrm{act}}^{t} T_A^{r}(s) \ ds/K_{T_0^8T_A^r}\right)\left(1+\int_{t-\tau_8^\mathrm{act}}^{t} Q^{8\mathrm{LN}}(s) \ ds/K_{T_0^8Q^{8\mathrm{LN}}}\right)}}_{\text{CD8+ T cell activation inhibited by $T_A^r$ and $Q^{8\mathrm{LN}}$}}.
\end{equation}
In particular, since effector Tregs and the PD-1/PD-L1 complex inhibit T cell activation during the whole activation process, it is not sufficient to consider point estimates of effector Treg and PD-1/PD-L1 concentration. Instead, we resort to considering the integrals of the concentrations of the relevant species throughout the entire $\tau_8^\mathrm{act}$ time that the CD8+ T cell takes to complete activation. This is because these integrals are proportional (with a proportionality constant of $1/{\tau_8^\mathrm{act}}$) to the average concentration of these species throughout activation, allowing us to properly incorporate their inhibition by effector Tregs and the PD-1/PD-L1 complex.
\subsubsection{Equation for Effector CD8+ T Cells in the TDLN ($T_A^8$)}
It is known that activated CD8+ T cells undergo clonal expansion in the TDLN and differentiate before they stop proliferating and migrate to the TS \cite{Harris2002supp, Catron2006supp}.\\~\\  
We assume that activated CD8+ T cells proliferate up to $n^8_\mathrm{max}$ times, upon which they stop dividing. For simplicity, we assume that the death rate of CD8+ T cells that have not completed their division program is equal to $d_{T_0^8}$, the death rate of naive CD8+ T cells, regardless of the number of cell divisions previously undergone. We also assume that only activated CD8+ T cells that have undergone $n^8_\mathrm{max}$ divisions become effector CD8+ T cells, which will leave the TDLN and migrate to the TS. Furthermore, we assume a constant cell cycle time of $\Delta_8$, except for the first cell division, which has a cycle time of $\Delta_8^0$. Thus, the duration of the activated CD8+ T cell division program to $n^8_\mathrm{max}$ divisions is given by
\begin{equation}
    \tau_{T_A^8} :=\Delta_8^0 + (n^8_\mathrm{max}-1)\Delta_8.
\end{equation}
In particular, we must take into account that some T cells will die before the division program is complete, so we must introduce a shrinkage factor of $e^{-d_{T_0^8}\tau_{T_A^8}}$. Furthermore, we must also take into account that effector Tregs and the PD-1/PD-L1 complex inhibit CD8+ T cell proliferation throughout the program \cite{Chen2022tregsupp, Li2020supp, Sakaguchi2008supp, Riley2009supp, Buchbinder2016supp}. We must also consider that some of these effector CD8+ T cells will migrate to the TS to perform effector functions. We finally assume that the death rate of CD8+ T cells that have completed their division program is equal to the death rate of CD8+ T cells in the TS. Taking this all into account leads to
\begin{equation}
\begin{split}
    \frac{dT_A^8}{dt} &= \underbrace{\frac{2^{n^8_\mathrm{max}}e^{-d_{T_0^8}\tau_{T_A^8}}R^8(t- \tau_{T_A^8})}{\left(1+\int_{t- \tau_{T_A^8}}^{t} T_A^{r}(s) \ ds/K_{T_A^8 T_A^{r}}\right)\left(1+\int_{t- \tau_{T_A^8}}^{t} Q^{8\mathrm{LN}}(s) \ ds/K_{T_A^8 Q^{8\mathrm{LN}}}\right)}}_{\text{CD8+ T cell proliferation inhibited by $T_A^r$ and $Q^{8\mathrm{LN}}$}} - \underbrace{\lambda_{T_A^8T_8}T_A^8}_{\substack{\text{$T_A^8$ migration} \\ \text{to the TS}}}-\underbrace{d_{T_8} T_A^8}_\text{death}.
\end{split}\label{TA8n8maxeqnsupp}
\end{equation}
\subsubsection{Equations for Effector and Exhausted CD8+ T Cells in the TS ($T_8$ and $T_\mathrm{ex}$)}
We assume that it takes time $\tau_a$ for effector CD8+ T cells in the TDLN to migrate to the TS. We must also account for CTL expansion due to IL-2 \cite{Rosenberg2014supp}, noting that this proliferation is inhibited by effector Tregs \cite{Chen2022tregsupp, Li2020supp, Sakaguchi2008supp}. Furthermore, the death of CD8+ T cells is resisted by IL-10 \cite{Oft2019supp, Qiu2017supp}. \\~\\
However, chronic antigen exposure can cause effector CD8+ T cells to enter a state of exhaustion, where they lose their ability to kill cancer cells, and the rate of cytokine secretion significantly decreases \cite{Blank2019supp, Lee2016supp, Shive2021supp}. We denote this exhausted CD8+ T cell population as $T_\mathrm{ex}(t)$. It has also been shown that pembrolizumab can ``reinvigorate'' these cells back into the effector state \cite{Pauken2015supp, Lee2015supp}. We model the re-invigoration and exhaustion using Michaelis-Menten terms in $A_1$ and $\int_{t-\tau_{l}}^t C(s) \ ds$ respectively, where $\tau_l$ is the median time that CD8+ T cells take to become exhausted after entering the TS. In particular, this has been shown to be more appropriate than simple mass-action kinetics as it accounts for extended antigen exposure \cite{DeBoer1995supp}. \\~\\
As such, remembering to take the volume change between the TDLN and the TS into account, this implies that
\begin{align}
\begin{split}
    \frac{d T_8}{dt} &= \frac{V_\mathrm{LN}}{V_\mathrm{TS}}\underbrace{\lambda_{T_A^8T_8}e^{-d_{T_8} \tau_a}T_A^8(t - \tau_a)}_{\text{$T_A^8$ migration to the TS}} + \underbrace{\lambda_{T_8 I_{2}}\frac{T_8 I_{2}}{K_{T_8 I_{2}}+ I_{2}}\frac{1}{1+T_r/K_{T_8T_r}}}_{\text{growth by $I_2$ inhibited by $T_r$}} \\
    &- \underbrace{\lambda_{T_8C}\frac{T_8\int_{t-\tau_{l}}^t C(s) \ ds}{K_{T_8C}+\int_{t-\tau_{l}}^t C(s) \ ds}}_{\text{$T_8 \to T_\mathrm{ex}$ from $C$ exposure}} + \underbrace{\lambda_{T_\mathrm{ex}A_1}\frac{T_\mathrm{ex}A_1}{K_{T_\mathrm{ex}A_1} + A_1}}_{\text{$T_\mathrm{ex} \to T_8$ by $A_1$}} - \underbrace{\frac{d_{T_8} T_8}{1+I_{10}/K_{T_8I_{10}}}}_{\substack{\text{death} \\ \text{inhibited by $I_{10}$}}},
    \end{split}\label{t8eqnsupp}\\
    \frac{dT_\mathrm{ex}}{dt} &= \underbrace{\lambda_{T_8C}\frac{T_8\int_{t-\tau_{l}}^t C(s) \ ds}{K_{T_8C}+\int_{t-\tau_{l}}^t C(s) \ ds}}_{\text{$T_8 \to T_\mathrm{ex}$ from $C$ exposure}} - \underbrace{\lambda_{T_\mathrm{ex}A_1}\frac{T_\mathrm{ex}A_1}{K_{T_\mathrm{ex}A_1} + A_1}}_{\text{$T_\mathrm{ex} \to T_8$ by $A_1$}} - \underbrace{\frac{d_{T_\mathrm{ex}} T_\mathrm{ex}}{1+I_{10}/K_{T_\mathrm{ex}I_{10}}}}_{\substack{\text{death} \\ \text{inhibited by $I_{10}$}}}. \label{Texeqnsupp}
\end{align}
\subsubsection{Equation for Naive CD4+ T Cells in the TDLN ($T_0^4$)}
For simplicity, we consider only the Th1 subtype that naive CD4+ T cells differentiate into upon activation, absorbing the influence of cytokines via the kinetic rate constant $\lambda_{T_0^4 T_A^1}$. Taking into account that effector Tregs and the PD-1/PD-L1 complex inhibit Th1 cell activation and some mature DCs migrate into the TDLN and activate naive CD4+ T cells, causing them to no longer be naive, and assuming that naive CD4+ T cells come into the TDLN at a rate $\mathcal{A}_{T_0^4}$, we can write a similar equation to \eqref{naivecd8eqnsupp}:
\begin{equation}
  \frac{dT_0^4}{dt} =\underbrace{\mathcal{A}_{T_0^4}}_{\text{source}} - \underbrace{R^1(t)}_{\text{Th1 cell activation}} - \underbrace{d_{T_0^4}T_0^4}_{\text{death}}, \label{naivecd4eqnsupp}
\end{equation}
where $R^1(t)$ is defined as
\begin{equation}
  R^1(t) := \underbrace{\frac{\lambda_{T_0^4 T_A^1}e^{-d_{T_0^4}\tau_4^\mathrm{act}} D^\mathrm{LN}(t-\tau_4^\mathrm{act})T_0^4(t-\tau_4^\mathrm{act})}{\left(1+\int_{t-\tau_4^\mathrm{act}}^{t} T_A^r(s) \ ds/K_{T_0^4 T_A^r}\right)\left(1+\int_{t-\tau_4^\mathrm{act}}^{t} Q^{1\mathrm{LN}}(s) \ ds/K_{T_0^4 Q^{1\mathrm{LN}}}\right)}}_{\text{Th1 cell activation inhibited by $T_A^r$ and $Q^{1\mathrm{LN}}$}}.
\end{equation}
\subsubsection{Equation for Effector Th1 Cells in the TDLN ($T_A^1$)}
We assume that Th1 cells proliferate up to $n^1_\mathrm{max}$ times, upon which they stop dividing and become effector cells. As before, we assume that the death rate of Th1 cells that have not completed their division program is equal to $d_{T_0^4}$, the death rate of naive CD4+ T cells, regardless of the number of cell divisions previously undergone. We assume a constant cell cycle time of $\Delta_1$, except for the first cell division, which has a cycle time of $\Delta_1^0$. Thus, the duration of the Th1 cell division program to $n^1_\mathrm{max}$ divisions is given by
\begin{equation}
    \tau_{T_A^1} :=\Delta_1^0 + (n^1_\mathrm{max}-1)\Delta_1.
\end{equation}
In particular, we must take into account that some Th1 cells will die before the division program is complete, so we must introduce a shrinkage factor of $e^{-d_{T_0^4}\tau_{T_A^1}}$. Furthermore, we must also take into account that effector Tregs and the PD-1/PD-L1 complex inhibit Th1 cell proliferation throughout their program. We also assume that the death rate of Th1 cells that have completed their division program is equal to the corresponding degradation rate in the TS. Taking this all into account, and incorporating effector Th1 cell migration to the TS, leads to
\begin{equation}
    \frac{dT_A^1}{dt} = \underbrace{\frac{2^{n^1_\mathrm{max}}e^{-d_{T_0^4}\tau_{T_A^1}} R^1(t- \tau_{T_A^1})}{\left(1+\int_{t- \tau_{T_A^1}}^{t} Q^{1\mathrm{LN}}(s) \ ds/K_{T_A^1 Q^{1\mathrm{LN}}}\right)\left(1+\int_{t- \tau_{T_A^1}}^{t} T_A^r(s) \ ds/K_{T_A^1 T_A^r}\right)}}_{\text{Th1 cell proliferation inhibited by $T_A^r$ and $Q^{1\mathrm{LN}}$}} - \underbrace{\lambda_{T_A^1T_1}T_A^1}_{\substack{\text{$T_A^1$ migration} \\ \text{to the TS}}}-\underbrace{d_{T_1} T_A^1}_\text{death} \label{TA1n1maxeqnsupp}.
\end{equation}
\subsubsection{Equation for Effector Th1 Cells in the TS ($T_1$)}
We assume that it takes time $\tau_a$ for these cells to migrate to the TS. We take into account the fact that IL-2 induces the growth of effector Th1 cells \cite{Choudhry2018supp}, noting that this proliferation is inhibited by effector Tregs \cite{Chen2022tregsupp, Li2020supp, Sakaguchi2008supp}. Furthermore, the PD-1/PD-L1 axis converts Th1 cells to Tregs \cite{Amarnath2011supp, Cai2019supp}, a process we consider to be mediated by the PD-1/PD-L1 complex on Th1 cells. Thus, we have that
\begin{align}
\frac{dT_1}{dt} &= \frac{V_\mathrm{LN}}{V_\mathrm{TS}}\underbrace{\lambda_{T_A^1T_1}e^{-d_{T_1} \tau_a}T_A^1(t-\tau_a)}_{\text{$T_A^1$ migration to the TS}} +\underbrace{\lambda_{T_1 I_{2}}\frac{T_1 I_{2}}{K_{T_1 I_{2}}+ I_{2}}\frac{1}{1+T_r/K_{T_1T_r}}}_{\text{growth by $I_2$ inhibited by $T_r$}} - \underbrace{\lambda_{T_1T_r}T_1\frac{Q^{T_1}}{K_{T_1Q^{T_1}} + Q^{T_1}}}_{\text{$T_1 \to T_r$ by $Q^{T_1}$}} - \underbrace{d_{T_1}T_1}_{\text{death}} \label{th1eqnsupp}.
\end{align}
\subsubsection{Equation for Naive Tregs in the TDLN ($T_0^r$)}
Finally, we consider the concentration of naive Tregs in the TDLN, following the same procedure as for CD8+ T cells and Th1 cells. We absorb the influence of cytokines on Treg activation via the kinetic rate constant $\lambda_{T_0^r T_A^r}$. We also take into account that some mature DCs migrate into the TDLN and activate naive Tregs, causing them to no longer be naive. Assuming that naive Tregs come into the TDLN at a rate $\mathcal{A}_{T_0^r}$, we can write a similar equation to \eqref{naivecd8eqnsupp} and \eqref{naivecd4eqnsupp}:
\begin{equation}
  \frac{dT_0^r}{dt} =\underbrace{\mathcal{A}_{T_0^r}}_{\text{source}} - \underbrace{R^r(t)}_{\text{Treg activation}} - \underbrace{d_{T_0^r}T_0^r}_{\text{death}}, \label{naivetregeqnsupp}
\end{equation}
where $R^r(t)$ is defined as
\begin{equation}
  R^r(t) := \underbrace{\lambda_{T_0^r T_A^r}e^{-d_{T_0^r}\tau_r^\mathrm{act}} D^\mathrm{LN}(t-\tau_r^\mathrm{act})T_0^r(t-\tau_r^\mathrm{act})}_{\text{Treg activation}}.
\end{equation}
\subsubsection{Equation for Effector Tregs in the TDLN ($T_A^r$)}
We assume that activated Tregs proliferate up to $n^r_\mathrm{max}$ times, upon which they stop dividing and become effector Tregs. As before, we assume that the death rate of Tregs that have not completed their division program is equal to $d_{T_0^r}$, the death rate of naive Tregs. We assume a constant cell cycle time of $\Delta_r$, except for the first cell division, which has a cycle time of $\Delta_r^0$. Thus, the duration of the activated Treg division program to $n^r_\mathrm{max}$ divisions is given by
\begin{equation}
    \tau_{T_A^r} :=\Delta_r^0 + (n^r_\mathrm{max}-1)\Delta_r.
\end{equation}
In particular, we must take into account that some T cells will die before the division program is complete, so we must introduce a shrinkage factor of $e^{-d_{T_0^r}\tau_{T_A^r}}$. We also assume that the death rate of effector Tregs in the TDLN is equal to the corresponding degradation rate in the TS. Taking this all into account, and incorporating effector Treg migration to the TS, leads to      
\begin{equation}
\begin{split}
    \frac{dT_A^r}{dt} &= \underbrace{2^{n^r_\mathrm{max}}e^{-d_{T_0^r}\tau_{T_A^r}} R^r(t- \tau_{T_A^r})}_{\text{Treg proliferation}} - \underbrace{\lambda_{T_A^rT_r}T_A^r}_{\substack{\text{$T_A^r$ migration} \\ \text{to the TS}}}-\underbrace{d_{T_r} T_A^r}_\text{death}.
\end{split}\label{TArnrmaxeqnsupp}
\end{equation}
\subsubsection{Equation for Effector Tregs in the TS ($T_r$)}
\begin{figure}[ht]
    \centering
    \fbox{\includegraphics[width=0.75\textwidth]{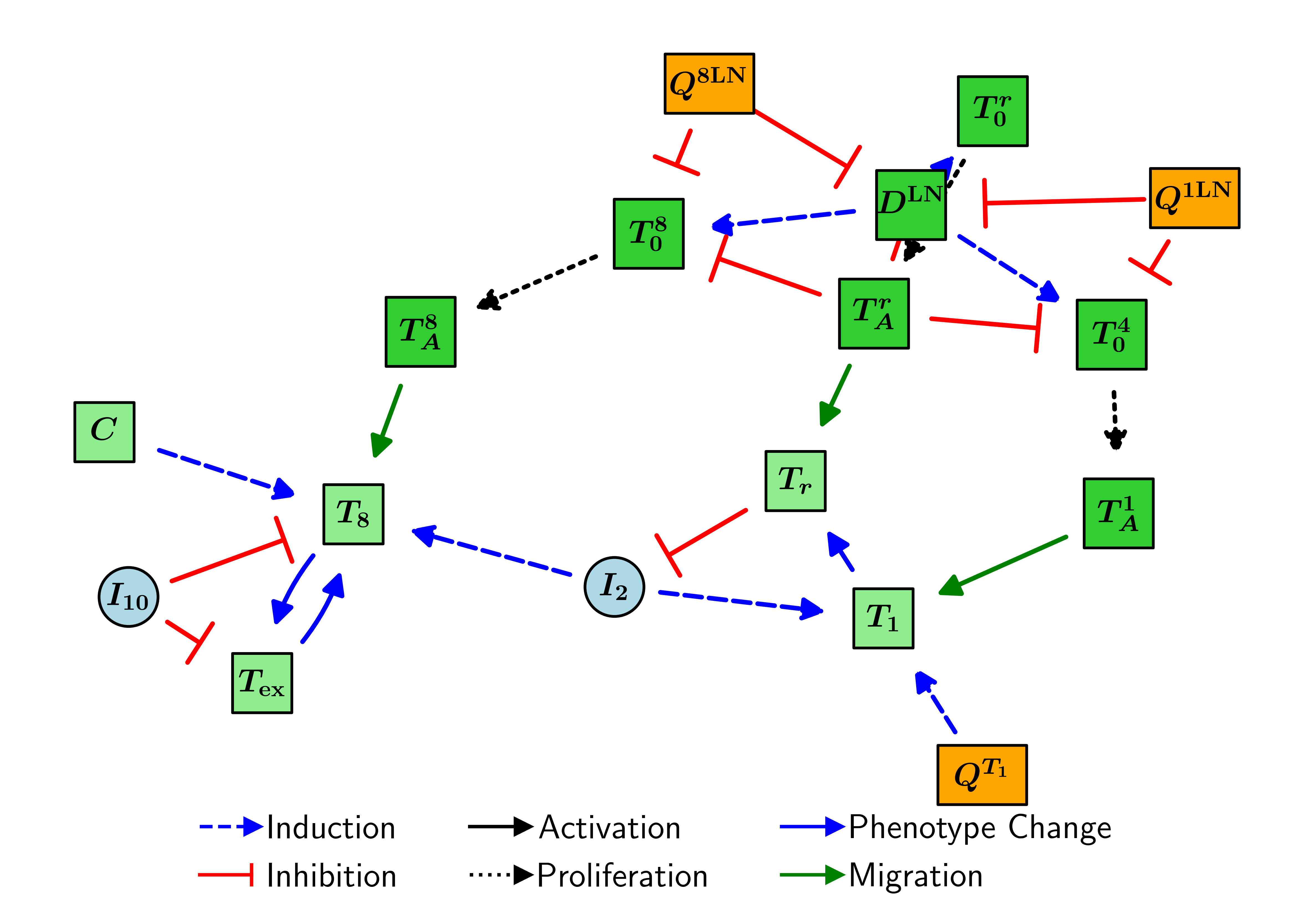}}
    \caption{\label{modeldiagramtcell}Schematic diagram of the interactions of T cells in the full model.}
\end{figure}

Assuming that it also takes time $\tau_a$ for Tregs to migrate to the TS, we have that
\begin{equation}
\frac{dT_r}{dt} = \frac{V_\mathrm{LN}}{V_\mathrm{TS}}\underbrace{\lambda_{T_A^rT_r}e^{-d_{T_r} \tau_a}T_A^r(t-\tau_a)}_{\text{$T_A^r$ migration to the TS}} + \underbrace{\lambda_{T_1T_r}T_1\frac{Q^{T_1}}{K_{T_1Q^{T_1}} + Q^{T_1}}}_{\text{$T_1 \to T_r$ by $Q^{T_1}$}} - \underbrace{d_{T_r}T_r}_{\text{death}}. \label{tregeqnsupp}
\end{equation}
A diagram encompassing the interactions of these components is shown in \autoref{modeldiagramtcell}.
\subsection{Equations for Other Immune Cells in the TS}
\subsubsection{Equations for Naive, M1, and M2 Macrophages ($M_0$, $M_1$, and $M_2$)}
TNF and IFN-$\upgamma$ polarise naive macrophages into M1 macrophages \cite{Kroner2014supp, Kratochvill2015supp, Nathan1983supp, Ivashkiv2018supp}, whilst IL-10 and TGF-$\upbeta$ polarise naive macrophages into the M2 phenotype \cite{Ambade2016supp, Chen2019TAMsupp, Zhang2016supp}. In addition, TGF-$\upbeta$ induces M1 macrophages to convert into M2 macrophages \cite{Zhang2016supp}. Furthermore, M2 macrophages change phenotype to M1 under the influence of TNF \cite{Kroner2014supp} and IFN-$\upgamma$ \cite{Ye2021supp}. Assuming a production rate $\mathcal{A}_{M_0}$ of naive macrophages, we thus have that
\begin{align}
 \begin{split}
 \frac{dM_0}{dt} &=\underbrace{\mathcal{A}_{M_0}}_{\text{source}} - \underbrace{\lambda_{M_1I_\upalpha}M_0\frac{I_\upalpha}{K_{M_1 I_\upalpha}+I_\upalpha}}_{\text{$M_0 \to M_1$ by $I_{\upalpha}$}} - \underbrace{\lambda_{M_1I_\upgamma}M_0\frac{I_\upgamma}{K_{M_1 I_\upgamma}+I_\upgamma}}_{\text{$M_0 \to M_1$ by $I_\upgamma$}} - \underbrace{\lambda_{M_2I_{10}}M_0\frac{I_{10}}{K_{M_2 I_{10}}+I_{10}}}_{\text{$M_0 \to M_2$ by $I_{10}$}} \\
 &- \underbrace{\lambda_{M_2I_\upbeta}M_0\frac{I_\upbeta}{K_{M_2 I_\upbeta}+I_\upbeta}}_{\text{$M_0 \to M_2$ by $I_\upbeta$}} - \underbrace{d_{M_0}M_0}_{\text{degradation}},
 \end{split} \label{M0eqnsupp} \\
 \begin{split}
 \frac{dM_1}{dt} &= \underbrace{\lambda_{M_1I_\upalpha}M_0\frac{I_\upalpha}{K_{M_1 I_\upalpha}+I_\upalpha}}_{\text{$M_0 \to M_1$ by $I_{\upalpha}$}} + \underbrace{\lambda_{M_1I_\upgamma}M_0\frac{I_\upgamma}{K_{M_1 I_\upgamma}+I_\upgamma}}_{\text{$M_0 \to M_1$ by $I_\upgamma$}} + \underbrace{\lambda_{MI_{\upgamma}}M_2\frac{I_{\upgamma}}{K_{MI_{\upgamma}}+I_{\upgamma}}}_{\text{$M_2 \to M_1$ by $I_{\upgamma}$}} + \underbrace{\lambda_{MI_{\upalpha}}M_2 \frac{I_{\upalpha}}{K_{MI_{\upalpha}}+I_{\upalpha}}}_{\text{$M_2 \to M_1$ by $I_{\upalpha}$}} \\
 &- \underbrace{\lambda_{MI_{\upbeta}}M_1 \frac{I_{\upbeta}}{K_{MI_{\upbeta}}+I_{\upbeta}}}_{\text{$M_1 \to M_2$ by $I_{\upbeta}$}} - \underbrace{d_{M_1}M_1}_{\text{degradation}},
 \end{split} \label{M1eqnsupp} \\
\begin{split}
 \frac{dM_2}{dt} &= \underbrace{\lambda_{M_2I_{10}}M_0\frac{I_{10}}{K_{M_2 I_{10}}+I_{10}}}_{\text{$M_0 \to M_2$ by $I_{10}$}} + \underbrace{\lambda_{M_2I_\upbeta}M_0\frac{I_\upbeta}{K_{M_2 I_\upbeta}+I_\upbeta}}_{\text{$M_0 \to M_2$ by $I_\upbeta$}} - \underbrace{\lambda_{MI_{\upgamma}}M_2\frac{I_{\upgamma}}{K_{MI_{\upgamma}}+I_{\upgamma}}}_{\text{$M_2 \to M_1$ by $I_{\upgamma}$}} - \underbrace{\lambda_{MI_{\upalpha}}M_2 \frac{I_{\upalpha}}{K_{MI_{\upalpha}}+I_{\upalpha}}}_{\text{$M_2 \to M_1$ by $I_{\upalpha}$}} \\
 &+ \underbrace{\lambda_{MI_{\upbeta}}M_1 \frac{I_{\upbeta}}{K_{MI_{\upbeta}}+I_{\upbeta}}}_{\text{$M_1 \to M_2$ by $I_{\upbeta}$}} - \underbrace{d_{M_2}M_2}_{\text{degradation}}.
 \end{split} \label{M2eqnsupp}
\end{align}
\subsubsection{Equations for Resting and Activated NK Cells ($K_0$ and $K$)}
\begin{figure}[ht]
    \centering
    \fbox{\includegraphics[width=0.75\textwidth]{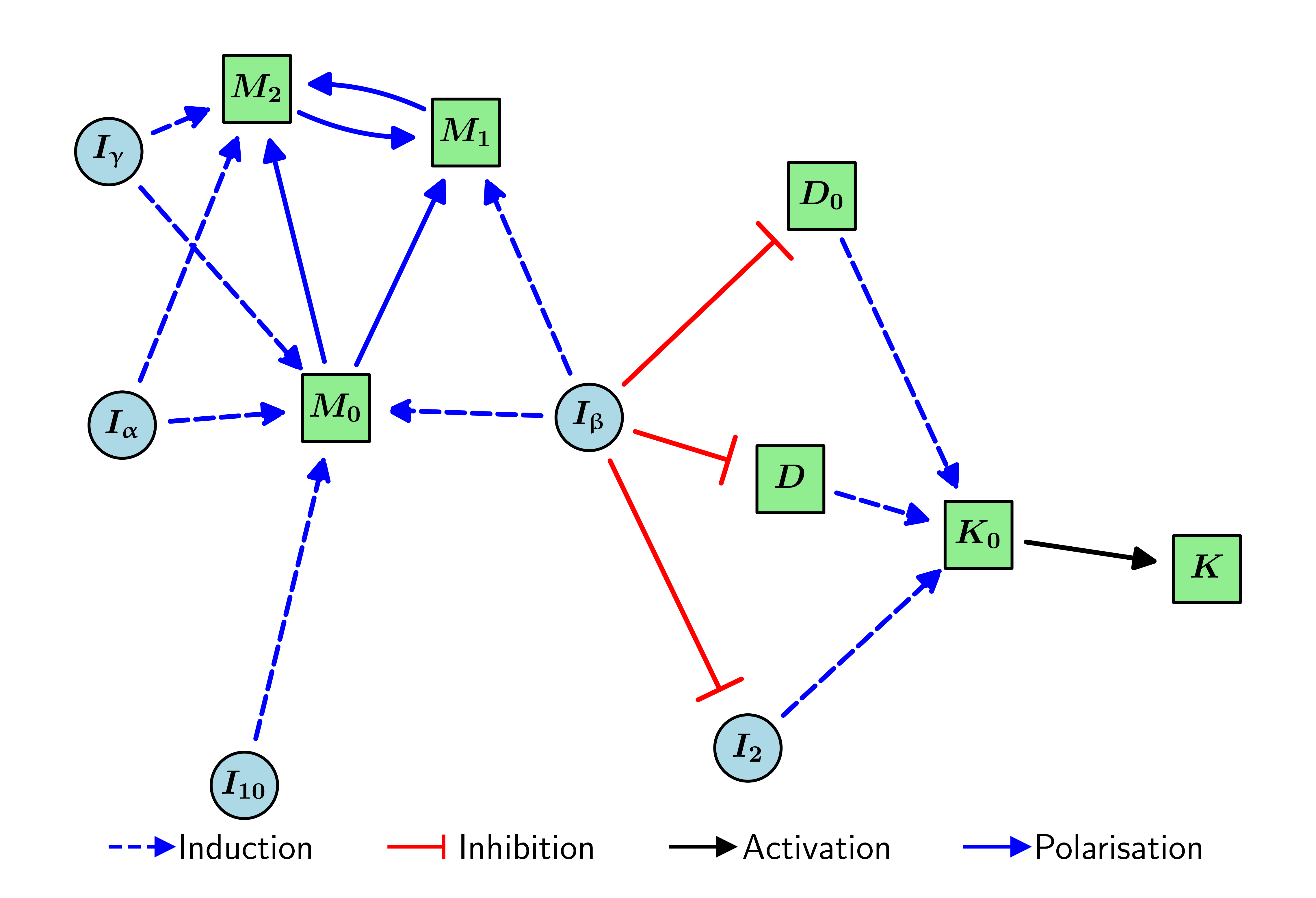}}
    \caption{\label{modeldiagramimmunecellTS}Schematic diagram of the interactions of macrophages and NK cells in the full model.}
\end{figure}

Resting NK cells are activated by IL-2 \cite{Konjevi2019supp, Widowati2020supp} and immature and mature DCs \cite{Ferlazzo2002supp}. However, NK cell activation is inhibited by TGF-$\upbeta$ \cite{Viel2016supp}. Thus, assuming a supply rate $\mathcal{A}_{K_0}$ of resting NK cells, we have that
\begin{align}
    \frac{dK_0}{dt} &= \underbrace{\mathcal{A}_{K_0}}_{\text{source}} - \left(\underbrace{\lambda_{KI_2}K_0\frac{I_2}{K_{KI_2}+I_2}}_{\text{$K_0 \to K$ by $I_2$}} +\underbrace{\lambda_{KD_{0}}K_0\frac{D_{0}}{K_{KD_0}+D_0}}_{\text{$K_0 \to K$ by $D_0$}}+\underbrace{\lambda_{KD}K_0\frac{D}{K_{KD}+D}}_{\text{$K_0 \to K$ by $D$}}\right)\underbrace{\frac{1}{1+I_\upbeta/K_{KI_\upbeta}}}_{\substack{\text{activation}\\ \text{inhibited by $I_\upbeta$}}} - \underbrace{d_{K_0}K_0}_{\text{degradation}}, \label{K0eqnsupp}\\
    \frac{dK}{dt} &= \left(\underbrace{\lambda_{KI_2}K_0\frac{I_2}{K_{KI_2}+I_2}}_{\text{$K_0 \to K$ by $I_2$}} + \underbrace{\lambda_{KD_{0}}K_0\frac{D_{0}}{K_{KD_0}+D_0}}_{\text{$K_0 \to K$ by $D_0$}}+\underbrace{\lambda_{KD}K_0\frac{D}{K_{KD}+D}}_{\text{$K_0 \to K$ by $D$}}\right)\underbrace{\frac{1}{1+I_\upbeta/K_{KI_\upbeta}}}_{\substack{\text{activation}\\ \text{inhibited by $I_\upbeta$}}} - \underbrace{d_{K}K}_{\text{degradation}}. \label{Keqnsupp}
  \end{align}
A diagram encompassing the interactions of these components is shown in \autoref{modeldiagramimmunecellTS}.
\subsection{Equations for Cytokines}
\subsubsection{Equation for IL-2 ($I_2$)}
IL-2 is produced by effector CD8+ T cells \cite{Kish2005supp, DSouza2004supp} and Th1 cells \cite{Hwang2005supp}, so that
\begin{equation}
 \frac{dI_2}{dt}= \underbrace{\lambda_{I_2 T_8}T_8}_{\text{production by $T_8$}} + \underbrace{\lambda_{I_2 T_1}T_1}_{\text{production by $T_1$}} - \underbrace{d_{I_2}I_2}_{\text{degradation}}. \label{il2eqnsupp}
\end{equation}
\subsubsection{Equation for IFN-$\upgamma$ ($I_\upgamma$)}
IFN-$\upgamma$ is produced by effector CD8+ T cells \cite{Bhat2017supp} and Th1 cells \cite{Szabo2002supp, Castro2018supp}, with both expressions being inhibited by Tregs \cite{Sojka2011supp}. Furthermore, activated NK cells also produce IFN-$\upgamma$ \cite{Cui2019supp}. Thus,
\begin{equation}
 \frac{dI_{\upgamma}}{dt}=\left(\underbrace{\lambda_{I_{\upgamma} T_8}T_8}_{\text{production by $T_8$}} + \underbrace{\lambda_{I_{\upgamma} T_1}T_1}_{\text{production by $T_1$}}\right)\underbrace{\frac{1}{1+T_r/K_{I_{\upgamma}T_r}}}_{\text{inhibition by $T_r$}} + \underbrace{\lambda_{I_{\upgamma} K}K}_{\text{production by $K$}} - \underbrace{d_{I_\upgamma}I_\upgamma}_{\text{degradation}}. \label{ifngammaeqnsupp}
\end{equation}
\subsubsection{Equation for TNF ($I_\upalpha$)}
TNF is produced by effector CD8+ T cells \cite{Hoekstra2021supp, Mehta2018supp} and Th1 cells \cite{Basu2021supp, Zhang2014supp}, M1 macrophages \cite{Chen2023supp}, and activated NK cells \cite{Fauriat2010supp, Wang2011supp}. Hence,
\begin{equation}
 \frac{dI_\upalpha}{dt}= \underbrace{\lambda_{I_{\upalpha}T_8}T_8}_{\text{production by $T_8$}} + \underbrace{\lambda_{I_{\upalpha}T_1}T_1}_{\text{production by $T_1$}} + \underbrace{\lambda_{I_{\upalpha}M_1}M_1}_{\text{production by $M_1$}} + \underbrace{\lambda_{I_{\upalpha}K}K}_{\text{production by $K$}}- \underbrace{d_{I_{\upalpha}}I_{\upalpha}}_{\text{degradation}}. \label{tnfeqnsupp}
\end{equation}
\subsubsection{Equation for TGF-$\upbeta$ ($I_\upbeta$)}
TGF-$\upbeta$ is produced by viable cancer cells \cite{Massagu2008supp}, effector Tregs \cite{Tang2008supp} and M2 macrophages \cite{Chen2019TAMsupp, Nuez2018supp}. Thus,
\begin{equation}
 \frac{dI_\upbeta}{dt}= \underbrace{\lambda_{I_{\upbeta}C}C}_{\text{production by $C$}} + \underbrace{\lambda_{I_{\upbeta}T_r}T_r}_{\text{production by $T_r$}} + \underbrace{\lambda_{I_{\upbeta}M_2}M_2}_{\text{production by $M_2$}} - \underbrace{d_{I_{\upbeta}}I_{\upbeta}}_{\text{degradation}}. \label{tgfbetaeqnsupp}
\end{equation}
\subsubsection{Equation for IL-10 ($I_{10}$)}
\begin{figure}[ht]
    \centering
    \fbox{\includegraphics[width=0.75\textwidth]{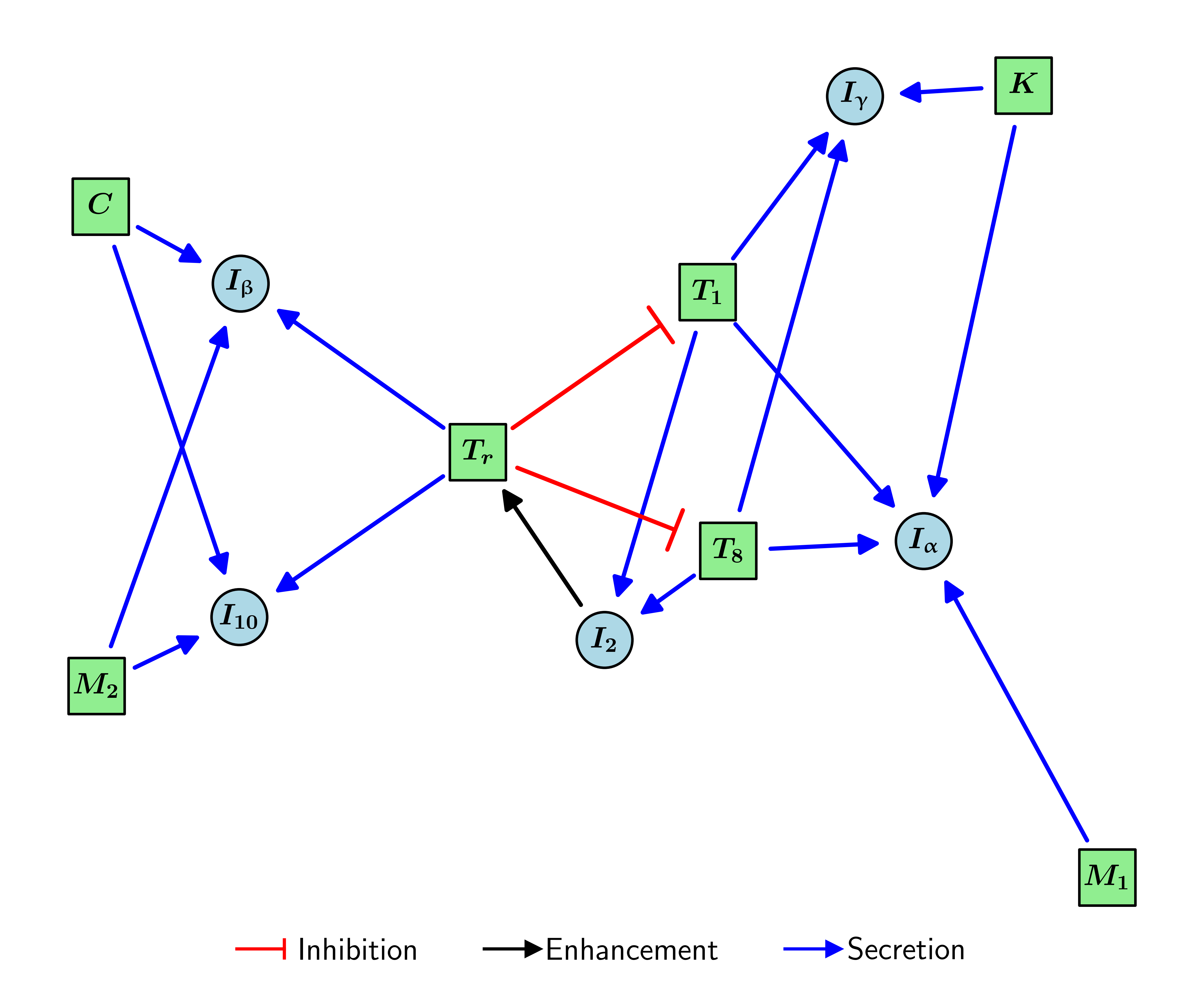}}
    \caption{\label{modeldiagramcytokines}Schematic diagram of the interactions of cytokines in the full model.}
\end{figure}

IL-10 is produced by viable cancer cells \cite{Itakura2011supp, KrgerKrasagakes1994supp} and M2 macrophages \cite{Chen2019supp, Qi2016supp}. Additionally, effector Tregs secrete IL-10 \cite{Moore2001supp} with IL-2 enhancing this production \cite{TsujiTakayama2008supp, TsujiTakayama2008n2supp}. Hence,
\begin{equation}
\frac{dI_{10}}{dt}= \underbrace{\lambda_{I_{10}C}C}_{\text{production by $C$}} + \underbrace{\lambda_{I_{10}M_2}M_2}_{\text{production by $M_2$}} + \underbrace{\lambda_{I_{10}T_{r}}T_r\left(1+\lambda_{I_{10}I_{2}}\frac{I_{2}}{K_{I_{10}I_{2}}+I_{2}}\right)}_{\text{production by $T_r$ enhanced by $I_2$}} - \underbrace{d_{I_{10}}I_{10}}_{\text{degradation}}. \label{il10eqnsupp}
\end{equation}
A diagram encompassing the interactions of cytokines is shown in \autoref{modeldiagramcytokines}.
\subsection{Equations for Immune Checkpoint-Associated Components in the TS}
\subsubsection{Equations for Unbound PD-1 receptors on Cells in the TS ($P_D^{T_8}$, $P_D^{T_1}$, $P_D^{K}$)}
It is known that PD-1 is expressed on the surface of effector CD8+ T cells \cite{Saito2012supp, Wu2014supp, Jiang2015supp}, effector Th1 cells \cite{LuzCrawford2012supp} and activated NK cells \cite{Liu2017supp, AlMterin2022supp, Hsu2018supp}. We assume that the rate of PD-1 synthesis is proportional to the concentration of the cell expressing it.
However, unbound PD-1 receptors on these PD-1-expressing cells can bind to either pembrolizumab or PD-L1, forming the PD-1/pembrolizumab and PD-1/PD-L1 complexes, respectively, resulting in the depletion of unbound PD-1 molecules \cite{han2020pdsupp, Lin2008supp}. For simplicity, we assume that the formation and dissociation rates of the PD-1/PD-L1 and PD-1/pembrolizumab complexes are invariant of the type of cell expressing PD-1. Considering unbound PD-1 receptors on effector CD8+ T cells in the TS at first, and taking into account the degradation of PD-1 receptors, this motivates the equation for $P_D^{T_8}$ to be
\begin{align}
    \frac{dP_D^{T_8}}{dt} &= \underbrace{\lambda_{P_D^{T_8}}T_8}_{\text{synthesis}} + \underbrace{\lambda_{Q_A}Q_A^{T_8}}_{\substack{\text{dissociation} \\ \text{of $Q_A^{T_8}$}}} + \underbrace{\lambda_{Q}Q^{T_8}}_{\substack{\text{dissociation} \\ \text{of $Q^{T_8}$}}} - \underbrace{\lambda_{P_DA_1}P_D^{T_8}A_1}_{\text{binding to $A_1$}} - \underbrace{\lambda_{P_DP_L}P_D^{T_8}P_L}_{\text{binding to $P_L$}} - \underbrace{d_{P_D}P_D^{T_8}}_{\text{degradation}}. \label{PD8eqntempsupp}
\intertext{Similarly, we have that}
    \frac{dP_D^{T_1}}{dt} &= \underbrace{\lambda_{P_D^{T_1}}T_1}_{\text{synthesis}} + \underbrace{\lambda_{Q_A}Q_A^{T_1}}_{\substack{\text{dissociation} \\ \text{of $Q_A^{T_1}$}}} + \underbrace{\lambda_{Q}Q^{T_1}}_{\substack{\text{dissociation} \\ \text{of $Q^{T_1}$}}} - \underbrace{\lambda_{P_DA_1}P_D^{T_1}A_1}_{\text{binding to $A_1$}} - \underbrace{\lambda_{P_DP_L}P_D^{T_1}P_L}_{\text{binding to $P_L$}} - \underbrace{d_{P_D}P_D^{T_1}}_{\text{degradation}}, \label{PD1eqntempsupp} \\
    \frac{dP_D^{K}}{dt} &= \underbrace{\lambda_{P_D^{K}}K}_{\text{synthesis}} + \underbrace{\lambda_{Q_A}Q_A^{K}}_{\substack{\text{dissociation} \\ \text{of $Q_A^{K}$}}} + \underbrace{\lambda_{Q}Q^{K}}_{\substack{\text{dissociation} \\ \text{of $Q^{K}$}}} - \underbrace{\lambda_{P_DA_1}P_D^{K}A_1}_{\text{binding to $A_1$}} - \underbrace{\lambda_{P_DP_L}P_D^{K}P_L}_{\text{binding to $P_L$}} - \underbrace{d_{P_D}P_D^{K}}_{\text{degradation}}. \label{PDKeqntempsupp}
\end{align}
\subsubsection{Equations for the PD-1/pembrolizumab Complex on Cells in the TS ($Q_A^{T_8}$, $Q_A^{T_1}$, $Q_A^{K}$)}
Pembrolizumab binds to unbound PD-1 on the surfaces of PD-1-expressing cells in a 1:1 ratio \cite{Na2016supp}, forming the PD-1/pembrolizumab complex in a reversible chemical process \cite{Tan2016supp, Wang2023pdsupp}. We must also account for loss due to the endocytosis and internalisation of the PD-1/pembrolizumab complex from the surface of cells \cite{Cowles2022supp, BenSaad2024supp}. We assume that the rates of PD-1/pembrolizumab complex internalisation and dissociation are invariant of the type of cell expressing PD-1, so that
\begin{align}
    \frac{dQ_A^{T_8}}{dt} &= \underbrace{\lambda_{P_DA_1}P_D^{T_8}A_1}_{\text{formation of $Q_A^{T_8}$}} - \underbrace{\lambda_{Q_A}Q_A^{T_8}}_{\text{dissociation of $Q_A^{T_8}$}} - \underbrace{d_{Q_A}Q_A^{T_8}}_{\text{internalisation}}, \label{QA8eqnsupp} \\
    \frac{dQ_A^{T_1}}{dt} &= \underbrace{\lambda_{P_DA_1}P_D^{T_1}A_1}_{\text{formation of $Q_A^{T_1}$}} - \underbrace{\lambda_{Q_A}Q_A^{T_1}}_{\text{dissociation of $Q_A^{T_1}$}} - \underbrace{d_{Q_A}Q_A^{T_1}}_{\text{internalisation}}, \label{QA1eqnsupp} \\
    \frac{dQ_A^{K}}{dt} &= \underbrace{\lambda_{P_DA_1}P_D^{K}A_1}_{\text{formation of $Q_A^{K}$}} - \underbrace{\lambda_{Q_A}Q_A^{K}}_{\text{dissociation of $Q_A^{K}$}} - \underbrace{d_{Q_A}Q_A^{K}}_{\text{internalisation}}. \label{QAKeqnsupp}
\end{align}
\subsubsection{Equation for Pembrolizumab in the TS ($A_1$)}
We assume that pembrolizumab is administered intravenously at times $t_1$, $t_2$, \dots, $t_n$ with doses $\xi_{1}$, $\xi_{2}$, \dots, $\xi_{n}$ respectively, and assume that the duration of infusion is negligible in comparison to the time period of interest. We also account for pembrolizumab depletion due to binding to unbound PD-1, replenishment due to PD-1/pembrolizumab complex dissociation, and elimination of pembrolizumab. It is important to note that the administered dose is not equal to the corresponding change in concentration in the TS. For simplicity, we assume linear pharmacokinetics so that, for some scaling factor $f_\mathrm{pembro}$, we have that
\begin{equation}
        \frac{dA_{1}}{dt} =\underbrace{\sum_{j=1}^{n} \xi_{j}f_\mathrm{pembro}\delta\left(t-t_j \right)}_{\text{infusion}} + \underbrace{\lambda_{Q_A}\left(Q_A^{T_8}+ Q_A^{T_1} + Q_A^{K}\right)}_{\text{dissociation of $Q_A^{T_8}$, $Q_A^{T_1}$, and $Q_A^{K}$}} - \underbrace{\lambda_{P_DA_1}\left(P_D^{T_8} + P_D^{T_1} + P_D^{K}\right)A_1}_{\text{formation of $Q_A^{T_8}$, $Q_A^{T_1}$, and $Q_A^{K}$}} - \underbrace{d_{A_1}A_{1}}_{\text{elimination}}. \label{A1eqnsupp}
\end{equation}
\subsubsection{Equation for Unbound PD-L1 in the TS ($P_L$)}
We also know that PD-L1 is expressed on the surface of viable cancer cells \cite{Zheng2019supp}, mature DCs \cite{Oh2020supp}, effector CD8+ T cells \cite{Zheng2022supp, Kowanetz2018supp}, effector Th1 cells \cite{Chen2004supp}, effector Tregs \cite{Gianchecchi2018supp}, and M2 macrophages \cite{Zhu2020supp}. For brevity, we denote $\mathcal{X}$ as the set of PD-L1-expressing cells in the TS, so that $\mathcal{X}:=\set{C, D, T_8, T_1, T_r, M_2}$. Furthermore, $\lambda_{P_LX}$ denotes the synthesis rate of unbound PD-L1 on the surface of $X \in \mathcal{X}$. We must take into account the synthesis of PD-L1, its depletion due to binding to unbound PD-1, replenishment due to PD-1/PD-L1 complex dissociation, and the degradation of PD-L1. Hence,
\begin{equation}
    \frac{dP_{L}}{dt} = \underbrace{\sum_{X \in \mathcal{X}}\lambda_{P_LX}X}_{\text{synthesis}} + \underbrace{\lambda_{Q}\left(Q^{T_8} + Q^{T_1} + Q^{K}\right)}_{\text{dissociation of $Q^{T_8}$, $Q^{T_1}$ and $Q^{K}$}} - \underbrace{\lambda_{P_DP_L}\left(P_D^{T_8} + P_D^{T_1} + P_D^{K}\right)P_L}_{\text{formation of $Q^{T_8}$, $Q^{T_1}$ and $Q^{K}$}} - \underbrace{d_{P_L}P_L}_{\text{degradation}}. \label{PLeqntempsupp}
\end{equation}
\subsubsection{Equations for the PD-1/PD-L1 Complex in the TS ($Q^{T_8}$, $Q^{T_1}$, and $Q^{K}$)}
\begin{figure}[ht]
    \centering
    \fbox{\includegraphics[width=0.75\textwidth]{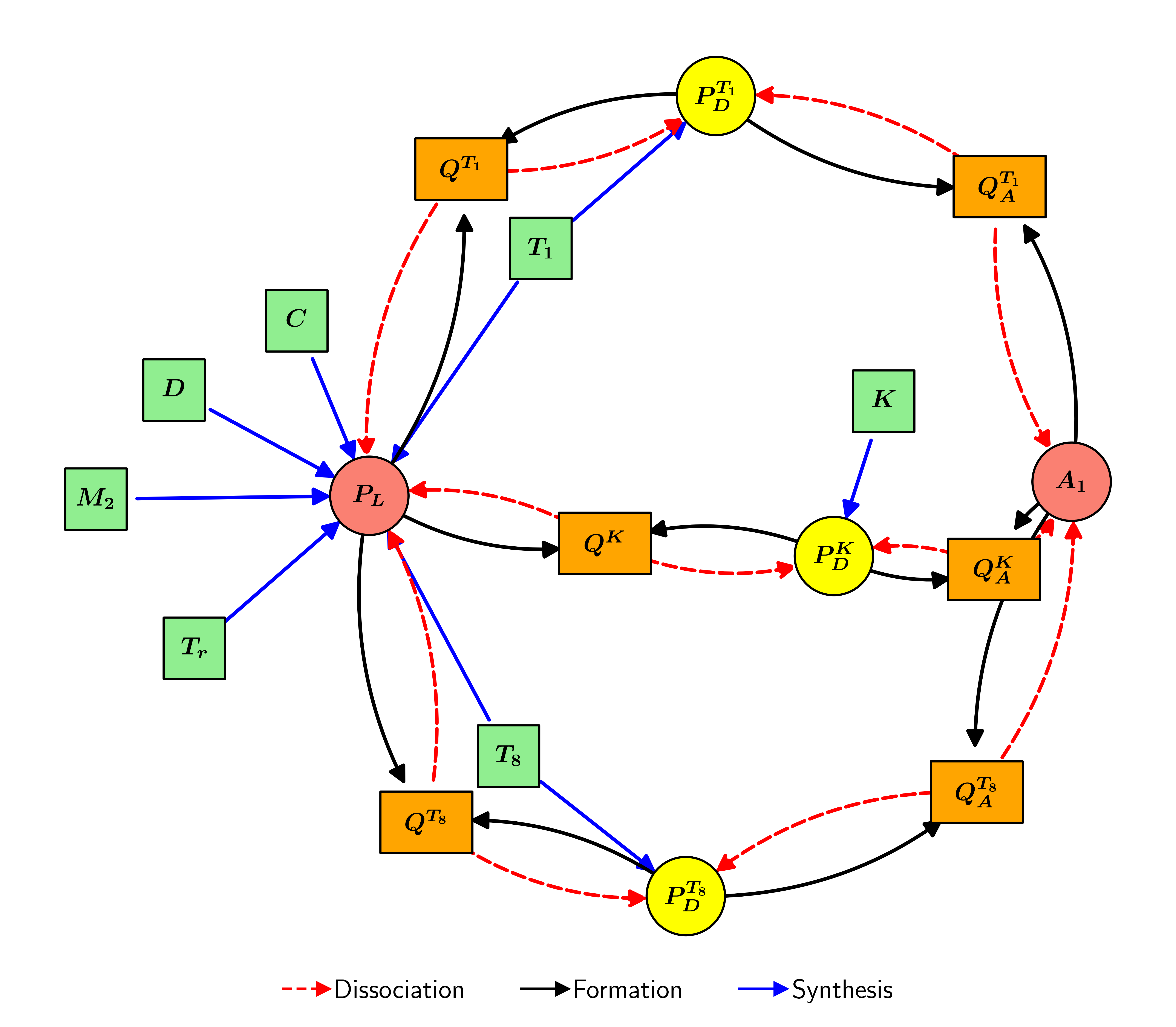}}
    \caption{\label{modeldiagramICITS}Schematic diagram of the interactions of immune checkpoint-associated components in the TS in the full model.}
\end{figure}

PD-L1 binds to unbound PD-1 receptors on the surfaces of PD-1-expressing cells in a 1:1 ratio \cite{Cheng2013supp}, forming the PD-1/PD-L1 complex in a reversible chemical process. Considering $Q^{T_8}$ as an example, we can express its formation and dissociation via the reaction $P_{D}^{T_8}+P_{L} \underset{\lambda_{Q}}{\stackrel{\lambda_{P_{D}P_{L}}}{\rightleftharpoons}} Q^{T_8}$. We assume that the degradation is negligible relative to the dissociation, so that
\begin{equation}
\frac{dQ^{T_8}}{dt} =\underbrace{\lambda_{P_{D} P_{L}}P_{D}^{T_8}P_{L}}_{\text{formation}} - \underbrace{\lambda_{Q}Q^{T_8}}_{\text{dissociation}}.
\end{equation}
However, the dissociation rate constant of the PD-1/PD-L1 complex is $1.44 \mathrm{~s^{-1}}$, corresponding to a mean lifetime of less than 1 second \cite{Cheng2013supp}. As such, we employ a quasi-steady-state approximation (QSSA) for $Q^{T_8}$, so that $\frac{dQ^{T_8}}{dt}=0$, so that
\begin{align}
Q^{T_8} &= \frac{\lambda_{P_{D}P_{L}}}{\lambda_{Q}} P_{D}^{T_8}P_{L}. \label{Q8eqnsupp}
\intertext{Similarly,}
Q^{T_1} &= \frac{\lambda_{P_{D}P_{L}}}{\lambda_{Q}} P_{D}^{T_1}P_{L}, \label{QT1eqnsupp} \\
Q^{K} &= \frac{\lambda_{P_{D}P_{L}}}{\lambda_{Q}} P_{D}^{K}P_{L}. \label{QKeqnsupp}
\end{align}
Furthermore, we can simplify \eqref{PD8eqntempsupp}~--~\eqref{PDKeqntempsupp} and \eqref{PLeqntempsupp} by substituting in \eqref{Q8eqnsupp}~--~\eqref{QKeqnsupp} so that
\begin{align}
    \frac{dP_D^{T_8}}{dt} &= \underbrace{\lambda_{P_D^{T_8}}T_8}_{\text{synthesis}} + \underbrace{\lambda_{Q_A}Q_A^{T_8}}_{\substack{\text{dissociation} \\ \text{of $Q_A^{T_8}$}}} - \underbrace{\lambda_{P_DA_1}P_D^{T_8}A_1}_{\text{binding to $A_1$}} - \underbrace{d_{P_D}P_D^{T_8}}_{\text{degradation}}, \label{PD8eqnsupp} \\
    \frac{dP_D^{T_1}}{dt} &= \underbrace{\lambda_{P_D^{T_1}}T_1}_{\text{synthesis}} + \underbrace{\lambda_{Q_A}Q_A^{T_1}}_{\substack{\text{dissociation} \\ \text{of $Q_A^{T_1}$}}} - \underbrace{\lambda_{P_DA_1}P_D^{T_1}A_1}_{\text{binding to $A_1$}} - \underbrace{d_{P_D}P_D^{T_1}}_{\text{degradation}}, \label{PD1eqnsupp} \\
    \frac{dP_D^{K}}{dt} &= \underbrace{\lambda_{P_D^{K}}K}_{\text{synthesis}} + \underbrace{\lambda_{Q_A}Q_A^{K}}_{\substack{\text{dissociation} \\ \text{of $Q_A^{K}$}}} - \underbrace{\lambda_{P_DA_1}P_D^{K}A_1}_{\text{binding to $A_1$}} - \underbrace{d_{P_D}P_D^{K}}_{\text{degradation}}, \label{PDKeqnsupp} \\
    \frac{dP_{L}}{dt} &= \underbrace{\sum_{X \in \mathcal{X}}\lambda_{P_L X}X}_{\text{synthesis}} - \underbrace{d_{P_L}P_L}_{\text{degradation}}. \label{PLeqnsupp}
\end{align}
A diagram encompassing the interactions of components is shown in \autoref{modeldiagramICITS}.
\subsection{Equations for Immune Checkpoint-Associated Components in the TDLN}
\subsubsection{Equations for Unbound PD-1 receptors on cells in the TDLN ($P_D^{8\mathrm{LN}}$ and $P_D^{1\mathrm{LN}}$)}
The equations for $P_D^{8\mathrm{LN}}$ and $P_D^{1\mathrm{LN}}$ follow identically to that of \eqref{PD8eqnsupp}~--~\eqref{PD1eqnsupp}. For simplicity, we assume that the formation and dissociation rates of the PD-1/pembrolizumab complex are identical in the TDLN and the TS, so that
\begin{align}
    \frac{dP_D^{8\mathrm{LN}}}{dt} &= \underbrace{\lambda_{P_D^{8\mathrm{LN}}}T_A^8}_{\text{synthesis}} + \underbrace{\lambda_{Q_A}Q_A^{8\mathrm{LN}}}_{\substack{\text{dissociation} \\ \text{of $Q_A^{8\mathrm{LN}}$}}} - \underbrace{\lambda_{P_DA_1}P_D^{8\mathrm{LN}}A_1^\mathrm{LN}}_{\text{binding to $A_1^\mathrm{LN}$}} - \underbrace{d_{P_D}P_D^{8\mathrm{LN}}}_{\text{degradation}}, \label{PD8LNeqnsupp} \\
    \frac{dP_D^{1\mathrm{LN}}}{dt} &= \underbrace{\lambda_{P_D^{1\mathrm{LN}}}T_A^1}_{\text{synthesis}} + \underbrace{\lambda_{Q_A}Q_A^{1\mathrm{LN}}}_{\substack{\text{dissociation} \\ \text{of $Q_A^{1\mathrm{LN}}$}}} - \underbrace{\lambda_{P_DA_1}P_D^{1\mathrm{LN}}A_1^\mathrm{LN}}_{\text{binding to $A_1^\mathrm{LN}$}} - \underbrace{d_{P_D}P_D^{1\mathrm{LN}}}_{\text{degradation}}. \label{PD1LNeqnsupp}
\end{align}
\subsubsection{Equations for the PD-1/pembrolizumab Complex on Cells in the TDLN ($Q_A^{8\mathrm{LN}}$ and $Q_A^{1\mathrm{LN}}$)}
The equations for $Q_A^{8\mathrm{LN}}$ and $Q_A^{1\mathrm{LN}}$ follow identically to that of \eqref{QA8eqnsupp}~--~\eqref{QA1eqnsupp}. For simplicity, we assume that the rates of PD-1 receptor internalisation are identical in the TDLN and the TS, so that
\begin{align}
    \frac{dQ_A^{8\mathrm{LN}}}{dt} = \underbrace{\lambda_{P_DA_1}P_D^{8\mathrm{LN}}A_1^\mathrm{LN}}_{\text{formation of $Q_A^{8\mathrm{LN}}$}} - \underbrace{\lambda_{Q_A}Q_A^{8\mathrm{LN}}}_{\text{dissociation of $Q_A^{8\mathrm{LN}}$}} - \underbrace{d_{Q_A}Q_A^{8\mathrm{LN}}}_{\text{internalisation}}, \label{QA8LNeqnsupp} \\
    \frac{dQ_A^{1\mathrm{LN}}}{dt} = \underbrace{\lambda_{P_DA_1}P_D^{1\mathrm{LN}}A_1^\mathrm{LN}}_{\text{formation of $Q_A^{1\mathrm{LN}}$}} - \underbrace{\lambda_{Q_A}Q_A^{1\mathrm{LN}}}_{\text{dissociation of $Q_A^{1\mathrm{LN}}$}} - \underbrace{d_{Q_A}Q_A^{1\mathrm{LN}}}_{\text{internalisation}}. \label{QA1LNeqnsupp}
\end{align}
\subsubsection{Equation for Pembrolizumab in the TDLN ($A_1^\mathrm{LN}$)}
The equation for $A_1^\mathrm{LN}$ follows identically to that of \eqref{A1eqnsupp} so that
\begin{equation}
        \frac{dA_{1}^\mathrm{LN}}{dt} =\underbrace{\sum_{j=1}^{n} \xi_{j}f_\mathrm{pembro}\delta\left(t-t_j \right)}_{\text{infusion}} + \underbrace{\lambda_{Q_A}\left(Q_A^{8\mathrm{LN}} + Q_A^{1\mathrm{LN}}\right)}_{\text{dissociation of $Q_A^{8\mathrm{LN}}$ and $Q_A^{1\mathrm{LN}}$}} - \underbrace{\lambda_{P_DA_1}\left(P_D^{8\mathrm{LN}} + P_D^{1\mathrm{LN}}\right)A_1^\mathrm{LN}}_{\text{formation of $Q_A^{8\mathrm{LN}}$ and $Q_A^{1\mathrm{LN}}$}} - \underbrace{d_{A_1}A_{1}^\mathrm{LN}}_{\text{elimination}}. \label{A1LNeqnsupp}
\end{equation}
\subsubsection{Equation for Unbound PD-L1 in the TDLN ($P_L^\mathrm{LN}$)}
We recall that PD-L1 is expressed on the surface of mature DCs, effector CD8+ T cells, effector Th1 cells, and effector Tregs. We denote $\mathcal{Y}$ as the set of PD-L1-expressing cells in the TDLN, so that $\mathcal{Y}:=\set{D^\mathrm{LN}, T_A^8, T_A^1, T_A^r}$, with $\lambda_{P_L^\mathrm{LN}Y}$ denoting the synthesis rate of unbound PD-L1 on the surface of $Y \in \mathcal{Y}$. The equation for $P_L^\mathrm{LN}$ follows identically to \eqref{PLeqnsupp} so that
\begin{equation}
\begin{split}
    \frac{dP_{L}^\mathrm{LN}}{dt} &= \underbrace{\sum_{Y \in \mathcal{Y}} \lambda_{P_L^\mathrm{LN}Y}Y}_{\text{synthesis}} - \underbrace{d_{P_L}P_L^\mathrm{LN}}_{\text{degradation}}. \label{PLLNeqnsupp}
\end{split}
\end{equation}
\subsubsection{Equations for the PD-1/PD-L1 Complex in the TDLN ($Q^{8\mathrm{LN}}$ and $Q^{1\mathrm{LN}}$)}
\begin{figure}[H]
    \centering
    \fbox{\includegraphics[width=0.75\textwidth]{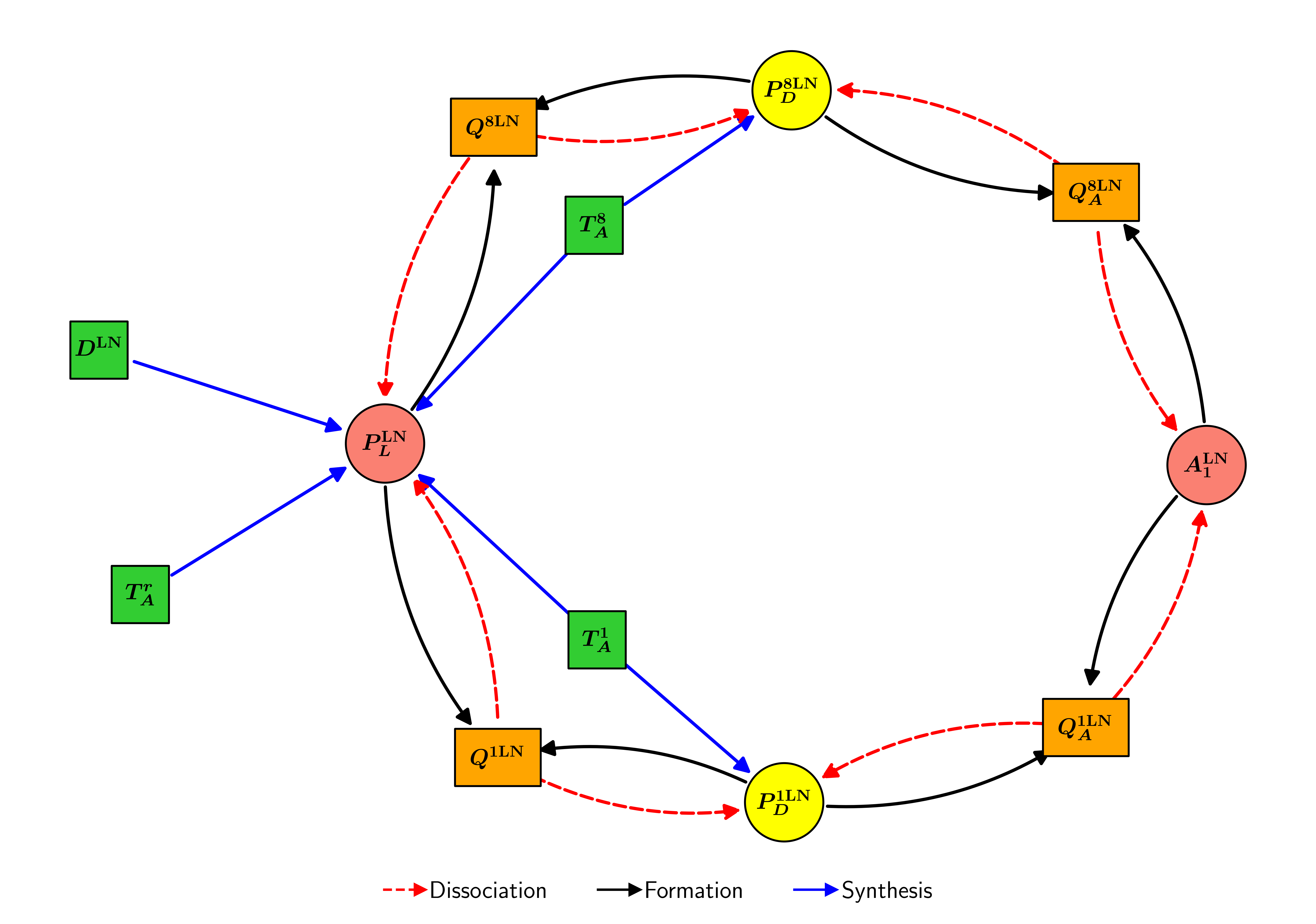}}
    \caption{\label{modeldiagramICITDLN}Schematic diagram of the interactions of immune checkpoint-associated components in the TDLN in the full model.}
\end{figure}

For simplicity, we assume that the formation and dissociation rates of the PD-1/PD-L1 complex are identical in the TDLN and the TS. The equations for $Q^{8\mathrm{LN}}$ and $Q^{1\mathrm{LN}}$ follow identically from \eqref{Q8eqnsupp}~--~\eqref{QT1eqnsupp} so that
\begin{align}
Q^{8\mathrm{LN}} &= \frac{\lambda_{P_{D}P_{L}}}{\lambda_{Q}} P_{D}^{8\mathrm{LN}}P_{L}^\mathrm{LN}, \label{Q8LNeqnsupp} \\
Q^{1\mathrm{LN}} &= \frac{\lambda_{P_{D}P_{L}}}{\lambda_{Q}} P_{D}^{1\mathrm{LN}}P_{L}^\mathrm{LN}. \label{Q1LNeqnsupp}
\end{align}
A diagram encompassing the interactions of these components is shown in \autoref{modeldiagramICITDLN}.
We note that throughout the model, the PD-1/PD-L1 complex appears only within an inhibition constant, making its absolute magnitude less important since it always appears as a ratio. One thing to note is that activated CD8+ T cells and Th1 cells also express PD-1 receptors and PD-L1 ligands, and we assume that effector and activated cells express these in equal amounts. However, as discussed in \Cref{TDLNssinitsupp}, the ratio between effector and activated T cells can be assumed to remain roughly constant. Since the PD-1/PD-L1 complex concentration is linearly proportional to the product of PD-L1 concentration and unbound PD-1 receptor concentration, and PD-1/PD-L1-mediated inhibition of T cell proliferation in the TDLN appears only as a ratio, it is sufficient to consider only PD-1, PD-L1, and PD-1/PD-L1 concentrations on effector cells, as this will be appropriately scaled by the corresponding inhibition constants. Furthermore, this also justifies using the PD-1/PD-L1 complex concentration on effector T cells as a proxy for its concentration on activated T cells that have not yet undergone division, given that their ratio to effector cells remains roughly constant and that PD-1/PD-L1-mediated inhibition of T cell activation in the TDLN appears only as a ratio.
\subsection{Model Reduction via QSSA}
We observe from \autoref{tableofparams} that the degradation rates of cytokines and DAMPs are, in general, orders of magnitude larger than those of immune and cancer cells. In particular, IL-2, IFN-$\upgamma$, TNF, and TGF-$\beta$ evolve on a very fast timescale, with degradation rates significantly higher than all other species in the model, causing them to equilibrate much more rapidly. As such, we perform a QSSA and reduce the model by setting \eqref{il2eqnsupp}~--~\eqref{tgfbetaeqn} to $0$ and solving for $I_2$, $I_\upgamma$, $I_\upalpha$, and $I_\upbeta$ in terms of the other parameters and variables in the model. This minimally affects the system's evolution after a very short period of transient behaviour \cite{Snowden2017supp}, and we justify this by observing that, empirically, the deviation in system trajectories remains negligible for nearby parameter choices. Performing the QSSA leads to
\begin{align}
    \frac{dI_2}{dt} &= 0 \implies I_2 = \frac{1}{d_{I_2}}\left(\lambda_{I_2 T_8}T_8 + \lambda_{I_2 T_1}T_1\right), \label{il2qssasupp}\\
    \frac{dI_\upgamma}{dt} &= 0 \implies I_\upgamma = \frac{1}{d_{I_\upgamma}}\left[\left(\lambda_{I_{\upgamma} T_8}T_8 + \lambda_{I_{\upgamma} T_1}T_1\right)\frac{1}{1+T_r/K_{I_{\upgamma}T_r}} + \lambda_{I_{\upgamma} K}K\right], \label{ifngammaqssasupp}\\
    \frac{dI_\upalpha}{dt} &= 0 \implies I_\upalpha = \frac{1}{d_{I_{\upalpha}}}\left(\lambda_{I_{\upalpha}T_8}T_8 + \lambda_{I_{\upalpha}T_1}T_1 + \lambda_{I_{\upalpha}M_1}M_1 + \lambda_{I_{\upalpha}K}K\right), \label{tnfqssasupp}\\
    \frac{dI_\upbeta}{dt} &= 0 \implies I_\upbeta = \frac{1}{d_{I_{\upbeta}}}\left(\lambda_{I_{\upbeta}C}C + \lambda_{I_{\upbeta}T_r}T_r + \lambda_{I_{\upbeta}M_2}M_2\right).\label{ibetaqssasupp}      
\end{align}
We note that this reduction is valid since the timescale of IFN-$\upgamma$, the slowest of the ``fast'' species, is significantly shorter than the timescales of all ``slow'' species in the model.
\subsection{Model Parameters for the Full Model}
The model parameter values are estimated in \cite{Hawi2026metasupp} and are listed in \autoref{tableofparams}.
\begin{center}
\begin{longtable}{|lp{160pt}lp{100pt}l|}
\caption{\label{tableofparams} Parameter values for the full model. TDLN denotes the tumour-draining lymph node, whilst TS denotes the tumour site. est. denotes estimated parameters.}\\
\hline \textbf{Parameter} & \textbf{Description} & \textbf{Value} & \textbf{Unit} & \textbf{References} \\
\hline
$f_\mathrm{pembro}$ & $A_1$/$A_1^\mathrm{LN}$ dose scaling factor & $1.17 \times 10^{12}$ & $\left(\mathrm{molec/cm^3}\right)/\mathrm{mg}$ & est.\\
$f_{C}$ & $C$ to $V_\mathrm{TS}$ scaling factor & $1.17 \times 10^{6}$ & $\mathrm{cell/(cm^3)^2}$ & est. \\
$f_{N_c}$ & $N_c$ to $V_\mathrm{TS}$ scaling factor & $6.16 \times 10^{4}$ & $\mathrm{cell/(cm^3)^2}$ & est. \\
\hline
$\mathcal{A}_{D_0}$ & Source of $D_0$ & $1.18 \times 10^6$ & $\left(\mathrm{cell/cm^{3}}\right)\mathrm{day^{-1}}$ & est. \\
$\mathcal{A}_{T_0^8}$ & Source of $T_0^8$ & $3.76 \times 10^5$ & $\left(\mathrm{cell/cm^{3}}\right)\mathrm{day^{-1}}$ & est. \\
$\mathcal{A}_{T_0^4}$ & Source of $T_0^4$ & $2.76 \times 10^5$ & $\left(\mathrm{cell/cm^{3}}\right)\mathrm{day^{-1}}$ & est. \\
$\mathcal{A}_{T_0^r}$ & Source of $T_0^r$ & $1.15 \times 10^5$ & $\left(\mathrm{cell/cm^{3}}\right)\mathrm{day^{-1}}$ & est. \\
$\mathcal{A}_{M_0}$ & Source of $M_0$ & $1.08 \times 10^6$ & $\left(\mathrm{cell/cm^{3}}\right)\mathrm{day^{-1}}$ & est. \\
$\mathcal{A}_{K_0}$ & Source of $K_0$ & $2.82 \times 10^5$ & $\left(\mathrm{cell/cm^{3}}\right)\mathrm{day^{-1}}$ & est. \\
\hline
$\lambda_{C}$ & Growth rate of $C$ & $5.25 \times 10^{-2}$ & $\mathrm{day^{-1}}$ & fitted \\
$\lambda_{CT_8}$ & Elimination rate of $C$ by $T_8$ & $8.01 \times 10^{-8}$ & $\left(\mathrm{cell/cm^{3}}\right)^{-1}\mathrm{day^{-1}}$ & fitted \\
$\lambda_{CK}$ & Elimination rate of $C$ by $K$ & $8.01 \times 10^{-8}$ & $\left(\mathrm{cell/cm^{3}}\right)^{-1}\mathrm{day^{-1}}$ & est. \\
$\lambda_{CI_{\upalpha}}$ & Necrosis rate of $C$ by $I_{\upalpha}$ & $6.02 \times 10^{-3}$ & $\mathrm{day^{-1}}$ & est. \\
$\lambda_{CI_{\upgamma}}$ & Necrosis rate of $C$ by $I_{\upgamma}$ & $1.20 \times 10^{-3}$ & $\mathrm{day^{-1}}$ & est. \\
$\lambda_{HN_c}$ & Production rate of $H$ by $N_c$ & $2.92 \times 10^{-14}$ & $\left(\mathrm{g/cell}\right)\mathrm{day^{-1}}$ & est. \\
$\lambda_{SN_c}$ & Production rate of $S$ by $N_c$ & $1.70 \times 10^{-14}$ & $\left(\mathrm{g/cell}\right)\mathrm{day^{-1}}$ & est. \\
$\lambda_{DH}$ & Maturation rate of $D_0$ by $H$ & $1.21$ & $\mathrm{day^{-1}}$ & est. \\
$\lambda_{DS}$ & Maturation rate of $D_0$ by $S$ & $1.21 \times 10^{-1}$ & $\mathrm{day^{-1}}$ & est. \\
$\lambda_{D_0K}$ & Killing rate of $D_0$ by $K$ & $5.49 \times 10^{-6}$ & $\left(\mathrm{cell/cm^3}\right)^{-1}\mathrm{day^{-1}}$ & est. \\
$\lambda_{DD^\mathrm{LN}}$ & Migration rate of $D$ to TDLN & $1.68 \times 10^{-2}$ & $\mathrm{day^{-1}}$ & est. \\
$\lambda_{T_0^8 T_A^8}$ & Kinetic rate constant for $T_0^8$ activation & $2.73 \times 10^{-11}$ & $\left(\mathrm{cell/cm^3}\right)^{-1}\mathrm{day^{-1}}$ & est. \\
$\lambda_{T_A^8T_8}$ & Kinetic rate constant for $T_A^8$ migration to the TS & $6.32 \times 10^{-1}$ & $\mathrm{day}^{-1}$ & est. \\
$\lambda_{T_8 I_2}$ & Growth rate of $T_8$ by $I_2$ & $1.53 \times 10^{-3}$ & $\mathrm{day}^{-1}$ & est. \\
$\lambda_{T_8 C}$ & Exhaustion rate of $T_8$ due to $C$ exposure & $6.31 \times 10^{-3}$ & $\mathrm{day}^{-1}$ & est. \\
$\lambda_{T_\mathrm{ex} A_1}$ & Reinvigoration rate of $T_\mathrm{ex}$ by $A_1$ & $2.25 \times 10^{-3}$ & $\mathrm{day}^{-1}$ & est. \\
$\lambda_{T_0^4 T_A^1}$ & Kinetic rate constant for $T_0^4$ activation into $T_A^1$ & $8.26 \times 10^{-11}$ & $\left(\mathrm{cell/cm^3}\right)^{-1}\mathrm{day^{-1}}$ & est. \\
$\lambda_{T_A^1 T_1}$ & Kinetic rate constant for $T_A^1$ migration to the TS & $6.17 \times 10^{-2}$ & $\mathrm{day}^{-1}$ & est. \\
$\lambda_{T_1 I_2}$ & Growth rate of $T_1$ by $I_2$ & $2.00 \times 10^{-3}$ & $\mathrm{day}^{-1}$ & est. \\
$\lambda_{T_1T_r}$ & Conversion rate of $T_1$ to $T_r$ by $Q^{T_1}$ & $4.00 \times 10^{-3}$ & $\mathrm{day}^{-1}$ & est. \\
$\lambda_{T_0^r T_A^r}$ & Kinetic rate constant for $T_0^r$ activation into $T_A^r$ & $1.08 \times 10^{-8}$ & $\left(\mathrm{cell/cm^3}\right)^{-1}\mathrm{day^{-1}}$ & est. \\
$\lambda_{T_A^r T_r}$ & Kinetic rate constant for $T_A^r$ migration to the TS & $4.88$ & $\mathrm{day}^{-1}$ & est. \\
$\lambda_{M_1I_\upalpha}$ & Polarisation rate of $M_0$ to $M_1$ by $I_{\upalpha}$ & $3.59 \times 10^{-1}$ & $\mathrm{day^{-1}}$ & est. \\
$\lambda_{M_1I_\upgamma}$ & Polarisation rate of $M_0$ to $M_1$ by $I_\upgamma$ & $4.18 \times 10^{-1}$ & $\mathrm{day^{-1}}$ & est. \\
$\lambda_{M_2I_{10}}$ & Polarisation rate of $M_0$ to $M_2$ by $I_{10}$ & $2.27 \times 10^{-1}$ & $\mathrm{day^{-1}}$ & est. \\
$\lambda_{M_2I_\upbeta}$ & Polarisation rate of $M_0$ to $M_2$ by $I_\upbeta$ & $2.54 \times 10^{-1}$ & $\mathrm{day^{-1}}$ & est. \\
$\lambda_{MI_{\upgamma}}$ & Polarisation rate of $M_2$ to $M_1$ by $I_{\upgamma}$ & $1.39 \times 10^{-2}$ & $\mathrm{day^{-1}}$ & est. \\
$\lambda_{MI_{\upalpha}}$ & Polarisation rate of $M_2$ to $M_1$ by $I_{\upalpha}$ & $1.15 \times 10^{-2}$ & $\mathrm{day^{-1}}$ & est. \\
$\lambda_{MI_{\upbeta}}$ & Polarisation rate of $M_1$ to $M_2$ by $I_{\upbeta}$ & $6.39 \times 10^{-3}$ & $\mathrm{day^{-1}}$ & est. \\
$\lambda_{KI_2}$ & Maturation rate of $K_0$ by $I_2$ & $4.62 \times 10^{-3}$ & $\mathrm{day^{-1}}$ & est. \\
$\lambda_{KD_0}$ & Maturation rate of $K_0$ by $D_0$ & $1.54 \times 10^{-3}$ & $\mathrm{day^{-1}}$ & est. \\
$\lambda_{KD}$ & Maturation rate of $K_0$ by $D$ & $7.70 \times 10^{-3}$ & $\mathrm{day^{-1}}$ & est. \\
$\lambda_{I_2 T_8}$ & Production rate of $I_2$ by $T_8$ & $5.95 \times 10^{-16}$ & $\left(\mathrm{g/cell}\right)\mathrm{day^{-1}}$ & est. \\
$\lambda_{I_2 T_1}$ & Production rate of $I_2$ by $T_1$ & $1.74 \times 10^{-15}$ & $\left(\mathrm{g/cell}\right)\mathrm{day^{-1}}$ & est. \\
$\lambda_{I_{\upgamma} T_8}$ & Production rate of $I_{\upgamma}$ by $T_8$ & $8.62 \times 10^{-16}$ & $\left(\mathrm{g/cell}\right)\mathrm{day^{-1}}$ & est. \\
$\lambda_{I_{\upgamma}T_1}$ & Production rate of $I_{\upgamma}$ by $T_1$ & $3.02 \times 10^{-16}$ & $\left(\mathrm{g/cell}\right)\mathrm{day^{-1}}$ & est. \\
$\lambda_{I_{\upgamma} K}$ & Production rate of $I_{\upgamma}$ by $K$ & $7.99 \times 10^{-15}$ & $\left(\mathrm{g/cell}\right)\mathrm{day^{-1}}$ & est. \\
$\lambda_{I_{\upalpha}T_8}$ & Production rate of $I_{\upalpha}$ by $T_8$ & $5.55 \times 10^{-15}$ & $\left(\mathrm{g/cell}\right)\mathrm{day^{-1}}$ & est. \\
$\lambda_{I_{\upalpha}T_1}$ & Production rate of $I_{\upalpha}$ by $T_1$ & $9.17 \times 10^{-15}$ & $\left(\mathrm{g/cell}\right)\mathrm{day^{-1}}$ & est. \\
$\lambda_{I_{\upalpha}M_1}$ & Production rate of $I_{\upalpha}$ by $M_1$ & $3.36 \times 10^{-15}$ & $\left(\mathrm{g/cell}\right)\mathrm{day^{-1}}$ & est. \\
$\lambda_{I_{\upalpha}K}$ & Production rate of $I_{\upalpha}$ by $K$ & $9.68 \times 10^{-15}$ & $\left(\mathrm{g/cell}\right)\mathrm{day^{-1}}$ & est. \\
$\lambda_{I_{\upbeta}C}$ & Production rate of $I_{\upbeta}$ by $C$ & $7.42 \times 10^{-12}$ & $\left(\mathrm{g/cell}\right)\mathrm{day^{-1}}$ & est. \\
$\lambda_{I_{\upbeta}T_r}$ & Production rate of $I_{\upbeta}$ by $T_r$ & $4.30 \times 10^{-11}$ & $\left(\mathrm{g/cell}\right)\mathrm{day^{-1}}$ & est. \\
$\lambda_{I_{\upbeta}M_2}$ & Production rate of $I_{\upbeta}$ by $M_2$ & $5.34 \times 10^{-11}$ & $\left(\mathrm{g/cell}\right)\mathrm{day^{-1}}$ & est. \\
$\lambda_{I_{10}C}$ & Production rate of $I_{10}$ by $C$ & $1.55 \times 10^{-17}$ & $\left(\mathrm{g/cell}\right)\mathrm{day^{-1}}$ & est. \\
$\lambda_{I_{10}M_2}$ & Production rate of $I_{10}$ by $M_2$ & $3.10 \times 10^{-17}$ & $\left(\mathrm{g/cell}\right)\mathrm{day^{-1}}$ & est. \\
$\lambda_{I_{10}T_r}$ & Production rate of $I_{10}$ by $T_r$ & $5.86 \times 10^{-18}$ & $\left(\mathrm{g/cell}\right)\mathrm{day^{-1}}$ & est. \\
$\lambda_{I_{10}I_{2}}$ & Production ratio of $I_{10}$ by $I_{2}$ & $3$ & dimensionless & \cite{Lo2016supp} est. \\
$\lambda_{P_D^{T_8}}$ & Synthesis rate of $P_D^{T_8}$ & $9.25 \times 10^2$ & $\left(\mathrm{molec/cell}\right)\mathrm{day^{-1}}$ & est. \\
$\lambda_{Q_A}$ & Dissociation rate of the PD-1/pembrolizumab complex & $2.6$ & $\mathrm{day}^{-1}$ & \cite{Li2021supp} \\
$\lambda_{Q}$ & Dissociation rate of the PD-1/PD-L1 complex & $1.24 \times 10^{5}$ & $\mathrm{day}^{-1}$ & \cite{Cheng2013supp} \\
$\lambda_{P_DA_1}$ & Formation rate of the PD-1/pembrolizumab complex & $4.63 \times 10^{-13}$ & $\left(\mathrm{molec/cm^3}\right)^{-1}\mathrm{day}^{-1}$ & fitted \\
$\lambda_{P_DP_L}$ & Formation rate of the PD-1/PD-L1 complex & $2.64 \times 10^{-11}$ & $\left(\mathrm{molec/cm^3}\right)^{-1}\mathrm{day}^{-1}$ & \cite{Cheng2013supp} \\
$\lambda_{P_D^{T_1}}$ & Synthesis rate of $P_D^{T_1}$ & $6.87 \times 10^2$ & $\left(\mathrm{molec/cell}\right)\mathrm{day^{-1}}$ & est. \\
$\lambda_{P_D^{K}}$ & Synthesis rate of $P_D^{K}$ & $1.85 \times 10^2$ & $\left(\mathrm{molec/cell}\right)\mathrm{day^{-1}}$ & est. \\
$\lambda_{P_LC}$ & Synthesis rate of $P_L$ by $C$ & $2.50 \times 10^{5}$ & $\left(\mathrm{molec/cell}\right)\mathrm{day^{-1}}$ & est. \\
$\lambda_{P_LD}$ & Synthesis rate of $P_L$ by $D$ & $2.46 \times 10^{4}$ & $\left(\mathrm{molec/cell}\right)\mathrm{day^{-1}}$ & est. \\
$\lambda_{P_LT_8}$ & Synthesis rate of $P_L$ by $T_8$ & $2.07 \times 10^{3}$ & $\left(\mathrm{molec/cell}\right)\mathrm{day^{-1}}$ & est. \\
$\lambda_{P_LT_1}$ & Synthesis rate of $P_L$ by $T_1$ & $2.89 \times 10^{3}$ & $\left(\mathrm{molec/cell}\right)\mathrm{day^{-1}}$ & est. \\
$\lambda_{P_LT_r}$ & Synthesis rate of $P_L$ by $T_r$ & $2.89 \times 10^{3}$ & $\left(\mathrm{molec/cell}\right)\mathrm{day^{-1}}$ & est. \\
$\lambda_{P_LM_2}$ & Synthesis rate of $P_L$ by $M_2$ & $3.71 \times 10^{5}$ & $\left(\mathrm{molec/cell}\right)\mathrm{day^{-1}}$ & est. \\
$\lambda_{P_D^{8\mathrm{LN}}}$ & Synthesis rate of $P_D^{8\mathrm{LN}}$ & $9.27 \times 10^{2}$ & $\left(\mathrm{molec/cell}\right)\mathrm{day^{-1}}$ & est. \\
$\lambda_{P_D^{1\mathrm{LN}}}$ & Synthesis rate of $P_D^{1\mathrm{LN}}$ & $6.89 \times 10^{2}$ & $\left(\mathrm{molec/cell}\right)\mathrm{day^{-1}}$ & est. \\
$\lambda_{P_L^\mathrm{LN}D^\mathrm{LN}}$ & Synthesis rate of $P_L^\mathrm{LN}$ by $D^\mathrm{LN}$ & $2.46 \times 10^{4}$ & $\left(\mathrm{molec/cell}\right)\mathrm{day^{-1}}$ & est. \\
$\lambda_{P_L^\mathrm{LN}T_A^8}$ & Synthesis rate of $P_L^\mathrm{LN}$ by $T_A^8$ & $2.07 \times 10^{3}$ & $\left(\mathrm{molec/cell}\right)\mathrm{day^{-1}}$ & est. \\
$\lambda_{P_L^\mathrm{LN}T_A^1}$ & Synthesis rate of $P_L^\mathrm{LN}$ by $T_A^1$ & $2.89 \times 10^{3}$ & $\left(\mathrm{molec/cell}\right)\mathrm{day^{-1}}$ & est. \\
$\lambda_{P_L^\mathrm{LN}T_A^r}$ & Synthesis rate of $P_L^\mathrm{LN}$ by $T_A^r$ & $2.89 \times 10^{3}$ & $\left(\mathrm{molec/cell}\right)\mathrm{day^{-1}}$ & est. \\
\hline
$K_{CI_{\upalpha}}$ & Half-saturation constant of $I_\upalpha$ for $C$ & $9.00 \times 10^{-11}$ & $\mathrm{g/cm^{3}}$ & est. \\
$K_{CI_{\upgamma}}$ & Half-saturation constant of $I_\upgamma$ for $C$ & $4.93 \times 10^{-11}$ & $\mathrm{g/cm^{3}}$ & est. \\
$K_{DH}$ & Half-saturation constant of $H$ for $D$ & $1.94 \times 10^{-8}$ & $\mathrm{g/cm^{3}}$ & est. \\
$K_{DS}$ & Half-saturation constant of $S$ for $D$ & $4.50 \times 10^{-8}$ & $\mathrm{g/cm^{3}}$ & est.\\
$K_{T_8I_2}$ & Half-saturation constant of $I_2$ for $T_8$ & $2.00 \times 10^{-12}$ & $\mathrm{g/cm^{3}}$ & est. \\
$K_{T_8C}$ & Half-saturation constant $T_8$ exhaustion due to $C$ exposure & $7.02 \times 10^{8}$ & $\left(\mathrm{cell}/\mathrm{cm^3}\right)\ \mathrm{day}$ & est. \\
$K_{T_\mathrm{ex} A_1}$ & Half-saturation constant of $T_\mathrm{ex}$ reinvigoration by $A_1$ & $2.05 \times 10^{14}$ & $\mathrm{molec/cm^3}$ & est. \\
$K_{T_1I_2}$ & Half-saturation constant of $I_2$ for $T_1$ & $2.00 \times 10^{-12}$ & $\mathrm{g/cm^{3}}$ & est. \\
$K_{T_1Q^{T_1}}$ & Half-saturation constant of $T_1$ conversion to $T_r$ by $Q^{T_1}$ & $6.01 \times 10^5$ & $\mathrm{molec/cm^{3}}$ & est. \\
$K_{M_1 I_\upalpha}$ & Half-saturation constant of $I_\upalpha$ for $M_1$ & $9.00 \times 10^{-11}$ & $\mathrm{g/cm^{3}}$ & est. \\
$K_{M_1 I_\upgamma}$ & Half-saturation constant of $I_\upgamma$ for $M_1$ & $4.93 \times 10^{-11}$ & $\mathrm{g/cm^{3}}$ & est. \\
$K_{M_2 I_{10}}$ & Half-saturation constant of $I_{10}$ for $M_{2}$ & $1.84 \times 10^{-10}$ & $\mathrm{g/cm^{3}}$ & est. \\
$K_{M_2 I_{\upbeta}}$ & Half-saturation constant of $I_{\upbeta}$ for $M_{2}$ & $1.51 \times 10^{-6}$ & $\mathrm{g/cm^{3}}$ & est. \\
$K_{MI_{\upgamma}}$ & Half-saturation constant of $I_{\upgamma}$ for $M_{1}/M_{2}$ & $4.93 \times 10^{-11}$ & $\mathrm{g/cm^{3}}$ & est.\\
$K_{MI_{\upalpha}}$ & Half-saturation constant of $I_{\upalpha}$ for $M_{1}/M_{2}$ & $9.00 \times 10^{-11}$ & $\mathrm{g/cm^{3}}$ & est. \\
$K_{MI_{\upbeta}}$ & Half-saturation constant of $I_{\upbeta}$ for $M_{1}/M_{2}$ & $1.51 \times 10^{-6}$ & $\mathrm{g/cm^{3}}$ & est. \\
$K_{KI_2}$ & Half-saturation constant of $I_2$ for $K$ & $2.00 \times 10^{-12}$ & $\mathrm{g/cm^{3}}$ & est. \\
$K_{KD_0}$ & Half-saturation constant of $D_0$ for $K$ & $9.55 \times 10^{5}$ & $\mathrm{cell/cm^{3}}$ & est. \\
$K_{KD}$ & Half-saturation constant of $D$ for $K$ & $1.91 \times 10^{6}$ & $\mathrm{cell/cm^{3}}$ & est. \\
$K_{I_{10}I_{2}}$ & Half-saturation constant of $I_{2}$ for $I_{10}$ & $2.00 \times 10^{-12}$ & $\mathrm{g/cm^{3}}$ & est. \\
\hline
$C_0$ & Carrying capacity of $C$ & $8.89 \times 10^{7}$ & $\mathrm{cell/cm^3}$ & fitted \\
$K_{CI_{\upbeta}}$ & Inhibition constant of $T_8$ and $K$ elimination of $C$ by $I_{\upbeta}$ & $1.51 \times 10^{-6}$ & $\mathrm{g/cm^{3}}$ & est. \\
$K_{CQ^{T_8}}$ & Inhibition constant of $T_8$ elimination of $C$ by $Q^{T_8}$ & $1.35 \times 10^6$ & $\mathrm{molec/cm^{3}}$ & est. \\
$K_{CQ^K}$ & Inhibition constant of $K$ elimination of $C$ by $Q^K$ & $2.96 \times 10^5$ & $\mathrm{molec/cm^3}$ & est. \\
$K_{D_0I_\upbeta}$ & Inhibition constant of $K$ elimination of $D_0$ by $I_\upbeta$ & $1.51 \times 10^{-6}$ & $\mathrm{g/cm^{3}}$ & est. \\
$V_\mathrm{LN}$ & Volume of the TDLN & $1.47 \times 10^{-1}$ & $\mathrm{cm^3}$ & \cite{Rssler2017supp} est. \\
$K_{T_0^8T_A^r}$ & Inhibition constant of $T_0^8$ activation by $T_A^r$ & $2.94 \times 10^{6}$ & $(\mathrm{cell/cm^3})\mathrm{~day}$ & est. \\
$K_{T_0^8 Q^{8\mathrm{LN}}}$ & Inhibition constant of $T_0^8$ activation by $Q^{8\mathrm{LN}}$ & $5.84 \times 10^{5}$ & $(\mathrm{molec/cm^3})$ & est. \\
$K_{T_A^8T_A^r}$ & Inhibition constant of $T_A^8$ proliferation by $T_A^r$ & $7.16 \times 10^{6}$ & $(\mathrm{cell/cm^3})\mathrm{~day}$ & est. \\
$K_{T_A^8 Q^{8\mathrm{LN}}}$ & Inhibition constant of $T_A^8$ proliferation by $Q^{8\mathrm{LN}}$ & $1.42 \times 10^{6}$ & $(\mathrm{molec/cm^3})\mathrm{~day}$ & est. \\
$K_{T_8T_r}$ & Inhibition constant of $I_2$-mediated growth of $T_8$ by $T_r$ & $2.78 \times 10^{5}$ & $\mathrm{cell/cm^{3}}$ & est. \\
$K_{T_8 I_{10}}$ & Inhibition constant of $T_8$ death by $I_{10}$ & $1.84 \times 10^{-10}$ & $\mathrm{g/cm^{3}}$ & est. \\
$K_{T_\mathrm{ex} I_{10}}$ & Inhibition constant of $T_\mathrm{ex}$ death by $I_{10}$ & $1.84 \times 10^{-10}$ & $\mathrm{g/cm^{3}}$ & est. \\
$K_{T_0^4 T_A^r}$ & Inhibition constant of $T_0^4$ activation by $T_A^r$ & $2.21 \times 10^{6}$ & $(\mathrm{cell/cm^3})\mathrm{~day}$ & est. \\
$K_{T_0^4 Q^{1\mathrm{LN}}}$ & Inhibition constant of $T_0^4$ activation by $Q^{1\mathrm{LN}}$ & $2.61 \times 10^{6}$ & $(\mathrm{molec/cm^3})\mathrm{~day}$ & est. \\
$K_{T_A^1 T_A^r}$ & Inhibition constant of $T_A^1$ proliferation by $T_A^r$ & $6.07 \times 10^{6}$ & $(\mathrm{cell/cm^3})\mathrm{~day}$ & est. \\
$K_{T_A^1 Q^{1\mathrm{LN}}}$ & Inhibition constant of $T_A^1$ proliferation by $Q^{1\mathrm{LN}}$ & $7.19 \times 10^{6}$ & $(\mathrm{molec/cm^3})\mathrm{~day}$ & est. \\
$K_{T_1T_r}$ & Inhibition constant of $I_2$-mediated growth of $T_1$ by $T_r$ & $2.78 \times 10^{5}$ & $\mathrm{cell/cm^{3}}$ & est. \\
$K_{KI_\upbeta}$ & Inhibition constant of NK cell activation by $I_\upbeta$ & $1.51 \times 10^{-6}$ & $\mathrm{g/cm^{3}}$ & est. \\
$K_{I_\upgamma T_r}$ & Inhibition constant of T cell production of $I_\upgamma$ by $T_r$ & $2.78 \times 10^{5}$ & $\mathrm{cell/cm^{3}}$ & est. \\
\hline
$d_{N_c}$ & Removal rate of $N_c$ & $6.88 \times 10^{-2}$ & $\mathrm{day}^{-1}$ & est. \\
$d_{H}$ & Degradation rate of $H$ & $5.55$ & $\mathrm{day}^{-1}$ & \cite{Zandarashvili2013supp} est. \\
$d_{S}$ & Degradation rate of $S$ & $1.39$ & $\mathrm{day}^{-1}$ & \cite{Goicoechea2003supp, Zhang2021calsupp} est. \\
$d_{D_0}$ & Death rate of $D_0$ & $3.57 \times 10^{-2}$ & $\mathrm{day}^{-1}$ & \cite{Ruedl2000supp} est. \\
$d_{D}$ & Death rate of $D$ & $3.15 \times 10^{-1}$ & $\mathrm{day}^{-1}$ & \cite{Kamath2002supp} est. \\
$d_{T_0^8}$ & Death rate of $T_0^8$ & $3.22 \times 10^{-2}$ & $\mathrm{day}^{-1}$ & \cite{Takada2009supp} est. \\
$d_{T_8}$ & Death rate of $T_8$ & $9\times 10^{-3}$ & $\mathrm{day}^{-1}$ & \cite{Hellerstein1999supp} \\
$d_{T_\mathrm{ex}}$ & Death rate of $T_\mathrm{ex}$ & $9\times 10^{-3}$ & $\mathrm{day}^{-1}$ & \cite{Hellerstein1999supp} \\
$d_{T_0^4}$ & Death rate of $T_0^4$ & $4.03\times 10^{-2}$ & $\mathrm{day}^{-1}$ & \cite{Takada2009supp} est. \\
$d_{T_1}$ & Death rate of $T_1$ & $8 \times 10^{-3}$ & $\mathrm{day}^{-1}$ & \cite{Hellerstein1999supp}\\
$d_{T_0^r}$ & Death rate of $T_0^r$ & $2.2 \times 10^{-3}$ & $\mathrm{day}^{-1}$ & \cite{Kumbhari2020n2supp} \\
$d_{T_r}$ & Death rate of $T_r$ & $6.30 \times 10^{-2}$ & $\mathrm{day}^{-1}$ & \cite{VukmanovicStejic2006supp} est. \\
$d_{M_0}$ & Death rate of $M_0$ & $0.73$ & $\mathrm{day}^{-1}$ & \cite{Patel2017supp} \\
$d_{M_1}$ & Death rate of $M_1$ & $0.99$ & $\mathrm{day}^{-1}$ & \cite{Patel2017supp} \\
$d_{M_2}$ & Death rate of $M_2$ & $1.35 \times 10^{-1}$ & $\mathrm{day}^{-1}$ & \cite{Patel2017supp} \\
$d_{K_0}$ & Death rate of $K_0$ & $6.93 \times 10^{-2}$ & $\mathrm{day}^{-1}$ & \cite{Wu2020supp, Vivier2008supp, Lowry2017supp} est. \\
$d_{K}$ & Death rate of $K$ & $6.93 \times 10^{-2}$ & $\mathrm{day}^{-1}$ & \cite{Wu2020supp, Vivier2008supp, Lowry2017supp} est. \\
$d_{I_2}$ & Degradation rate of $I_2$ & $1.45 \times 10^2$ & $\mathrm{day}^{-1}$ & \cite{Lotze1985n1supp} est. \\
$d_{I_\upgamma}$ & Degradation rate of $I_\upgamma$ & $3.33 \times 10^1$ & $\mathrm{day}^{-1}$ & \cite{Balachandran2013supp} est. \\
$d_{I_{\upalpha}}$ & Degradation rate of $I_{\upalpha}$ & $5.48 \times 10^{1}$ & $\mathrm{day}^{-1}$ & \cite{Ma2015supp, Oliver1993-vlsupp} est. \\
$d_{I_{\upbeta}}$ & Degradation rate of $I_{\upbeta}$ & $3.99 \times 10^{2}$ & $\mathrm{day}^{-1}$ & \cite{TiradoRodriguez2014supp} est. \\
$d_{I_{10}}$ & Degradation rate of $I_{10}$ & $6.16$ & $\mathrm{day}^{-1}$ & \cite{Huhn1997supp} est. \\
$d_{P_D}$ & Degradation rate of unbound PD-1 receptors & $3.36 \times 10^{-1}$ & $\mathrm{day}^{-1}$ & \cite{Lassman2021supp} \\
$d_{Q_A}$ & Internalisation rate of the PD-1/pembrolizumab complex & $0.43$ & $\mathrm{day}^{-1}$ & \cite{Li2021supp} \\
$d_{A_1}$ & Elimination rate of $A_1/A_1^\mathrm{LN}$ & $2.92 \times 10^{-2}$ & $\mathrm{day}^{-1}$ & \cite{Li2017supp, Li2019pembrosupp, Ahamadi2016supp} est. \\
$d_{P_L}$ & Degradation rate of unbound PD-L1 & $1.39$ & $\mathrm{day}^{-1}$ & \cite{Cha2019supp} \\
\hline
$\tau_m$ & DC migration time from TDLN to the TS & $0.75$ & $\mathrm{day}$ & \cite{Catron2004supp} est. \\
$\tau_8^\mathrm{act}$ & CD8+ T cell activation time & $2$ & $\mathrm{day}$ & \cite{Kinjyo2015supp} \\
$\Delta_8^0$ & Time taken for first CTL division & $1.63$ & $\mathrm{day}$ & \cite{Plambeck2022supp} \\
$n^8_\mathrm{max}$ & Maximal number of CTL divisions in the TDLN & $10$ & dimensionless & \cite{Kaech2001supp, Masopust2007supp} est. \\
$\Delta_8$ & Time taken for successive CTL divisions & $0.36$ & $\mathrm{day}$ & \cite{Kaech2001supp} \\
$\tau_{T_A^8}$ & Time taken for CTL division program & $4.87$ & $\mathrm{day}$ & est. \\
$\tau_a$ & T cell migration time from the TDLN to the TS & $0.27$ & $\mathrm{day}$ & est. \\
$\tau_l$ & Time for CTL to become exhausted in TS & $10$ & $\mathrm{day}$ & \cite{Blank2019supp, McLane2019supp} est. \\
$\tau_4^\mathrm{act}$ & CD4+ T cell activation time & $1.5$& $\mathrm{day}$ & \cite{JelleyGibbs2000supp} est. \\
$\Delta_1^0$ & Time taken for first Th1 cell division & $0.77$ & $\mathrm{day}$ & \cite{Kaech2002supp} est. \\
$n^1_\mathrm{max}$ & Maximal number of Th1 cell divisions in the TDLN & $9$ & dimensionless & \cite{Homann2001supp} est. \\
$\Delta_1$ & Time taken for successive Th1 cell divisions & $0.42$ & $\mathrm{day}$ & \cite{Kaech2002supp} est. \\
$\tau_{T_A^1}$ & Time taken for Th1 cell division program & $4.13$ & $\mathrm{day}$ & est. \\
$\tau_r^\mathrm{act}$ & Treg activation time & $1.5$ & $\mathrm{day}$ & \cite{JelleyGibbs2000supp} est. \\
$\Delta_r^0$ & Time taken for first Treg division & $0.77$ & $\mathrm{day}$ & \cite{Kaech2002supp} est. \\
$n^r_\mathrm{max}$ & Maximal number of Treg divisions in the TDLN & $6$ & dimensionless & \cite{DarrasseJze2009supp} est. \\
$\Delta_r$ & Time taken for successive Treg divisions & $0.42$ & $\mathrm{day}$ & \cite{Kaech2002supp} est. \\
$\tau_{T_A^r}$ & Time taken for Treg division program & $2.87$ & $\mathrm{day}$ & est. \\
\hline
\end{longtable}
\end{center}
\subsection{Steady States and Initial Conditions \label{initcond section supp}}
All steady states and initial conditions are as in \cite{Hawi2026metasupp}, but are repeated here for completeness. 
\subsubsection{Steady States and Initial Conditions for Cells in the TS\label{tsitesssupp}}
\begin{table}[H]
\centering
\begin{tabular}{|c|c|c|c|c|c|c|}
\hline
$C$ & $N_c$ & $D_0$ & $D$ & $T_8$ & $T_\mathrm{ex}$ & $T_1$ \\ 
\hline
$7.02 \times 10^{7}$ & $3.69 \times 10^{6}$ & $9.55 \times 10^{5}$ & $1.91 \times 10^{6}$ & $1.77 \times 10^{5}$ & $1.24 \times 10^{5}$ & $1.06 \times 10^{5}$ \\ 
\hline
$T_r$ & $M_0$ & $M_1$ & $M_2$ & $K_0$ & $K$ & \\ 
\hline
$2.78 \times 10^{5}$ & $7.93 \times 10^{5}$ & $3.27 \times 10^{5}$ & $1.30 \times 10^{6}$ & $3.88 \times 10^{6}$ & $1.94 \times 10^{5}$ & \\ 
\hline
\end{tabular}
\caption{\label{firstmodelsteadysupp}TS steady-state cell densities for the full model. All values are in $\mathrm{cell/cm^3}$.}
\end{table}
\begin{table}[H]
\centering
\begin{tabular}{|c|c|c|c|c|c|c|}
\hline
$C$ & $N_c$ & $D_0$ & $D$ & $T_8$ & $T_\mathrm{ex}$ & $T_1$ \\ 
\hline
$3.90 \times 10^{7}$ & $2.05 \times 10^{6}$ & $1.04 \times 10^{6}$ & $1.04 \times 10^{6}$ & $1.61 \times 10^{5}$ & $1.93 \times 10^{5}$ & $1.01 \times 10^{5}$ \\ 
\hline
$T_r$ & $M_0$ & $M_1$ & $M_2$ & $K_0$ & $K$ & \\ 
\hline
$2.02 \times 10^{5}$ & $2.50 \times 10^{5}$ & $2.09 \times 10^{5}$ & $1.29 \times 10^{6}$ & $4.47 \times 10^{6}$ & $4.47 \times 10^{5}$ & \\ 
\hline
\end{tabular}
\caption{\label{firstmodelinitcondsupp}TS initial condition cell densities for the full model. All values are in $\mathrm{cell/cm^3}$.}
\end{table}
\subsubsection{Steady States and Initial Conditions for Cells in the TDLN\label{TDLNssinitsupp}}
\begin{table}[H]
\centering
\begin{tabular}{|c|c|c|c|c|c|c|}
\hline
$D^\mathrm{LN}$ & $T_0^8$ & $T_A^8$ & $T_0^4$ & $T_A^1$ & $T_0^r$ & $T_A^r$ \\
\hline
$3.28 \times 10^{7}$ & $1.16 \times 10^{7}$ & $8.31 \times 10^{5}$ & $6.73 \times 10^{6}$ & $6.66 \times 10^{6}$ & $3.23 \times 10^{5}$ & $1.47 \times 10^{6}$ \\
\hline
\end{tabular}
\caption{\label{TDLNfirstmodelsteadysupp}TDLN steady-state cell densities for the full model. All values are in $\mathrm{cell/cm^3}$.}
\end{table}
\begin{table}[H]
\centering
\begin{tabular}{|c|c|c|c|c|c|c|}
\hline
$D^\mathrm{LN}$ & $T_0^8$ & $T_A^8$ & $T_0^4$ & $T_A^1$ & $T_0^r$ & $T_A^r$ \\
\hline
$1.78 \times 10^7$ & $1.20 \times 10^{7}$ & $8.60 \times 10^{5}$ & $4.31 \times 10^{6}$ & $7.76 \times 10^{6}$ & $1.72 \times 10^{5}$ & $7.81 \times 10^{5}$ \\
\hline
\end{tabular}
\caption{\label{TDLNfirstmodelinitcondsupp}TDLN initial condition cell densities for the full model. All values are in $\mathrm{cell/cm^3}$.}
\end{table}
\subsubsection{Steady States and Initial Conditions for DAMPs\label{DAMPssinitsecsupp}}
\begin{table}[H]
    \centering
    \begin{tabular}{|c|c|c|}
    \hline
    \textbf{DAMP} & \textbf{Steady State} & \textbf{Initial Condition} \\
    \hline 
    $H$ & $1.94 \times 10^{-8}$ & $1.33 \times 10^{-8}$ \\
    $S$ & $4.50 \times 10^{-8}$ & $3.25 \times 10^{-8}$ \\
    \hline
    \end{tabular}
    \caption{\label{DAMPtablesupp}DAMP steady states and initial conditions for the full model. All values are in units of $\mathrm{g/cm^3}$.}
\end{table}
\subsubsection{Steady States and Initial Conditions for Cytokines}
\begin{table}[H]
    \centering
    \begin{tabular}{|c|c|c|}
    \hline
    \textbf{Cytokine} & \textbf{Steady State} & \textbf{Initial Condition} \\
    \hline 
    $I_2$ & $2.00 \times 10^{-12}$ & $1.87 \times 10^{-12}$\\
    $I_\upgamma$ & $4.93 \times 10^{-11}$ & $1.10 \times 10^{-10}$ \\
    $I_\upalpha$ & $9.00 \times 10^{-11}$ & $1.25 \times 10^{-10}$\\
    $I_\upbeta$ & $1.51 \times 10^{-6}$ & $9.20 \times 10^{-7}$ \\
    $I_{10}$ & $1.84 \times 10^{-10}$ & $1.15 \times 10^{-10}$\\
    \hline
    \end{tabular}
    \caption{\label{cytokinetablesupp}Cytokine steady states and initial conditions for the full model. All values are in units of $\mathrm{g/cm^3}$.}
\end{table}
\subsubsection{Steady States and Initial Conditions for Immune Checkpoint-Associated Components in the TS}
\begin{table}[H]
    \centering
    \begin{tabular}{|c|c|c|}
    \hline
    \textbf{Protein} & \textbf{Steady State} & \textbf{Initial Condition} \\
    \hline 
    $P_D^{T_8}$ & $4.87 \times 10^8$ & $4.44 \times 10^8$ \\
    $P_D^{T_1}$ & $2.17 \times 10^8$ & $2.07 \times 10^8$ \\
    $P_D^{K}$ & $1.07 \times 10^8$ & $2.46 \times 10^8$ \\
    $P_L$ & $1.30 \times 10^{13}$ & $7.40 \times 10^{12}$ \\
    $Q^{T_8}$ & $1.35 \times 10^6$ & $6.99 \times 10^5$ \\
    $Q^{T_1}$ & $6.01 \times 10^5$ & $3.26 \times 10^5$ \\
    $Q^{K}$ & $2.96 \times 10^5$ & $3.88 \times 10^5$ \\
    \hline
    \end{tabular}
    \caption{\label{TSICItablesupp}TS immune checkpoint-associated component steady states and initial conditions for the full model. All values are in units of $\mathrm{molec/cm^3}$.}
\end{table}
\subsubsection{Steady States and Initial Conditions for Immune Checkpoint-Associated Components in the TDLN}
\begin{table}[H]
    \centering
    \begin{tabular}{|c|c|c|}
    \hline
    \textbf{Protein} & \textbf{Steady State} & \textbf{Initial Condition} \\
    \hline 
    $P_{D}^{8\mathrm{LN}}$ & $2.29 \times 10^9$ & $2.37 \times 10^9$ \\
    $P_{D}^{1\mathrm{LN}}$ & $1.37 \times 10^{10}$ & $1.59 \times 10^{10}$ \\
    $P_{L}^{\mathrm{LN}}$ & $5.99 \times 10^{11}$ & $3.34 \times 10^{11}$ \\
    $Q^{8\mathrm{LN}}$ & $2.92 \times 10^5$ & $1.69 \times 10^5$ \\
    $Q^{1\mathrm{LN}}$ & $1.74 \times 10^6$ & $1.13 \times 10^6$ \\
    \hline
    \end{tabular}
    \caption{\label{TDLNICItablesupp}TDLN immune checkpoint-associated component steady states and initial conditions for the full model. All values are in units of $\mathrm{molec/cm^3}$.}
\end{table}
\subsubsection{Steady States and Initial Conditions for Pembrolizumab-Associated Components in the TS}
\begin{table}[H]
    \centering
    \begin{tabular}{|c|c|c|}
    \hline
    \textbf{Protein} & \textbf{Steady State} & \textbf{Initial Condition} \\
    \hline 
    $Q_A^{T_8}$ & $0$ & $0$ \\
    $Q_A^{T_1}$ & $0$ & $0$ \\
    $Q_A^{K}$ & $0$ & $0$ \\
    $A_1$ & $0$ & $0$ \\
    \hline
    \end{tabular}
    \caption{\label{TSpembrotablesupp}Steady states and initial conditions for pembrolizumab-associated components in the TS in the full model. All values are in units of $\mathrm{molec/cm^3}$.}
\end{table}
\subsubsection{Steady States and Initial Conditions for Pembrolizumab-Associated Components in the TDLN}
\begin{table}[H]
    \centering
    \begin{tabular}{|c|c|c|}
    \hline
    \textbf{Protein} & \textbf{Steady State} & \textbf{Initial Condition} \\
    \hline 
    $Q_A^{8\mathrm{LN}}$ & $0$ & $0$ \\
    $Q_A^{1\mathrm{LN}}$ & $0$ & $0$ \\
    $A_1^\mathrm{LN}$ & $0$ & $0$ \\
    \hline
    \end{tabular}
    \caption{\label{TDLNpembrotablesupp}Steady states and initial conditions for pembrolizumab-associated components in the TDLN in the full model. All values are in units of $\mathrm{molec/cm^3}$.}
\end{table}
\putbib[References.bib]
\end{bibunit}
\newpage
\title{Appendix B: Sensitivity analysis-guided model reduction of a mathematical model of pembrolizumab therapy for de novo metastatic MSI-H/dMMR colorectal cancer}
\maketitle
\setcounter{page}{1}
\begin{bibunit}[vancouver]
\section{Reduced Model Parameter Estimation\label{reducedmodelparamestsection}}
We estimate all parameters, where possible, under the assumption that no pembrolizumab has been or will be administered. The exceptions to this are the parameters directly related to pembrolizumab treatment, for which the assumptions are explicitly stated during estimation. Many of the assumptions and techniques in this section are adopted from \cite{Hawi2026metasupp} and \cite{Hawi2025localsupp}.\\~\\
For the sake of brevity, we present parameter estimation only for those parameters that differ between the full and the reduced model. The absence of a parameter in this section implies that its value and derivation are identical to those provided in \cite{Hawi2026metasupp} and \autoref{tableofparams}.
\subsection{Half-Saturation Constants}
We recall that for some species $X$, $K_X$ is denoted the half-saturation constant of $X$ in a term of the form 
\begin{equation*}
    \frac{X}{K_X +X}.
\end{equation*}
For simplicity, we assume that if $\overline{X}$ denotes the steady state value of $X$, then
\begin{equation}
    \frac{\overline{X}}{K_X + \overline{X}} = \frac{1}{2} \implies K_X = \overline{X}. \label{halfsateqn}
\end{equation}
Using \eqref{halfsateqn}, we have that
\begin{align*}
    K_{T_8C} &= \overline{C} = 7.02 \times 10^7 ~\mathrm{cell/cm^3}, \\
    K_{T_1Q^{T_1}}&= \overline{Q^{T_1}} = 6.01 \times 10^5 ~\mathrm{molec/cm^3}.
\end{align*}
\subsection{Inhibition Constants}
We recall that for some species $X$, $K_X$ is denoted as the inhibition constant of $X$ in a term of the form
\begin{equation*}
    \frac{1}{1 + X/K_X}. \label{inhibitionconsteqn}
\end{equation*}
For simplicity, we assume that if $\overline{X}$ denotes the steady state value of $X$, then
\begin{equation}
    \frac{1}{1 + \overline{X}/K_X} = \frac{1}{2} \implies K_X = \overline{X}.
\end{equation}
Using \eqref{inhibitionconsteqn}, we have that
\begin{align*}
K_{CQ^{T_8}} &= \overline{Q^{T_8}} = 1.35 \times 10^{6} ~\mathrm{molec/cm^3}, \\
K_{CQ^{K}} &= \overline{Q^{K}} = 2.95 \times 10^{6} ~\mathrm{molec/cm^3}, \\
K_{T_0^8T_A^r} = K_{T_A^8T_A^r} = K_{T_0^4 T_A^r} = K_{T_A^1 T_A^r} &= \overline{T_A^{r}} = 1.47 \times 10^{6} ~\mathrm{cell/cm^3}, \\
K_{T_0^8 Q^{8\mathrm{LN}}} = K_{T_A^8 Q^{8\mathrm{LN}}} &= \overline{Q^{8\mathrm{LN}}} = 2.90 \times 10^5 ~ \mathrm{molec/cm^3}, \\
K_{T_0^4 Q^{1\mathrm{LN}}} = K_{T_A^1 Q^{1\mathrm{LN}}} &= \overline{Q^{1\mathrm{LN}}} = 1.73 \times 10^6 ~\mathrm{molec/cm^3}.
\end{align*}
\subsection{Cytokine Production Parameters\label{reducedcytokinessic}}
\subsubsection{Estimates for $I_\upgamma$}
Considering \eqref{reducedifngammaeqn} at steady state, or equivalently considering \eqref{reducedifngammaqssa}, leads to
\begin{equation*}
\lambda_{I_{\upgamma} K}\overline{K} -d_{I_\upgamma}\overline{I_\upgamma}=0.
\end{equation*}
Solving this leads to
\begin{align*}
    \lambda_{I_{\upgamma} K} &= 8.46 \times 10^{-15} ~\left(\mathrm{g/cell}\right)\mathrm{day^{-1}}.
\end{align*}
Consequently, considering \eqref{reducedifngammaqssa}, we have that
\begin{equation*}
    I_\upgamma(0) = \frac{\lambda_{I_{\upgamma} K}}{d_{I_\upgamma}} K(0) = 1.14 \times 10^{-10} \mathrm{~g/cm^3}.
\end{equation*}
\subsubsection{Estimates for $I_{10}$}
Amongst $48$ different cell lines tested, it was found in \cite{Gastl1993supp} that cancer IL-10 production was maximised in cell lines derived from colon carcinomas. As such, we assume that at steady state, cancer production of IL-10 is equal to half of that by $M_2$ macrophages. This, in conjunction with considering \eqref{reducedil10eqn} at steady state, leads to the equations
\begin{equation*}
    \frac{\lambda_{I_{10}C}}{1} = \frac{\lambda_{I_{10}M_2}}{2},
\end{equation*}
and
\begin{equation*}
\lambda_{I_{10}C}\overline{C} + \lambda_{I_{10}M_2}\overline{M_2} - d_{I_{10}}\overline{I_{10}} = 0.
\end{equation*}
Solving these simultaneously leads to
\begin{align*}
    \lambda_{I_{10} C} &= 1.56 \times 10^{-17} ~\left(\mathrm{g/cell}\right)\mathrm{day^{-1}},\\
    \lambda_{I_{10} M_2} &= 3.11 \times 10^{-17} ~\left(\mathrm{g/cell}\right)\mathrm{day^{-1}}.
\end{align*}
\subsection{Parameters for DCs, Macrophages, and NK Cells}
\subsubsection{Estimates for $D_0$ and $D$\label{reduceddcestsection}}
Adding \eqref{reducedD0eqn} and \eqref{reducedDeqn} at steady state leads to
\begin{equation*}
    \mathcal{A}_{D_0} - \frac{\lambda_{D_0K} \overline{D_0}\overline{K}}{2} - d_{D_0}\overline{D_0} - \lambda_{DD^\mathrm{LN}}\overline{D} - d_{D}\overline{D} =0.
\end{equation*}
In \cite{Piccioli2002supp}, it was also shown that the percentage of immature DCs that were lysed as a result of NK cells is roughly linear in the ratio of NK cells to immature DCs. When a 1:1 ratio of activated NK cells to immature DCs is present, after 24 hours, roughly $35.5\%$ of immature DCs are lysed, whereas if a 5:1 ratio is present, $85.5\%$ of immature DCs are lysed. At steady state, the ratio of NK cells to immature DCs is $\approx 2.39:1$, corresponding to an approximate $52.85\%$ being lysed. However, if we consider only immature DC loss due to degradation, after $24$ hours, only $ 1-\exp\left(-d_{D_0}\right) \approx 3.54\%$ is lost to it. Thus, we assume at steady state that
\begin{equation*}
    \frac{\lambda_{D_0K} \overline{D_0}\overline{K}}{2 \times 0.5285} = \frac{d_{D_0}\overline{D_0}}{0.0354} \implies \lambda_{D_0K} = 5.49 \times 10^{-6} ~\left(\mathrm{cell/cm^3}\right)^{-1}\mathrm{day^{-1}}.
\end{equation*}
Considering \eqref{reducedDeqn} at steady state leads to
\begin{equation*}
    \frac{\lambda_{DH}\overline{D_0}}{2} - \lambda_{DD^\mathrm{LN}}\overline{D} - d_{D}\overline{D} =0.
\end{equation*}
Finally, it was found in \cite{Verdijk2009supp} that only a limited number of DCs migrate up to the TDLN, with at most 4\% of DCs reaching the TDLN in melanoma patients when DCs were injected intradermally. We assume at steady state that this holds, too, for MSI-H/dMMR CRC. Taking into account that only $\exp\left(-d_D\tau_m\right)$ of mature DCs that leave the TS survive their migration to the TDLN, we have that
\begin{equation*}
    \frac{\lambda_{DD^\mathrm{LN}}}{0.04\exp\left(d_D\tau_m\right)} = \frac{d_D}{1-0.04\exp\left(d_D\tau_m\right)}.
\end{equation*}
This leads to
\begin{align*}
    \mathcal{A}_{D_0} &= 1.18 \times 10^{6} ~\left(\mathrm{cell/cm^{3}}\right)\mathrm{day^{-1}}, \\
    \lambda_{DH} &= 1.33 ~\mathrm{day^{-1}}, \\
    \lambda_{DD^\mathrm{LN}} &= 1.68 \times 10^{-2} ~\mathrm{day^{-1}}.
\end{align*}
\subsubsection{Estimates for $M_0$, $M_1$, and $M_2$}
To estimate the macrophage production constants, we consider \eqref{reducedM0eqn}, \eqref{reducedM1eqn}, \eqref{reducedM2eqn} at steady state, and use the data from \cite{Cui2023supp}. We assume that the magnitude of response to a specific cytokine is proportional to its corresponding macrophage polarisation rate, where the response is defined as the Euclidean distance between the centroid vectors of cytokine-treated macrophages and phosphate-buffered saline (PBS)-treated macrophages. \\~\\
Adding \eqref{reducedM0eqn}, \eqref{reducedM1eqn}, and \eqref{reducedM2eqn} at steady state leads to
\begin{equation*}
    \mathcal{A}_{M_0} - d_{M_0}\overline{M_0} - d_{M_1}\overline{M_1}-d_{M_2}\overline{M_2} = 0 \implies \mathcal{A}_{M_0} = 1.08 \times 10^6 ~\left(\mathrm{cell/cm^{3}}\right)\mathrm{day^{-1}}.       
\end{equation*}
Using values from \cite{Cui2023supp} and considering \eqref{reducedM0eqn} and \eqref{reducedM1eqn} at steady state leads to the equations
\begin{align*}
    &\mathcal{A}_{M_0} - \frac{\lambda_{M_1 I_\upalpha}\overline{M_0}}{2} - \frac{\lambda_{M_1 I_\upgamma}\overline{M_0}}{2} - \frac{\lambda_{M_2 I_{10}}\overline{M_0}}{2} - \frac{\lambda_{M_2 I_\upbeta}\overline{M_0}}{2} - d_{M_0}\overline{M_0} = 0,\\
    & \frac{\lambda_{M_1 I_\upalpha}\overline{M_0}}{2} + \frac{\lambda_{M_1 I_\upgamma}\overline{M_0}}{2} - d_{M_1}\overline{M_1} = 0, \\
    &\frac{\lambda_{M_1 I_\upalpha}}{2 \times 10.77} = \frac{\lambda_{M_1 I_\upgamma}}{2 \times 12.54},\\
    &\frac{\lambda_{M_2 I_{10}}}{2 \times 6.81} = \frac{\lambda_{M_2I_\upbeta}}{2 \times 7.63}.
\end{align*}
Solving these simultaneously leads to
\begin{align*}
\lambda_{M_1I_\upalpha} & = 3.77 \times 10^{-1} \mathrm{~day^{-1}}, \\
\lambda_{M_1I_\upgamma} & = 4.39 \times 10^{-1} \mathrm{~day^{-1}}, \\
\lambda_{M_2I_{10}} & = 2.09 \times 10^{-1} \mathrm{~day^{-1}}, \\
\lambda_{M_2I_\upbeta} & = 2.34 \times 10^{-1} \mathrm{~day^{-1}}.
\end{align*}
\subsubsection{Estimates for $K_0$ and $K$\label{reducednkestsection}}
To estimate NK cell production parameters, we do a similar process to that of macrophages. Adding \eqref{reducedK0eqn} and \eqref{reducedKeqn} at steady state leads to
\begin{equation*}
    \mathcal{A}_{K_0} - d_{K_0}\overline{K_0} - d_{K}\overline{K}= 0 \implies \mathcal{A}_{K_0} = 2.82 \times 10^5 ~\left(\mathrm{cell/cm^{3}}\right)\mathrm{day^{-1}}.
\end{equation*}
Considering \eqref{reducedKeqn} at steady state leads to
\begin{equation*}
    \frac{1}{2}\left(\frac{\lambda_{KI_2}\overline{K_0}}{2} + \frac{\lambda_{KD}\overline{K_0}}{2}\right) - d_{K}\overline{K} = 0.
\end{equation*}
We finally assume that DC-mediated NK-cell activation is twice as potent as cytokine-induced activation at steady state, so that
\begin{equation*}
    \frac{\lambda_{KD}/2}{2} = \frac{\lambda_{KI_2}/2}{1}.
\end{equation*}
Solving these simultaneously leads to
\begin{align*}
    \lambda_{KI_2} &= 4.62 \times 10^{-3} \mathrm{~day^{-1}}, \\
    \lambda_{KD} &= 9.24 \times 10^{-3} \mathrm{~day^{-1}}.
\end{align*}
\subsection{T Cell Parameters and Estimates}
\subsubsection{Estimates for $T_0^8$, $T_A^8$, $T_8$, and $T_\mathrm{ex}$\label{reducedcd8estsection}}
Considering \eqref{reducednaivecd8eqn} at steady state leads to
\begin{equation*}
    \mathcal{A}_{T_0^8} - \overline{R^8} - d_{T_0^8}\overline{T_0^8} = 0,
\end{equation*}
and in particular,
\begin{equation*}
    \overline{R^8} = \frac{\lambda_{T_0^8 T_A^8}\overline{D^\mathrm{LN}}\overline{T_0^8}}{4}.
\end{equation*}
Considering \eqref{reducedTA8n8maxeqn} at steady state leads to
\begin{equation*}
    \frac{2^{n^8_\mathrm{max}}\exp\left(-d_{T_0^8}\tau_{T_A^8}\right)\overline{R^8}}{4} - \lambda_{T_A^8T_8}\overline{T_A^8} - d_{T_8} \overline{T_A^8} = 0.
\end{equation*}
We first consider the case where no pembrolizumab is present. Considering \eqref{reducedt8eqn} and \eqref{reducedTexeqn} at steady state leads to
\begin{align*}
    \frac{V_\mathrm{LN}}{\overline{V_\mathrm{TS}}}\lambda_{T_A^8T_8}\overline{T_A^8} - \frac{\lambda_{T_8C}\overline{T_8}}{2} - \frac{d_{T_8}\overline{T_8}}{2}&=0, \\
    \frac{\lambda_{T_8C}\overline{T_8}}{2} - \frac{d_{T_\mathrm{ex}}\overline{T_\mathrm{ex}}}{2}&=0.
\end{align*}
To determine $\lambda_{T_\mathrm{ex}A_1}$, we assume that when pembrolizumab is present, at steady state, $20\%$ of exhausted CD8+ T cells are reinvigorated. That is, we assume that
\begin{equation*}
\frac{\lambda_{T_\mathrm{ex}A_1}\overline{T_\mathrm{ex}}/2}{0.2}=\frac{d_{T_\mathrm{ex}}\overline{T_\mathrm{ex}}/2}{0.8}.
\end{equation*}
Solving these equations simultaneously leads to
\begin{align*}
\mathcal{A}_{T_0^8} &= 3.76 \times 10^5 ~\left(\mathrm{cell/cm^{3}}\right)\mathrm{day^{-1}}, \\
\lambda_{T_0^8 T_A^8} &= 2.69 \times 10^{-11} ~\left(\mathrm{cell/cm^3}\right)^{-1}\mathrm{day^{-1}}, \\
\overline{R^8} &= 2.18 \times 10^3 ~\left(\mathrm{cell/cm^{3}}\right)\mathrm{day^{-1}}, \\
\lambda_{T_A^8T_8} &= 6.64 \times 10^{-1} ~\mathrm{day}^{-1}, \\
\lambda_{T_8C} &= 6.31 \times 10^{-3} ~\mathrm{day}^{-1}, \\
\lambda_{T_\mathrm{ex}A_1} &= 2.25 \times 10^{-3} ~\mathrm{day}^{-1}.
\end{align*}
\subsubsection{Estimates for $T_0^4$, $T_A^1$, and $T_1$}
Considering \eqref{reducednaivecd4eqn} at steady state leads to
\begin{equation*}
    \mathcal{A}_{T_0^4} - \overline{R^1} - d_{T_0^4}\overline{T_0^4} = 0,
\end{equation*}
where
\begin{equation*}
    \overline{R^1} = \frac{\lambda_{T_0^4 T_A^1}\overline{D^\mathrm{LN}}\overline{T_0^4}}{4}.
\end{equation*}
Considering \eqref{reducedTA1n1maxeqn} at steady state leads to
\begin{equation*}
    \frac{2^{n^1_\mathrm{max}}\exp\left(-d_{T_0^4}\tau_{T_A^1}\right) \overline{R^1}}{4} - \lambda_{T_A^1T_1}\overline{T_A^1}-d_{T_1}\overline{T_A^1}=0.
\end{equation*}
Based on murine data from \cite{Tan2025supp}, we assume that at steady state, $20\%$ of Th1 cells are converted to Tregs. That is, we assume that
\begin{equation*}
\frac{\lambda_{T_1T_r}\overline{T_1}/2}{0.2} = \frac{d_{T_1}\overline{T_1}}{0.8}.
\end{equation*}
Finally, considering \eqref{reducedth1eqn} at steady state leads to
\begin{align*}
    \frac{V_\mathrm{LN}}{\overline{V_\mathrm{TS}}}\lambda_{T_A^1 T_1}\overline{T_A^1} - \frac{\lambda_{T_1T_r}\overline{T_1}}{2}- d_{T_1}\overline{T_1} &= 0.
\end{align*}
Solving these equations simultaneously leads to
\begin{align*}
\mathcal{A}_{T_0^4} &= 2.76 \times 10^5 ~\left(\mathrm{cell/cm^{3}}\right)\mathrm{day^{-1}}, \\
\lambda_{T_0^4 T_A^1} &= 8.12 \times 10^{-11} ~\left(\mathrm{cell/cm^3}\right)^{-1}\mathrm{day^{-1}}, \\
\overline{R^1} &= 3.79 \times 10^3 ~\left(\mathrm{cell/cm^{3}}\right)\mathrm{day^{-1}}, \\
\lambda_{T_A^1 T_1} &= 6.48 \times 10^{-2} ~\mathrm{day}^{-1}, \\
\lambda_{T_1T_r} &= 4.00 \times 10^{-3} ~\mathrm{day}^{-1}.
\end{align*}
\subsubsection{Estimates for $T_0^r$, $T_A^r$, and $T_r$}
Considering \eqref{reducednaivetregeqn} at steady state leads to
\begin{equation*}
    \mathcal{A}_{T_0^r} - \overline{R^r} - d_{T_0^r}\overline{T_0^r} = 0,
\end{equation*}
where
\begin{equation*}
    \overline{R^r} = \lambda_{T_0^r T_A^r}\overline{D^\mathrm{LN}}\overline{T_0^r}.
\end{equation*}
Considering \eqref{reducedTArnrmaxeqn} at steady state leads to
\begin{equation*}
    2^{n^r_\mathrm{max}}\exp\left(-d_{T_0^r}\tau_{T_A^r}\right) \overline{R^r} - \lambda_{T_A^rT_r}\overline{T_A^r}-d_{T_r}\overline{T_A^r}=0.
\end{equation*}
Finally, considering \eqref{reducedtregeqn} at steady state leads to
\begin{align*}
    \frac{V_\mathrm{LN}}{\overline{V_\mathrm{TS}}}\lambda_{T_A^r T_r}\overline{T_A^r} + \frac{\lambda_{T_1T_r}\overline{T_1}}{2} - d_{T_r}\overline{T_r} &= 0.
\end{align*}
Solving these equations simultaneously leads to
\begin{align*}
\mathcal{A}_{T_0^r} &= 1.15 \times 10^5 ~\left(\mathrm{cell/cm^{3}}\right)\mathrm{day^{-1}}, \\
\lambda_{T_0^r T_A^r} &= 1.06 \times 10^{-8} ~\left(\mathrm{cell/cm^3}\right)^{-1}\mathrm{day^{-1}}, \\
\overline{R^r} &= 1.12 \times 10^5 ~\left(\mathrm{cell/cm^{3}}\right)\mathrm{day^{-1}}, \\
\lambda_{T_A^r T_r} &= 4.79 ~\mathrm{day}^{-1}.
\end{align*}
\subsection{Estimates for Cancer Cells}
Considering \eqref{reducedcancereqn} at steady state leads to
\begin{equation*}
    \lambda_{C}\left(1-\frac{\overline{C}}{C_0}\right)-\frac{\lambda_{CT_8}}{4}\overline{T_8} - \frac{\lambda_{CK}}{4}\overline{K}-\frac{\lambda_{CI_\upalpha}}{2} = 0.
\end{equation*}
We assume that CD8+ T cells and NK cells kill cancer cells with similar potency, so we approximate
\begin{equation*}
\lambda_{CK}/{4} = \lambda_{CT_8}/4 \implies \lambda_{CK} = \lambda_{CT_8}.
\end{equation*}
Considering \eqref{reducednecroticcelleqn} at steady state leads to the equation
\begin{equation*}
\frac{\lambda_{CI_\upalpha}\overline{C}}{2} - d_{N_c}\overline{N_c} = 0.
\end{equation*}
Solving these simultaneously, and using the same $\lambda_C$, $\lambda_{CT_8}$, and $C_0$ as the full model, leads to
\begin{align*}
    \lambda_C &= 5.25 \times 10^{-2} ~\mathrm{day}^{-1}, \\
    \lambda_{CT_8}&= 8.01 \times 10^{-8} ~\left(\mathrm{cell/cm^{3}}\right)^{-1}\mathrm{day^{-1}}, \\
    C_0 &= 8.89 \times 10^{7} ~\mathrm{cell/cm^3}, \\
    \lambda_{CK} &= 8.01 \times 10^{-8} ~\left(\mathrm{cell/cm^{3}}\right)^{-1}\mathrm{day^{-1}}, \\
    \lambda_{CI_\upalpha}&= 7.23 \times 10^{-3} ~ \mathrm{day}^{-1}, \\
    d_{N_c} &= 6.88 \times 10^{-2} \mathrm{~day^{-1}}.
\end{align*}
\subsection{Estimates for Immune Checkpoint-Associated Components in the TS\label{reducedtsicissinitappendix}}
\subsubsection{Estimates for Synthesis Rates and Steady States}
We first denote $\rho_{P_D^{T_8}}$, $\rho_{P_D^{T_1}}$, and $\rho_{P_D^{K}}$ as the number of PD-1 molecules expressed on the surface of CD8+ T cells, Th1 cells, and activated NK cells in the TS, respectively. To determine these parameters, we used the baseline data collected in \cite{Pluim2019supp} on $5$ advanced cancer patients before their pembrolizumab infusions. The net number of PD-1 molecules on the surface of CD4+ T cells was $2053~\mathrm{molec/cell}$, and so we set $\rho_{P_D^{T_1}} = 2.05 \times 10^{3} ~\mathrm{molec/cell}$. The net number of PD-1 molecules on the surface of CD8+ T cells was $2761~\mathrm{molec/cell}$, and so we set $\rho_{P_D^{T_8}} = 2.76 \times 10^{3}~ \mathrm{molec/cell}$. Despite the net number of PD-1 molecules on the surface of NK cells being below the lower limit of quantification in \cite{Pluim2019supp}, NK cells substantially express PD-1 \cite{Liu2017supp} in CRC, and so we set $\rho_{P_D^{K}} = \rho_{P_D^{T_8}}/5 = 5.52 \times 10^{2}~ \mathrm{molec/cell}$.\\~\\
We next denote $\rho_{P_LC}$ and $\rho_{P_LM_2}$ as the number of PD-L1 molecules expressed on cancer cells and M2 macrophages, respectively. In their quantitative systems pharmacology model of colorectal cancer, Anbari et al.\ estimated the baseline numbers of PD-L1 molecules per cancer cell and per APC to be $180,000 ~\mathrm{molec/cell}$ and $266,666 ~\mathrm{molec/cell}$, respectively \cite{Anbari2023supp}. This makes sense, noting that PD-L1 expression in macrophages is stronger and more continuous than that in cancer cells \cite{Saito2022supp}. As such, we set $\rho_{P_LC} = 1.8 \times 10^{5} \mathrm{~molec/cell}$ and $\rho_{P_L{M_2}} = 2.67 \times 10^{5} \mathrm{~molec/cell}$. \\~\\
Considering \eqref{reducedPD8eqn}~--~\eqref{reducedPDKeqn}, and \eqref{reducedPLeqn}~--~\eqref{reducedQKeqn} at steady state in the absence of pembrolizumab leads to
\begin{align*}
    \lambda_{P_D^{T_8}}\overline{T_8} - d_{P_D}\overline{P_D^{T_8}} &=0, \\
    \lambda_{P_D^{T_1}}\overline{T_1} - d_{P_D}\overline{P_D^{T_1}} &=0, \\
    \lambda_{P_D^{K}}\overline{K} - d_{P_D}\overline{P_D^{K}} &=0, \\
    \lambda_{P_L C}\overline{C} + \lambda_{P_L M_2}\overline{M_2} - d_{P_L}\overline{P_L} &=0, \\
    \overline{Q^{T_8}} - \frac{\lambda_{P_DP_L}}{\lambda_Q}\overline{P_D^{T_8}}\overline{P_L}&=0, \\
    \overline{Q^{T_1}} - \frac{\lambda_{P_DP_L}}{\lambda_Q}\overline{P_D^{T_1}}\overline{P_L}&=0, \\
    \overline{Q^{K}} - \frac{\lambda_{P_DP_L}}{\lambda_Q}\overline{P_D^{K}}\overline{P_L}&=0.
    \intertext{By considering the total number of PD-1 receptors expressed on each PD-1-expressing cell at steady state, we expect in the absence of pembrolizumab that}
    \overline{P_D^{T_8}} + \overline{Q^{T_8}} &=\rho_{P_D^{T_8}}\overline{T_8}, \\
    \overline{P_D^{T_1}} + \overline{Q^{T_1}} &=\rho_{P_D^{T_1}}\overline{T_1}, \\
    \overline{P_D^{K}} + \overline{Q^{K}} &=\rho_{P_D^{K}}\overline{K}.
    \intertext{We can also consider the total number of PD-L1 ligands at steady state so that}
    \overline{P_L} + \overline{Q^{T_8}} + \overline{Q^{T_1}} + \overline{Q^{K}} &= \rho_{P_L C}\overline{C} + \rho_{P_L M_2}\overline{M_2}.
    \intertext{Finally, we expect the synthesis rates of PD-1 and PD-L1 to be proportional to the total number of PD-1 molecules expressed per PD-1- and PD-L1-expressing cell, so that}
    \frac{\lambda_{P_D^{T_8}}}{\rho_{P_D^{T_8}}} =     \frac{\lambda_{P_D^{T_1}}}{\rho_{P_D^{T_1}}} &=     \frac{\lambda_{P_D^{K}}}{\rho_{P_D^{K}}}, \\
    \frac{\lambda_{P_LC}}{\rho_{P_LC}} &= \frac{\lambda_{P_LM_2}}{\rho_{P_LM_2}}.
\end{align*}
Solving these simultaneously and ensuring all model parameters are positive leads to
\begin{align*}
    \lambda_{P_D^{T_8}} &= 9.25 \times 10^2 ~\left(\mathrm{molec/cell}\right)\mathrm{day^{-1}}, \\
    \lambda_{P_D^{T_1}} &= 6.87 \times 10^2 ~\left(\mathrm{molec/cell}\right)\mathrm{day^{-1}}, \\
    \lambda_{P_D^{K}} &= 1.85 \times 10^2 ~\left(\mathrm{molec/cell}\right)\mathrm{day^{-1}}, \\
    \lambda_{P_LC} &= 2.50 \times 10^{5} ~\left(\mathrm{molec/cell}\right)\mathrm{day^{-1}}, \\
    \lambda_{P_LM_2} &= 3.71 \times 10^{5} ~\left(\mathrm{molec/cell}\right)\mathrm{day^{-1}}.
\end{align*}
This leads to
\begin{align*}
    \overline{P_D^{T_8}} &= 4.87 \times 10^8 \mathrm{~molec/cm^3}, \\
    \overline{P_D^{T_1}} &= 2.17 \times 10^8 \mathrm{~molec/cm^3}, \\
    \overline{P_D^{K}} &= 1.07 \times 10^8 \mathrm{~molec/cm^3}, \\
    \overline{P_L} &= 1.30 \times 10^{13} \mathrm{~molec/cm^3}, \\
    \overline{Q^{T_8}} &= 1.35 \times 10^6 \mathrm{~molec/cm^3}, \\
    \overline{Q^{T_1}} &= 5.99 \times 10^5 \mathrm{~molec/cm^3}, \\
    \overline{Q^{K}} &= 2.95 \times 10^5 \mathrm{~molec/cm^3}.
\end{align*}
\subsubsection{Estimates for Initial Conditions}
To determine the relevant initial conditions, we can simply consider the total number of PD-1 receptors on each PD-1-expressing cell and PD-L1 ligands in the absence of pembrolizumab, so that
\begin{align*}
P_D^{T_8}(0) + Q^{T_8}(0) &=\rho_{P_D^{T_8}}T_8(0), \\
P_D^{T_1}(0) + Q^{T_1}(0) &=\rho_{P_D^{T_1}}T_1(0), \\
P_D^{K}(0) + Q^{K}(0) &=\rho_{P_D^{K}}K(0), \\
P_L(0) + Q^{T_8}(0) + Q^{T_1}(0) + Q^{K}(0) &= \rho_{P_L C}C(0) + \rho_{P_L M_2}M_2(0).
\intertext{We can also consider \eqref{reducedQ8eqn}~--~\eqref{reducedQKeqn} initially, so that}
    Q^{T_8}(0) - \frac{\lambda_{P_DP_L}}{\lambda_Q}P_D^{T_8}(0)P_L(0)&=0, \\
    Q^{T_1}(0) - \frac{\lambda_{P_DP_L}}{\lambda_Q}P_D^{T_1}(0)P_L(0)&=0, \\
    Q^{K}(0) - \frac{\lambda_{P_DP_L}}{\lambda_Q}P_D^{K}(0)P_L(0)&=0.
\end{align*}
Solving these simultaneously leads to
\begin{align*}
    P_D^{T_8}(0) &= 4.44 \times 10^8 \mathrm{~molec/cm^3}, \\
    P_D^{T_1}(0) &= 2.07 \times 10^8 \mathrm{~molec/cm^3}, \\
    P_D^{K}(0) &= 2.46 \times 10^8 \mathrm{~molec/cm^3}, \\
    P_L(0) &= 7.36 \times 10^{12} \mathrm{~molec/cm^3}, \\
    Q^{T_8}(0) &= 6.96 \times 10^5 \mathrm{~molec/cm^3}, \\
    Q^{T_1}(0) &= 3.24 \times 10^5 \mathrm{~molec/cm^3}, \\
    Q^{K}(0) &= 3.86 \times 10^5 \mathrm{~molec/cm^3}.
\end{align*}
We note that excluding bound PD-1 receptors when considering the total number of PD-1 receptors on PD-1-expressing cells does not affect the parameter estimates, steady states, or initial conditions at this level of precision, since the number of unbound PD-1 receptors is several orders of magnitude larger than the number of bound PD-1 receptors on PD-1-expressing cells. Furthermore, this also applies when considering the total number of PD-L1 ligands.
\subsection{Estimates for Immune Checkpoint-Associated Components in the TDLN\label{reducedtdlnicissinitappendix}}
\subsubsection{Estimates for Synthesis Rates and Steady States}
For simplicity, we assume that the total number of PD-1 receptors on cells in the TDLN is equal to the number on the corresponding cells in the TS. Thus, denoting $\rho_{P_D^{8\mathrm{LN}}}$ and $\rho_{P_D^{1\mathrm{LN}}}$ as the number of PD-1 molecules expressed on the surface of CD8+ T cells and Th1 cells in the TDLN, respectively, we have that $\rho_{P_D^{8\mathrm{LN}}} = \rho_{P_D^{T_8}}$ and $\rho_{P_D^{1\mathrm{LN}}} = \rho_{P_D^{T_1}}$. \\~\\
We denote $\rho_{P_L^\mathrm{LN}D^\mathrm{LN}}$ and $\rho_{P_L^\mathrm{LN}T_A^1}$ as the number of PD-L1 molecules expressed on mature DCs, and effector Th1 cells in the TDLN, respectively.
It was found in \cite{Cheng2013supp} that the PD-L1 expression on activated CD3+ PD-L1+ T cells was $9282 ~\mathrm{molec/cell}$, whilst the PD-L1 expression on mature DCs was $80,372 ~\mathrm{molec/cell}$. However, amongst advanced CRC patients, only $22.4\%$ of CD4+ T cells were PD-L1+ \cite{Saito2021supp}. Moreover, only $22\%$ of colonic DCs were PD-L1+ in \cite{Moreira2021supp}. We thus assumed that $\rho_{P_L^\mathrm{LN}T_A^1} = 2.08 \times 10^{3} ~\mathrm{molec/cell}$, and $\rho_{P_L^\mathrm{LN}D^\mathrm{LN}} = 1.77 \times 10^{4} ~\mathrm{molec/cell}$. \\~\\
The procedure for estimating parameters, steady states, and initial conditions for PD-1, PD-L1, and the PD-1/PD-L1 complex in the TDLN is the same as in the TS. Considering \eqref{reducedPD8LNeqn}~--~\eqref{reducedPD1LNeqn} and \eqref{reducedPLLNeqn}~--~\eqref{reducedQ1LNeqn} at steady state in the absence of pembrolizumab, and making the same assumptions for estimation as in the TS, we obtain        
\begin{align*}
    \lambda_{P_D^{8\mathrm{LN}}}\overline{T_A^8} - d_{P_D}\overline{P_D^{8\mathrm{LN}}} &=0, \\
    \lambda_{P_D^{1\mathrm{LN}}}\overline{T_A^1} - d_{P_D}\overline{P_D^{1\mathrm{LN}}} &=0, \\
    \lambda_{P_L^\mathrm{LN}D^\mathrm{LN}}\overline{D^\mathrm{LN}} + \lambda_{P_L^\mathrm{LN}T_A^1}\overline{T_A^1} - d_{P_L}\overline{P_L^\mathrm{LN}} &=0, \\
    \overline{Q^{8\mathrm{LN}}} - \frac{\lambda_{P_DP_L}}{\lambda_Q}\overline{P_D^{8\mathrm{LN}}}\overline{P_L^\mathrm{LN}}&=0, \\
    \overline{Q^{1\mathrm{LN}}} - \frac{\lambda_{P_DP_L}}{\lambda_Q}\overline{P_D^{1\mathrm{LN}}}\overline{P_L^\mathrm{LN}}&=0, \\
    \overline{P_D^{8\mathrm{LN}}} + \overline{Q^{8\mathrm{LN}}} &=\rho_{P_D^{8\mathrm{LN}}}\overline{T_A^8}, \\
    \overline{P_D^{1\mathrm{LN}}} + \overline{Q^{1\mathrm{LN}}} &=\rho_{P_D^{1\mathrm{LN}}}\overline{T_A^1}, \\
    \overline{P_L^\mathrm{LN}} + \overline{Q^{8\mathrm{LN}}} + \overline{Q^{1\mathrm{LN}}} &= \rho_{P_L^\mathrm{LN}D^\mathrm{LN}} \overline{D^\mathrm{LN}} + \rho_{P_L^\mathrm{LN}T_A^1}\overline{T_A^1},\\
    \frac{\lambda_{P_D^{8\mathrm{LN}}}}{\rho_{P_D^{8\mathrm{LN}}}} &=     \frac{\lambda_{P_D^{1\mathrm{LN}}}}{\rho_{P_D^{1\mathrm{LN}}}}, \\
    \frac{\lambda_{P_L^\mathrm{LN}D^\mathrm{LN}}}{\rho_{P_L^\mathrm{LN}D^\mathrm{LN}}} &= \frac{\lambda_{P_L^\mathrm{LN}T_A^1}}{\rho_{P_L^\mathrm{LN}T_A^1}}.
\end{align*}
Solving these simultaneously and ensuring all model parameters are positive leads to
\begin{align*}
    \lambda_{P_D^{8\mathrm{LN}}} &= 9.27 \times 10^2 ~\left(\mathrm{molec/cell}\right)\mathrm{day^{-1}}, \\
    \lambda_{P_D^{1\mathrm{LN}}} &= 6.89 \times 10^2 ~\left(\mathrm{molec/cell}\right)\mathrm{day^{-1}}, \\
    \lambda_{P_L^\mathrm{LN}D^\mathrm{LN}} &= 2.46 \times 10^{4} ~\left(\mathrm{molec/cell}\right)\mathrm{day^{-1}}, \\
    \lambda_{P_L^\mathrm{LN}T_A^1} &= 2.89 \times 10^{3} ~\left(\mathrm{molec/cell}\right)\mathrm{day^{-1}}.
\end{align*}
This leads to
\begin{align*}
\overline{P_D^{8\mathrm{LN}}} &= 2.29 \times 10^9 \mathrm{~molec/cm^3}, \\
\overline{P_D^{1\mathrm{LN}}} &= 1.37 \times 10^{10} \mathrm{~molec/cm^3}, \\
\overline{P_L^\mathrm{LN}} &= 5.94 \times 10^{11} \mathrm{~molec/cm^3}, \\
\overline{Q^{8\mathrm{LN}}} &= 2.90 \times 10^5 \mathrm{~molec/cm^3}, \\
\overline{Q^{1\mathrm{LN}}} &= 1.73 \times 10^6 \mathrm{~molec/cm^3}.
\end{align*}
\subsubsection{Estimates for Initial Conditions}
To determine the relevant immune checkpoint initial conditions, we can simply consider the total number of PD-1 receptors on each PD-1-expressing cell and PD-L1 ligands in the absence of pembrolizumab, so that
\begin{align*}
P_D^{8\mathrm{LN}}(0) + Q^{8\mathrm{LN}}(0) &=\rho_{P_D^{8\mathrm{LN}}}T_A^8(0), \\
P_D^{1\mathrm{LN}}(0) + Q^{1\mathrm{LN}}(0) &=\rho_{P_D^{1\mathrm{LN}}}T_A^1(0), \\
P_L^\mathrm{LN}(0) + Q^{8\mathrm{LN}}(0) + Q^{1\mathrm{LN}}(0) &= \rho_{P_L^\mathrm{LN}D^\mathrm{LN}} D^\mathrm{LN}(0) + \rho_{P_L^\mathrm{LN}T_A^1}T_A^1(0).
\intertext{We can also consider \eqref{reducedQ8LNeqn} and \eqref{reducedQ1LNeqn} initially, so that}
    Q^{8\mathrm{LN}}(0) - \frac{\lambda_{P_DP_L}}{\lambda_Q}P_D^{8\mathrm{LN}}(0)P_L^\mathrm{LN}(0)&=0, \\
    Q^{1\mathrm{LN}}(0) - \frac{\lambda_{P_DP_L}}{\lambda_Q}P_D^{1\mathrm{LN}}(0)P_L^\mathrm{LN}(0)&=0.
\end{align*}
Solving these simultaneously leads to
\begin{align*}
    P_D^{8\mathrm{LN}}(0) &= 2.37 \times 10^9 \mathrm{~molec/cm^3}, \\
    P_D^{1\mathrm{LN}}(0) &= 1.59 \times 10^{10} \mathrm{~molec/cm^3}, \\
    P_L^\mathrm{LN}(0) &= 3.31 \times 10^{11} \mathrm{~molec/cm^3}, \\
    Q^{8\mathrm{LN}}(0) &= 1.67 \times 10^5 \mathrm{~molec/cm^3}, \\
    Q^{1\mathrm{LN}}(0) &= 1.12 \times 10^6 \mathrm{~molec/cm^3}.
\end{align*}
We note again that excluding bound PD-1 receptors when considering the total number of PD-1 receptors on PD-1-expressing cells does not affect the parameter estimates, steady states, or initial conditions at this level of precision, since the number of unbound PD-1 receptors is several orders of magnitude larger than the number of bound PD-1 receptors on PD-1-expressing cells. Furthermore, this also applies when considering the total number of PD-L1 ligands.
\putbib[References.bib]
\end{bibunit}
\newpage
\title{Appendix C: Sensitivity analysis-guided model reduction of a mathematical model of pembrolizumab therapy for de novo metastatic MSI-H/dMMR colorectal cancer}
\maketitle
\setcounter{page}{1}
\begin{bibunit}[vancouver]
\section{Model Parameters for the Reduced Model}
The model parameter values are estimated in \Cref{reducedmodelparamestsection} and are listed in \autoref{reducedtableofparams}.
\begin{center}
\begin{longtable}{|lp{160pt}lp{100pt}l|}
\caption{\label{reducedtableofparams}Parameter values for the reduced model. TDLN denotes the tumour-draining lymph node, whilst TS denotes the tumour site. est. denotes estimated parameters.}\\       
\hline \textbf{Parameter} & \textbf{Description} & \textbf{Value} & \textbf{Unit} & \textbf{References} \\
\hline
$f_\mathrm{pembro}$ & $A_1$/$A_1^\mathrm{LN}$ dose scaling factor & $1.17 \times 10^{12}$ & $\left(\mathrm{molec/cm^3}\right)/\mathrm{mg}$ & est.\\
$f_{C}$ & $C$ to $V_\mathrm{TS}$ scaling factor & $1.17 \times 10^{6}$ & $\mathrm{cell/(cm^3)^2}$ & est. \\
$f_{N_c}$ & $N_c$ to $V_\mathrm{TS}$ scaling factor & $6.16 \times 10^{4}$ & $\mathrm{cell/(cm^3)^2}$ & est. \\
\hline
$\mathcal{A}_{D_0}$ & Source of $D_0$ & $1.18 \times 10^6$ & $\left(\mathrm{cell/cm^{3}}\right)\mathrm{day^{-1}}$ & est. \\
$\mathcal{A}_{T_0^8}$ & Source of $T_0^8$ & $3.76 \times 10^5$ & $\left(\mathrm{cell/cm^{3}}\right)\mathrm{day^{-1}}$ & est. \\
$\mathcal{A}_{T_0^4}$ & Source of $T_0^4$ & $2.76 \times 10^5$ & $\left(\mathrm{cell/cm^{3}}\right)\mathrm{day^{-1}}$ & est. \\
$\mathcal{A}_{T_0^r}$ & Source of $T_0^r$ & $1.15 \times 10^5$ & $\left(\mathrm{cell/cm^{3}}\right)\mathrm{day^{-1}}$ & est. \\
$\mathcal{A}_{M_0}$ & Source of $M_0$ & $1.08 \times 10^6$ & $\left(\mathrm{cell/cm^{3}}\right)\mathrm{day^{-1}}$ & est. \\
$\mathcal{A}_{K_0}$ & Source of $K_0$ & $2.82 \times 10^5$ & $\left(\mathrm{cell/cm^{3}}\right)\mathrm{day^{-1}}$ & est. \\
\hline
$\lambda_{C}$ & Growth rate of $C$ & $5.25 \times 10^{-2}$ & $\mathrm{day^{-1}}$ & fitted \\
$\lambda_{CT_8}$ & Elimination rate of $C$ by $T_8$ & $8.01 \times 10^{-8}$ & $\left(\mathrm{cell/cm^{3}}\right)^{-1}\mathrm{day^{-1}}$ & fitted \\
$\lambda_{CK}$ & Elimination rate of $C$ by $K$ & $8.01 \times 10^{-8}$ & $\left(\mathrm{cell/cm^{3}}\right)^{-1}\mathrm{day^{-1}}$ & est. \\
$\lambda_{CI_{\upalpha}}$ & Necrosis rate of $C$ by $I_{\upalpha}$ & $7.23 \times 10^{-3}$ & $\mathrm{day^{-1}}$ & est. \\
$\lambda_{HN_c}$ & Production rate of $H$ by $N_c$ & $2.92 \times 10^{-14}$ & $\left(\mathrm{g/cell}\right)\mathrm{day^{-1}}$ & est. \\
$\lambda_{SN_c}$ & Production rate of $S$ by $N_c$ & $1.70 \times 10^{-14}$ & $\left(\mathrm{g/cell}\right)\mathrm{day^{-1}}$ & est. \\
$\lambda_{DH}$ & Maturation rate of $D_0$ by $H$ & $1.33$ & $\mathrm{day^{-1}}$ & est. \\
$\lambda_{D_0K}$ & Killing rate of $D_0$ by $K$ & $5.49 \times 10^{-6}$ & $\left(\mathrm{cell/cm^3}\right)^{-1}\mathrm{day^{-1}}$ & est. \\
$\lambda_{DD^\mathrm{LN}}$ & Migration rate of $D$ to TDLN & $1.68 \times 10^{-2}$ & $\mathrm{day^{-1}}$ & est. \\
$\lambda_{T_0^8 T_A^8}$ & Kinetic rate constant for $T_0^8$ activation & $2.69 \times 10^{-11}$ & $\left(\mathrm{cell/cm^3}\right)^{-1}\mathrm{day^{-1}}$ & est. \\
$\lambda_{T_A^8T_8}$ & Kinetic rate constant for $T_A^8$ migration to the TS & $6.64 \times 10^{-1}$ & $\mathrm{day}^{-1}$ & est. \\
$\lambda_{T_8 C}$ & Exhaustion rate of $T_8$ due to $C$ exposure & $6.31 \times 10^{-3}$ & $\mathrm{day}^{-1}$ & est. \\
$\lambda_{T_\mathrm{ex} A_1}$ & Reinvigoration rate of $T_\mathrm{ex}$ by $A_1$ & $2.25 \times 10^{-3}$ & $\mathrm{day}^{-1}$ & est. \\
$\lambda_{T_0^4 T_A^1}$ & Kinetic rate constant for $T_0^4$ activation into $T_A^1$ & $8.12 \times 10^{-11}$ & $\left(\mathrm{cell/cm^3}\right)^{-1}\mathrm{day^{-1}}$ & est. \\
$\lambda_{T_A^1 T_1}$ & Kinetic rate constant for $T_A^1$ migration to the TS & $6.48 \times 10^{-2}$ & $\mathrm{day}^{-1}$ & est. \\
$\lambda_{T_1T_r}$ & Conversion rate of $T_1$ to $T_r$ by $Q^{T_1}$ & $4.00 \times 10^{-3}$ & $\mathrm{day^{-1}}$ & est. \\
$\lambda_{T_0^r T_A^r}$ & Kinetic rate constant for $T_0^r$ activation into $T_A^r$ & $1.06 \times 10^{-8}$ & $\left(\mathrm{cell/cm^3}\right)^{-1}\mathrm{day^{-1}}$ & est. \\
$\lambda_{T_A^r T_r}$ & Kinetic rate constant for $T_A^r$ migration to the TS & $4.79$ & $\mathrm{day}^{-1}$ & est. \\
$\lambda_{M_1I_\upalpha}$ & Polarisation rate of $M_0$ to $M_1$ by $I_{\upalpha}$ & $3.77 \times 10^{-1}$ & $\mathrm{day^{-1}}$ & est. \\
$\lambda_{M_1I_\upgamma}$ & Polarisation rate of $M_0$ to $M_1$ by $I_\upgamma$ & $4.39 \times 10^{-1}$ & $\mathrm{day^{-1}}$ & est. \\
$\lambda_{M_2I_{10}}$ & Polarisation rate of $M_0$ to $M_2$ by $I_{10}$ & $2.09 \times 10^{-1}$ & $\mathrm{day^{-1}}$ & est. \\
$\lambda_{M_2I_\upbeta}$ & Polarisation rate of $M_0$ to $M_2$ by $I_\upbeta$ & $2.34 \times 10^{-1}$ & $\mathrm{day^{-1}}$ & est. \\
$\lambda_{KI_2}$ & Maturation rate of $K_0$ by $I_2$ & $4.62 \times 10^{-3}$ & $\mathrm{day^{-1}}$ & est. \\
$\lambda_{KD}$ & Maturation rate of $K_0$ by $D$ & $9.24 \times 10^{-3}$ & $\mathrm{day^{-1}}$ & est. \\
$\lambda_{I_2 T_8}$ & Production rate of $I_2$ by $T_8$ & $5.95 \times 10^{-16}$ & $\left(\mathrm{g/cell}\right)\mathrm{day^{-1}}$ & est. \\
$\lambda_{I_2 T_1}$ & Production rate of $I_2$ by $T_1$ & $1.74 \times 10^{-15}$ & $\left(\mathrm{g/cell}\right)\mathrm{day^{-1}}$ & est. \\
$\lambda_{I_{\upgamma} K}$ & Production rate of $I_{\upgamma}$ by $K$ & $8.46 \times 10^{-15}$ & $\left(\mathrm{g/cell}\right)\mathrm{day^{-1}}$ & est. \\
$\lambda_{I_{\upalpha}T_8}$ & Production rate of $I_{\upalpha}$ by $T_8$ & $5.55 \times 10^{-15}$ & $\left(\mathrm{g/cell}\right)\mathrm{day^{-1}}$ & est. \\
$\lambda_{I_{\upalpha}T_1}$ & Production rate of $I_{\upalpha}$ by $T_1$ & $9.17 \times 10^{-15}$ & $\left(\mathrm{g/cell}\right)\mathrm{day^{-1}}$ & est. \\
$\lambda_{I_{\upalpha}M_1}$ & Production rate of $I_{\upalpha}$ by $M_1$ & $3.36 \times 10^{-15}$ & $\left(\mathrm{g/cell}\right)\mathrm{day^{-1}}$ & est. \\
$\lambda_{I_{\upalpha}K}$ & Production rate of $I_{\upalpha}$ by $K$ & $9.68 \times 10^{-15}$ & $\left(\mathrm{g/cell}\right)\mathrm{day^{-1}}$ & est. \\
$\lambda_{I_{\upbeta}C}$ & Production rate of $I_{\upbeta}$ by $C$ & $7.42 \times 10^{-12}$ & $\left(\mathrm{g/cell}\right)\mathrm{day^{-1}}$ & est. \\
$\lambda_{I_{\upbeta}T_r}$ & Production rate of $I_{\upbeta}$ by $T_r$ & $4.30 \times 10^{-11}$ & $\left(\mathrm{g/cell}\right)\mathrm{day^{-1}}$ & est. \\
$\lambda_{I_{\upbeta}M_2}$ & Production rate of $I_{\upbeta}$ by $M_2$ & $5.34 \times 10^{-11}$ & $\left(\mathrm{g/cell}\right)\mathrm{day^{-1}}$ & est. \\
$\lambda_{I_{10}C}$ & Production rate of $I_{10}$ by $C$ & $1.56 \times 10^{-17}$ & $\left(\mathrm{g/cell}\right)\mathrm{day^{-1}}$ & est. \\
$\lambda_{I_{10}M_2}$ & Production rate of $I_{10}$ by $M_2$ & $3.11 \times 10^{-17}$ & $\left(\mathrm{g/cell}\right)\mathrm{day^{-1}}$ & est. \\
$\lambda_{P_D^{T_8}}$ & Synthesis rate of $P_D^{T_8}$ & $9.25 \times 10^2$ & $\left(\mathrm{molec/cell}\right)\mathrm{day^{-1}}$ & est. \\
$\lambda_{Q_A}$ & Dissociation rate of the PD-1/pembrolizumab complex & $2.6$ & $\mathrm{day}^{-1}$ & \cite{Li2021supp} \\
$\lambda_{Q}$ & Dissociation rate of the PD-1/PD-L1 complex & $1.24 \times 10^{5}$ & $\mathrm{day^{-1}}$ & \cite{Cheng2013supp} \\
$\lambda_{P_DA_1}$ & Formation rate of the PD-1/pembrolizumab complex & $4.63 \times 10^{-13}$ & $\left(\mathrm{molec/cm^3}\right)^{-1}\mathrm{day}^{-1}$ & fitted \\
$\lambda_{P_DP_L}$ & Formation rate of the PD-1/PD-L1 complex & $2.64 \times 10^{-11}$ & $\left(\mathrm{molec/cm^3}\right)^{-1}\mathrm{day}^{-1}$ & \cite{Cheng2013supp} \\
$\lambda_{P_D^{T_1}}$ & Synthesis rate of $P_D^{T_1}$ & $6.87 \times 10^2$ & $\left(\mathrm{molec/cell}\right)\mathrm{day^{-1}}$ & est. \\
$\lambda_{P_D^{K}}$ & Synthesis rate of $P_D^{K}$ & $1.85 \times 10^2$ & $\left(\mathrm{molec/cell}\right)\mathrm{day^{-1}}$ & est. \\
$\lambda_{P_LC}$ & Synthesis rate of $P_L$ by $C$ & $2.50 \times 10^5$ & $\left(\mathrm{molec/cell}\right)\mathrm{day^{-1}}$ & est. \\
$\lambda_{P_LM_2}$ & Synthesis rate of $P_L$ by $M_2$ & $3.71 \times 10^5$ & $\left(\mathrm{molec/cell}\right)\mathrm{day^{-1}}$ & est. \\
$\lambda_{P_D^{8\mathrm{LN}}}$ & Synthesis rate of $P_D^{8\mathrm{LN}}$ & $9.27 \times 10^{2}$ & $\left(\mathrm{molec/cell}\right)\mathrm{day^{-1}}$ & est. \\
$\lambda_{P_D^{1\mathrm{LN}}}$ & Synthesis rate of $P_D^{1\mathrm{LN}}$ & $6.89 \times 10^{2}$ & $\left(\mathrm{molec/cell}\right)\mathrm{day^{-1}}$ & est. \\
$\lambda_{P_L^\mathrm{LN}D^\mathrm{LN}}$ & Synthesis rate of $P_L^\mathrm{LN}$ by $D^\mathrm{LN}$ & $2.46 \times 10^{4}$ & $\left(\mathrm{molec/cell}\right)\mathrm{day^{-1}}$ & est. \\
$\lambda_{P_L^\mathrm{LN}T_A^1}$ & Synthesis rate of $P_L^\mathrm{LN}$ by $T_A^1$ & $2.89 \times 10^{3}$ & $\left(\mathrm{molec/cell}\right)\mathrm{day^{-1}}$ & est. \\
\hline
$K_{CI_{\upalpha}}$ & Half-saturation constant of $I_\upalpha$ for $C$ & $9.00 \times 10^{-11}$ & $\mathrm{g/cm^{3}}$ & est. \\
$K_{DH}$ & Half-saturation constant of $H$ for $D$ & $1.94 \times 10^{-8}$ & $\mathrm{g/cm^{3}}$ & est. \\
$K_{T_8C}$ & Half-saturation constant $T_8$ exhaustion due to $C$ exposure & $7.02 \times 10^{7}$ & $\mathrm{cell}/\mathrm{cm^3}$ & est. \\
$K_{T_\mathrm{ex} A_1}$ & Half-saturation constant of $T_\mathrm{ex}$ reinvigoration by $A_1$ & $2.05 \times 10^{14}$ & $\mathrm{molec/cm^3}$ & est. \\
$K_{T_1Q^{T_1}}$ & Half-saturation constant of $T_1$ conversion to $T_r$ by $Q^{T_1}$ & $5.99 \times 10^5$ & $\mathrm{molec/cm^{3}}$ & est. \\
$K_{M_1 I_\upalpha}$ & Half-saturation constant of $I_\upalpha$ for $M_1$ & $9.00 \times 10^{-11}$ & $\mathrm{g/cm^{3}}$ & est. \\
$K_{M_1 I_\upgamma}$ & Half-saturation constant of $I_\upgamma$ for $M_1$ & $4.93 \times 10^{-11}$ & $\mathrm{g/cm^{3}}$ & est. \\
$K_{M_2 I_{10}}$ & Half-saturation constant of $I_{10}$ for $M_{2}$ & $1.84 \times 10^{-10}$ & $\mathrm{g/cm^{3}}$ & est. \\
$K_{M_2 I_{\upbeta}}$ & Half-saturation constant of $I_{\upbeta}$ for $M_{2}$ & $1.51 \times 10^{-6}$ & $\mathrm{g/cm^{3}}$ & est. \\
$K_{KI_2}$ & Half-saturation constant of $I_2$ for $K$ & $2.00 \times 10^{-12}$ & $\mathrm{g/cm^{3}}$ & est. \\
$K_{KD}$ & Half-saturation constant of $D$ for $K$ & $1.91 \times 10^{6}$ & $\mathrm{cell/cm^{3}}$ & est. \\
\hline
$C_0$ & Carrying capacity of $C$ & $8.89 \times 10^{7}$ & $\mathrm{cell/cm^3}$ & fitted \\
$K_{CI_{\upbeta}}$ & Inhibition constant of $T_8$ and $K$ elimination of $C$ by $I_{\upbeta}$ & $1.51 \times 10^{-6}$ & $\mathrm{g/cm^{3}}$ & est. \\
$K_{CQ^{T_8}}$ & Inhibition constant of $T_8$ elimination of $C$ by $Q^{T_8}$ & $1.35 \times 10^6$ & $\mathrm{molec/cm^{3}}$ & est. \\
$K_{CQ^K}$ & Inhibition constant of $K$ elimination of $C$ by $Q^K$ & $2.95 \times 10^5$ & $\mathrm{molec/cm^3}$ & est. \\
$K_{D_0I_\upbeta}$ & Inhibition constant of $K$ elimination of $D_0$ by $I_\upbeta$ & $1.51 \times 10^{-6}$ & $\mathrm{g/cm^{3}}$ & est. \\
$V_\mathrm{LN}$ & Volume of the TDLN & $1.47 \times 10^{-1}$ & $\mathrm{cm^3}$ & \cite{Rssler2017supp} est. \\
$K_{T_0^8T_A^r}$ & Inhibition constant of $T_0^8$ activation by $T_A^r$ & $1.47 \times 10^{6}$ & $\mathrm{cell/cm^3}$ & est. \\
$K_{T_0^8 Q^{8\mathrm{LN}}}$ & Inhibition constant of $T_0^8$ activation by $Q^{8\mathrm{LN}}$ & $2.90 \times 10^5$ & $\mathrm{molec/cm^3}$ & est. \\
$K_{T_A^8T_A^r}$ & Inhibition constant of $T_A^8$ proliferation by $T_A^r$ & $1.47 \times 10^{6}$ & $\mathrm{cell/cm^3}$ & est. \\
$K_{T_A^8 Q^{8\mathrm{LN}}}$ & Inhibition constant of $T_A^8$ proliferation by $Q^{8\mathrm{LN}}$ & $2.90 \times 10^5$ & $\mathrm{molec/cm^3}$ & est. \\
$K_{T_8 I_{10}}$ & Inhibition constant of $T_8$ death by $I_{10}$ & $1.84 \times 10^{-10}$ & $\mathrm{g/cm^{3}}$ & est. \\
$K_{T_\mathrm{ex} I_{10}}$ & Inhibition constant of $T_\mathrm{ex}$ death by $I_{10}$ & $1.84 \times 10^{-10}$ & $\mathrm{g/cm^{3}}$ & est. \\
$K_{T_0^4 T_A^r}$ & Inhibition constant of $T_0^4$ activation by $T_A^r$ & $1.47 \times 10^{6}$ & $\mathrm{cell/cm^3}$ & est. \\
$K_{T_0^4 Q^{1\mathrm{LN}}}$ & Inhibition constant of $T_0^4$ activation by $Q^{1\mathrm{LN}}$ & $1.73 \times 10^6$ & $\mathrm{molec/cm^3}$ & est. \\
$K_{T_A^1 T_A^r}$ & Inhibition constant of $T_A^1$ proliferation by $T_A^r$ & $1.47 \times 10^{6}$ & $\mathrm{cell/cm^3}$ & est. \\
$K_{T_A^1 Q^{1\mathrm{LN}}}$ & Inhibition constant of $T_A^1$ proliferation by $Q^{1\mathrm{LN}}$ & $1.73 \times 10^6$ & $\mathrm{molec/cm^3}$ & est. \\
$K_{KI_\upbeta}$ & Inhibition constant of NK cell activation by $I_\upbeta$ & $1.51 \times 10^{-6}$ & $\mathrm{g/cm^{3}}$ & est. \\
\hline
$d_{N_c}$ & Removal rate of $N_c$ & $6.88 \times 10^{-2}$ & $\mathrm{day}^{-1}$ & est. \\
$d_{H}$ & Degradation rate of $H$ & $5.55$ & $\mathrm{day}^{-1}$ & \cite{Zandarashvili2013supp} est. \\
$d_{S}$ & Degradation rate of $S$ & $1.39$ & $\mathrm{day}^{-1}$ & \cite{Goicoechea2003supp, Zhang2021calsupp} est. \\
$d_{D_0}$ & Death rate of $D_0$ & $3.57 \times 10^{-2}$ & $\mathrm{day}^{-1}$ & \cite{Ruedl2000supp} est. \\
$d_{D}$ & Death rate of $D$ & $3.15 \times 10^{-1}$ & $\mathrm{day}^{-1}$ & \cite{Kamath2002supp} est. \\
$d_{T_0^8}$ & Death rate of $T_0^8$ & $3.22 \times 10^{-2}$ & $\mathrm{day}^{-1}$ & \cite{Takada2009supp} est. \\
$d_{T_8}$ & Death rate of $T_8$ & $9\times 10^{-3}$ & $\mathrm{day}^{-1}$ & \cite{Hellerstein1999supp} \\
$d_{T_\mathrm{ex}}$ & Death rate of $T_\mathrm{ex}$ & $9\times 10^{-3}$ & $\mathrm{day}^{-1}$ & \cite{Hellerstein1999supp} \\
$d_{T_0^4}$ & Death rate of $T_0^4$ & $4.03\times 10^{-2}$ & $\mathrm{day}^{-1}$ & \cite{Takada2009supp} est. \\
$d_{T_1}$ & Death rate of $T_1$ & $8 \times 10^{-3}$ & $\mathrm{day}^{-1}$ & \cite{Hellerstein1999supp}\\
$d_{T_0^r}$ & Death rate of $T_0^r$ & $2.2 \times 10^{-3}$ & $\mathrm{day}^{-1}$ & \cite{Kumbhari2020n2supp} \\
$d_{T_r}$ & Death rate of $T_r$ & $6.30 \times 10^{-2}$ & $\mathrm{day}^{-1}$ & \cite{VukmanovicStejic2006supp} est. \\
$d_{M_0}$ & Death rate of $M_0$ & $0.73$ & $\mathrm{day}^{-1}$ & \cite{Patel2017supp} \\
$d_{M_1}$ & Death rate of $M_1$ & $0.99$ & $\mathrm{day}^{-1}$ & \cite{Patel2017supp} \\
$d_{M_2}$ & Death rate of $M_2$ & $1.35 \times 10^{-1}$ & $\mathrm{day}^{-1}$ & \cite{Patel2017supp} \\
$d_{K_0}$ & Death rate of $K_0$ & $6.93 \times 10^{-2}$ & $\mathrm{day}^{-1}$ & \cite{Wu2020supp, Vivier2008supp, Lowry2017supp} est. \\
$d_{K}$ & Death rate of $K$ & $6.93 \times 10^{-2}$ & $\mathrm{day}^{-1}$ & \cite{Wu2020supp, Vivier2008supp, Lowry2017supp} est. \\
$d_{I_2}$ & Degradation rate of $I_2$ & $1.45 \times 10^2$ & $\mathrm{day}^{-1}$ & \cite{Lotze1985n1supp} est. \\
$d_{I_\upgamma}$ & Degradation rate of $I_\upgamma$ & $3.33 \times 10^1$ & $\mathrm{day}^{-1}$ & \cite{Balachandran2013supp} est. \\
$d_{I_{\upalpha}}$ & Degradation rate of $I_{\upalpha}$ & $5.48 \times 10^{1}$ & $\mathrm{day}^{-1}$ & \cite{Ma2015supp, Oliver1993-vlsupp} est. \\
$d_{I_{\upbeta}}$ & Degradation rate of $I_{\upbeta}$ & $3.99 \times 10^{2}$ & $\mathrm{day}^{-1}$ & \cite{TiradoRodriguez2014supp} est. \\
$d_{I_{10}}$ & Degradation rate of $I_{10}$ & $6.16$ & $\mathrm{day}^{-1}$ & \cite{Huhn1997supp} est. \\
$d_{P_D}$ & Degradation rate of unbound PD-1 receptors & $3.36 \times 10^{-1}$ & $\mathrm{day}^{-1}$ & \cite{Lassman2021supp} \\
$d_{Q_A}$ & Internalisation rate of the PD-1/pembrolizumab complex & $0.43$ & $\mathrm{day}^{-1}$ & \cite{Li2021supp} \\
$d_{A_1}$ & Elimination rate of $A_1/A_1^\mathrm{LN}$ & $2.92 \times 10^{-2}$ & $\mathrm{day}^{-1}$ & \cite{Li2017supp, Li2019pembrosupp, Ahamadi2016supp} est. \\
$d_{P_L}$ & Degradation rate of unbound PD-L1 & $1.39$ & $\mathrm{day}^{-1}$ & \cite{Cha2019supp} \\
\hline
$\tau_m$ & DC migration time from TDLN to the TS & $0.75$ & $\mathrm{day}$ & \cite{Catron2004supp} est. \\
$\Delta_8^0$ & Time taken for first CTL division & $1.63$ & $\mathrm{day}$ & \cite{Plambeck2022supp} \\
$n^8_\mathrm{max}$ & Maximal number of CTL divisions in the TDLN & $10$ & dimensionless & \cite{Kaech2001supp, Masopust2007supp} est. \\
$\Delta_8$ & Time taken for successive CTL divisions & $0.36$ & $\mathrm{day}$ & \cite{Kaech2001supp} \\
$\tau_{T_A^8}$ & Time taken for CTL division program & $4.87$ & $\mathrm{day}$ & est. \\
$\Delta_1^0$ & Time taken for first Th1 cell division & $0.77$ & $\mathrm{day}$ & \cite{Kaech2002supp} est. \\
$n^1_\mathrm{max}$ & Maximal number of Th1 cell divisions in the TDLN & $9$ & dimensionless & \cite{Homann2001supp} est. \\
$\Delta_1$ & Time taken for successive Th1 cell divisions & $0.42$ & $\mathrm{day}$ & \cite{Kaech2002supp} est. \\
$\tau_{T_A^1}$ & Time taken for Th1 cell division program & $4.13$ & $\mathrm{day}$ & est. \\
$\Delta_r^0$ & Time taken for first Treg division & $0.77$ & $\mathrm{day}$ & \cite{Kaech2002supp} est. \\
$n^r_\mathrm{max}$ & Maximal number of Treg divisions in the TDLN & $6$ & dimensionless & \cite{DarrasseJze2009supp} est. \\
$\Delta_r$ & Time taken for successive Treg divisions & $0.42$ & $\mathrm{day}$ & \cite{Kaech2002supp} est. \\
$\tau_{T_A^r}$ & Time taken for Treg division program & $2.87$ & $\mathrm{day}$ & est. \\
\hline
\end{longtable}
\end{center}
\putbib[References.bib]
\end{bibunit}
\newpage
\title{Appendix D: Sensitivity analysis-guided model reduction of a mathematical model of pembrolizumab therapy for de novo metastatic MSI-H/dMMR colorectal cancer}
\maketitle
\setcounter{page}{1}
\begin{bibunit}[vancouver]
\section{Minimal Model Parameter Estimation\label{minimalmodelparamestsection}}
We estimate all parameters, where possible, under the assumption that no pembrolizumab has been or will be administered. The exceptions to this are the parameters directly related to pembrolizumab treatment, for which the assumptions are explicitly stated during estimation. Many of the assumptions and techniques in this section are adopted from \cite{Hawi2026metasupp} and \cite{Hawi2025localsupp}.\\~\\
For the sake of brevity, we present parameter estimation only for those parameters that differ between the reduced and the minimal model. The absence of a parameter in this section implies that its value and derivation are identical to those provided in \Cref{reducedmodelparamestsection} and \autoref{reducedtableofparams}.
\subsection{Half-Saturation Constants}
Using \eqref{halfsateqn}, we have that
\begin{align*}
    K_{T_8T_\mathrm{ex}} &= \overline{C} = 7.02 \times 10^7 ~\mathrm{cell/cm^3}, \\
    K_{DN_c} &= \overline{N_c} = 3.69 \times 10^6 ~\mathrm{cell/cm^3}, \\
    K_{KT_8} &= \overline{T_8} = 1.77 \times 10^5 ~\mathrm{cell/cm^3}.
\end{align*}
\subsection{Inhibition Constants}
Using \eqref{inhibitionconsteqn}, we have that
\begin{align*}
K_{CQ^{T_8}} &= \overline{Q^{T_8}} = 1.31 \times 10^{6} ~\mathrm{molec/cm^3}, \\
K_{CQ^{K}} &= \overline{Q^{K}} = 2.87 \times 10^{6} ~\mathrm{molec/cm^3}, \\
K_{T_0^8T_A^r} = K_{T_A^8T_A^r}  &= \overline{T_A^{r}} = 1.47 \times 10^{6} ~\mathrm{cell/cm^3}, \\
K_{T_8C} = K_{T_\mathrm{ex}C} &= \overline{C} = 7.02 \times 10^7 ~\mathrm{cell/cm^3}, \\
K_{T_0^8 Q^{8\mathrm{LN}}} = K_{T_A^8 Q^{8\mathrm{LN}}} &= \overline{Q^{8\mathrm{LN}}} = 2.83 \times 10^5 ~ \mathrm{molec/cm^3}.
\end{align*}
\subsection{Cytokine Production Parameters\label{minimalcytokinessic}}
To estimate the cytokine production constants, we consider \eqref{minimaltnfeqn} and \eqref{minimaltgfbetaeqn} at steady state and use the data from \cite{Cui2023supp}. For each immune cell, we assume that each cytokine's corresponding gene expression is proportional to its production rate by that cell.
\subsubsection{Estimates for $I_\upalpha$}
Using values from \cite{Cui2023supp} and considering \eqref{minimaltnfeqn} at steady state, or equivalently considering \eqref{minimaltnfqssa}, leads to the equations
\begin{equation*}
\frac{\lambda_{I_{\upalpha}T_8}}{0.0654443776961264} = \frac{\lambda_{I_{\upalpha}K}}{0.114108294134927},
\end{equation*}
and
\begin{equation*}
\lambda_{I_{\upalpha}T_8}\overline{T_8} + \lambda_{I_{\upalpha}K}\overline{K} - d_{I_{\upalpha}}\overline{I_{\upalpha}} = 0.
\end{equation*}
Solving these simultaneously leads to
\begin{align*}
    \lambda_{I_{\upalpha} T_8} &= 9.57 \times 10^{-15} ~\left(\mathrm{g/cell}\right)\mathrm{day^{-1}}, \\
    \lambda_{I_{\upalpha} K} &= 1.67 \times 10^{-14} ~\left(\mathrm{g/cell}\right)\mathrm{day^{-1}}.
\end{align*}
Consequently, considering \eqref{minimaltnfqssa}, we have that
\begin{equation*}
    I_\upalpha(0) = \frac{1}{d_{I_{\upalpha}}}\left(\lambda_{I_{\upalpha}T_8}T_8(0) + \lambda_{I_{\upalpha}K}K(0)\right) = 1.64 \times 10^{-10} \mathrm{~g/cm^3}.
\end{equation*}
\subsubsection{Estimates for $I_\upbeta$}
Estimating the production constants for TGF-$\upbeta$ is slightly more complicated compared to other cytokines. We assume that the results for fibroblastic reticular cells in \cite{Zhang2023supp} translate directly to results for cancer-associated fibroblasts (CAFs), which are considered to be all fibroblasts found in the TME \cite{Zhang2023supp}. We assume that at steady state, CAFs produce twice as much TGF-$\upbeta$ as cancer cells in the TME, and denote the production rate of TGF-$\upbeta$ by CAFs as $\lambda_{I_\upbeta C_F}$. This, in conjunction with values from \cite{Cui2023supp}, and considering \eqref{minimaltgfbetaeqn} at steady state, or equivalently considering \eqref{minimalibetaqssa}, leads to the equations
\begin{equation*}
    \frac{\lambda_{I_{\upbeta}C_F}}{2} = \frac{\lambda_{I_{\upbeta}C}}{1},
\end{equation*}
and
\begin{equation*}
\frac{\lambda_{I_{\upbeta}C_F}}{0.175283003265127} = \frac{\lambda_{I_{\upbeta}T_r}}{0.507677682409403},
\end{equation*}
and
\begin{equation*}
\lambda_{I_{\upbeta}C}\overline{C} + \lambda_{I_{\upbeta}T_r}\overline{T_r} - d_{I_{\upbeta}}\overline{I_{\upbeta}} = 0.
\end{equation*}
Solving these simultaneously leads to
\begin{align*}
    \lambda_{I_{\upbeta} C} &= 8.39 \times 10^{-12} ~\left(\mathrm{g/cell}\right)\mathrm{day^{-1}},\\
    \lambda_{I_{\upbeta} T_r} &= 4.86 \times 10^{-11} ~\left(\mathrm{g/cell}\right)\mathrm{day^{-1}}.
\end{align*}
Consequently, considering \eqref{minimalibetaqssa}, we have that
\begin{equation*}
    I_\upbeta(0) = \frac{1}{d_{I_{\upbeta}}}\left(\lambda_{I_{\upbeta}C}C(0) + \lambda_{I_{\upbeta}T_r}T_r(0)\right) = 8.45 \times 10^{-7} \mathrm{~g/cm^3}.
\end{equation*}

\subsection{Parameters for DCs and NK Cells}
\subsubsection{Estimates for $D_0$ and $D$}
Using the values from \Cref{reduceddcestsection}, as well as substituting $\lambda_{DN_c}=\lambda_{DH}$, we get that
\begin{align*}
    \mathcal{A}_{D_0} &= 1.18 \times 10^{6} ~\left(\mathrm{cell/cm^{3}}\right)\mathrm{day^{-1}}, \\
    \lambda_{DN_c} &= 1.33 ~\mathrm{day^{-1}}, \\
    \lambda_{DD^\mathrm{LN}} &= 1.68 \times 10^{-2} ~\mathrm{day^{-1}}.
\end{align*}
\subsubsection{Estimates for $K$}
Using the values from \Cref{reducednkestsection} and applying the transformations $\lambda_{KI_2}\overline{K_0} \mapsto \lambda_{KI_2}$ and $\lambda_{KD}\overline{K_0} \mapsto \lambda_{KD}$, where $\overline{K_0}$ is as in both the full and reduced models, leads to
\begin{align*}
    \lambda_{KT_8} &= 1.79 \times 10^{4} ~\left(\mathrm{cell/cm^3}\right)\mathrm{day^{-1}}, \\
    \lambda_{KD} &= 3.59 \times 10^{4} ~\left(\mathrm{cell/cm^3}\right)\mathrm{day^{-1}}.
\end{align*}
\subsection{T Cell Parameters and Estimates}
\subsubsection{Estimates for $T_A^8$, $T_8$ and $T_\mathrm{ex}$}
Using the values from \Cref{reducedcd8estsection} and applying the transformation $\lambda_{T_0^8 T_A^8}\overline{T_0^8} \mapsto \lambda_{T_0^8 T_A^8}$, where $\overline{T_0^8}$ is as in both the full and reduced models, leads to
\begin{align*}
\lambda_{T_0^8 T_A^8} &= 3.12 \times 10^{-4} ~\mathrm{day^{-1}}, \\
\overline{R^8} &= 2.55 \times 10^3 ~\left(\mathrm{cell/cm^{3}}\right)\mathrm{day^{-1}}, \\
\lambda_{T_A^8T_8} &= 6.64 \times 10^{-1} ~\mathrm{day}^{-1}, \\
\lambda_{T_8C} &= 6.31 \times 10^{-3} ~\mathrm{day}^{-1}, \\
\lambda_{T_\mathrm{ex}A_1} &= 2.25 \times 10^{-3} ~\mathrm{day}^{-1}.
\end{align*}
\subsubsection{Estimates for $T_0^r$, $T_A^r$, and $T_r$}
Considering \eqref{minimalnaivetregeqn} at steady state leads to
\begin{equation*}
    \mathcal{A}_{T_0^r} - \overline{R^r} - d_{T_0^r}\overline{T_0^r} = 0,
\end{equation*}
where
\begin{equation*}
    \overline{R^r} = \lambda_{T_0^r T_A^r}\overline{D^\mathrm{LN}}\overline{T_0^r}.
\end{equation*}
Considering \eqref{minimalTArnrmaxeqn} at steady state leads to
\begin{equation*}
    2^{n^r_\mathrm{max}}\exp\left(-d_{T_0^r}\tau_{T_A^r}\right) \overline{R^r} - \lambda_{T_A^rT_r}\overline{T_A^r}-d_{T_r}\overline{T_A^r}=0.
\end{equation*}
Finally, considering \eqref{minimaltregeqn} at steady state leads to
\begin{align*}
    \frac{V_\mathrm{LN}}{\overline{V_\mathrm{TS}}}\lambda_{T_A^r T_r}\overline{T_A^r} - d_{T_r}\overline{T_r} &= 0.
\end{align*}
Solving these equations simultaneously leads to
\begin{align*}
\mathcal{A}_{T_0^r} &= 1.14 \times 10^5 ~\left(\mathrm{cell/cm^{3}}\right)\mathrm{day^{-1}}, \\
\lambda_{T_0^r T_A^r} &= 1.07 \times 10^{-8} ~\left(\mathrm{cell/cm^3}\right)^{-1}\mathrm{day^{-1}}, \\
\overline{R^r} &= 1.14 \times 10^5 ~\left(\mathrm{cell/cm^{3}}\right)\mathrm{day^{-1}}, \\
\lambda_{T_A^r T_r} &= 4.85 ~\mathrm{day}^{-1}.
\end{align*}
\subsection{Estimates for Immune Checkpoint-Associated Components in the TS \label{minimaltsicissinitappendix}}
\subsubsection{Estimates for Synthesis Rates and Steady States}
As in \Cref{reducedtsicissinitappendix}, we denote $\rho_{P_D^{T_8}}$ and $\rho_{P_D^{K}}$ as the number of PD-1 molecules expressed on the surface of CD8+ T cells and activated NK cells in the TS, respectively, with $\rho_{P_D^{T_8}} = 2.76 \times 10^{3}~ \mathrm{molec/cell}$ and $\rho_{P_D^{K}} = 5.52 \times 10^{2}~ \mathrm{molec/cell}$. Similarly, we denote $\rho_{P_LC}$ as the number of PD-L1 molecules expressed on cancer cells with $\rho_{P_LC} = 1.8 \times 10^{5} \mathrm{~molec/cell}$.\\~\\
Considering \eqref{minimalPD8eqn}~--~\eqref{minimalPDKeqn}, and \eqref{minimalPLeqn}~--~\eqref{minimalQKeqn} at steady state in the absence of pembrolizumab leads to
\begin{align*}
    \lambda_{P_D^{T_8}}\overline{T_8} - d_{P_D}\overline{P_D^{T_8}} &=0, \\
    \lambda_{P_D^{K}}\overline{K} - d_{P_D}\overline{P_D^{K}} &=0, \\
    \lambda_{P_L C}\overline{C} - d_{P_L}\overline{P_L} &=0, \\
    \overline{Q^{T_8}} - \frac{\lambda_{P_DP_L}}{\lambda_Q}\overline{P_D^{T_8}}\overline{P_L}&=0, \\
    \overline{Q^{K}} - \frac{\lambda_{P_DP_L}}{\lambda_Q}\overline{P_D^{K}}\overline{P_L}&=0.
    \intertext{By considering the total number of PD-1 receptors expressed on each PD-1-expressing cell at steady state, we expect in the absence of pembrolizumab that}
    \overline{P_D^{T_8}} + \overline{Q^{T_8}} &=\rho_{P_D^{T_8}}\overline{T_8}, \\
    \overline{P_D^{K}} + \overline{Q^{K}} &=\rho_{P_D^{K}}\overline{K}.
    \intertext{We can also consider the total number of PD-L1 ligands at steady state so that}
    \overline{P_L} + \overline{Q^{T_8}} + \overline{Q^{K}} &= \rho_{P_L C}\overline{C}.
    \intertext{Finally, we expect the synthesis rates of PD-1 to be proportional to the total number of PD-1 molecules expressed per PD-1-expressing cell so that}
    \frac{\lambda_{P_D^{T_8}}}{\rho_{P_D^{T_8}}} &=     \frac{\lambda_{P_D^{K}}}{\rho_{P_D^{K}}}.
\end{align*}
Solving these simultaneously and ensuring all model parameters are positive leads to
\begin{align*}
    \lambda_{P_D^{T_8}} &= 9.25 \times 10^2 ~\left(\mathrm{molec/cell}\right)\mathrm{day^{-1}}, \\
    \lambda_{P_D^{K}} &= 1.85 \times 10^2 ~\left(\mathrm{molec/cell}\right)\mathrm{day^{-1}}, \\
    \lambda_{P_LC} &= 2.50 \times 10^5 ~\left(\mathrm{molec/cell}\right)\mathrm{day^{-1}}.
\end{align*}
This leads to
\begin{align*}
    \overline{P_D^{T_8}} &= 4.87 \times 10^8 \mathrm{~molec/cm^3}, \\
    \overline{P_D^{K}} &= 1.07 \times 10^8 \mathrm{~molec/cm^3}, \\
    \overline{P_L} &= 1.26 \times 10^{13} \mathrm{~molec/cm^3}, \\
    \overline{Q^{T_8}} &= 1.31 \times 10^6 \mathrm{~molec/cm^3}, \\
    \overline{Q^{K}} &= 2.87 \times 10^5 \mathrm{~molec/cm^3}.
\end{align*}
\subsubsection{Estimates for Initial Conditions}
To determine the relevant initial conditions, we can simply consider the total number of PD-1 receptors on each PD-1-expressing cell and PD-L1 ligands in the absence of pembrolizumab, so that
\begin{align*}
P_D^{T_8}(0) + Q^{T_8}(0) &=\rho_{P_D^{T_8}}T_8(0), \\
P_D^{K}(0) + Q^{K}(0) &=\rho_{P_D^{K}}K(0), \\
P_L(0) + Q^{T_8}(0) + Q^{K}(0) &= \rho_{P_L C}C(0).
\intertext{We can also consider \eqref{minimalQ8eqn}~--~\eqref{minimalQKeqn} initially, so that}
    Q^{T_8}(0) - \frac{\lambda_{P_DP_L}}{\lambda_Q}P_D^{T_8}(0)P_L(0)&=0, \\
    Q^{K}(0) - \frac{\lambda_{P_DP_L}}{\lambda_Q}P_D^{K}(0)P_L(0)&=0.
\end{align*}
Solving these simultaneously leads to
\begin{align*}
    P_D^{T_8}(0) &= 4.44 \times 10^8 \mathrm{~molec/cm^3}, \\
    P_D^{K}(0) &= 2.46 \times 10^8 \mathrm{~molec/cm^3}, \\
    P_L(0) &= 7.02 \times 10^{12} \mathrm{~molec/cm^3}, \\
    Q^{T_8}(0) &= 6.63 \times 10^5 \mathrm{~molec/cm^3}, \\
    Q^{K}(0) &= 3.68 \times 10^5 \mathrm{~molec/cm^3}.
\end{align*}
We note that excluding bound PD-1 receptors when considering the total number of PD-1 receptors on PD-1-expressing cells does not affect the parameter estimates, steady states, or initial conditions at this level of precision, since the number of unbound PD-1 receptors is several orders of magnitude larger than the number of bound PD-1 receptors on PD-1-expressing cells. Furthermore, this also applies when considering the total number of PD-L1 ligands.
\subsection{Estimates for Immune Checkpoint-Associated Components in the TDLN\label{minimaltdlnicissinitappendix}}
\subsubsection{Estimates for Synthesis Rates and Steady States}
Again, for simplicity, we assume that the total number of PD-1 receptors on cells in the TDLN is equal to the number on the corresponding cells in the TS. Thus, denoting $\rho_{P_D^{8\mathrm{LN}}}$ as the number of PD-1 molecules expressed on the surface of CD8+ T cells in the TDLN, we have that $\rho_{P_D^{8\mathrm{LN}}} = \rho_{P_D^{T_8}} = 2.76 \times 10^{3}~ \mathrm{molec/cell}$. Similarly, as in \Cref{reducedtdlnicissinitappendix}, we denote $\rho_{P_L^\mathrm{LN}D^\mathrm{LN}}$ as the number of PD-L1 molecules expressed on the surface of mature DCs in the TDLN, with $\rho_{P_L^\mathrm{LN}D^\mathrm{LN}} = 1.77 \times 10^{4} ~\mathrm{molec/cell}$.\\~\\
Considering \eqref{minimalPD8LNeqn} and \eqref{minimalPLLNeqn}~--~\eqref{minimalQ8LNeqn} at steady state in the absence of pembrolizumab, and making the same assumptions for estimation as in the TS, we obtain
\begin{align*}
    \lambda_{P_D^{8\mathrm{LN}}}\overline{T_A^8} - d_{P_D}\overline{P_D^{8\mathrm{LN}}} &=0, \\    \lambda_{P_L^\mathrm{LN}D^\mathrm{LN}}\overline{D^\mathrm{LN}} - d_{P_L}\overline{P_L^\mathrm{LN}} &=0, \\
    \overline{Q^{8\mathrm{LN}}} - \frac{\lambda_{P_DP_L}}{\lambda_Q}\overline{P_D^{8\mathrm{LN}}}\overline{P_L^\mathrm{LN}}&=0, \\
    \overline{P_D^{8\mathrm{LN}}} + \overline{Q^{8\mathrm{LN}}} &=\rho_{P_D^{8\mathrm{LN}}}\overline{T_A^8}, \\
    \overline{P_L^\mathrm{LN}} + \overline{Q^{8\mathrm{LN}}} &= \rho_{P_L^\mathrm{LN}D^\mathrm{LN}} \overline{D^\mathrm{LN}}.
\end{align*}
Solving these simultaneously and ensuring all model parameters are positive leads to
\begin{align*}
    \lambda_{P_D^{8\mathrm{LN}}} &= 9.27 \times 10^2 ~\left(\mathrm{molec/cell}\right)\mathrm{day^{-1}}, \\
    \lambda_{P_L^\mathrm{LN}D^\mathrm{LN}} &= 2.46 \times 10^{4} ~\left(\mathrm{molec/cell}\right)\mathrm{day^{-1}}.
\end{align*}
This leads to
\begin{align*}
\overline{P_D^{8\mathrm{LN}}} &= 2.29 \times 10^9 \mathrm{~molec/cm^3}, \\
\overline{P_L^\mathrm{LN}} &= 5.81 \times 10^{11} \mathrm{~molec/cm^3}, \\
\overline{Q^{8\mathrm{LN}}} &= 2.83 \times 10^5 \mathrm{~molec/cm^3}.
\end{align*}
\subsubsection{Estimates for Initial Conditions}
To determine the relevant immune checkpoint initial conditions, we can simply consider the total number of PD-1 receptors on each PD-1-expressing cell and PD-L1 ligands in the absence of pembrolizumab, so that
\begin{align*}
P_D^{8\mathrm{LN}}(0) + Q^{8\mathrm{LN}}(0) &=\rho_{P_D^{8\mathrm{LN}}}T_A^8(0), \\
P_L^\mathrm{LN}(0) + Q^{8\mathrm{LN}}(0)&= \rho_{P_L^\mathrm{LN}D^\mathrm{LN}} D^\mathrm{LN}(0).
\intertext{We can also consider \eqref{minimalQ8LNeqn} initially so that}
    Q^{8\mathrm{LN}}(0) - \frac{\lambda_{P_DP_L}}{\lambda_Q}P_D^{8\mathrm{LN}}(0)P_L^\mathrm{LN}(0)&=0.
\end{align*}
Solving these simultaneously leads to
\begin{align*}
    P_D^{8\mathrm{LN}}(0) &= 2.37 \times 10^9 \mathrm{~molec/cm^3}, \\
    P_L^\mathrm{LN}(0) &= 3.15 \times 10^{11} \mathrm{~molec/cm^3}, \\
    Q^{8\mathrm{LN}}(0) &= 1.59 \times 10^5 \mathrm{~molec/cm^3}.
\end{align*}
We note again that excluding bound PD-1 receptors when considering the total number of PD-1 receptors on PD-1-expressing cells does not affect the parameter estimates, steady states, or initial conditions at this level of precision, since the number of unbound PD-1 receptors is several orders of magnitude larger than the number of bound PD-1 receptors on PD-1-expressing cells. Furthermore, this also applies when considering the total number of PD-L1 ligands.
\putbib[References.bib]
\end{bibunit}
\newpage
\title{Appendix E: Sensitivity analysis-guided model reduction of a mathematical model of pembrolizumab therapy for de novo metastatic MSI-H/dMMR colorectal cancer}
\maketitle
\setcounter{page}{1}
\begin{bibunit}[vancouver]
\section{Model Parameters for the Minimal Model}
The model parameter values are estimated in \Cref{minimalmodelparamestsection} and are listed in \autoref{minimaltableofparams}.
\begin{center}
\begin{longtable}{|lp{160pt}lp{100pt}l|}
\caption{\label{minimaltableofparams}Parameter values for the minimal model. TDLN denotes the tumour-draining lymph node, whilst TS denotes the tumour site. est. denotes estimated parameters.}\\       
\hline \textbf{Parameter} & \textbf{Description} & \textbf{Value} & \textbf{Unit} & \textbf{References} \\
\hline
$f_\mathrm{pembro}$ & $A_1$/$A_1^\mathrm{LN}$ dose scaling factor & $1.17 \times 10^{12}$ & $\left(\mathrm{molec/cm^3}\right)/\mathrm{mg}$ & est.\\
$f_{C}$ & $C$ to $V_\mathrm{TS}$ scaling factor & $1.17 \times 10^{6}$ & $\mathrm{cell/(cm^3)^2}$ & est. \\
$f_{N_c}$ & $N_c$ to $V_\mathrm{TS}$ scaling factor & $6.16 \times 10^{4}$ & $\mathrm{cell/(cm^3)^2}$ & est. \\
\hline
$\mathcal{A}_{D_0}$ & Source of $D_0$ & $1.18 \times 10^6$ & $\left(\mathrm{cell/cm^{3}}\right)\mathrm{day^{-1}}$ & est. \\
$\mathcal{A}_{T_0^r}$ & Source of $T_0^r$ & $1.14 \times 10^5$ & $\left(\mathrm{cell/cm^{3}}\right)\mathrm{day^{-1}}$ & est. \\
\hline
$\lambda_{C}$ & Growth rate of $C$ & $5.25 \times 10^{-2}$ & $\mathrm{day^{-1}}$ & fitted \\
$\lambda_{CT_8}$ & Elimination rate of $C$ by $T_8$ & $8.01 \times 10^{-8}$ & $\left(\mathrm{cell/cm^{3}}\right)^{-1}\mathrm{day^{-1}}$ & fitted \\
$\lambda_{CK}$ & Elimination rate of $C$ by $K$ & $8.01 \times 10^{-8}$ & $\left(\mathrm{cell/cm^{3}}\right)^{-1}\mathrm{day^{-1}}$ & est. \\
$\lambda_{CI_{\upalpha}}$ & Necrosis rate of $C$ by $I_{\upalpha}$ & $7.23 \times 10^{-3}$ & $\mathrm{day^{-1}}$ & est. \\
$\lambda_{DN_c}$ & Maturation rate of $D_0$ by DAMPs & $1.33$ & $\mathrm{day^{-1}}$ & est. \\
$\lambda_{D_0K}$ & Killing rate of $D_0$ by $K$ & $5.49 \times 10^{-6}$ & $\left(\mathrm{cell/cm^3}\right)^{-1}\mathrm{day^{-1}}$ & est. \\
$\lambda_{DD^\mathrm{LN}}$ & Migration rate of $D$ to TDLN & $1.68 \times 10^{-2}$ & $\mathrm{day^{-1}}$ & est. \\
$\lambda_{T_0^8 T_A^8}$ & Kinetic rate constant for naive CD8+ T cell activation & $3.12 \times 10^{-4}$ & $\mathrm{day^{-1}}$ & est. \\
$\lambda_{T_A^8T_8}$ & Kinetic rate constant for $T_A^8$ migration to the TS & $6.64 \times 10^{-1}$ & $\mathrm{day}^{-1}$ & est. \\
$\lambda_{T_8 C}$ & Exhaustion rate of $T_8$ due to $C$ exposure & $6.31 \times 10^{-3}$ & $\mathrm{day}^{-1}$ & est. \\
$\lambda_{T_\mathrm{ex} A_1}$ & Reinvigoration rate of $T_\mathrm{ex}$ by $A_1$ & $2.25 \times 10^{-3}$ & $\mathrm{day}^{-1}$ & est. \\
$\lambda_{T_0^r T_A^r}$ & Kinetic rate constant for $T_0^r$ activation into $T_A^r$ & $1.07 \times 10^{-8}$ & $\left(\mathrm{cell/cm^3}\right)^{-1}\mathrm{day^{-1}}$ & est. \\
$\lambda_{T_A^r T_r}$ & Kinetic rate constant for $T_A^r$ migration to the TS & $4.85$ & $\mathrm{day}^{-1}$ & est. \\
$\lambda_{KT_8}$ & Maturation rate of $K_0$ by $I_2$ & $1.79 \times 10^{4}$ & $\left(\mathrm{cell/cm^3}\right)\mathrm{day^{-1}}$ & est. \\
$\lambda_{KD}$ & Maturation rate of $K_0$ by $D$ & $3.59 \times 10^{4}$ & $\left(\mathrm{cell/cm^3}\right)\mathrm{day^{-1}}$ & est. \\
$\lambda_{I_{\upalpha}T_8}$ & Production rate of $I_{\upalpha}$ by $T_8$ & $9.57 \times 10^{-15}$ & $\left(\mathrm{g/cell}\right)\mathrm{day^{-1}}$ & est. \\
$\lambda_{I_{\upalpha}K}$ & Production rate of $I_{\upalpha}$ by $K$ & $1.67 \times 10^{-14}$ & $\left(\mathrm{g/cell}\right)\mathrm{day^{-1}}$ & est. \\
$\lambda_{I_{\upbeta}C}$ & Production rate of $I_{\upbeta}$ by $C$ & $8.39 \times 10^{-12}$ & $\left(\mathrm{g/cell}\right)\mathrm{day^{-1}}$ & est. \\
$\lambda_{I_{\upbeta}T_r}$ & Production rate of $I_{\upbeta}$ by $T_r$ & $4.86 \times 10^{-11}$ & $\left(\mathrm{g/cell}\right)\mathrm{day^{-1}}$ & est. \\
$\lambda_{P_D^{T_8}}$ & Synthesis rate of $P_D^{T_8}$ & $9.25 \times 10^2$ & $\left(\mathrm{molec/cell}\right)\mathrm{day^{-1}}$ & est. \\
$\lambda_{Q_A}$ & Dissociation rate of the PD-1/pembrolizumab complex & $2.6$ & $\mathrm{day}^{-1}$ & \cite{Li2021supp} \\
$\lambda_{Q}$ & Dissociation rate of the PD-1/PD-L1 complex & $1.24 \times 10^{5}$ & $\mathrm{day^{-1}}$ & \cite{Cheng2013supp} \\
$\lambda_{P_DA_1}$ & Formation rate of the PD-1/pembrolizumab complex & $4.63 \times 10^{-13}$ & $\left(\mathrm{molec/cm^3}\right)^{-1}\mathrm{day}^{-1}$ & fitted \\
$\lambda_{P_DP_L}$ & Formation rate of the PD-1/PD-L1 complex & $2.64 \times 10^{-11}$ & $\left(\mathrm{molec/cm^3}\right)^{-1}\mathrm{day}^{-1}$ & \cite{Cheng2013supp} \\
$\lambda_{P_D^{K}}$ & Synthesis rate of $P_D^{K}$ & $1.85 \times 10^2$ & $\left(\mathrm{molec/cell}\right)\mathrm{day^{-1}}$ & est. \\
$\lambda_{P_LC}$ & Synthesis rate of $P_L$ by $C$ & $2.50 \times 10^{5}$ & $\left(\mathrm{molec/cell}\right)\mathrm{day^{-1}}$ & est. \\
$\lambda_{P_D^{8\mathrm{LN}}}$ & Synthesis rate of $P_D^{8\mathrm{LN}}$ & $9.27 \times 10^{2}$ & $\left(\mathrm{molec/cell}\right)\mathrm{day^{-1}}$ & est. \\
$\lambda_{P_L^\mathrm{LN}D^\mathrm{LN}}$ & Synthesis rate of $P_L^\mathrm{LN}$ by $D^\mathrm{LN}$ & $2.46 \times 10^{4}$ & $\left(\mathrm{molec/cell}\right)\mathrm{day^{-1}}$ & est. \\
\hline
$K_{CI_{\upalpha}}$ & Half-saturation constant of $I_\upalpha$ for $C$ & $9.00 \times 10^{-11}$ & $\mathrm{g/cm^{3}}$ & est. \\
$K_{DN_c}$ & Half-saturation constant of $N_c$ for $D$ & $3.69 \times 10^{6}$ & $\mathrm{cell/cm^{3}}$ & est. \\
$K_{T_8 T_\mathrm{ex}}$ & Half-saturation constant $T_8$ exhaustion due to $C$ exposure & $7.02 \times 10^{7}$ & $\mathrm{cell}/\mathrm{cm^3}$ & est. \\
$K_{T_\mathrm{ex} A_1}$ & Half-saturation constant of $T_\mathrm{ex}$ reinvigoration by $A_1$ & $2.05 \times 10^{14}$ & $\mathrm{molec/cm^3}$ & est. \\
$K_{KT_8}$ & Half-saturation constant of $T_8$ for $K$ & $1.77 \times 10^5$ & $\mathrm{cell/cm^{3}}$ & est. \\
$K_{KD}$ & Half-saturation constant of $D$ for $K$ & $1.91 \times 10^{6}$ & $\mathrm{cell/cm^{3}}$ & est. \\
\hline
$C_0$ & Carrying capacity of $C$ & $8.89 \times 10^{7}$ & $\mathrm{cell/cm^3}$ & fitted \\
$K_{CI_{\upbeta}}$ & Inhibition constant of $T_8$ and $K$ elimination of $C$ by $I_{\upbeta}$ & $1.51 \times 10^{-6}$ & $\mathrm{g/cm^{3}}$ & est. \\
$K_{CQ^{T_8}}$ & Inhibition constant of $T_8$ elimination of $C$ by $Q^{T_8}$ & $1.31 \times 10^6$ & $\mathrm{molec/cm^{3}}$ & est. \\
$K_{CQ^K}$ & Inhibition constant of $K$ elimination of $C$ by $Q^K$ & $2.87 \times 10^5$ & $\mathrm{molec/cm^3}$ & est. \\
$K_{D_0I_\upbeta}$ & Inhibition constant of $K$ elimination of $D_0$ by $I_\upbeta$ & $1.51 \times 10^{-6}$ & $\mathrm{g/cm^{3}}$ & est. \\
$V_\mathrm{LN}$ & Volume of the TDLN & $1.47 \times 10^{-1}$ & $\mathrm{cm^3}$ & \cite{Rssler2017supp} est. \\
$K_{T_0^8T_A^r}$ & Inhibition constant of $T_0^8$ activation by $T_A^r$ & $1.47 \times 10^{6}$ & $\mathrm{cell/cm^3}$ & est. \\
$K_{T_0^8 Q^{8\mathrm{LN}}}$ & Inhibition constant of $T_0^8$ activation by $Q^{8\mathrm{LN}}$ & $2.83 \times 10^5$ & $\mathrm{molec/cm^3}$ & est. \\
$K_{T_A^8T_A^r}$ & Inhibition constant of $T_A^8$ proliferation by $T_A^r$ & $1.47 \times 10^{6}$ & $\mathrm{cell/cm^3}$ & est. \\
$K_{T_A^8 Q^{8\mathrm{LN}}}$ & Inhibition constant of $T_A^8$ proliferation by $Q^{8\mathrm{LN}}$ & $2.83 \times 10^5$ & $\mathrm{molec/cm^3}$ & est. \\
$K_{T_8C}$ & Inhibition constant of $T_8$ death by IL-10 & $7.02 \times 10^{7}$ & $\mathrm{cell/cm^{3}}$ & est. \\
$K_{T_\mathrm{ex} C}$ & Inhibition constant of $T_\mathrm{ex}$ death by IL-10 & $7.02 \times 10^{7}$ & $\mathrm{cell/cm^{3}}$ & est. \\
$K_{KI_\upbeta}$ & Inhibition constant of NK cell activation by $I_\upbeta$ & $1.51 \times 10^{-6}$ & $\mathrm{g/cm^{3}}$ & est. \\
\hline
$d_{N_c}$ & Removal rate of $N_c$ & $6.88 \times 10^{-2}$ & $\mathrm{day}^{-1}$ & est. \\
$d_{D_0}$ & Death rate of $D_0$ & $3.57 \times 10^{-2}$ & $\mathrm{day}^{-1}$ & \cite{Ruedl2000supp} est. \\
$d_{D}$ & Death rate of $D$ & $3.15 \times 10^{-1}$ & $\mathrm{day}^{-1}$ & \cite{Kamath2002supp} est. \\
$d_{T_0^8}$ & Death rate of naive CD8+ T cells & $3.22 \times 10^{-2}$ & $\mathrm{day}^{-1}$ & \cite{Takada2009supp} est. \\
$d_{T_8}$ & Death rate of $T_8$ & $9\times 10^{-3}$ & $\mathrm{day}^{-1}$ & \cite{Hellerstein1999supp} \\
$d_{T_\mathrm{ex}}$ & Death rate of $T_\mathrm{ex}$ & $9\times 10^{-3}$ & $\mathrm{day}^{-1}$ & \cite{Hellerstein1999supp} \\
$d_{T_0^r}$ & Death rate of $T_0^r$ & $2.2 \times 10^{-3}$ & $\mathrm{day}^{-1}$ & \cite{Kumbhari2020n2supp} \\
$d_{T_r}$ & Death rate of $T_r$ & $6.30 \times 10^{-2}$ & $\mathrm{day}^{-1}$ & \cite{VukmanovicStejic2006supp} est. \\
$d_{K}$ & Death rate of $K$ & $6.93 \times 10^{-2}$ & $\mathrm{day}^{-1}$ & \cite{Wu2020supp, Vivier2008supp, Lowry2017supp} est. \\
$d_{I_{\upalpha}}$ & Degradation rate of $I_{\upalpha}$ & $5.48 \times 10^{1}$ & $\mathrm{day}^{-1}$ & \cite{Ma2015supp, Oliver1993-vlsupp} est. \\
$d_{I_{\upbeta}}$ & Degradation rate of $I_{\upbeta}$ & $3.99 \times 10^{2}$ & $\mathrm{day}^{-1}$ & \cite{TiradoRodriguez2014supp} est. \\
$d_{P_D}$ & Degradation rate of unbound PD-1 receptors & $3.36 \times 10^{-1}$ & $\mathrm{day}^{-1}$ & \cite{Lassman2021supp} \\
$d_{Q_A}$ & Internalisation rate of the PD-1/pembrolizumab complex & $0.43$ & $\mathrm{day}^{-1}$ & \cite{Li2021supp} \\
$d_{A_1}$ & Elimination rate of $A_1/A_1^\mathrm{LN}$ & $2.92 \times 10^{-2}$ & $\mathrm{day}^{-1}$ & \cite{Li2017supp, Li2019pembrosupp, Ahamadi2016supp} est. \\
$d_{P_L}$ & Degradation rate of unbound PD-L1 & $1.39$ & $\mathrm{day}^{-1}$ & \cite{Cha2019supp} \\
\hline
$\tau_m$ & DC migration time from TDLN to the TS & $0.75$ & $\mathrm{day}$ & \cite{Catron2004supp} est. \\
$\Delta_8^0$ & Time taken for first CTL division & $1.63$ & $\mathrm{day}$ & \cite{Plambeck2022supp} \\
$n^8_\mathrm{max}$ & Maximal number of CTL divisions in the TDLN & $10$ & dimensionless & \cite{Kaech2001supp, Masopust2007supp} est. \\
$\Delta_8$ & Time taken for successive CTL divisions & $0.36$ & $\mathrm{day}$ & \cite{Kaech2001supp} \\
$\tau_{T_A^8}$ & Time taken for CTL division program & $4.87$ & $\mathrm{day}$ & est. \\
$\Delta_r^0$ & Time taken for first Treg division & $0.77$ & $\mathrm{day}$ & \cite{Kaech2002supp} est. \\
$n^r_\mathrm{max}$ & Maximal number of Treg divisions in the TDLN & $6$ & dimensionless & \cite{DarrasseJze2009supp} est. \\
$\Delta_r$ & Time taken for successive Treg divisions & $0.42$ & $\mathrm{day}$ & \cite{Kaech2002supp} est. \\
$\tau_{T_A^r}$ & Time taken for Treg division program & $2.87$ & $\mathrm{day}$ & est. \\
\hline
\end{longtable}
\end{center}
\putbib[References.bib]
\end{bibunit}
\newpage
\title{Appendix F: Sensitivity analysis-guided model reduction of a mathematical model of pembrolizumab therapy for de novo metastatic MSI-H/dMMR colorectal cancer}
\maketitle
\setcounter{page}{1}
\section{Full Model Sensitivity Analysis Indices\label{fullmodelSAresultsection}}
\begin{center}
{\small
\begin{longtable}{|c|c|c|c|c|c|c|c|c|c|}
\caption{\label{fullmodelSAresults}Maximum first-order indices ($S_i$) and total-order indices ($S_{Ti}$) for the RMSRE of variables in the full model at 180.9 days under the standard regimen, along with the indices associated with the RMSRE of $V_\mathrm{TS}$ --- the primary tumour volume. In particular, ``maximum first-order indices'' refers to the largest value of $S_i$ across the RMSRE for all variables in \autoref{modelvars}, while ``maximum total-order indices'' is defined analogously using $S_{Ti}$.}\\
\hline 
\textbf{Param} & \textbf{Max} & \textbf{Max} & $\bm{S_i}$ \textbf{for} & $\bm{S_{Ti}}$ \textbf{for} & \textbf{Param} & \textbf{Max} & \textbf{Max} & $\bm{S_i}$ \textbf{for} & $\bm{S_{Ti}}$ \textbf{for}\\
 & $\bm{S_i}$ & $\bm{S_{Ti}}$ & $\bm{V_\mathrm{TS}}$ & $\bm{V_\mathrm{TS}}$ & & $\bm{S_i}$ & $\bm{S_{Ti}}$ & $\bm{V_\mathrm{TS}}$ & $\bm{V_\mathrm{TS}}$\\
\hline
$\bm{f_\mathrm{pembro}}$ & 0.196455 & 0.781830 & 0.000142 & 0.030835 & $\bm{f_{C}}$ & 0.154395 & 0.689673 & 0.154395 & 0.468892 \\
$\bm{f_{N_c}}$ & 0.000572 & 0.286282 & 0.000572 & 0.028513 & $\bm{\mathcal{A}_{D_0}}$ & 0.098118 & 0.361952 & 0.000924 & 0.032853 \\
$\bm{\mathcal{A}_{T_0^8}}$ & 0.098988 & 0.745633 & 0.000013 & 0.026813 & $\bm{\mathcal{A}_{T_0^4}}$ & 0.113653 & 0.867688 & 0.000033 & 0.029270 \\
$\bm{\mathcal{A}_{T_0^r}}$ & 0.053728 & 0.868902 & 0.000023 & 0.033829 & $\bm{\mathcal{A}_{M_0}}$ & 0.240664 & 0.836605 & 0.000052 & 0.024638 \\
$\bm{\mathcal{A}_{K_0}}$ & 0.111575 & 0.841465 & 0.001600 & 0.051894 & $\bm{\lambda_{C}}$ & 0.045830 & 0.756570 & 0.044835 & 0.312093 \\
$\bm{\lambda_{CT_8}}$ & 0.020103 & 0.487583 & 0.001134 & 0.063549 & $\bm{\lambda_{CK}}$ & 0.013353 & 0.826863 & 0.003405 & 0.084351 \\
$\bm{\lambda_{CI_{\upalpha}}}$ & 0.039973 & 0.581582 & 0.000587 & 0.036730 & $\bm{\lambda_{CI_{\upgamma}}}$ & 0.001039 & 0.527882 & 0.000025 & 0.031523 \\
$\bm{\lambda_{HN_c}}$ & 0.053316 & 0.696267 & 0.000025 & 0.029590 & $\bm{\lambda_{SN_c}}$ & 0.051485 & 0.557700 & 0.000011 & 0.025580 \\
$\bm{\lambda_{DH}}$ & 0.007218 & 0.428341 & 0.000022 & 0.030887 & $\bm{\lambda_{DS}}$ & 0.000717 & 0.821022 & 0.000016 & 0.030449 \\
$\bm{\lambda_{D_0K}}$ & 0.010631 & 0.803754 & 0.000049 & 0.028679 & $\bm{\lambda_{DD^\mathrm{LN}}}$ & 0.016469 & 0.440553 & 0.000095 & 0.032598 \\
$\bm{\lambda_{T_0^8 T_A^8}}$ & 0.009913 & 0.815172 & 0.000031 & 0.038289 & $\bm{\lambda_{T_A^8T_8}}$ & 0.025050 & 0.683439 & 0.000005 & 0.025455 \\
$\bm{\lambda_{T_8 I_2}}$ & 0.000398 & 0.693214 & 0.000009 & 0.023958 & $\bm{\lambda_{T_8 C}}$ & 0.002766 & 0.715876 & 0.000102 & 0.028300 \\
$\bm{\lambda_{T_\mathrm{ex} A_1}}$ & 0.000595 & 0.686386 & 0.000022 & 0.031097 & $\bm{\lambda_{T_0^4 T_A^1}}$ & 0.008659 & 0.611239 & 0.000006 & 0.030416 \\
$\bm{\lambda_{T_A^1 T_1}}$ & 0.006964 & 0.867024 & 0.000005 & 0.024661 & $\bm{\lambda_{T_1 I_2}}$ & 0.000277 & 0.367976 & 0.000025 & 0.031124 \\
$\bm{\lambda_{T_1T_r}}$ & 0.000419 & 0.549419 & 0.000006 & 0.026430 & $\bm{\lambda_{T_0^r T_A^r}}$ & 0.011332 & 0.529358 & 0.000002 & 0.025710 \\
$\bm{\lambda_{T_A^r T_r}}$ & 0.037037 & 0.694783 & 0.000098 & 0.034798 & $\bm{\lambda_{M_1I_\upalpha}}$ & 0.005957 & 0.383106 & 0.000036 & 0.029591 \\
$\bm{\lambda_{M_1I_\upgamma}}$ & 0.006690 & 0.544117 & 0.000008 & 0.023943 & $\bm{\lambda_{M_2I_{10}}}$ & 0.012403 & 0.528181 & 0.000009 & 0.027346 \\
$\bm{\lambda_{M_2I_\upbeta}}$ & 0.015300 & 0.443515 & 0.000018 & 0.025119 & $\bm{\lambda_{MI_{\upgamma}}}$ & 0.000516 & 0.626537 & 0.000015 & 0.028707 \\
$\bm{\lambda_{MI_{\upalpha}}}$ & 0.000453 & 0.425568 & 0.000008 & 0.022884 & $\bm{\lambda_{MI_{\upbeta}}}$ & 0.000508 & 0.654224 & 0.000008 & 0.028281 \\
$\bm{\lambda_{KI_2}}$ & 0.019479 & 0.379776 & 0.000115 & 0.033761 & $\bm{\lambda_{KD_0}}$ & 0.000408 & 0.587977 & 0.000036 & 0.025974 \\
$\bm{\lambda_{KD}}$ & 0.002305 & 0.477241 & 0.000489 & 0.033624 & $\bm{\lambda_{I_2 T_8}}$ & 0.000498 & 0.640375 & 0.000022 & 0.028599 \\
$\bm{\lambda_{I_2 T_1}}$ & 0.000511 & 0.621754 & 0.000053 & 0.029137 & $\bm{\lambda_{I_{\upgamma} T_8}}$ & 0.002761 & 0.505682 & 0.000008 & 0.027038 \\
$\bm{\lambda_{I_{\upgamma}T_1}}$ & 0.000435 & 0.549770 & 0.000012 & 0.023860 & $\bm{\lambda_{I_{\upgamma} K}}$ & 0.006003 & 0.709358 & 0.000017 & 0.031138 \\
$\bm{\lambda_{I_{\upalpha}T_8}}$ & 0.000473 & 0.592204 & 0.000007 & 0.029306 & $\bm{\lambda_{I_{\upalpha}T_1}}$ & 0.000619 & 0.538387 & 0.000025 & 0.024793 \\
$\bm{\lambda_{I_{\upalpha}M_1}}$ & 0.000474 & 0.585327 & 0.000012 & 0.026705 & $\bm{\lambda_{I_{\upalpha}K}}$ & 0.000667 & 0.560243 & 0.000044 & 0.025397 \\
$\bm{\lambda_{I_{\upbeta}C}}$ & 0.009754 & 0.519025 & 0.004696 & 0.050936 & $\bm{\lambda_{I_{\upbeta}T_r}}$ & 0.001067 & 0.418672 & 0.000016 & 0.023998 \\
$\bm{\lambda_{I_{\upbeta}M_2}}$ & 0.000497 & 0.628144 & 0.000080 & 0.027637 & $\bm{\lambda_{I_{10}C}}$ & 0.002527 & 0.535806 & 0.000109 & 0.022529 \\
$\bm{\lambda_{I_{10}M_2}}$ & 0.000316 & 0.438253 & 0.000009 & 0.021582 & $\bm{\lambda_{I_{10}T_r}}$ & 0.000871 & 0.471220 & 0.000009 & 0.024644 \\
$\bm{\lambda_{I_{10}I_{2}}}$ & 0.000289 & 0.393170 & 0.000012 & 0.029589 & $\bm{\lambda_{P_D^{T_8}}}$ & 0.002471 & 0.395530 & 0.000016 & 0.029102 \\
$\bm{\lambda_{Q_A}}$ & 0.025001 & 0.710132 & 0.000012 & 0.027026 & $\bm{\lambda_{Q}}$ & 0.018023 & 0.588070 & 0.000009 & 0.028938 \\
$\bm{\lambda_{P_DA_1}}$ & 0.060459 & 0.572133 & 0.000013 & 0.023879 & $\bm{\lambda_{P_DP_L}}$ & 0.017739 & 0.678943 & 0.000063 & 0.024731 \\
$\bm{\lambda_{P_D^{T_1}}}$ & 0.000676 & 0.520618 & 0.000013 & 0.026382 & $\bm{\lambda_{P_D^{K}}}$ & 0.049163 & 0.563250 & 0.000030 & 0.023153 \\
$\bm{\lambda_{P_LC}}$ & 0.009473 & 0.380163 & 0.000027 & 0.028253 & $\bm{\lambda_{P_LD}}$ & 0.000349 & 0.476055 & 0.000010 & 0.022271 \\
$\bm{\lambda_{P_LT_8}}$ & 0.000285 & 0.422810 & 0.000014 & 0.024339 & $\bm{\lambda_{P_LT_1}}$ & 0.000479 & 0.605555 & 0.000019 & 0.027530 \\
$\bm{\lambda_{P_LT_r}}$ & 0.000416 & 0.558997 & 0.000018 & 0.024703 & $\bm{\lambda_{P_LM_2}}$ & 0.000253 & 0.492355 & 0.000020 & 0.027700 \\
$\bm{\lambda_{P_D^{8\mathrm{LN}}}}$ & 0.016826 & 0.373082 & 0.000015 & 0.028454 & $\bm{\lambda_{P_D^{1\mathrm{LN}}}}$ & 0.009439 & 0.350992 & 0.000013 & 0.024811 \\
$\bm{\lambda_{P_L^\mathrm{LN}D^\mathrm{LN}}}$ & 0.018080 & 0.550259 & 0.000011 & 0.026725 & $\bm{\lambda_{P_L^\mathrm{LN}T_A^8}}$ & 0.000469 & 0.606482 & 0.000028 & 0.025046 \\
$\bm{\lambda_{P_L^\mathrm{LN}T_A^1}}$ & 0.000423 & 0.558748 & 0.000023 & 0.023104 & $\bm{\lambda_{P_L^\mathrm{LN}T_A^r}}$ & 0.000224 & 0.344181 & 0.000018 & 0.027921 \\
$\bm{K_{CI_{\upalpha}}}$ & 0.007051 & 0.571664 & 0.000069 & 0.024656 & $\bm{K_{CI_{\upgamma}}}$ & 0.000331 & 0.502299 & 0.000007 & 0.027936 \\
$\bm{K_{DH}}$ & 0.001836 & 0.536396 & 0.000028 & 0.025420 & $\bm{K_{DS}}$ & 0.000297 & 0.435087 & 0.000024 & 0.023824 \\
$\bm{K_{T_8I_2}}$ & 0.000450 & 0.482466 & 0.000026 & 0.023693 & $\bm{K_{T_8C}}$ & 0.001754 & 0.352477 & 0.000073 & 0.029270 \\
$\bm{K_{T_\mathrm{ex} A_1}}$ & 0.000355 & 0.489704 & 0.000055 & 0.023480 & $\bm{K_{T_1I_2}}$ & 0.000275 & 0.523019 & 0.000012 & 0.029960 \\
$\bm{K_{T_1Q^{T_1}}}$ & 0.000471 & 0.594317 & 0.000021 & 0.027960 & $\bm{K_{M_1 I_\upalpha}}$ & 0.000603 & 0.335303 & 0.000007 & 0.029736 \\
$\bm{K_{M_1 I_\upgamma}}$ & 0.000815 & 0.503111 & 0.000024 & 0.028087 & $\bm{K_{M_2 I_{10}}}$ & 0.004164 & 0.562269 & 0.000011 & 0.029065 \\
$\bm{K_{M_2 I_{\upbeta}}}$ & 0.002805 & 0.385447 & 0.000021 & 0.025002 & $\bm{K_{MI_{\upgamma}}}$ & 0.000314 & 0.443723 & 0.000022 & 0.023068 \\
$\bm{K_{MI_{\upalpha}}}$ & 0.000196 & 0.334591 & 0.000016 & 0.029593 & $\bm{K_{MI_{\upbeta}}}$ & 0.000363 & 0.517795 & 0.000009 & 0.029048 \\
$\bm{K_{KI_2}}$ & 0.000389 & 0.386053 & 0.000047 & 0.026070 & $\bm{K_{KD_0}}$ & 0.000261 & 0.397275 & 0.000006 & 0.028063 \\
$\bm{K_{KD}}$ & 0.001280 & 0.532152 & 0.000094 & 0.026022 & $\bm{K_{I_{10}I_{2}}}$ & 0.000222 & 0.375698 & 0.000011 & 0.028945 \\
$\bm{C_0}$ & 0.216866 & 0.663083 & 0.141903 & 0.449168 & $\bm{K_{CI_{\upbeta}}}$ & 0.006226 & 0.397751 & 0.004442 & 0.064300 \\
$\bm{K_{CQ^{T_8}}}$ & 0.000393 & 0.526433 & 0.000019 & 0.025382 & $\bm{K_{CQ^K}}$ & 0.000254 & 0.390986 & 0.000025 & 0.028895 \\
$\bm{K_{D_0I_\upbeta}}$ & 0.002976 & 0.372823 & 0.000036 & 0.026400 & $\bm{V_\mathrm{LN}}$ & 0.032220 & 0.761538 & 0.000011 & 0.035396 \\
$\bm{K_{T_0^8T_A^r}}$ & 0.001215 & 0.360980 & 0.000058 & 0.029030 & $\bm{K_{T_0^8 Q^{8\mathrm{LN}}}}$ & 0.000647 & 0.353112 & 0.000020 & 0.028440 \\
$\bm{K_{T_A^8T_A^r}}$ & 0.000978 & 0.507490 & 0.000015 & 0.032324 & $\bm{K_{T_A^8 Q^{8\mathrm{LN}}}}$ & 0.000543 & 0.590254 & 0.000013 & 0.031324 \\
$\bm{K_{T_8T_r}}$ & 0.000327 & 0.440936 & 0.000010 & 0.026860 & $\bm{K_{T_8 I_{10}}}$ & 0.000221 & 0.361707 & 0.000109 & 0.027530 \\
$\bm{K_{T_\mathrm{ex} I_{10}}}$ & 0.000386 & 0.515569 & 0.000003 & 0.025007 & $\bm{K_{T_0^4 T_A^r}}$ & 0.000906 & 0.416674 & 0.000013 & 0.027251 \\
$\bm{K_{T_0^4 Q^{1\mathrm{LN}}}}$ & 0.000607 & 0.601432 & 0.000017 & 0.027814 & $\bm{K_{T_A^1 T_A^r}}$ & 0.001141 & 0.390313 & 0.000013 & 0.027740 \\
$\bm{K_{T_A^1 Q^{1\mathrm{LN}}}}$ & 0.000659 & 0.663544 & 0.000003 & 0.028727 & $\bm{K_{T_1T_r}}$ & 0.000358 & 0.483177 & 0.000013 & 0.025290 \\
$\bm{K_{KI_\upbeta}}$ & 0.003117 & 0.367399 & 0.000659 & 0.033125 & $\bm{K_{I_\upgamma T_r}}$ & 0.001390 & 0.547457 & 0.000007 & 0.025100 \\
$\bm{d_{N_c}}$ & 0.091396 & 0.868058 & 0.000264 & 0.024513 & $\bm{d_{H}}$ & 0.091265 & 0.449674 & 0.000004 & 0.026243 \\
$\bm{d_{S}}$ & 0.076757 & 0.567976 & 0.000005 & 0.025488 & $\bm{d_{D_0}}$ & 0.000779 & 0.598566 & 0.000005 & 0.023806 \\
$\bm{d_{D}}$ & 0.147321 & 0.776825 & 0.000978 & 0.045696 & $\bm{d_{T_0^8}}$ & 0.250308 & 0.896970 & 0.000010 & 0.036000 \\
$\bm{d_{T_8}}$ & 0.001293 & 0.486334 & 0.000195 & 0.025046 & $\bm{d_{T_\mathrm{ex}}}$ & 0.000777 & 0.524852 & 0.000009 & 0.029467 \\
$\bm{d_{T_0^4}}$ & 0.242511 & 0.882519 & 0.000010 & 0.027031 & $\bm{d_{T_1}}$ & 0.000718 & 0.657060 & 0.000023 & 0.022276 \\
$\bm{d_{T_0^r}}$ & 0.000798 & 0.540833 & 0.000013 & 0.026578 & $\bm{d_{T_r}}$ & 0.003225 & 0.534336 & 0.000023 & 0.025321 \\
$\bm{d_{M_0}}$ & 0.055325 & 0.586456 & 0.000029 & 0.028598 & $\bm{d_{M_1}}$ & 0.226859 & 0.644728 & 0.000023 & 0.026919 \\
$\bm{d_{M_2}}$ & 0.125127 & 0.460048 & 0.000069 & 0.030346 & $\bm{d_{K_0}}$ & 0.264315 & 0.882659 & 0.001545 & 0.051803 \\
$\bm{d_{K}}$ & 0.095537 & 0.514093 & 0.003873 & 0.068345 & $\bm{d_{I_2}}$ & 0.000631 & 0.437762 & 0.000063 & 0.024649 \\
$\bm{d_{I_\upgamma}}$ & 0.034021 & 0.601873 & 0.000043 & 0.026979 & $\bm{d_{I_{\upalpha}}}$ & 0.006243 & 0.611070 & 0.000176 & 0.027942 \\
$\bm{d_{I_{\upbeta}}}$ & 0.019153 & 0.509755 & 0.007105 & 0.079089 & $\bm{d_{I_{10}}}$ & 0.016709 & 0.602077 & 0.000054 & 0.026421 \\
$\bm{d_{P_D}}$ & 0.000559 & 0.501914 & 0.000007 & 0.025763 & $\bm{d_{Q_A}}$ & 0.095190 & 0.522846 & 0.000045 & 0.026934 \\
$\bm{d_{A_1}}$ & 0.217898 & 0.800762 & 0.000073 & 0.027830 & $\bm{d_{P_L}}$ & 0.118656 & 0.751497 & 0.000022 & 0.029878 \\
$\bm{\tau_m}$ & 0.000804 & 0.724374 & 0.000017 & 0.027370 & $\bm{\tau_8^\mathrm{act}}$ & 0.001235 & 0.713127 & 0.000059 & 0.027581 \\
$\bm{\Delta_8^0}$ & 0.000563 & 0.669466 & 0.000037 & 0.026410 & $\bm{n^8_\mathrm{max}}$ & 0.342734 & 0.873702 & 0.004335 & 0.287343 \\
$\bm{\Delta_8}$ & 0.000841 & 0.608572 & 0.000032 & 0.024798 & $\bm{\tau_a}$ & 0.000816 & 0.774151 & 0.000017 & 0.025973 \\
$\bm{\tau_l}$ & 0.020278 & 0.644328 & 0.000062 & 0.025358 & $\bm{\tau_4^\mathrm{act}}$ & 0.000787 & 0.756608 & 0.000016 & 0.025606 \\
$\bm{\Delta_1^0}$ & 0.000723 & 0.766798 & 0.000025 & 0.027289 & $\bm{n^1_\mathrm{max}}$ & 0.245710 & 0.786345 & 0.000248 & 0.028258 \\
$\bm{\Delta_1}$ & 0.001106 & 0.767379 & 0.000011 & 0.026659 & $\bm{\tau_r^\mathrm{act}}$ & 0.000608 & 0.689967 & 0.000025 & 0.024408 \\
$\bm{\Delta_r^0}$ & 0.000549 & 0.700840 & 0.000026 & 0.025698 & $\bm{n^r_\mathrm{max}}$ & 0.455148 & 0.872521 & 0.001873 & 0.123402 \\
$\bm{\Delta_r}$ & 0.000208 & 0.296629 & 0.000014 & 0.026179 & & & & & \\
\hline
\end{longtable}
}
\end{center}
\newpage
\title{Appendix G: Sensitivity analysis-guided model reduction of a mathematical model of pembrolizumab therapy for de novo metastatic MSI-H/dMMR colorectal cancer}
\maketitle
\setcounter{page}{1}
\section{Reduced Model Sensitivity Analysis Indices\label{reducedmodelSAresultsection}}
\begin{center}
{\small
\begin{longtable}{|c|c|c|c|c|c|c|c|c|c|}
\caption{\label{reducedmodelSAresults}Maximum first-order indices ($S_i$) and total-order indices ($S_{Ti}$) for the RMSRE of variables in the reduced model at 180.9 days under the standard regimen, along with the indices associated with the RMSRE of $V_\mathrm{TS}$ --- the primary tumour volume. In particular, ``maximum first-order indices'' refers to the largest value of $S_i$ across the RMSRE for all variables in \autoref{modelvars}, while ``maximum total-order indices'' is defined analogously using $S_{Ti}$.}\\
\hline
\textbf{Param} & \textbf{Max} & \textbf{Max} & $\bm{S_i}$ \textbf{for} & $\bm{S_{Ti}}$ \textbf{for} & \textbf{Param} & \textbf{Max} & \textbf{Max} & $\bm{S_i}$ \textbf{for} & $\bm{S_{Ti}}$ \textbf{for}\\
& $\bm{S_i}$ & $\bm{S_{Ti}}$ & $\bm{V_\mathrm{TS}}$ & $\bm{V_\mathrm{TS}}$ & & $\bm{S_i}$ & $\bm{S_{Ti}}$ & $\bm{V_\mathrm{TS}}$ & $\bm{V_\mathrm{TS}}$\\
\hline
$\bm{f_\mathrm{pembro}}$ & 0.196699 & 0.782368 & 0.000018 & 0.028101 & $\bm{f_{C}}$ & 0.156199 & 0.865079 & 0.156199 & 0.479742 \\
$\bm{f_{N_c}}$ & 0.000394 & 0.463306 & 0.000394 & 0.031755 & $\bm{\mathcal{A}_{D_0}}$ & 0.074493 & 0.511470 & 0.000396 & 0.038541 \\
$\bm{\mathcal{A}_{T_0^8}}$ & 0.097871 & 0.834404 & 0.000070 & 0.031580 & $\bm{\mathcal{A}_{T_0^4}}$ & 0.110457 & 0.750074 & 0.000039 & 0.029241 \\
$\bm{\mathcal{A}_{T_0^r}}$ & 0.054063 & 0.609495 & 0.000258 & 0.038541 & $\bm{\mathcal{A}_{M_0}}$ & 0.255852 & 0.847846 & 0.000108 & 0.029145 \\
$\bm{\mathcal{A}_{K_0}}$ & 0.112343 & 0.726756 & 0.001923 & 0.049679 & $\bm{\lambda_{C}}$ & 0.043134 & 0.457370 & 0.042920 & 0.300684 \\
$\bm{\lambda_{CT_8}}$ & 0.016052 & 0.625139 & 0.001379 & 0.066753 & $\bm{\lambda_{CK}}$ & 0.010201 & 0.578997 & 0.003086 & 0.077898 \\
$\bm{\lambda_{CI_{\upalpha}}}$ & 0.044683 & 0.645409 & 0.000704 & 0.037197 & $\bm{\lambda_{HN_c}}$ & 0.042463 & 0.594310 & 0.000017 & 0.029839 \\
$\bm{\lambda_{SN_c}}$ & 0.054364 & 0.730376 & 0.000006 & 0.026103 & $\bm{\lambda_{DH}}$ & 0.009758 & 0.492553 & 0.000063 & 0.030330 \\
$\bm{\lambda_{D_0K}}$ & 0.018260 & 0.789082 & 0.000037 & 0.029910 & $\bm{\lambda_{DD^\mathrm{LN}}}$ & 0.013301 & 0.533852 & 0.000061 & 0.030025 \\
$\bm{\lambda_{T_0^8 T_A^8}}$ & 0.009162 & 0.567546 & 0.000013 & 0.037923 & $\bm{\lambda_{T_A^8T_8}}$ & 0.020333 & 0.626416 & 0.000035 & 0.029267 \\
$\bm{\lambda_{T_8 C}}$ & 0.010465 & 0.533543 & 0.000185 & 0.031121 & $\bm{\lambda_{T_\mathrm{ex} A_1}}$ & 0.000618 & 0.727339 & 0.000080 & 0.032953 \\
$\bm{\lambda_{T_0^4 T_A^1}}$ & 0.006429 & 0.843575 & 0.000081 & 0.032684 & $\bm{\lambda_{T_A^1 T_1}}$ & 0.010808 & 0.860397 & 0.000009 & 0.028221 \\
$\bm{\lambda_{T_1T_r}}$ & 0.000201 & 0.413099 & 0.000005 & 0.028087 & $\bm{\lambda_{T_0^r T_A^r}}$ & 0.010996 & 0.438753 & 0.000015 & 0.030774 \\
$\bm{\lambda_{T_A^r T_r}}$ & 0.046559 & 0.543385 & 0.000053 & 0.033947 & $\bm{\lambda_{M_1I_\upalpha}}$ & 0.007544 & 0.505262 & 0.000016 & 0.028446 \\
$\bm{\lambda_{M_1I_\upgamma}}$ & 0.005616 & 0.673982 & 0.000016 & 0.027777 & $\bm{\lambda_{M_2I_{10}}}$ & 0.012041 & 0.564238 & 0.000017 & 0.027264 \\
$\bm{\lambda_{M_2I_\upbeta}}$ & 0.015363 & 0.711958 & 0.000008 & 0.027330 & $\bm{\lambda_{KI_2}}$ & 0.012993 & 0.434006 & 0.000167 & 0.035123 \\
$\bm{\lambda_{KD}}$ & 0.004344 & 0.530308 & 0.000986 & 0.037752 & $\bm{\lambda_{I_2 T_8}}$ & 0.000346 & 0.437036 & 0.000021 & 0.030972 \\
$\bm{\lambda_{I_2 T_1}}$ & 0.000505 & 0.470982 & 0.000015 & 0.031061 & $\bm{\lambda_{I_{\upgamma} K}}$ & 0.047020 & 0.469126 & 0.000022 & 0.030584 \\
$\bm{\lambda_{I_{\upalpha}T_8}}$ & 0.000358 & 0.483409 & 0.000015 & 0.032958 & $\bm{\lambda_{I_{\upalpha}T_1}}$ & 0.000577 & 0.609750 & 0.000019 & 0.026065 \\
$\bm{\lambda_{I_{\upalpha}M_1}}$ & 0.000272 & 0.448167 & 0.000013 & 0.027329 & $\bm{\lambda_{I_{\upalpha}K}}$ & 0.000744 & 0.587610 & 0.000011 & 0.026171 \\
$\bm{\lambda_{I_{\upbeta}C}}$ & 0.009140 & 0.476749 & 0.005372 & 0.065933 & $\bm{\lambda_{I_{\upbeta}T_r}}$ & 0.001208 & 0.582373 & 0.000009 & 0.027140 \\
$\bm{\lambda_{I_{\upbeta}M_2}}$ & 0.000482 & 0.470403 & 0.000111 & 0.034241 & $\bm{\lambda_{I_{10}C}}$ & 0.070524 & 0.653060 & 0.000028 & 0.025077 \\
$\bm{\lambda_{I_{10}M_2}}$ & 0.000420 & 0.469329 & 0.000010 & 0.033561 & $\bm{\lambda_{P_D^{T_8}}}$ & 0.001438 & 0.364811 & 0.000026 & 0.027597 \\
$\bm{\lambda_{Q_A}}$ & 0.033563 & 0.533773 & 0.000037 & 0.033103 & $\bm{\lambda_{Q}}$ & 0.023865 & 0.550601 & 0.000021 & 0.030076 \\
$\bm{\lambda_{P_DA_1}}$ & 0.065135 & 0.449660 & 0.000093 & 0.032422 & $\bm{\lambda_{P_DP_L}}$ & 0.029281 & 0.714233 & 0.000068 & 0.034821 \\
$\bm{\lambda_{P_D^{T_1}}}$ & 0.000702 & 0.443972 & 0.000009 & 0.031971 & $\bm{\lambda_{P_D^{K}}}$ & 0.048038 & 0.365372 & 0.000022 & 0.031603 \\
$\bm{\lambda_{P_LC}}$ & 0.072411 & 0.394143 & 0.000033 & 0.027409 & $\bm{\lambda_{P_LM_2}}$ & 0.000299 & 0.462689 & 0.000007 & 0.032642 \\
$\bm{\lambda_{P_D^{8\mathrm{LN}}}}$ & 0.020138 & 0.416773 & 0.000019 & 0.034631 & $\bm{\lambda_{P_D^{1\mathrm{LN}}}}$ & 0.008002 & 0.558867 & 0.000007 & 0.031067 \\
$\bm{\lambda_{P_L^\mathrm{LN}D^\mathrm{LN}}}$ & 0.018164 & 0.654417 & 0.000011 & 0.030028 & $\bm{\lambda_{P_L^\mathrm{LN}T_A^1}}$ & 0.000198 & 0.380404 & 0.000012 & 0.027096 \\        
$\bm{K_{CI_{\upalpha}}}$ & 0.008810 & 0.384462 & 0.000181 & 0.027225 & $\bm{K_{DH}}$ & 0.002113 & 0.528954 & 0.000023 & 0.030381 \\
$\bm{K_{T_8C}}$ & 0.004342 & 0.616622 & 0.000008 & 0.027834 & $\bm{K_{T_\mathrm{ex} A_1}}$ & 0.000512 & 0.617239 & 0.000034 & 0.027696 \\
$\bm{K_{T_1Q^{T_1}}}$ & 0.000550 & 0.627109 & 0.000013 & 0.031656 & $\bm{K_{M_1 I_\upalpha}}$ & 0.000714 & 0.599344 & 0.000014 & 0.028523 \\
$\bm{K_{M_1 I_\upgamma}}$ & 0.000974 & 0.406854 & 0.000019 & 0.029125 & $\bm{K_{M_2 I_{10}}}$ & 0.002916 & 0.492126 & 0.000013 & 0.032249 \\
$\bm{K_{M_2 I_{\upbeta}}}$ & 0.002589 & 0.380563 & 0.000012 & 0.030998 & $\bm{K_{KI_2}}$ & 0.000460 & 0.568078 & 0.000047 & 0.032673 \\
$\bm{K_{KD}}$ & 0.002210 & 0.535704 & 0.000190 & 0.035304 & $\bm{C_0}$ & 0.216418 & 0.659926 & 0.134699 & 0.438727 \\
$\bm{K_{CI_{\upbeta}}}$ & 0.006842 & 0.413199 & 0.005062 & 0.071254 & $\bm{K_{CQ^{T_8}}}$ & 0.000264 & 0.402001 & 0.000011 & 0.031511 \\
$\bm{K_{CQ^K}}$ & 0.000451 & 0.559173 & 0.000002 & 0.031582 & $\bm{K_{D_0I_\upbeta}}$ & 0.002991 & 0.542746 & 0.000048 & 0.028396 \\
$\bm{V_\mathrm{LN}}$ & 0.031432 & 0.491612 & 0.000034 & 0.037856 & $\bm{K_{T_0^8T_A^r}}$ & 0.001054 & 0.419920 & 0.000010 & 0.031579 \\
$\bm{K_{T_0^8 Q^{8\mathrm{LN}}}}$ & 0.000495 & 0.666322 & 0.000009 & 0.028865 & $\bm{K_{T_A^8T_A^r}}$ & 0.001104 & 0.440874 & 0.000017 & 0.025948 \\
$\bm{K_{T_A^8 Q^{8\mathrm{LN}}}}$ & 0.000647 & 0.756078 & 0.000006 & 0.027987 & $\bm{K_{T_8 I_{10}}}$ & 0.000205 & 0.352087 & 0.000074 & 0.030958 \\
$\bm{K_{T_\mathrm{ex} I_{10}}}$ & 0.000451 & 0.624676 & 0.000007 & 0.029233 & $\bm{K_{T_0^4 T_A^r}}$ & 0.000717 & 0.534408 & 0.000034 & 0.032599 \\
$\bm{K_{T_0^4 Q^{1\mathrm{LN}}}}$ & 0.000567 & 0.609289 & 0.000008 & 0.029015 & $\bm{K_{T_A^1 T_A^r}}$ & 0.000968 & 0.464576 & 0.000026 & 0.028965 \\
$\bm{K_{T_A^1 Q^{1\mathrm{LN}}}}$ & 0.000707 & 0.557800 & 0.000018 & 0.031212 & $\bm{K_{KI_\upbeta}}$ & 0.002623 & 0.436641 & 0.000541 & 0.036278 \\
$\bm{d_{N_c}}$ & 0.072744 & 0.872253 & 0.000139 & 0.033711 & $\bm{d_{H}}$ & 0.074184 & 0.512795 & 0.000003 & 0.029891 \\
$\bm{d_{S}}$ & 0.074398 & 0.598835 & 0.000018 & 0.031017 & $\bm{d_{D_0}}$ & 0.000585 & 0.572300 & 0.000026 & 0.030700 \\
$\bm{d_{D}}$ & 0.106839 & 0.443295 & 0.001198 & 0.061665 & $\bm{d_{T_0^8}}$ & 0.252751 & 0.900001 & 0.000015 & 0.034062 \\
$\bm{d_{T_8}}$ & 0.000577 & 0.355399 & 0.000213 & 0.032954 & $\bm{d_{T_\mathrm{ex}}}$ & 0.021297 & 0.399114 & 0.000014 & 0.033962 \\
$\bm{d_{T_0^4}}$ & 0.240662 & 0.881627 & 0.000017 & 0.033046 & $\bm{d_{T_1}}$ & 0.000397 & 0.363997 & 0.000011 & 0.032331 \\
$\bm{d_{T_0^r}}$ & 0.000822 & 0.479558 & 0.000019 & 0.034144 & $\bm{d_{T_r}}$ & 0.000768 & 0.461821 & 0.000009 & 0.029399 \\
$\bm{d_{M_0}}$ & 0.074767 & 0.547764 & 0.000007 & 0.032210 & $\bm{d_{M_1}}$ & 0.220101 & 0.671529 & 0.000009 & 0.033616 \\
$\bm{d_{M_2}}$ & 0.142668 & 0.514810 & 0.000058 & 0.029554 & $\bm{d_{K_0}}$ & 0.268048 & 0.882146 & 0.001720 & 0.043581 \\
$\bm{d_{K}}$ & 0.083896 & 0.508186 & 0.002192 & 0.066429 & $\bm{d_{I_2}}$ & 0.000554 & 0.393571 & 0.000077 & 0.033173 \\
$\bm{d_{I_\upgamma}}$ & 0.100467 & 0.455116 & 0.000013 & 0.023677 & $\bm{d_{I_{\upalpha}}}$ & 0.010932 & 0.465299 & 0.000106 & 0.024842 \\
$\bm{d_{I_{\upbeta}}}$ & 0.020297 & 0.454115 & 0.007799 & 0.083064 & $\bm{d_{I_{10}}}$ & 0.129228 & 0.422049 & 0.000133 & 0.030982 \\
$\bm{d_{P_D}}$ & 0.000295 & 0.369661 & 0.000011 & 0.032970 & $\bm{d_{Q_A}}$ & 0.080240 & 0.386906 & 0.000012 & 0.032232 \\
$\bm{d_{A_1}}$ & 0.217378 & 0.800610 & 0.000037 & 0.027867 & $\bm{d_{P_L}}$ & 0.142276 & 0.535291 & 0.000053 & 0.035635 \\
$\bm{\tau_m}$ & 0.000950 & 0.406262 & 0.000021 & 0.023889 & $\bm{\Delta_8^0}$ & 0.000140 & 0.324964 & 0.000017 & 0.030383 \\
$\bm{n^8_\mathrm{max}}$ & 0.341779 & 0.865224 & 0.005149 & 0.312416 & $\bm{\Delta_8}$ & 0.000211 & 0.324558 & 0.000015 & 0.038173 \\
$\bm{\Delta_1^0}$ & 0.000129 & 0.432236 & 0.000018 & 0.038985 & $\bm{n^1_\mathrm{max}}$ & 0.259140 & 0.761432 & 0.000279 & 0.039751 \\
$\bm{\Delta_1}$ & 0.000230 & 0.332084 & 0.000006 & 0.026368 & $\bm{\Delta_r^0}$ & 0.000199 & 0.478516 & 0.000021 & 0.037171 \\
$\bm{n^r_\mathrm{max}}$ & 0.537028 & 0.885371 & 0.002362 & 0.140304 & $\bm{\Delta_r}$ & 0.000116 & 0.274462 & 0.000030 & 0.023804 \\
\hline
\end{longtable}
}
\end{center}
\newpage
\title{Appendix H: Sensitivity analysis-guided model reduction of a mathematical model of pembrolizumab therapy for de novo metastatic MSI-H/dMMR colorectal cancer}
\maketitle
\setcounter{page}{1}
\section{Minimal Model Sensitivity Analysis Indices\label{minimalmodelSAresultsection}}
\begin{center}
{\small
\begin{longtable}{|c|c|c|c|c|c|c|c|c|c|}
\caption{\label{minimalmodelSAresults}Maximum first-order indices ($S_i$) and total-order indices ($S_{Ti}$) for the RMSRE of variables in the minimal model at 180.9 days under the standard regimen, along with the indices associated with the RMSRE of $V_\mathrm{TS}$ --- the primary tumour volume. In particular, ``maximum first-order indices'' refers to the largest value of $S_i$ across the RMSRE for all variables in \autoref{modelvars}, while ``maximum total-order indices'' is defined analogously using $S_{Ti}$.}\\
\hline
\textbf{Param} & \textbf{Max} & \textbf{Max} & $\bm{S_i}$ \textbf{for} & $\bm{S_{Ti}}$ \textbf{for} & \textbf{Param} & \textbf{Max} & \textbf{Max} & $\bm{S_i}$ \textbf{for} & $\bm{S_{Ti}}$ \textbf{for}\\
& $\bm{S_i}$ & $\bm{S_{Ti}}$ & $\bm{V_\mathrm{TS}}$ & $\bm{V_\mathrm{TS}}$ & & $\bm{S_i}$ & $\bm{S_{Ti}}$ & $\bm{V_\mathrm{TS}}$ & $\bm{V_\mathrm{TS}}$\\
\hline
$\bm{f_\mathrm{pembro}}$ & 0.196714 & 0.781397 & 0.000053 & 0.029181 & $\bm{f_{C}}$ & 0.155495 & 0.477257 & 0.155495 & 0.466771 \\
$\bm{f_{N_c}}$ & 0.000802 & 0.230561 & 0.000513 & 0.034046 & $\bm{\mathcal{A}_{D_0}}$ & 0.080313 & 0.353478 & 0.000795 & 0.044875 \\
$\bm{\mathcal{A}_{T_0^r}}$ & 0.052635 & 0.405867 & 0.000072 & 0.029519 & $\bm{\lambda_{C}}$ & 0.045353 & 0.358603 & 0.045353 & 0.290649 \\
$\bm{\lambda_{CT_8}}$ & 0.018126 & 0.266294 & 0.001267 & 0.069703 & $\bm{\lambda_{CK}}$ & 0.008360 & 0.289387 & 0.003218 & 0.064299 \\
$\bm{\lambda_{CI_{\upalpha}}}$ & 0.036782 & 0.334747 & 0.000788 & 0.038577 & $\bm{\lambda_{DN_c}}$ & 0.008558 & 0.295177 & 0.000127 & 0.031737 \\
$\bm{\lambda_{D_0K}}$ & 0.019602 & 0.288701 & 0.000070 & 0.029169 & $\bm{\lambda_{DD^\mathrm{LN}}}$ & 0.011157 & 0.381432 & 0.000080 & 0.029407 \\
$\bm{\lambda_{T_0^8 T_A^8}}$ & 0.008600 & 0.505163 & 0.000020 & 0.038178 & $\bm{\lambda_{T_A^8T_8}}$ & 0.024229 & 0.369096 & 0.000002 & 0.026556 \\
$\bm{\lambda_{T_8 C}}$ & 0.003473 & 0.328211 & 0.000182 & 0.027518 & $\bm{\lambda_{T_\mathrm{ex} A_1}}$ & 0.001145 & 0.332576 & 0.000028 & 0.025827 \\
$\bm{\lambda_{T_0^r T_A^r}}$ & 0.010304 & 0.310825 & 0.000057 & 0.028596 & $\bm{\lambda_{T_A^r T_r}}$ & 0.046368 & 0.355147 & 0.000009 & 0.037796 \\
$\bm{\lambda_{KT_8}}$ & 0.011316 & 0.343820 & 0.000199 & 0.029113 & $\bm{\lambda_{KD}}$ & 0.007381 & 0.462180 & 0.001108 & 0.037246 \\
$\bm{\lambda_{I_{\upalpha}T_8}}$ & 0.002358 & 0.364446 & 0.000016 & 0.025512 & $\bm{\lambda_{I_{\upalpha}K}}$ & 0.002799 & 0.285920 & 0.000096 & 0.029224 \\
$\bm{\lambda_{I_{\upbeta}C}}$ & 0.013157 & 0.240479 & 0.005137 & 0.063172 & $\bm{\lambda_{I_{\upbeta}T_r}}$ & 0.001449 & 0.405260 & 0.000025 & 0.029763 \\
$\bm{\lambda_{P_D^{T_8}}}$ & 0.014701 & 0.299428 & 0.000022 & 0.027847 & $\bm{\lambda_{Q_A}}$ & 0.029650 & 0.285242 & 0.000027 & 0.028960 \\
$\bm{\lambda_{Q}}$ & 0.027488 & 0.400916 & 0.000044 & 0.029495 & $\bm{\lambda_{P_DA_1}}$ & 0.062137 & 0.348826 & 0.000006 & 0.026113 \\
$\bm{\lambda_{P_DP_L}}$ & 0.015567 & 0.326772 & 0.000048 & 0.029725 & $\bm{\lambda_{P_D^{K}}}$ & 0.046745 & 0.407042 & 0.000006 & 0.026605 \\
$\bm{\lambda_{P_LC}}$ & 0.078813 & 0.366148 & 0.000029 & 0.025288 & $\bm{\lambda_{P_D^{8\mathrm{LN}}}}$ & 0.019039 & 0.374169 & 0.000014 & 0.025930 \\
$\bm{\lambda_{P_L^\mathrm{LN}D^\mathrm{LN}}}$ & 0.015530 & 0.356287 & 0.000005 & 0.026136 & $\bm{K_{CI_{\upalpha}}}$ & 0.009615 & 0.319686 & 0.000210 & 0.034703 \\
$\bm{K_{DN_c}}$ & 0.002715 & 0.307489 & 0.000077 & 0.030232 & $\bm{K_{T_8 T_\mathrm{ex}}}$ & 0.002102 & 0.311059 & 0.000034 & 0.033769 \\
$\bm{K_{T_\mathrm{ex} A_1}}$ & 0.000490 & 0.240533 & 0.000021 & 0.029109 & $\bm{K_{KT_8}}$ & 0.001318 & 0.338152 & 0.000106 & 0.027441 \\
$\bm{K_{KD}}$ & 0.002650 & 0.327721 & 0.000186 & 0.030883 & $\bm{C_0}$ & 0.239215 & 0.669053 & 0.146308 & 0.447836 \\
$\bm{K_{CI_{\upbeta}}}$ & 0.005655 & 0.319877 & 0.003165 & 0.050119 & $\bm{K_{CQ^{T_8}}}$ & 0.001067 & 0.345712 & 0.000031 & 0.028128 \\
$\bm{K_{CQ^K}}$ & 0.000720 & 0.306517 & 0.000032 & 0.031147 & $\bm{K_{D_0I_\upbeta}}$ & 0.003973 & 0.314481 & 0.000058 & 0.028029 \\
$\bm{V_\mathrm{LN}}$ & 0.035060 & 0.458468 & 0.000025 & 0.033560 & $\bm{K_{T_0^8T_A^r}}$ & 0.002032 & 0.501240 & 0.000033 & 0.030690 \\
$\bm{K_{T_0^8 Q^{8\mathrm{LN}}}}$ & 0.000644 & 0.317083 & 0.000027 & 0.032202 & $\bm{K_{T_A^8T_A^r}}$ & 0.001591 & 0.409334 & 0.000050 & 0.029279 \\
$\bm{K_{T_A^8 Q^{8\mathrm{LN}}}}$ & 0.000545 & 0.285883 & 0.000021 & 0.031257 & $\bm{K_{T_8C}}$ & 0.000763 & 0.307931 & 0.000143 & 0.030565 \\
$\bm{K_{T_\mathrm{ex} C}}$ & 0.000644 & 0.263686 & 0.000021 & 0.027768 & $\bm{K_{KI_\upbeta}}$ & 0.004027 & 0.289919 & 0.000639 & 0.033093 \\
$\bm{d_{N_c}}$ & 0.074221 & 0.789225 & 0.000226 & 0.031497 & $\bm{d_{D_0}}$ & 0.000524 & 0.315449 & 0.000007 & 0.027834 \\
$\bm{d_{D}}$ & 0.108263 & 0.423138 & 0.000884 & 0.054999 & $\bm{d_{T_0^8}}$ & 0.000541 & 0.522774 & 0.000008 & 0.026585 \\
$\bm{d_{T_8}}$ & 0.000626 & 0.447007 & 0.000372 & 0.031383 & $\bm{d_{T_\mathrm{ex}}}$ & 0.034663 & 0.369677 & 0.000012 & 0.029355 \\
$\bm{d_{T_0^r}}$ & 0.000937 & 0.458592 & 0.000049 & 0.026591 & $\bm{d_{T_r}}$ & 0.001711 & 0.384029 & 0.000011 & 0.030659 \\
$\bm{d_{K}}$ & 0.109494 & 0.721891 & 0.002838 & 0.055667 & $\bm{d_{I_{\upalpha}}}$ & 0.009238 & 0.431077 & 0.000228 & 0.027491 \\
$\bm{d_{I_{\upbeta}}}$ & 0.023981 & 0.468133 & 0.007949 & 0.078904 & $\bm{d_{P_D}}$ & 0.000298 & 0.508645 & 0.000006 & 0.026462 \\
$\bm{d_{Q_A}}$ & 0.107907 & 0.522369 & 0.000024 & 0.028929 & $\bm{d_{A_1}}$ & 0.218158 & 0.801259 & 0.000054 & 0.027654 \\
$\bm{d_{P_L}}$ & 0.125013 & 0.425971 & 0.000047 & 0.031199 & $\bm{\tau_m}$ & 0.000586 & 0.379641 & 0.000021 & 0.030613 \\
$\bm{\Delta_8^0}$ & 0.000207 & 0.340482 & 0.000006 & 0.028781 & $\bm{n^8_\mathrm{max}}$ & 0.366052 & 0.842425 & 0.008491 & 0.361357 \\
$\bm{\Delta_8}$ & 0.000365 & 0.391673 & 0.000030 & 0.032536 & $\bm{\Delta_r^0}$ & 0.000203 & 0.319411 & 0.000036 & 0.031947 \\
$\bm{n^r_\mathrm{max}}$ & 0.561872 & 0.894193 & 0.002088 & 0.132968 & $\bm{\Delta_r}$ & 0.000199 & 0.279566 & 0.000006 & 0.028821 \\
\hline
\end{longtable}
}
\end{center}
\newpage
\title{Appendix I: Sensitivity analysis-guided model reduction of a mathematical model of pembrolizumab therapy for de novo metastatic MSI-H/dMMR colorectal cancer}
\maketitle
\setcounter{page}{1}
\section{MRE and RMSRE Tables\label{metricstablesection}}
The MRE and RMSRE at 180.9 days and 672 days, both without treatment and under the standard regimen, are shown in \autoref{reducedperformancemetricssupp} and \autoref{minimalperformancemetricssupp} for variables in the reduced and minimal models, respectively.
\begin{table}[H]
    \centering
    \resizebox{\columnwidth}{!}{%
    \begin{tabular}{|c|c|c|c|c|c|c|c|c|c|}
    \hline
    & \multicolumn{4}{|c|}{\textbf{No Treatment}} & \multicolumn{4}{|c|}{\textbf{Standard Regimen}} \\
    \hline
    \textbf{Variable} & \textbf{MRE at} & \textbf{RMSRE at} & \textbf{MRE at} & \textbf{RMSRE at} & \textbf{MRE at} & \textbf{RMSRE at} & \textbf{MRE at} & \textbf{RMSRE at} \\
    & \textbf{180.9 days} & \textbf{180.9 days} & \textbf{672 days} & \textbf{672 days} & \textbf{180.9 days} & \textbf{180.9 days} & \textbf{672 days} & \textbf{672 days} \\
    \hline
$\bm{V_\mathrm{TS}}$ & 0.0023734 & 0.0011311 & 0.0023734 & 0.0008632 & 0.0121151 & 0.0082577 & 0.0393465 & 0.0238138 \\
$\bm{C}$ & 0.0028276 & 0.0014299 & 0.0028276 & 0.0010730 & 0.0119563 & 0.0077838 & 0.0406321 & 0.0237730 \\
$\bm{N_c}$ & 0.0116307 & 0.0070559 & 0.0116307 & 0.0050830 & 0.0239736 & 0.0193486 & 0.0532521 & 0.0306637 \\
$\bm{D_0}$ & 0.0111911 & 0.0060997 & 0.0111911 & 0.0044845 & 0.0130414 & 0.0069654 & 0.0199002 & 0.0116856 \\
$\bm{D}$ & 0.0050825 & 0.0024155 & 0.0050825 & 0.0017696 & 0.0212771 & 0.0162837 & 0.0583087 & 0.0323700 \\
$\bm{D^\mathrm{LN}}$ & 0.0074041 & 0.0032461 & 0.0074041 & 0.0024157 & 0.0292734 & 0.0239427 & 0.0976282 & 0.0530018 \\
$\bm{T_0^8}$ & 0.0003015 & 0.0001725 & 0.0003036 & 0.0002443 & 0.0002914 & 0.0002265 & 0.0002914 & 0.0001807 \\
$\bm{T_A^8}$ & 0.1176080 & 0.0554367 & 0.1176080 & 0.0402138 & 0.3381936 & 0.0426412 & 0.3381936 & 0.0490073 \\
$\bm{T_8}$ & 0.0118881 & 0.0046049 & 0.0118881 & 0.0056179 & 0.0215579 & 0.0073560 & 0.0418209 & 0.0209546 \\
$\bm{T_{\mathrm{ex}}}$ & 0.0041987 & 0.0023023 & 0.0041987 & 0.0023141 & 0.0023742 & 0.0007896 & 0.0315004 & 0.0150487 \\
$\bm{T_0^4}$ & 0.0006140 & 0.0003765 & 0.0006197 & 0.0005077 & 0.0006537 & 0.0005110 & 0.0006537 & 0.0003966 \\
$\bm{T_A^1}$ & 0.0509770 & 0.0246850 & 0.0509770 & 0.0189467 & 0.0752486 & 0.0134399 & 0.0772959 & 0.0403713 \\
$\bm{T_1}$ & 0.0058466 & 0.0025310 & 0.0095796 & 0.0066107 & 0.0074193 & 0.0048965 & 0.0349808 & 0.0163249 \\
$\bm{T_0^r}$ & 0.0579932 & 0.0337331 & 0.0579932 & 0.0262837 & 0.0564815 & 0.0158846 & 0.0729100 & 0.0417866 \\
$\bm{T_A^r}$ & 0.3607671 & 0.0932472 & 0.3607671 & 0.0682983 & 0.3575233 & 0.0428688 & 0.3575233 & 0.0288222 \\
$\bm{T_r}$ & 0.0426825 & 0.0245072 & 0.0426825 & 0.0208654 & 0.0396430 & 0.0129175 & 0.0542403 & 0.0223937 \\
$\bm{M_0}$ & 0.0090110 & 0.0054764 & 0.0090110 & 0.0039405 & 0.0092214 & 0.0062718 & 0.0096419 & 0.0077581 \\
$\bm{M_1}$ & 0.0175630 & 0.0050505 & 0.0175630 & 0.0037992 & 0.0175664 & 0.0057575 & 0.0175664 & 0.0118202 \\
$\bm{M_2}$ & 0.0234309 & 0.0143249 & 0.0234309 & 0.0105519 & 0.0252418 & 0.0137304 & 0.0252418 & 0.0118795 \\
$\bm{K_0}$ & 0.0008137 & 0.0004406 & 0.0008137 & 0.0003211 & 0.0010401 & 0.0008128 & 0.0020765 & 0.0013507 \\
$\bm{K}$ & 0.0128971 & 0.0066352 & 0.0128971 & 0.0048838 & 0.0158519 & 0.0124569 & 0.0305380 & 0.0200647 \\
$\bm{H}$ & 0.0116299 & 0.0070071 & 0.0116299 & 0.0050480 & 0.0239736 & 0.0193350 & 0.0532520 & 0.0306621 \\
$\bm{S}$ & 0.0116177 & 0.0068527 & 0.0116177 & 0.0049372 & 0.0239730 & 0.0192928 & 0.0532508 & 0.0306566 \\
$\bm{I_2}$ & 0.0071315 & 0.0031239 & 0.0086286 & 0.0061209 & 0.0102981 & 0.0054271 & 0.0369606 & 0.0175178 \\
$\bm{I_\upgamma}$ & 0.0305339 & 0.0138417 & 0.0305339 & 0.0101214 & 0.0305339 & 0.0183497 & 0.0305339 & 0.0143329 \\
$\bm{I_\upalpha}$ & 0.0069435 & 0.0034038 & 0.0069435 & 0.0037210 & 0.0095717 & 0.0035390 & 0.0178490 & 0.0087006 \\
$\bm{I_\upbeta}$ & 0.0050327 & 0.0034577 & 0.0050327 & 0.0025952 & 0.0116172 & 0.0085044 & 0.0277417 & 0.0165819 \\
$\bm{I_{10}}$ & 0.0043210 & 0.0033474 & 0.0043210 & 0.0032899 & 0.0182637 & 0.0080897 & 0.0589050 & 0.0322900 \\
$\bm{P_D^{T_8}}$ & 0.0106262 & 0.0038416 & 0.0106262 & 0.0053039 & 0.0203469 & 0.0073149 & 0.0418084 & 0.0209389 \\
$\bm{P_D^{T_1}}$ & 0.0056780 & 0.0022927 & 0.0095794 & 0.0065533 & 0.0074107 & 0.0048876 & 0.0349697 & 0.0163154 \\
$\bm{P_D^{K}}$ & 0.0127414 & 0.0060668 & 0.0127414 & 0.0044844 & 0.0158096 & 0.0123657 & 0.0305355 & 0.0200480 \\
$\bm{Q_A^{T_8}}$ & 0.0000000 & 0.0000000 & 0.0000000 & 0.0000000 & 0.0203293 & 0.0073279 & 0.0418054 & 0.0209380 \\
$\bm{Q_A^{T_1}}$ & 0.0000000 & 0.0000000 & 0.0000000 & 0.0000000 & 0.0074095 & 0.0048879 & 0.0349663 & 0.0163140 \\
$\bm{Q_A^{K}}$ & 0.0000000 & 0.0000000 & 0.0000000 & 0.0000000 & 0.0158079 & 0.0123532 & 0.0305338 & 0.0200462 \\
$\bm{P_L}$ & 0.0054054 & 0.0018101 & 0.0054054 & 0.0020073 & 0.0083754 & 0.0048989 & 0.0437430 & 0.0224439 \\
$\bm{Q^{T_8}}$ & 0.0106080 & 0.0043049 & 0.0106080 & 0.0067769 & 0.0178932 & 0.0086467 & 0.0449959 & 0.0240326 \\
$\bm{Q^{T_1}}$ & 0.0082186 & 0.0031159 & 0.0116895 & 0.0081573 & 0.0134638 & 0.0085448 & 0.0330700 & 0.0185309 \\
$\bm{Q^{K}}$ & 0.0125800 & 0.0066470 & 0.0125800 & 0.0054160 & 0.0155286 & 0.0091175 & 0.0560795 & 0.0265618 \\
$\bm{A_{1}}$ & 0.0000000 & 0.0000000 & 0.0000000 & 0.0000000 & 0.3901297 & 0.0041671 & 0.3901297 & 0.0031769 \\
$\bm{P_D^{8\mathrm{LN}}}$ & 0.0970748 & 0.0451112 & 0.0970748 & 0.0328601 & 0.2211528 & 0.0317253 & 0.2211528 & 0.0466997 \\
$\bm{P_D^{1\mathrm{LN}}}$ & 0.0445350 & 0.0206293 & 0.0445350 & 0.0162430 & 0.0458369 & 0.0125530 & 0.0772799 & 0.0401409 \\
$\bm{Q_A^{8\mathrm{LN}}}$ & 0.0000000 & 0.0000000 & 0.0000000 & 0.0000000 & 0.2054100 & 0.0314589 & 0.2054100 & 0.0466267 \\
$\bm{Q_A^{1\mathrm{LN}}}$ & 0.0000000 & 0.0000000 & 0.0000000 & 0.0000000 & 0.0457394 & 0.0125347 & 0.0772585 & 0.0401058 \\
$\bm{P_L^\mathrm{LN}}$ & 0.0232450 & 0.0130925 & 0.0232450 & 0.0104517 & 0.0373812 & 0.0092167 & 0.0738867 & 0.0346097 \\
$\bm{Q^{8\mathrm{LN}}}$ & 0.1165856 & 0.0540520 & 0.1165856 & 0.0400645 & 0.2362887 & 0.0372978 & 0.2362887 & 0.0757193 \\
$\bm{Q^{1\mathrm{LN}}}$ & 0.0639956 & 0.0315627 & 0.0639956 & 0.0252690 & 0.0776799 & 0.0212030 & 0.1309049 & 0.0698038 \\
$\bm{A_{1}^\mathrm{LN}}$ & 0.0000000 & 0.0000000 & 0.0000000 & 0.0000000 & 0.3905869 & 0.0041715 & 0.3905869 & 0.0031790 \\
\hline
\end{tabular}
\caption{\label{reducedperformancemetricssupp}MRE and RMSRE between the reduced model and the full model at 180.9 days and 672 days, in the cases of no treatment and the standard regimen.}%
}
\end{table}
\begin{table}[H]
    \centering
    \resizebox{\columnwidth}{!}{%
    \begin{tabular}{|c|c|c|c|c|c|c|c|c|c|}
    \hline
    & \multicolumn{4}{|c|}{\textbf{No Treatment}} & \multicolumn{4}{|c|}{\textbf{Standard Regimen}} \\
    \hline
    \textbf{Variable} & \textbf{MRE at} & \textbf{RMSRE at} & \textbf{MRE at} & \textbf{RMSRE at} & \textbf{MRE at} & \textbf{RMSRE at} & \textbf{MRE at} & \textbf{RMSRE at} \\
    & \textbf{180.9 days} & \textbf{180.9 days} & \textbf{672 days} & \textbf{672 days} & \textbf{180.9 days} & \textbf{180.9 days} & \textbf{672 days} & \textbf{672 days} \\    
    \hline
$\bm{V_\mathrm{TS}}$ & 0.0024068 & 0.0015786 & 0.0024068 & 0.0011702 & 0.0393053 & 0.0218701 & 0.0836609 & 0.0454335 \\
$\bm{C}$ & 0.0034820 & 0.0022233 & 0.0034820 & 0.0016458 & 0.0390739 & 0.0221238 & 0.0856075 & 0.0461282 \\
$\bm{N_c}$ & 0.0180043 & 0.0110073 & 0.0180043 & 0.0081071 & 0.0425986 & 0.0202978 & 0.0785886 & 0.0386240 \\
$\bm{D_0}$ & 0.0173109 & 0.0091230 & 0.0173109 & 0.0066168 & 0.0151593 & 0.0093197 & 0.0324771 & 0.0165763 \\
$\bm{D}$ & 0.0240020 & 0.0117156 & 0.0240020 & 0.0084664 & 0.0322543 & 0.0198373 & 0.0627882 & 0.0340109 \\
$\bm{D^\mathrm{LN}}$ & 0.0212806 & 0.0087804 & 0.0212806 & 0.0064094 & 0.0698406 & 0.0393707 & 0.1515504 & 0.0796794 \\
$\bm{T_A^8}$ & 0.1205968 & 0.0600061 & 0.1205968 & 0.0432066 & 0.3483231 & 0.0577471 & 0.3483231 & 0.0823991 \\
$\bm{T_8}$ & 0.0123520 & 0.0056882 & 0.0123520 & 0.0046167 & 0.0293080 & 0.0088689 & 0.0491172 & 0.0247417 \\
$\bm{T_{\mathrm{ex}}}$ & 0.0037352 & 0.0017204 & 0.0037427 & 0.0025960 & 0.0251028 & 0.0112476 & 0.0458564 & 0.0273572 \\
$\bm{T_0^r}$ & 0.0663734 & 0.0423494 & 0.0663734 & 0.0304796 & 0.0647645 & 0.0384575 & 0.1339163 & 0.0715741 \\
$\bm{T_A^r}$ & 0.3475644 & 0.0869672 & 0.3475644 & 0.0625942 & 0.3443561 & 0.0384203 & 0.3443561 & 0.0271378 \\
$\bm{T_r}$ & 0.0372989 & 0.0203915 & 0.0372989 & 0.0149035 & 0.0409380 & 0.0210639 & 0.0687808 & 0.0368607 \\
$\bm{K}$ & 0.0394841 & 0.0222479 & 0.0394841 & 0.0161036 & 0.0349595 & 0.0118913 & 0.0472535 & 0.0276112 \\
$\bm{I_\upalpha}$ & 0.3160494 & 0.1203937 & 0.3160494 & 0.0866372 & 0.3160355 & 0.0712895 & 0.3160355 & 0.0875083 \\
$\bm{I_\upbeta}$ & 0.0815496 & 0.0346187 & 0.0815496 & 0.0249066 & 0.0815496 & 0.0544669 & 0.0930705 & 0.0545702 \\
$\bm{P_D^{T_8}}$ & 0.0109792 & 0.0049939 & 0.0109792 & 0.0041829 & 0.0279134 & 0.0088848 & 0.0490822 & 0.0247239 \\
$\bm{P_D^{K}}$ & 0.0386324 & 0.0194238 & 0.0386324 & 0.0140869 & 0.0343874 & 0.0117107 & 0.0472614 & 0.0275892 \\
$\bm{Q_A^{T_8}}$ & 0.0000000 & 0.0000000 & 0.0000000 & 0.0000000 & 0.0278835 & 0.0089090 & 0.0490940 & 0.0247264 \\
$\bm{Q_A^{K}}$ & 0.0000000 & 0.0000000 & 0.0000000 & 0.0000000 & 0.0343683 & 0.0117090 & 0.0472444 & 0.0275741 \\
$\bm{P_L}$ & 0.0513514 & 0.0380499 & 0.0513514 & 0.0343658 & 0.0840797 & 0.0593228 & 0.1107315 & 0.0614505 \\
$\bm{Q^{T_8}}$ & 0.0513514 & 0.0385474 & 0.0513514 & 0.0334537 & 0.0881127 & 0.0597162 & 0.0995980 & 0.0584293 \\
$\bm{Q^{K}}$ & 0.0729991 & 0.0525336 & 0.0729991 & 0.0423383 & 0.0909687 & 0.0627461 & 0.1257442 & 0.0743049 \\
$\bm{A_{1}}$ & 0.0000000 & 0.0000000 & 0.0000000 & 0.0000000 & 0.4868756 & 0.0048969 & 0.4868756 & 0.0036682 \\
$\bm{P_D^{8\mathrm{LN}}}$ & 0.0963670 & 0.0494992 & 0.0963670 & 0.0356555 & 0.2311865 & 0.0501352 & 0.2311865 & 0.0809391 \\
$\bm{Q_A^{8\mathrm{LN}}}$ & 0.0000000 & 0.0000000 & 0.0000000 & 0.0000000 & 0.2156159 & 0.0496510 & 0.2156159 & 0.0808541 \\
$\bm{P_L^\mathrm{LN}}$ & 0.0990899 & 0.0632076 & 0.0990899 & 0.0505471 & 0.1931496 & 0.1386761 & 0.2195455 & 0.1367663 \\
$\bm{Q^{8\mathrm{LN}}}$ & 0.1798825 & 0.0966435 & 0.1798825 & 0.0724442 & 0.3020117 & 0.1689593 & 0.3020117 & 0.1771299 \\
$\bm{A_{1}^\mathrm{LN}}$ & 0.0000000 & 0.0000000 & 0.0000000 & 0.0000000 & 0.4882693 & 0.0049396 & 0.4882693 & 0.0036959 \\
\hline
\end{tabular}
\caption{\label{minimalperformancemetricssupp}MRE and RMSRE between the minimal model and the full model at 180.9 days and at 672 days, in the cases of no treatment and the standard regimen.}%
}
\end{table}
\section{MRE and RMSRE Time Traces\label{metricstimetracessection}}
Time traces for the MRE and RMSRE of model variables, both without treatment and under the standard regimen, from 0 to 672 days for the reduced and minimal models are shown in \autoref{MREplots} and \autoref{RMSREplots}, respectively.
\subsection{MRE Time Traces}
\begin{figure}[H]
\begin{subfigure}{\textwidth}
    \centering
    \includegraphics[width=\textwidth]{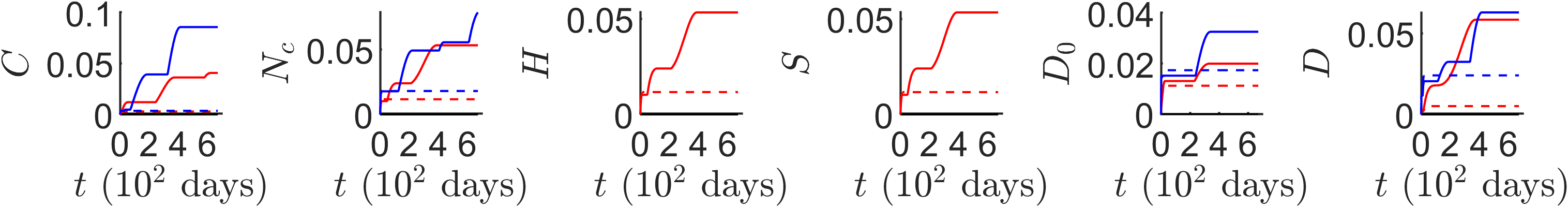}
\end{subfigure}
\end{figure}%

\begin{figure}[H]\ContinuedFloat
\begin{subfigure}{\textwidth}
    \centering
    \includegraphics[width=\textwidth]{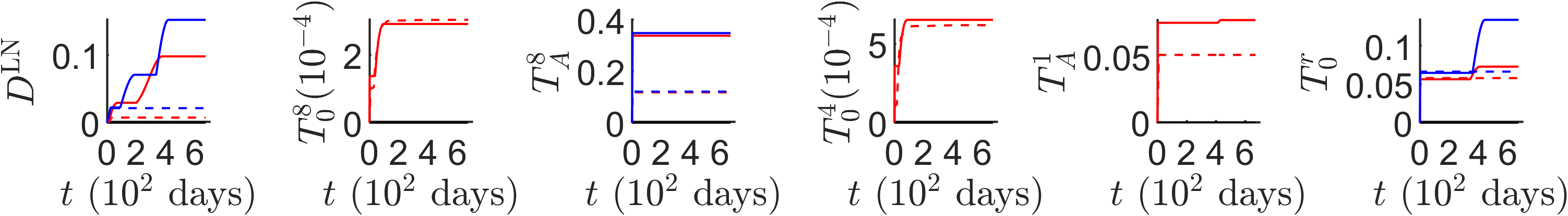}
\end{subfigure}
\end{figure}%

\begin{figure}[H]\ContinuedFloat
\begin{subfigure}{\textwidth}
    \centering
    \includegraphics[width=\textwidth]{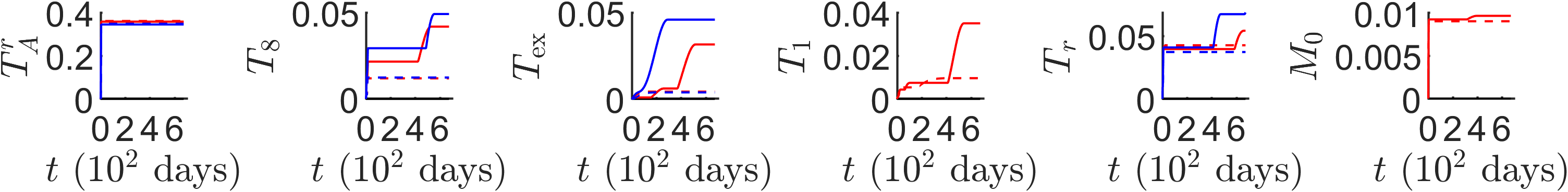}
\end{subfigure}
\end{figure}%

\begin{figure}[H]\ContinuedFloat
\begin{subfigure}{\textwidth}
    \centering
    \includegraphics[width=\textwidth]{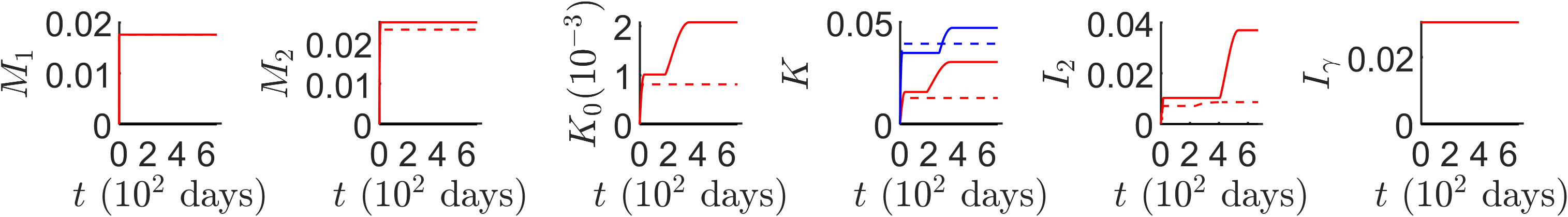}
\end{subfigure}
\end{figure}%

\begin{figure}[H]\ContinuedFloat
\begin{subfigure}{\textwidth}
    \centering
    \includegraphics[width=\textwidth]{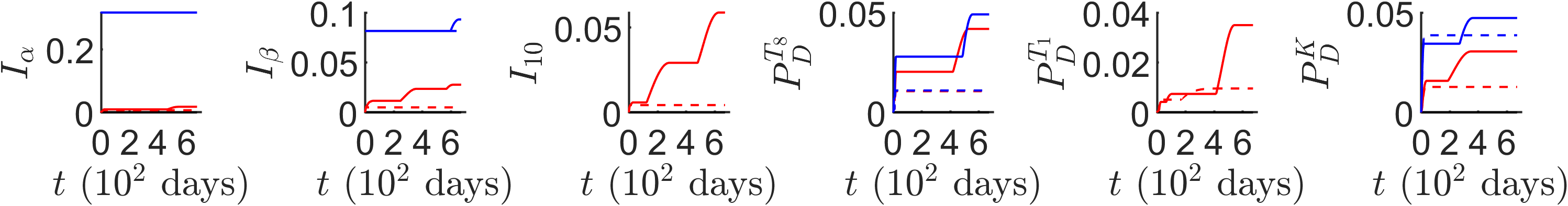}
\end{subfigure}
\end{figure}%

\begin{figure}[H]\ContinuedFloat
\begin{subfigure}{\textwidth}
    \centering
    \includegraphics[width=\textwidth]{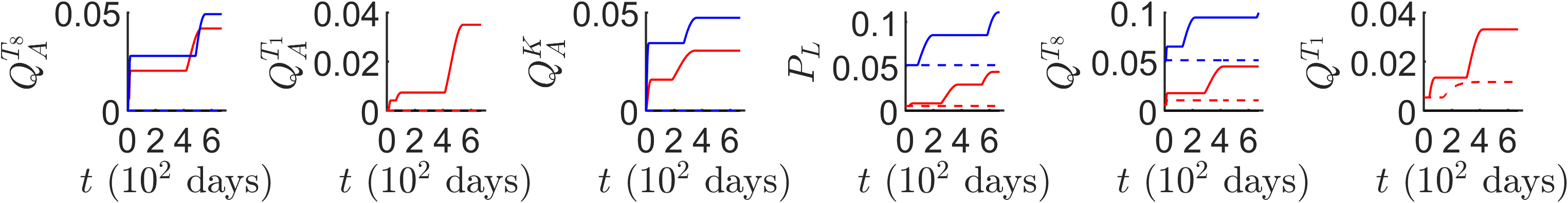}
\end{subfigure}
\end{figure}%

\begin{figure}[H]\ContinuedFloat
\begin{subfigure}{\textwidth}
    \centering
    \includegraphics[width=\textwidth]{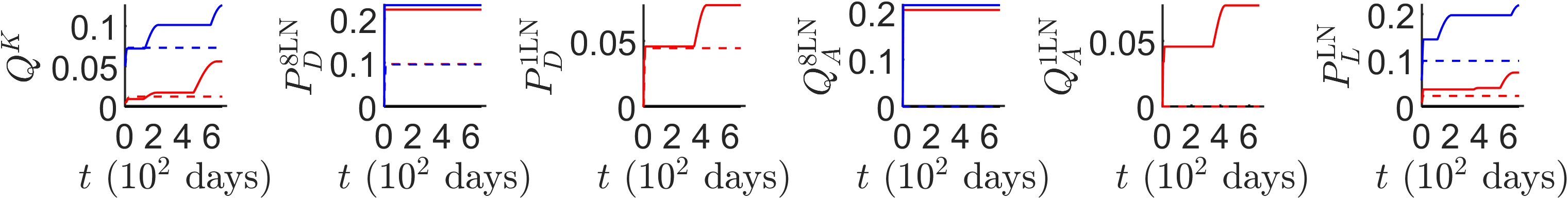}
\end{subfigure}
\end{figure}%

\begin{figure}[H]\ContinuedFloat
\begin{subfigure}{\textwidth}
    \centering
    \includegraphics[width=\textwidth]{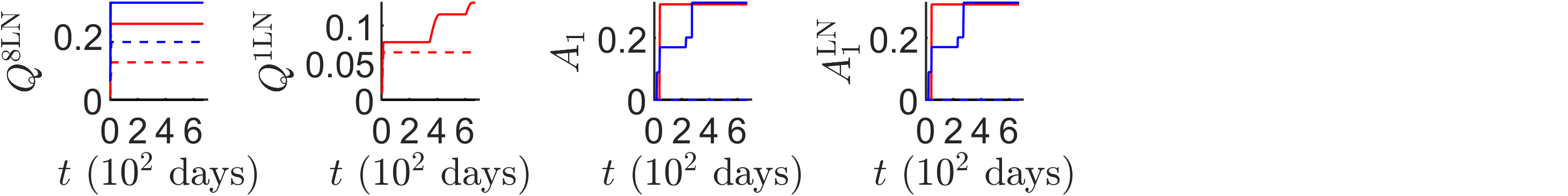}
\end{subfigure}
    \caption{\label{MREplots}Time traces of MRE for model variables between the reduced model and the full model (red) and the minimal model and full model (blue), from 0 days up to 672 days from commencement, with and without treatment. Dashed lines indicate no treatment, and solid lines indicate treatment with the standard regimen.}
\end{figure}
\subsection{RMSRE Time Traces}
\begin{figure}[H]
\begin{subfigure}{\textwidth}
    \centering
    \includegraphics[width=\textwidth]{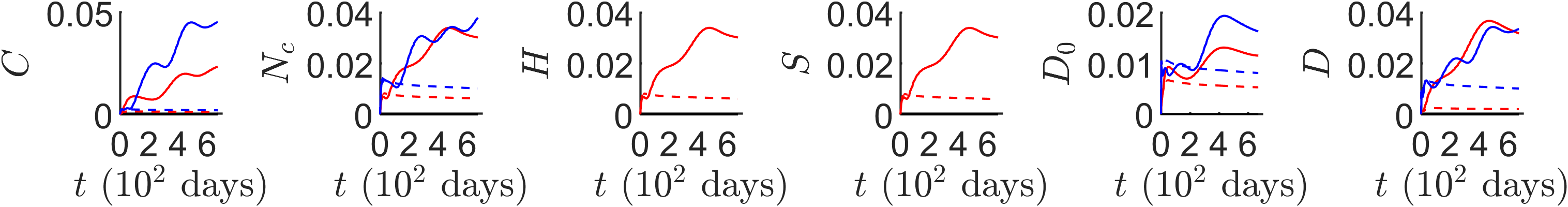}
\end{subfigure}
\end{figure}%

\begin{figure}[H]\ContinuedFloat
\begin{subfigure}{\textwidth}
    \centering
    \includegraphics[width=\textwidth]{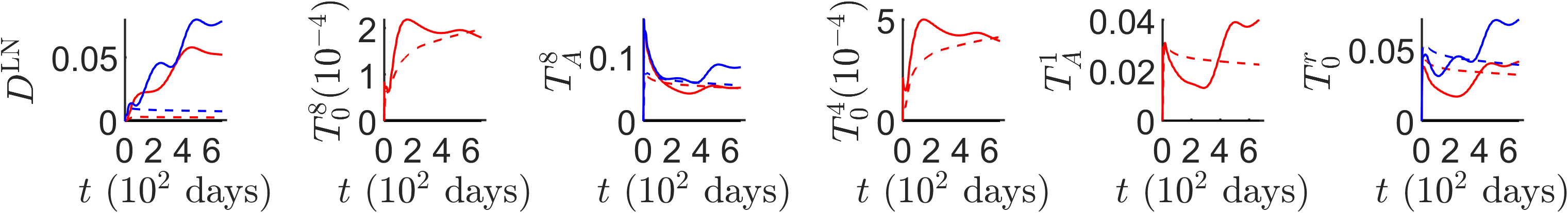}
\end{subfigure}
\end{figure}%

\begin{figure}[H]\ContinuedFloat
\begin{subfigure}{\textwidth}
    \centering
    \includegraphics[width=\textwidth]{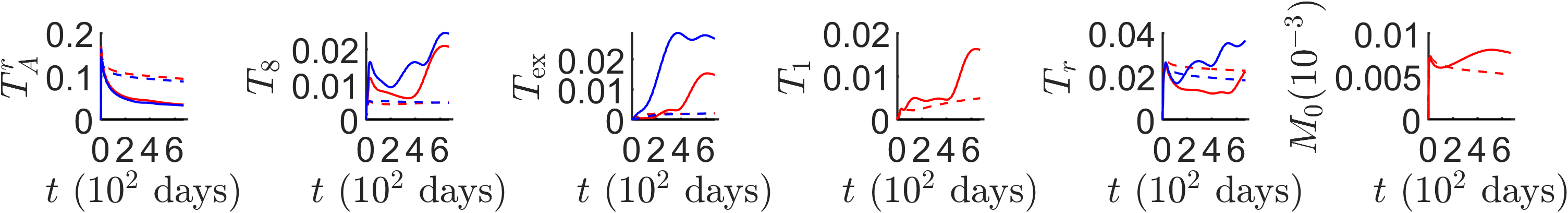}
\end{subfigure}
\end{figure}%

\begin{figure}[H]\ContinuedFloat
\begin{subfigure}{\textwidth}
    \centering
    \includegraphics[width=\textwidth]{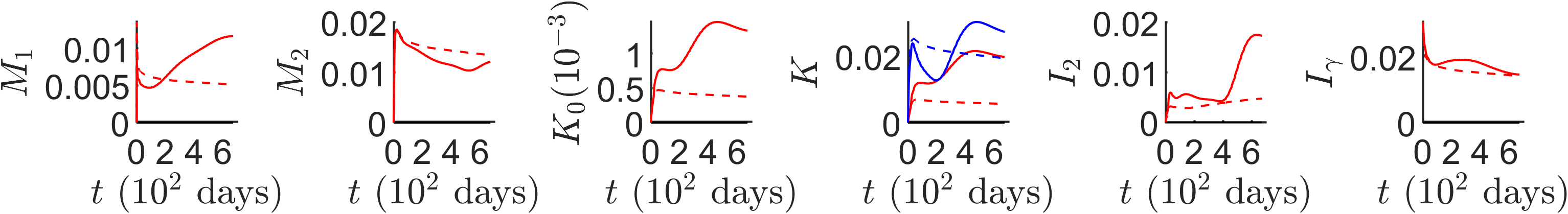}
\end{subfigure}
\end{figure}%

\begin{figure}[H]\ContinuedFloat
\begin{subfigure}{\textwidth}
    \centering
    \includegraphics[width=\textwidth]{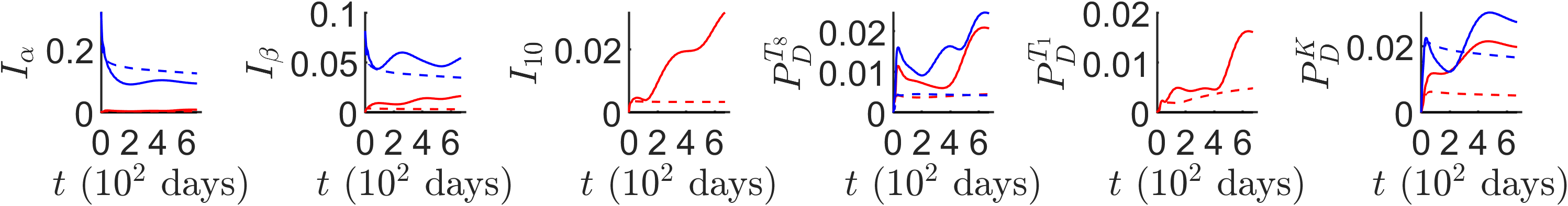}
\end{subfigure}
\end{figure}%

\begin{figure}[H]\ContinuedFloat
\begin{subfigure}{\textwidth}
    \centering
    \includegraphics[width=\textwidth]{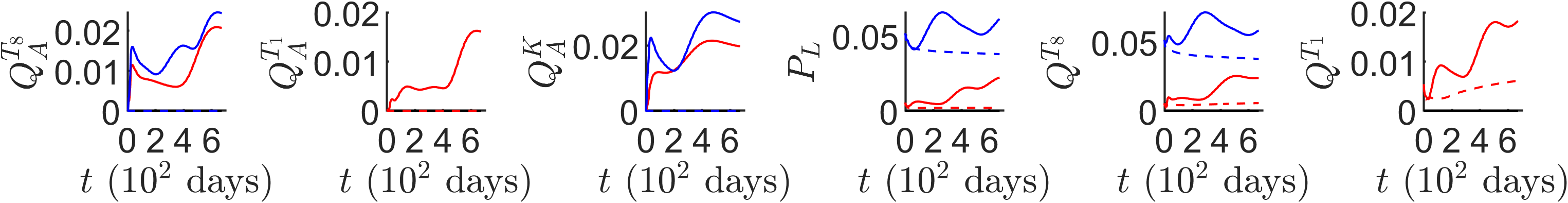}
\end{subfigure}
\end{figure}%

\begin{figure}[H]\ContinuedFloat
\begin{subfigure}{\textwidth}
    \centering
    \includegraphics[width=\textwidth]{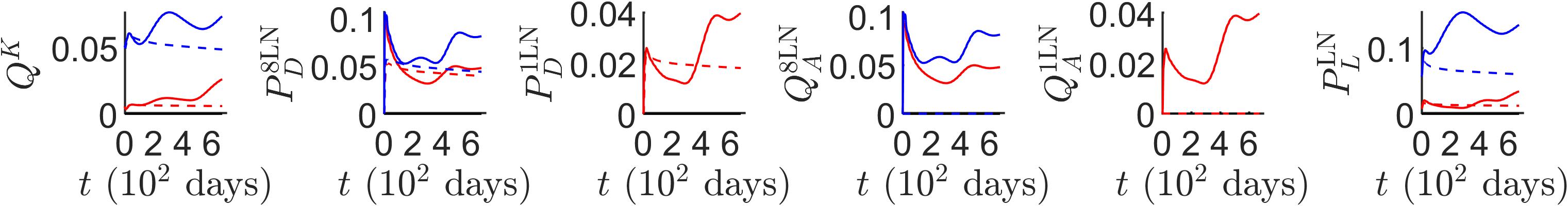}
\end{subfigure}
\end{figure}%

\begin{figure}[H]\ContinuedFloat
\begin{subfigure}{\textwidth}
    \centering
    \includegraphics[width=\textwidth]{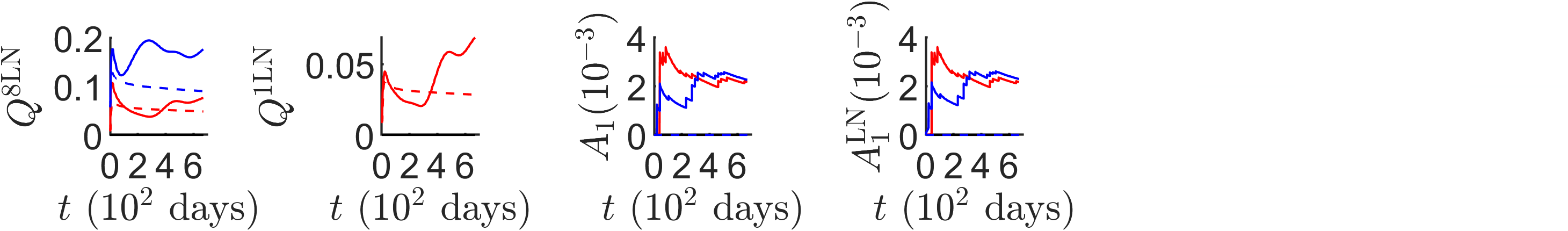}
\end{subfigure}
    \caption{\label{RMSREplots}Time traces of RMSRE for model variables between the reduced model and the full model (red) and the minimal model and full model (blue), from 0 days up to 672 days from commencement, with and without treatment. Dashed lines indicate no treatment, and solid lines indicate treatment with the standard regimen.}
\end{figure}
\end{document}